\documentclass[
submitting
]{nst}

\usepackage{graphicx}
\usepackage{amsmath}
\usepackage{amsfonts}
\usepackage{amssymb}
\usepackage{mathrsfs}
\usepackage{xcolor}
\usepackage{xspace}
\usepackage[capitalise]{cleveref}
\usepackage{threeparttable}
\usepackage{multirow}
\usepackage[compat=1.1.0]{tikz-feynman}
\usepackage{enumerate}
\usepackage{subfig}
\usepackage{textgreek}
\usepackage{url}
\usepackage{hyperref}

\tikzfeynmanset{warn luatex=false}
\newcommand{\ket}[1]{|#1\rangle}
\newcommand{\dd}{{\mathrm{d}}}
\newcommand{\norm}[1]{\left|#1\right|}
\newcommand{\mmbar}{\text{M}-\overline{\text{M}}}
\newcommand{\mmu}{{\text{M}}}
\newcommand{\mmup}{{\text{M}^P}}
\newcommand{\mmuv}{{\text{M}^V}}
\newcommand{\ammu}{{\overline{\text{M}}}}

\begin{document}

\title{Conceptual Design of the Muonium-to-Antimuonium Conversion Experiment (MACE)}

\author{Ai-Yu~Bai}
\affiliation{School of Physics, Sun Yat-sen University, Guangzhou 510275, China}
\affiliation{Platform for Muon Science and Technology, Sun Yat-sen University, Guangzhou 510275, China}
\author{Hanjie~Cai}
\affiliation{Institute of Modern Physics, Chinese Academy of Sciences, Lanzhou 730000, China}
\affiliation{University of Chinese Academy of Sciences, Beijing 100049, China}
\author{Chang-Lin~Chen}
\affiliation{School of Physics and Electronics, Hunan University, Changsha 410082, China}
\author{Siyuan~Chen}
\affiliation{School of Physics, Sun Yat-sen University, Guangzhou 510275, China}
\affiliation{Platform for Muon Science and Technology, Sun Yat-sen University, Guangzhou 510275, China}
\author{Xurong~Chen}
\affiliation{Institute of Modern Physics, Chinese Academy of Sciences, Lanzhou 730000, China}
\affiliation{University of Chinese Academy of Sciences, Beijing 100049, China}
\affiliation{Southern Center for Nuclear Science Theory (SCNT), Institute of Modern Physics, Chinese Academy of Sciences, Huizhou 516000, Guangdong Province, China}
\author{Yu~Chen}
\affiliation{School of Physics, Sun Yat-sen University, Guangzhou 510275, China}
\affiliation{Platform for Muon Science and Technology, Sun Yat-sen University, Guangzhou 510275, China}
\author{Weibin~Cheng}
\affiliation{School of Physics, Liaoning University, Shenyang 110036, China}
\author{Ling-Yun~Dai}
\affiliation{School of Physics and Electronics, Hunan University, Changsha 410082, China}
\affiliation{Hunan Provincial Key Laboratory of High-Energy Scale Physics and Applications, Hunan University, Changsha 410082, China}
\author{Rui-Rui~Fan}
\affiliation{Institute of High Energy Physics, Chinese Academy of Sciences, Beijing 100049, China}
\affiliation{China Spallation Neutron Source, Dongguan 523803, China}
\affiliation{State Key Laboratory of Particle Detection and Electronics, Beijing, 100049, China}
\author{Li~Gong}
\affiliation{School of Physics, Liaoning University, Shenyang 110036, China}
\author{Zihao~Guo}
\affiliation{School of Physics, Southeast University, Nanjing 211189, China}
\author{Yuan~He}
\affiliation{Institute of Modern Physics, Chinese Academy of Sciences, Lanzhou 730000, China}
\affiliation{University of Chinese Academy of Sciences, Beijing 100049, China}
\author{Zhilong~Hou}
\affiliation{Institute of High Energy Physics, Chinese Academy of Sciences, Beijing 100049, China}
\author{Yinyuan~Huang}
\affiliation{School of Physics, Sun Yat-sen University, Guangzhou 510275, China}
\affiliation{Platform for Muon Science and Technology, Sun Yat-sen University, Guangzhou 510275, China}
\author{Huan~Jia}
\affiliation{Institute of Modern Physics, Chinese Academy of Sciences, Lanzhou 730000, China}
\affiliation{University of Chinese Academy of Sciences, Beijing 100049, China}
\author{Hao~Jiang}
\affiliation{School of Physics, Sun Yat-sen University, Guangzhou 510275, China}
\affiliation{Platform for Muon Science and Technology, Sun Yat-sen University, Guangzhou 510275, China}
\author{Han-Tao~Jing}
\affiliation{Institute of High Energy Physics, Chinese Academy of Sciences, Beijing 100049, China}
\author{Xiaoshen~Kang}
\affiliation{School of Physics, Liaoning University, Shenyang 110036, China}
\author{Hai-Bo~Li}
\affiliation{Institute of High Energy Physics, Chinese Academy of Sciences, Beijing 100049, China}
\affiliation{University of Chinese Academy of Sciences, Beijing 100049, China}
\author{Jincheng~Li}
\affiliation{Institute of Modern Physics, Chinese Academy of Sciences, Lanzhou 730000, China}
\affiliation{University of Chinese Academy of Sciences, Beijing 100049, China}
\author{Yang~Li}
\affiliation{Institute of High Energy Physics, Chinese Academy of Sciences, Beijing 100049, China}
\author{Daming~Liu}
\affiliation{Key Laboratory of Particle and Radiation Imaging, Department of Engineering Physics, Tsinghua University, Beĳing 100084, China}
\author{Shulin~Liu}
\affiliation{Institute of High Energy Physics, Chinese Academy of Sciences, Beijing 100049, China}
\affiliation{University of Chinese Academy of Sciences, Beijing 100049, China}
\affiliation{State Key Laboratory of Particle Detection and Electronics, Beijing 100049, China}
\author{Guihao~Lu}
\affiliation{School of Physics, Sun Yat-sen University, Guangzhou 510275, China}
\affiliation{Platform for Muon Science and Technology, Sun Yat-sen University, Guangzhou 510275, China}
\author{Han~Miao}
\affiliation{Institute of High Energy Physics, Chinese Academy of Sciences, Beijing 100049, China}
\affiliation{University of Chinese Academy of Sciences, Beijing 100049, China}
\author{Yunsong~Ning}
\affiliation{School of Physics, Sun Yat-sen University, Guangzhou 510275, China}
\affiliation{Platform for Muon Science and Technology, Sun Yat-sen University, Guangzhou 510275, China}
\author{Jianwei~Niu}
\affiliation{Institute of Modern Physics, Chinese Academy of Sciences, Lanzhou 730000, China}
\affiliation{School of Nuclear Science and Technology, Lanzhou University, Lanzhou 730000, China}
\author{Huaxing~Peng}
\affiliation{Institute of High Energy Physics, Chinese Academy of Sciences, Beijing 100049, China}
\affiliation{University of Chinese Academy of Sciences, Beijing 100049, China}
\affiliation{State Key Laboratory of Particle Detection and Electronics, Beijing 100049, China}
\author{Alexey~A.~Petrov}
\affiliation{Department of Physics and Astronomy, University of South Carolina, Columbia, South Carolina 29208, USA}
\author{Yuanshuai~Qin}
\affiliation{Institute of Modern Physics, Chinese Academy of Sciences, Lanzhou 730000, China}
\author{Mingchen~Sun}
\affiliation{School of Physics, Sun Yat-sen University, Guangzhou 510275, China}
\affiliation{Platform for Muon Science and Technology, Sun Yat-sen University, Guangzhou 510275, China}
\author{Jian~Tang}
\email{Corresponding author: tangjian5@mail.sysu.edu.cn}
\affiliation{School of Physics, Sun Yat-sen University, Guangzhou 510275, China}
\affiliation{Platform for Muon Science and Technology, Sun Yat-sen University, Guangzhou 510275, China}
\author{Jing-Yu~Tang}
\affiliation{School of Nuclear Science and Technology, University of Science and Technology of China, Hefei 230026, China}
\author{Ye~Tian}
\affiliation{Institute of Modern Physics, Chinese Academy of Sciences, Lanzhou 730000, China}
\author{Rong~Wang}
\affiliation{Institute of Modern Physics, Chinese Academy of Sciences, Lanzhou 730000, China}
\affiliation{University of Chinese Academy of Sciences, Beijing 100049, China}
\author{Xiaodong~Wang}
\affiliation{School of Nuclear Science and Technology, University of South China, Hengyang 421001, China}
\affiliation{Key Laboratory of Advanced Nuclear Energy Design and Safety (MOE), University of South China, Hengyang 421001, China}
\author{Yi~Wang}
\affiliation{Key Laboratory of Particle and Radiation Imaging, Department of Engineering Physics, Tsinghua University, Beĳing 100084, China}
\author{Zhichao~Wang}
\affiliation{School of Physics, Sun Yat-sen University, Guangzhou 510275, China}
\affiliation{Platform for Muon Science and Technology, Sun Yat-sen University, Guangzhou 510275, China}
\author{Chen~Wu}
\affiliation{Institute of High Energy Physics, Chinese Academy of Sciences, Beijing 100049, China}
\affiliation{China Spallation Neutron Source, Dongguan 523803, China}
\author{Tian-Yu~Xing}
\affiliation{INFN Sezione di Milano, Milano 20133, Italy}
\affiliation{Universita degli Studi di Milano, Milano 20122, Italy}
\author{Weizhi~Xiong}
\affiliation{Key Laboratory of Particle Physics and Particle Irradiation (MOE), Institute of Frontier and Interdisciplinary Science, Shandong University, Qingdao 266237, China}
\author{Yu~Xu}
\affiliation{Advanced energy science and technology Guangdong laboratory, Huizhou 516007, China}
\author{Baojun~Yan}
\affiliation{Institute of High Energy Physics, Chinese Academy of Sciences, Beijing 100049, China}
\affiliation{State Key Laboratory of Particle Detection and Electronics, Beijing 100049, China}
\author{De-Liang~Yao}
\affiliation{School of Physics and Electronics, Hunan University, Changsha 410082, China}
\affiliation{Hunan Provincial Key Laboratory of High-Energy Scale Physics and Applications, Hunan University, Changsha 410082, China}
\author{Tao~Yu}
\affiliation{School of Physics, Sun Yat-sen University, Guangzhou 510275, China}
\affiliation{Platform for Muon Science and Technology, Sun Yat-sen University, Guangzhou 510275, China}
\author{Ye~Yuan}
\affiliation{Institute of High Energy Physics, Chinese Academy of Sciences, Beijing 100049, China}
\affiliation{University of Chinese Academy of Sciences, Beijing 100049, China}
\author{Yi~Yuan}
\affiliation{School of Physics, Sun Yat-sen University, Guangzhou 510275, China}
\affiliation{Platform for Muon Science and Technology, Sun Yat-sen University, Guangzhou 510275, China}
\author{Yao~Zhang}
\affiliation{Institute of High Energy Physics, Chinese Academy of Sciences, Beijing 100049, China}
\author{Yongchao~Zhang}
\affiliation{School of Physics, Southeast University, Nanjing 211189, China}
\author{Zhilv~Zhang}
\affiliation{Institute of Modern Physics, Chinese Academy of Sciences, Lanzhou 730000, China}
\author{Guang~Zhao}
\affiliation{Institute of High Energy Physics, Chinese Academy of Sciences, Beijing 100049, China}
\author{Shihan~Zhao}
\affiliation{School of Physics, Sun Yat-sen University, Guangzhou 510275, China}
\affiliation{Platform for Muon Science and Technology, Sun Yat-sen University, Guangzhou 510275, China}

\date{\today}

\begin{abstract}
The spontaneous conversion of muonium to antimuonium is one of the interesting charged lepton flavor violation phenomena offering a sensitive probe of potential new physics and serving as a tool to constrain the parameter space beyond the Standard Model. The Muonium-to-Antimuonium Conversion Experiment (MACE) is designed to utilize a high-intensity muon beam, a Michel electron magnetic spectrometer, a positron transport system, and a positron detection system, to either discover or constrain this rare process with a conversion probability of $\mathcal{O}(10^{-13})$. This article presents an overview of the theoretical framework as well as a detailed description of the experimental design for the search for muonium-to-antimuonium conversion.
\end{abstract}


\maketitle

\nolinenumbers

\tableofcontents

\section{Introduction}

Neutrino oscillation is a neutral lepton flavor violation process with profound implications on particle physics. On one hand, it points out that neutrinos have mass and it is therefore the first direct evidence of physics beyond standard model (BSM). The existence of neutral lepton flavor violation also leads to ask whether there can be also charged lepton flavor violation (cLFV). On the other hand, the origin of neutrino masses remains one of the unsolved mysteries in particle physics. The traditional way to explain the neutrino masses is the seesaw mechanism, which is often predicted together with cLFV effects~\cite{Ilakovac:1994kj}. For example, in the type-II seesaw model a Higgs triplet is introduced under a $SU(2)$ symmetry. After the spontaneous symmetry breaking, the massive Higgs boson can contribute to cLFV processes. Therefore, searching for cLFV is an important quest in BSM physics. It is well motivated to push forward the experimental efforts to look for BSM physics by cLFV processes.

There are many experimental efforts to search for cLFV. Such muon-electron-sector cLFV experiments as COMET~\cite{Adamov:2018vin} in Japan and Mu2e~\cite{Mu2e:2014fns} in the USA are currently under construction to search for the coherent muon to electron conversion $\mu^-N\to e^- N$. The accelerator muon beam experiments at Paul Scherrer Institute (PSI) are also searching for cLFV processes, with Mu3e~\cite{Mu3e:2020gyw} looking for $\mu^+\to e^+ e^- e^+$ and MEG~II~\cite{Baldini:2018nnn} for $\mu^+ \to e^+ \gamma$, respectively.

Muonium is an atom consists of a muon and an electron, firstly discovered by Vernon W. Hughes \textit{et al.} in 1960~\cite{PhysRevLett.5.63}. Another interesting approach to probe BSM physics via cLFV is to take the muonium atom and see whether there is a spontaneous conversion from muonium to antimuonium. The original idea was first brought by Pontecorvo in 1957~\cite{Pontecorvo:1957cp}. The latest upper limit on the muonium-to-antimuonium conversion probability, established at $P \lesssim 8.3 \times 10^{-11}$ at 90\% confidence level, was reported in the MACS experiment at PSI in 1999~\cite{Willmann:1998gd}. This result has remained unchallenged in any experiment over the past two decades. The Muonium-to-Antimuonium Conversion Experiment (MACE) aims to discover or further constrain this rare process at an unprecedented level.

As a $\Delta L_\ell=2$ process, the muonium-to-antimuonium conversion is distinct from $\Delta L_\ell=1$ charged lepton flavor violation (cLFV) processes such as $\mu^+ \to e^+\gamma$, $\mu^+ \to e^+e^-e^+$, or $\mu^- N \to e^- N$. From the perspective of Standard Model effective field theory (SMEFT)~\cite{Grzadkowski:2010es,Conlin:2020veq,Petrov:2022wau,Fernandez-Martinez:2024bxg}, effective operators that generate $\Delta L_\ell=1$ processes cannot directly account for $\Delta L_\ell=2$ cLFV phenomena. Consequently, the sensitivity of these processes to $\Delta L_\ell=2$ operators is significantly reduced, indicating that even if $\Delta L_\ell=1$ cLFV processes are observed, they give little rise to $\Delta L_\ell=2$ operators. Furthermore, $\Delta L_\ell=2$ operators can be fundamentally distinct from $\Delta L_\ell=1$ operators, as they carry different lepton numbers. It is feasible to impose a symmetry in the charged-lepton sector that forbids $\Delta L_\ell=1$ processes while permitting $\Delta L_\ell=2$~\cite{Heeck:2024uiz}. This suggests the possibility that $\Delta L_\ell=1$ processes may not exist fundamentally. Therefore, searches targeting $\Delta L_\ell=2$ processes can investigate phenomena that are inaccessible to $\Delta L_\ell=1$ processes. As a result, muonium-to-antimuonium conversion is decorrelated from $\Delta L_\ell=1$ processes, making it a promising target for experimental searches.

Specific theoretical models, including neutral or doubly charged Higgs boson in a type-II hybrid seesaw~\cite{BhupalDev:2018vpr,Han:2021nod}, axion-like particles (ALPs)~\cite{Calibbi:2024rcm}, and flavored gauge bosons such as $Z'$~\cite{Foldenauer:2016rpi,Kriewald:2022erk,Liu:2024gui,Ding:2024zaj}, predict muonium-to-antimuonium conversion at the tree level. Additionally, heavy neutral leptons, such as Majorana neutrinos, can induce this conversion at the one-loop level~\cite{Abada:2018nio,Zhang:2022cyr,Chen:2022gmk,Kriewald:2024zou}. A comprehensive theoretical study has provided a complete list of models allow for this conversion process~\cite{Fukuyama:2021iyw}. In many of these models, searches for muonium-to-antimuonium conversion can complement high-energy collider experiments as well as other low-energy experiments such as the muon $g-2$. Along with all these experimental efforts, we can further constrain these models and reveal the mystery of many BSM phenomena including neutrino masses, matter-antimatter asymmetry, and dark matter.

In addition to the $\mmu$-to-$\ammu$ conversion process, the flavor violating muonium decay is also of interest. In 1959, C. M. York et al. measured the upper limit of the reaction rate for muon-electron annihilation ($\mu^+e^-\to\gamma\gamma$) in a copper target~\cite{PhysRevLett.3.288}. Since then, no other analogous experimental results have been published. Besides, other Muonium cLFV decays such as pair-final-state decay ($\mathrm{M}\to e^+e^-$) and invisible decay ($\mmu\to invisible$) have not yet been constrained by any experiment.
A search for these muonium rare decay modes will complement other existing muon cLFV experiments to provide more evidence for possible new physics~\cite{PhysRev.126.375,Li:1988xb,Czarnecki:1999yj,Eeg:2001cca,Cvetic:2006yg,Gninenko:2012nt,Shkerin:2013pia,Conlin:2022sga}.
Current upper limits of similar processes such as $\mu^+\to e^+\gamma\gamma$ ($\mathcal{BR}<7.2\times 10^{-11}$) and $\mu^+\to e^+e^-e^+$ ($\mathcal{BR}<1.0\times 10^{-12}$) are respectively reported by the Crystal Box~\cite{Bolton:1988af} and the SINDRUM~\cite{SINDRUM:1987nra}.
The Mu3e experiment is proposed to constrain the limit of the latter process to the level of $10^{-16}$ ~\cite{Mu3e:2020gyw}.
However, no experimental search plan for $\mu^+ \to e^+ \gamma \gamma$ has been reported since the 1980s.

The advent of new intense and slow muon sources and the significant advances in modern particle detection technologies will lead to new possibilities in the design of new lepton flavor violation experiments. Regarding the muonium-to-antimuonium conversion, we intend to improve the present bound by more than two orders of magnitude in the proposed Muonium-to-Antimuonium Conversion Experiment (MACE). For the barely touched process $\mathrm{M}\to\gamma\gamma$, a sensitivity of $\mathcal{O}(10^{-12})$ or better is pursued in the pilot MACE \textsc{Phase-I} experiment. The limit of $\mu^+\to e^+\gamma\gamma$ is also expected to be improved by one order of magnitude in this experiment stage. The experimental plan of searching for the aforementioned processes by optimizing the detector system and changing the event selection criteria is also discussed at the end of this article.

The conceptual design study focuses on the following aspects: the physics motivation, the development of a high-intensity muon beam, the high-efficiency muonium production in a vacuum, an optimized design of detector system and requirements in physics performance. This article summarizes the theoretical and experimental aspects of MACE.

\section{Overview of theoretical framework}

The main decay channel for the muonium is determined by the weak decay of the muon, $\mmu \to e^+e^- \bar \nu_\mu \nu_e$. The average lifetime of a muonium state $\tau_\mmu$ is expected to be almost the same as that of the muon,
\begin{equation}\label{MuoniumWidth}
\begin{aligned}
    1/\tau_\mmu &= \Gamma\left(\mmu \to e^+e^- \nu_e \bar\nu_\mu \right) \\
    &\approx \Gamma\left(\mu^+ \to e^+ \nu_e \bar\nu_\mu \right) = \frac{G_F^2 m_\mu^5}{192 \pi^3} =1/\tau_\mu ,
\end{aligned}
\end{equation}
with $\tau_\mu = (2.1969811 \pm 0.0000022)\times 10^{-6}$ s~\cite{ParticleDataGroup:2024cfk}, apart from the tiny effect due to time dilation~\cite{Czarnecki:1999yj}. Note that \cref{MuoniumWidth} represents the leading-order result. The results including subleading corrections are available~\cite{Czarnecki:1999yj}.

Like a hydrogen atom, muonium could be formed in two spin configurations. A spin-one triplet state $\ket{\mmuv}$ is called ortho-muonium, while a spin-zero singlet state $\ket{\mmup}$ is called para-muonium. In what follows, we will drop the superscript and employ the notation $\ket{\mmu}$ if the spin of the muonium state is not important for the discussion.

Since the interactions with $\Delta L_\ell=2$ can change the muonium state into the antimuonium one, the possibility to study muonium--antimuonium oscillations arises. Theoretical analyses of conversion probability from a muonium to an antimuonium have been performed, both in particular new physics models~\cite{Pontecorvo:1957cp,Feinberg:1961zza,ClarkLove:2004,CDKK:2005,Li:2019xvv,Endo:2020mev}, and using the framework of effective field theory~\cite{Conlin:2020veq}, where all possible BSM models are encoded in a few Wilson coefficients of effective operators. Observation of muonium converting into anti-muonium provides clean probes of new physics in the leptonic sector~\cite{Bernstein:2013hba,Willmann:1998gd}.

\subsection{Phenomenology of muonium conversion}\label{sec:pheno of m conv}
In order to determine experimental observables related to $\mmbar$ oscillations, we recall that the treatment of the two-level system that represents muonium and antimuonium is similar to that of meson-antimeson oscillations~\cite{Voloshin:1986dir,Burdman:2003rs,Lenz:2006hd,Golowich:2007ka,Nierste,Donoghue,Petrov:2021idw,Song:2024jjn}. There are, however, several important differences. First, both ortho- and para-muonium can oscillate. Second, the SM oscillation probability is tiny, as it is related to a function of neutrino masses. Any experimental indication of oscillation would represent a sign of new physics.

In the presence of the interactions coupling $\mmu$ and $\ammu$, the time development of a muonium and anti-muonium states would be coupled, so it would be appropriate to consider their combined evolution,
\begin{equation}
|\psi(t)\rangle =
\left( {\begin{array}{c}
 a(t) \\
 b(t) \\
 \end{array} } \right) =
 a(t) |\mmu\rangle + b(t) |\ammu\rangle.
\end{equation}
The time evolution of $|\psi(t)\rangle$ evolution is governed by a Schr\"odinger-like equation,
\begin{equation}\label{two state time evolution}
i\frac{\dd}{\dd t}
\left(
\begin{array}{c}
\ket{\mmu(t)} \\ \ket{\ammu(t)}
\end{array}
\right)
=
\left(M-i\frac{\Gamma}{2}\right)
\left(
\begin{array}{c}
\ket{\mmu(t)} \\ \ket{\ammu(t)}
\end{array}
\right)~,
\end{equation}
where $\left(M-i\frac{\Gamma}{2}\right)_{ik}$ is a $2\times 2$ Hamiltonian (mass matrix) with non-zero off-diagonal terms originating from the $\Delta L_\ell=2$ interactions. CPT-invariance dictates that the masses and widths of the muonium and anti-muonium are the same, so $M_{11}= M_{22}$, $\Gamma_{11}=\Gamma_{22}$. In what follows, we assume CP-invariance of the $\Delta L_\ell = 2$ interaction. A more general formalism without this assumption follows the same steps as that for the $B\bar B$ or $K\bar K$ mixing~\cite{Petrov:2021idw,Donoghue}. Then,
\begin{equation}\label{off_diagonal_elements}
M_{12}=M^{*}_{21}, \qquad \Gamma_{12}=\Gamma^{*}_{21}.
\end{equation}
The off-diagonal matrix elements in \cref{off_diagonal_elements} can be related to the matrix elements of the effective operators, as discussed in~\cite{Petrov:2021idw,Donoghue}. The related Lagrangian expressed in effective operators can be found in \cref{sec:muonium conversion EFT}.
\begin{equation}\label{OffDiagonal}
\begin{aligned}
    \left(M-\frac{i}{2} \Gamma\right)_{12}&=\frac{1}{2 m}\left\langle\ammu\left|{\cal H}_{\rm eff} \right| \mmu\right\rangle \\
    &+ \frac{1}{2 m} \sum_{n} \frac{\left\langle\ammu
\left|{\cal H}_{\rm eff} \right| n\right\rangle\left\langle n\left|{\cal H}_{\rm eff} \right| \mmu\right\rangle}{m-E_{n}+i \epsilon}~,
\end{aligned}
\end{equation}
where $m = \left(m_1+m_2\right)/2$ is the muonium mass, and $m_i$ are the masses of the physical mass eigenstates $ |\mmu_{1,2} \rangle$ as to be discussed below.

To find the propagating states, the mass matrix needs to be diagonalized. The basis in which the mass matrix is diagonal is represented by the mass eigenstates $ |\mmu_{1,2} \rangle$, which are related to the flavor eigenstates $\mmu$ and $\ammu$ as
\begin{equation}
 |\mmu_{1,2} \rangle = \frac{1}{\sqrt{2}} \left(|\mmu \rangle \mp |\ammu \rangle
 \right) ,
\end{equation}
where we employed a convention where $CP |\mmu_\pm \rangle = \mp |\mmu_\pm \rangle$. The mass and the width differences of the mass eigenstates are
\begin{equation}
    \Delta m \equiv m_{1}-m_{2}, \qquad \Delta \Gamma \equiv \Gamma_{2}-\Gamma_{1}.
\end{equation}
Here, $m_i$ ($\Gamma_i$) are the masses (widths) of the physical mass eigenstates $ |\mmu_{1,2} \rangle$. It is interesting to see how the \cref{OffDiagonal} defines the mass and the lifetime differences. Since the first term in \cref{OffDiagonal} is defined by a local operator, its matrix element does not develop an absorptive part, so it contributes to $M_{12}$, i.e., the mass difference. The second term contains bi-local contributions connected by physical intermediate states. This term has both real and imaginary parts and thus contributes to both $M_{12}$ and $\Gamma_{12}$.

It is often convenient to introduce dimensionless quantities,
\begin{equation}\label{XandY}
x = \frac{\Delta m}{\Gamma}, \qquad y = \frac{\Delta \Gamma}{2\Gamma},
\end{equation}
where the average width $\Gamma=(\Gamma_1+\Gamma_2)/2$. Noting that $\Gamma$ is defined by the standard model decay rate of the muon, and $x$ and $y$ are driven by the lepton-flavor violating interactions, we should expect that both $x,y \ll 1$.

The time evolution of flavor eigenstates follows from \cref{two state time evolution}~\cite{Donoghue,Nierste,Conlin:2020veq},
\begin{eqnarray}
\ket{\mmu(t)} &=& g_+(t) \ket{\mmu} + g_-(t) \ket{\ammu},
\nonumber \\
\ket{\ammu(t)} &=& g_-(t) \ket{\mmu} + g_+(t) \ket{\ammu},
\end{eqnarray}
where the coefficients $g_\pm(t)$ are defined as
\begin{equation}\label{TimeDep}
g_\pm(t) = \frac{1}{2} e^{-\Gamma_1 t/2} e^{-im_1t} \left(1 \pm e^{\Delta\Gamma t/2} e^{i \Delta m t} \right).
\end{equation}

As $x,y \ll 1$ we can expand \cref{TimeDep} in power series in $x$ and $y$ to obtain

\begin{eqnarray}
g_+(t) &=& e^{-\Gamma_1 t/2} e^{-im_1t} \left(1 + \frac{1}{8} \left(y-ix\right)^2 \left(\Gamma t\right)^2 \right),
\nonumber \\
g_-(t) &=& \frac{1}{2} e^{-\Gamma_1 t/2} e^{-im_1t} \left(y-ix\right) \left(\Gamma t\right).
\end{eqnarray}

The most natural way to detect $\mmbar$ oscillations experimentally is by producing $\mmu$ state and looking for the decay products of the CP-conjugated state $\ammu$. Denoting an amplitude for the $\mmu$ decay into a final state $f$ as $A_f = \langle f|{\cal H} |\mmu\rangle$ and an amplitude for its decay into a CP-conjugated final state $\bar{f}$ as $A_{\bar f} = \langle \bar{f}|{\cal H} |\mmu\rangle$, we can write the time-dependent decay rate of $\mmu$ into the $\bar{f}$,

\begin{equation}
\Gamma(\mmu \to \bar{f})(t) = \frac{1}{2} N_f \left|A_f\right|^2 e^{-\Gamma t} \left(\Gamma t\right)^2 R_M(x,y),
\end{equation}
where $N_f$ is a phase-space factor and we defined the oscillation rate $R_M(x,y)$ as
\begin{equation}
R_M(x,y) = \frac{1}{2} \left(x^2+y^2\right).
\end{equation}
Integrating over time and normalizing to $\Gamma(\mmu \to f)$ we get the probability of $\mmu$ decaying as $\ammu$ at some time $t > 0$,
\begin{equation}\label{Prob_osc}
P(\mmu \rightarrow \ammu) = \frac{\Gamma(\mmu \to \bar{f})}{\Gamma(\mmu \to f)} = R_M(x,y).
\end{equation}
The \cref{Prob_osc} generalizes oscillation probability found in the papers~\cite{Feinberg:1961zza,CDKK:2005,Conlin:2020veq} by allowing for a non-zero lifetime difference in $\mmbar$ oscillations.

\subsection{Muonium conversion and new physics beyond the Standard Model}\label{sec:muonium conversion EFT}
The Standard Model Effective Field Theory (SMEFT) serves as a powerful framework for providing a model-independent characterization of heavy new physics beyond the Standard Model~\cite{Grzadkowski:2010es,Fernandez-Martinez:2024bxg}. If we believe that the Standard Model acts as a low-energy approximation of a complete theory, the Standard Model Lagrangian can be interpreted as the leading term in a series expansion of the complete theory. Higher-order terms in this series expansion are the effective forms of new physics beyond the Standard Model. The Standard Model operators have a mass dimension of 4, whereas higher-dimensional operators, which are scaled by inverse powers of the unknown new physics scale (NP scale), are incorporated into the SMEFT Lagrangian. Thus, the SMEFT Lagrangian can be expressed as a series expansion in terms of the inverse power of NP scale,
\begin{equation}
\begin{aligned}
    \mathcal{L}_\text{eff}
    & =\mathcal{L}_\text{SM} + \sum_{n>4}\frac{1}{\Lambda^{n-4}}\sum_i C^{(n)}_i Q^{(n)}_i \\
    & =\mathcal{L}_\text{SM} + \frac{1}{\Lambda}\sum_i C^{(5)}_i Q^{(5)}_i + \frac{1}{\Lambda^2}\sum_i C^{(6)}_i Q^{(6)}_i + \cdots.
\end{aligned}
\end{equation}
The leading term in the series is the Standard Model Lagrangian, denoted as $\mathcal{L}_\text{SM}$. In the higher-order terms, each dimension-$n$ effective operator $Q^{(n)}_i$ is associated with a Wilson coefficient $C^{(n)}_i$ and scaled by the new physics scale $\Lambda$. Qualitatively, these higher-order terms encapsulate the potential new physics models through the effective operators, while the Wilson coefficients represent the coupling strengths. The new physics scale $\Lambda$ can be interpreted as a characteristic energy scale in new physics, such as a combination of quantities with dimension of mass involved in interactions beyond the Standard Model. Both the NP scale and the Wilson coefficients will ultimately manifest in observables, allowing experiments to constrain the new physics scale or the coupling strengths in a model-independent way.

\begin{figure}[t]
    \nolinenumbers
    \centering
    \begin{tikzpicture}
        \begin{feynman}
            \vertex (Mu) {$\mu^+$};
            \vertex [right=1.75cm of Mu] (Mu1);
            \vertex [above=1.5cm of Mu] (e) {$e^-$};
            \vertex [right=1.75cm of e] (e1);
            \vertex [above=0.75cm of Mu1] (tmp0);
            \vertex [right=0.75cm of tmp0, empty dot] (EFT) {};
            \vertex [right=0.75cm of EFT] (text0) {$\frac{C_{\bar{\mu}e\bar{\mu}e}}{\Lambda^2}$};
            \vertex [right=0.75cm of EFT] (tmp1);
            \vertex [below=0.75cm of tmp1] (mu1);
            \vertex [right=1.5cm of mu1] (mu) {$\mu^-$};
            \vertex [above=0.75cm of tmp1] (E1);
            \vertex [right=1.5cm of E1] (E) {$e^+\,$};
            \diagram* {
            (e) -- [fermion] (e1) -- [quarter left] (EFT) -- [quarter left] (E1) -- [anti fermion] (E),
            (Mu) -- [anti fermion] (Mu1) -- [quarter right] (EFT) -- [quarter right] (mu1) -- [fermion] (mu),
            };
            \draw [decoration={brace}, decorate] (Mu.west) -- (e.west) node [pos=0.5, left] {$\mmu$};
            \draw [decoration={brace}, decorate] (E.east) -- (mu.east) node [pos=0.5, right] {$\ammu$};
        \end{feynman}
    \end{tikzpicture}
    \caption{\label{fig:deltaL2-SMEFT-Mconv}The SMEFT tree-level diagram for muonium-to-antimuonium conversion with one $\Delta L_\mu=2$ four-fermion effective vertex. The conversion probability is proportional to $1/\Lambda^4$.}
\end{figure}
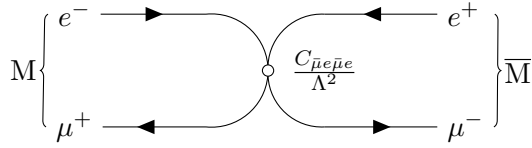

Dimension-6 operators are the lowest-order operators where $\Delta L_\mu=2$ muonium-to-antimuonium conversion can be introduced at the tree level with single SMEFT vertex~\cite{Conlin:2020veq}. There are five independent dimension-6 $\bar{\mu}e\bar{\mu}e$ four-fermion operators that leads to muonium-to-antimuonium conversion~\cite{Conlin:2020veq,Fukuyama:2021iyw,Petrov:2022wau,Conlin:2022sga},
\begin{equation}
\begin{gathered}
    Q^{LL}_V=\left(\bar\mu \gamma_\alpha(1-\gamma_5) e\right)\left(\bar\mu \gamma^\alpha(1-\gamma_5) e\right),\\
    Q^{RR}_V=\left(\bar\mu \gamma_\alpha(1+\gamma_5) e\right)\left(\bar\mu \gamma^\alpha(1+\gamma_5) e\right),\\
    Q^{LR}_V=\left(\bar\mu \gamma_\alpha(1-\gamma_5) e\right)\left(\bar\mu \gamma^\alpha(1+\gamma_5) e\right),\\
    Q^{LR}_S=\left(\bar\mu (1+\gamma_5) e\right)\left(\bar\mu (1+\gamma_5) e\right),\\
    Q^{RL}_S=\left(\bar\mu (1-\gamma_5) e\right)\left(\bar\mu (1-\gamma_5) e\right).
\end{gathered}
\end{equation}
These five operators contribute to the mass difference as defined in \cref{sec:pheno of m conv}. The width difference can arise from the following $\bar{\mu}e\bar{\nu}_\mu\nu_e$ four-fermion operators,
\begin{equation}
\begin{gathered}
    Q^{L\nu}_V=\left(\bar\mu \gamma_\alpha(1-\gamma_5) e\right)\left(\bar\nu_{\mu} \gamma^\alpha(1-\gamma_5) \nu_{e}\right),\\
    Q^{R\nu}_V=\left(\bar\mu \gamma_\alpha(1+\gamma_5) e\right)\left(\bar\nu_{\mu} \gamma^\alpha(1-\gamma_5) \nu_{e}\right).
\end{gathered}
\end{equation}
However, the contributions of the width difference to the conversion probability are suppressed by a factor of $G_F m^2 \approx 10^{-7}$~\cite{Conlin:2020veq,Conlin:2022sga}, which is orders of magnitude smaller than the contributions from the mass difference as long as $|C^{L\nu}_V /C_\text{other}|$ and $|C^{R\nu}_V /C_\text{other}|$ are not too large. We will therefore consider only the mass difference contributions in the following calculations.

The effective Lagrangian for muonium-to-antimuonium conversion can be written as
\begin{equation}
\begin{aligned}
    \mathcal{L}_\text{eff}=\frac{1}{\Lambda^2}\big(
    & C^{LL}_V Q^{LL}_V + C^{RR}_V Q^{RR}_V + C^{LR}_V Q^{LR}_V + \\
    & C^{LR}_S Q^{LR}_S + C^{RL}_S Q^{RL}_S
    \big)~.
\end{aligned}
\end{equation}
The corresponding effective Hamiltonian in \cref{OffDiagonal} is $\mathcal{H}_\text{eff}=-\mathcal{L}_\text{eff}$, which generates muonium-to-antimuonium conversion as shown in \cref{fig:deltaL2-SMEFT-Mconv}.

\begin{table*}[t]
\nolinenumbers
\centering
\caption{\label{tab:deltaL2 processes} $\Delta L_\ell=-\Delta L_{\ell'}=2$ cLFV processes.}
\begin{tabular}{cccc}
\hline\hline
Process &
  Type &
  Experiment &
  Current bound \\ \hline
$\mmu\to\ammu$ &
  $\mmbar$ mixing &
  MACS~\cite{Willmann:1998gd}, MACE &
  $P_{0.1\,\text{T}}(\mmu\to\ammu)<8.3\times 10^{-11}$~\cite{Willmann:1998gd} \\ \hline
$\mu^+e^-\to\mu^-e^+$ &
  \multirow{3}{*}{Scattering} &
  \multirow{3}{*}{$\mu$TRISTAN~\cite{Hamada:2022mua}} &
  \multirow{3}{*}{None} \\ \cline{1-1}
$\mu^+\mu^+\to e^+e^+$ &
   &
   &
   \\ \cline{1-1}
$\mu^+\mu^+\to \tau^+\tau^+$ &
   &
   &
   \\ \hline
$\mu^+\to e^+\bar{\nu}_e\nu_\mu$ & \multirow{2}{*}{Decay} & $\tau_\mu$ measurement & $\Delta\tau_\mu/\tau_\mu=1\times10^{-6}$~\cite{ParticleDataGroup:2024cfk} \\ \cline{1-1} \cline{3-4}
$Z\to\ell'^\pm\ell'^\pm\ell^\mp\ell^\mp$ &
   &
  CEPC~\cite{CEPCStudyGroup:2018ghi}, FCC-ee~\cite{FCC:2018evy} &
  None \\ \hline
\hline
\end{tabular}
\end{table*}

According to the SMEFT Lagrangian, we can calculate the amplitudes for muonium-to-antimuonium conversion. In experiments, muonium atoms are produced in two spin states: spin-zero singlet states (para-muonium, denoted as $\mmu_P$) and spin-one triplet states (ortho-muonium, denoted as $\mmu_{V,m}$, where $m=0,\pm 1$). The conversion probability is expressed in terms of the dimensionless mass differences, i.e. $x$ in \cref{sec:pheno of m conv}, under the assumption that the width differences are negligible. Each spin state has its own corresponding values for $x$. They are given by~\cite{Fukuyama:2021iyw,Fukuyama:2023drl}
\begin{equation}\label{eq:muonium mass differences}
\begin{gathered}
    x_P=\frac{16(\alpha\mu)^3}{\pi\Gamma\Lambda^2}\left(C_0-\frac{3}{2}C^{LR}_V\right)~,\\
    x_{V,0}=x_{V,\pm 1}=\frac{16(\alpha\mu)^3}{\pi\Gamma\Lambda^2}\left(C_0+\frac{1}{2}C^{LR}_V\right)~,
\end{gathered}
\end{equation}
where
\begin{equation}
    C_0 = C^{LL}_V + C^{RR}_V - \frac{1}{4}\left(C^{LR}_S + C^{RL}_S\right)~.
\end{equation}
Here, $\mu=m_\mu m_e/(m_\mu+m_e)$ is the reduced mass of muonium, $\alpha$ is the fine-structure constant, and $\Gamma$ is the muonium width.

It is important to note that a non-zero external magnetic field will introduce a suppression to the conversion probability~\cite{Hou:1995np,Fukuyama:2021iyw,Fukuyama:2023drl}. Since the decay kinematics of different spin states are indistinguishable in experiments like MACE, the total conversion probability is expressed as a weighted sum over all spin states, according to their populations,
\begin{equation}
\begin{aligned}
    P_B(\mmu\to\ammu) & = f_P P_B(\mmu_P\to\ammu_P) \\
    & \quad + f_{V,0} P_B(\mmu_{V,0}\to\ammu_{V,0}) \\  
    & \quad + \sum_{m=\pm 1} f_{V,m} P_B(\mmu_{V,m}\to\ammu_{V,m})~,
\end{aligned}
\end{equation}
where $f_P$ and $f_{V,m}$ denote the populations of para- and ortho-muonium states, respectively. $P_B$ denotes the conversion probability in an external magnetic field with a magnitude of $B$.

The conversion probabilities for different spin states exhibit distinct magnetic field dependencies. For the $m = \pm 1$ ortho-muonium states, the external magnetic field induces an energy splitting $\Delta E_B$, which suppresses their conversion probability
\begin{equation}
    P_B(\mmu_{V,\pm 1}\to\ammu_{V,\pm 1}) = \frac{x_{V,\pm 1}^2/2}{1 + \left(\Delta E_B/\Gamma\right)^2}~,
\end{equation}
where $\Delta E_B/\Gamma \approx 3.85 \times 10^5 \times B/\text{T}$~\cite{Fukuyama:2021iyw}. The $(\Delta E_B/\Gamma)^2$ term suppresses the conversion probability by $\mathcal{O}(10^{-9})$ in a 0.1~T magnetic field. Consequently, contributions from the $m = \pm 1$ states are quenched in MACE.

On the contrary, $m=0$ states are free from energy splitting so the conversion probabilities are still sizeable in a practical external magnetic field. Instead, the $m=0$ para- and ortho-muonium states are mixed due to the magnetic field, and the dimensionless mass differences are written as
\begin{equation}
\begin{gathered}
    x_P^B=\frac{1}{2}\left(x_P-x_{V,0}+\frac{x_P+x_{V,0}}{\sqrt{1+X^2}}\right)~,\\
    x_{V,0}^B=\frac{1}{2}\left(x_{V,0}-x_P+\frac{x_P+x_{V,0}}{\sqrt{1+X^2}}\right)~,
\end{gathered}
\end{equation}
where $X\approx 6.31\times B/\text{T}$~\cite{Fukuyama:2021iyw}. Here, $x_P^B$ and $x_{V,0}^B$ represents the dimensionless mass differences in an external magnetic field with a magnitude of $B$.

The total magnetic-field-dependent conversion probability reads
\begin{equation}
\begin{aligned}
    P_B(\mmu\to\ammu) = \frac{1}{2}\bigg( & f_P\left(x_P^B\right)^2 + f_{V,0}\left(x_{V,0}^B\right)^2 \\
    & + \sum_{m=\pm 1} \frac{f_{V,m} x_{V,\pm 1}^2}{1 + \left(\Delta E_B/\Gamma\right)^2}\bigg)~.
\end{aligned}
\end{equation}
If there exists some sizable magnetic field (typically larger than the geomagnetic field) and contributions from $m=\pm1$ states are negligible as in MACE's case, the total conversion probability can be simplified as
\begin{equation}
     P_B(\mmu\to\ammu) = \frac{1}{2}\left(f_P\left(x_P^B\right)^2 + f_{V,0}\left(x_{V,0}^B\right)^2\right)~.
\end{equation}
We can now estimate the new physics scale that MACE could probe in a $B=0.1$~T magnetic field by assuming $f_P = 0.32$, $f_{V,0} = 0.18$ , and taking all Wilson coefficients to be 1. This leads to the result
\begin{equation}\label{eq:MACE NP scale}
    \Lambda \gtrsim \frac{0.02~\text{TeV}}{P_{0.1\,\text{T}}^\text{up}(\mmu \to \ammu)^{1/4}}~.
\end{equation}
If the experimental upper limit $P_{0.1\,\text{T}}^\text{up}(\mmu \to \ammu)$ is set at $\mathcal{O}(10^{-13})$, we conclude that MACE could probe new physics at the scale of 10--100~TeV.

Since $\Delta L_\ell=1$ processes are distinct from $\Delta L_\ell=2$ physics and they are not directly comparable to muonium-to-antimuonium conversion, it is essential to figure out which processes are sensitive to $\Delta L_\ell=2$ physics. They should involve fermions from two different generations and violate lepton flavor by 2 units. An incomplete list of candidate processes for experimental searches is provided in \cref{tab:deltaL2 processes}. The current most stringent limit is set by the MACS experiment~\cite{Willmann:1998gd}, while most of the other processes remain unconstrained. These unconstrained processes are likely to be targeted for experimental searches or measurements in the coming decades.

\begin{figure}[htbp]
    \nolinenumbers
    \centering
    \subfloat[\label{fig:mup em to mum ep scattering}$\mu^+e^-\to\mu^-e^+$]{
        \begin{tikzpicture}
            \begin{feynman}
                \vertex (left);
                \vertex [right=2cm of left, empty dot] (EFT) {};
                \vertex [right=2cm of EFT] (right);

                \vertex [above=1cm of left] (e) {$e^-$};
                \vertex [below=1cm of left] (Mu) {$\mu^+$};
                \vertex [above=1cm of right] (E) {$e^+$};
                \vertex [below=1cm of right] (mu) {$\mu^-$};

                \vertex [above=0.5cm of EFT] {$\frac{C_{\bar{\mu}e\bar{\mu}e}}{\Lambda^2}$};

                \diagram* {
                (e) -- [fermion] (EFT) -- [fermion] (mu),
                (Mu) -- [anti fermion] (EFT) -- [anti fermion] (E),
                };
            \end{feynman}
        \end{tikzpicture}
    }\\
    \subfloat[\label{fig:mup mup to ep ep scattering}$\mu^+\mu^+\to \ell^+\ell^+~(\ell=e,\tau)$]{
        \begin{tikzpicture}
            \begin{feynman}
                \vertex (left);
                \vertex [right=2cm of left, empty dot] (EFT) {};
                \vertex [right=2cm of EFT] (right);

                \vertex [above=1cm of left] (Mu1) {$\mu^+$};
                \vertex [below=1cm of left] (Mu2) {$\mu^+$};
                \vertex [above=1cm of right] (E1) {$\ell^+$};
                \vertex [below=1cm of right] (E2) {$\ell^+$};

                \vertex [above=0.5cm of EFT] {$\frac{C_{\bar{\mu}\ell\bar{\mu}\ell}}{\Lambda^2}$};

                \diagram* {
                (Mu1) -- [anti fermion] (EFT) -- [anti fermion] (E1),
                (Mu2) -- [anti fermion] (EFT) -- [anti fermion] (E2),
                };
            \end{feynman}
        \end{tikzpicture}
    }\qquad
    \caption{Charged lepton flavor violating scattering $\mu^+e^-\to\mu^-e^+$ and $\mu^+\mu^+\to \ell^+\ell^+~(\ell=e,\tau)$.}
\end{figure}
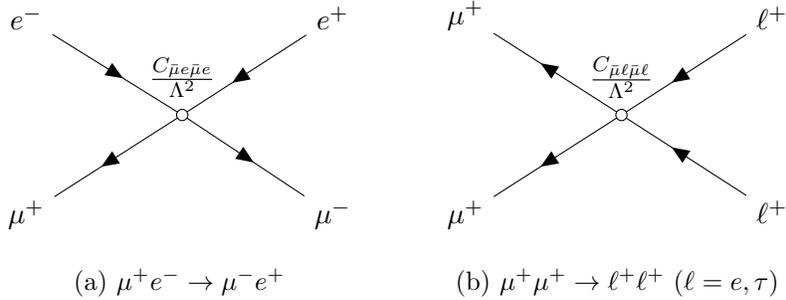

The muonium-to-antimuonium conversion generating $\bar{\mu}e\bar{\mu}e$ four-fermion operators can also lead to cLFV scattering processes $\mu^+e^- \to \mu^-e^+$ and $\mu^+\mu^+ \to \ell^+\ell^+~(\ell=e,\tau)$, which could be searched for in future lepton colliders. Recently, a muon collider named $\mu$TRISTAN was proposed~\cite{Hamada:2022mua}. The conceptual design utilizes an intensive muonium source produced from a multi-layer silica aerogel target to provide an ultra-cold muon source by ionizing muonium atoms. The collider-quality muon beam is produced by reaccelerating the ultra-cold muons. The center-of-mass energies for $\mu^+e^-$ and $\mu^+\mu^+$ collisions are 346~GeV and 2~TeV, respectively, with estimated instantaneous luminosities of $5.7 \times 10^{32}$~cm$^{-2}$~s$^{-1}$ and $4.6 \times 10^{33}$~cm$^{-2}$~s$^{-1}$. This high luminosity gives $\mu$TRISTAN the potential to provide stringent bounds on these two cLFV scattering processes. Based on previous phenomenological calculations on similar four-fermion operators, $\mu$TRISTAN could potentially push the new physics scale limit on $\bar{\mu}e\bar{\mu}e$ four-fermion operators up to 50~TeV~\cite{Hamada:2022uyn}. Substantial efforts are still required before the $\mu^+e^-$ and $\mu^+\mu^+$ collider can be operational and accessible for physicists.

\begin{figure*}[t]
    \nolinenumbers
    \centering
    \subfloat{\includegraphics[width=0.45\textwidth]{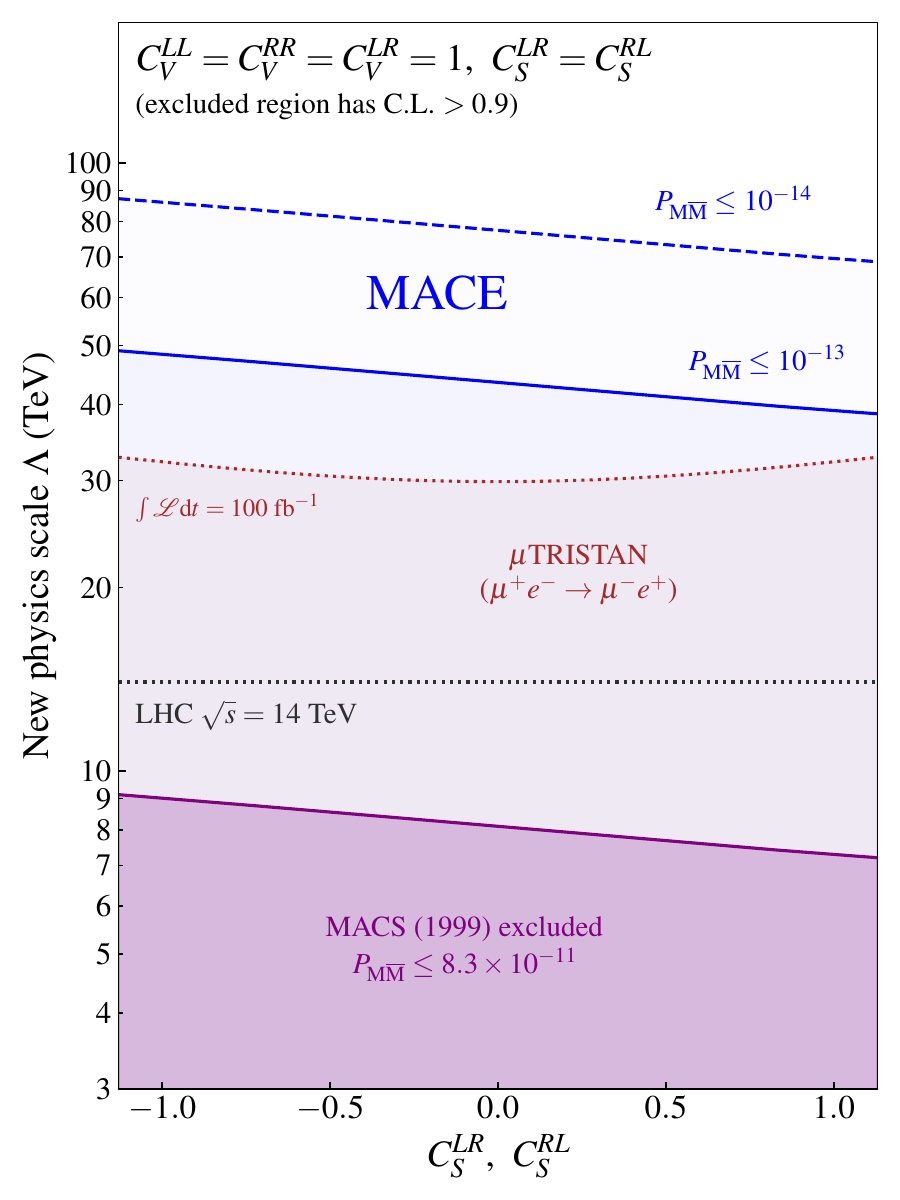}}
    \hfill
    \subfloat{\includegraphics[width=0.45\textwidth]{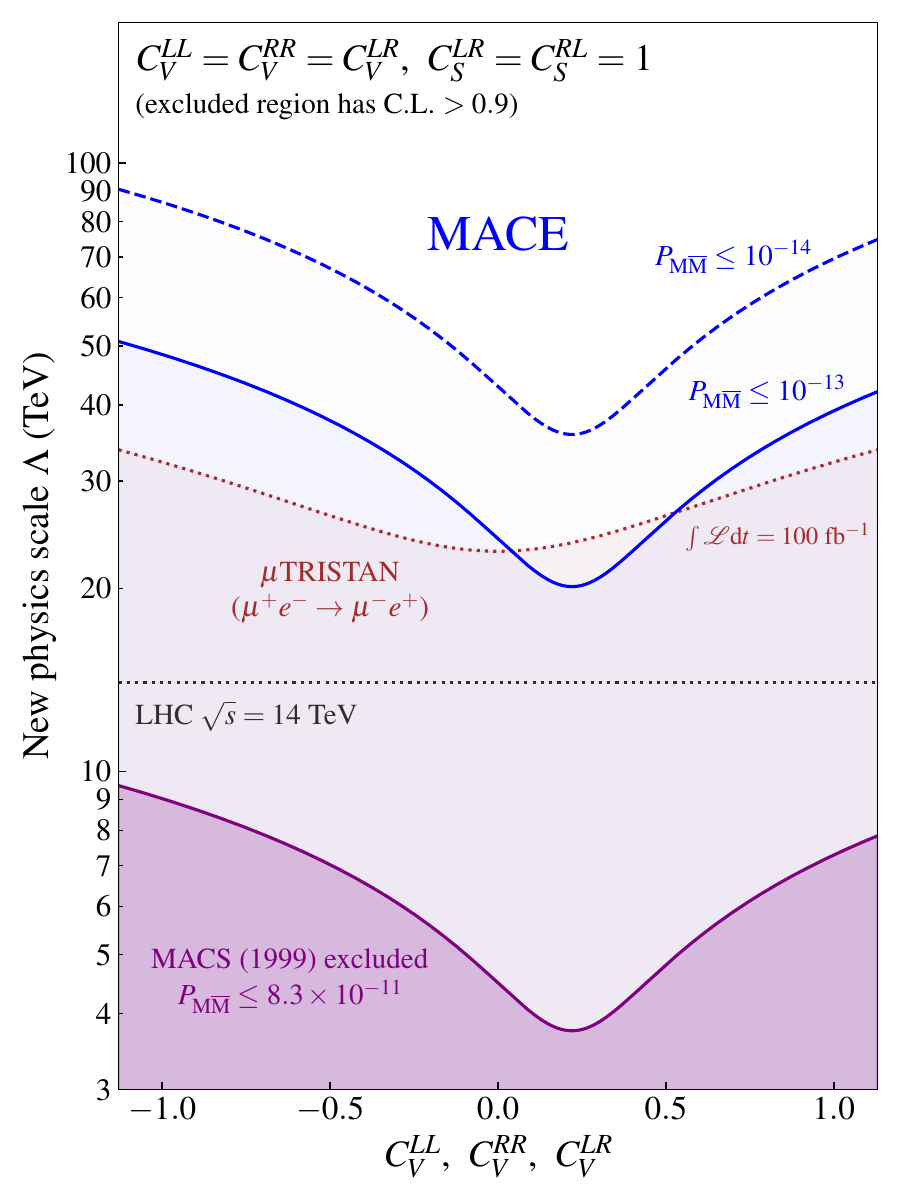}}
    \caption{\label{fig:MACE NP scale}The new physics scale accessible to experiments searching for $\Delta L_\ell=2$ processes.}
\end{figure*}

The MUonE experiment is a fixed-target experiment aimed at measuring the $\mu-e$ scattering differential cross section with unprecedented precision of $\mathcal{O}(10^{-5})$~\cite{Spedicato:2024sjh}. MUonE is designed to utilize the intense 150~GeV muon beam available at CERN, incident on beryllium targets, to measure the scattering of muons on atomic electrons. The design aims to achieve the target statistical sensitivity within three years of running time, with an integrated luminosity of 15~fb$^{-1}$. Limit on the $\mu^+e^- \to \mu^-e^+$ cross section could be set by measuring anomalies in the $\mu-e$ scattering cross section. Unfortunately, this approach may not be sensitive enough to heavy new physics to supersede the existing MACS limit due to the low center-of-mass energy~\cite{Masiero:2020vxk}.

\begin{figure*}[t]
    \nolinenumbers
    \centering
    \subfloat[$e^-$ energy spectrum.\label{fig:antimuounium-electron-spectrum}]{
        \includegraphics[width=0.3\textwidth]{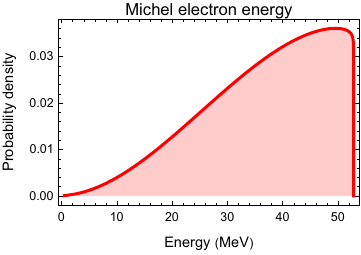}
    }\vspace{1pt}
    \subfloat[Leading-order $\ammu$ decay.]{
        \begin{tikzpicture}
        \begin{feynman}
            \vertex (Mu) {$\mu^-$};
            \vertex [right=1.75cm of Mu] (MuWnuMu);
            \vertex [below right= of MuWnuMu] (nuMu) {$\nu_\mu$};
            \vertex [above right= of MuWnuMu] (EWnue);
            \vertex [below right= of EWnue] (nue) {$\bar{\nu}_e$};
            \vertex [right= of EWnue] (E) {\textcolor{red}{$e^-$}};
            \vertex [above=1em of Mu] (e) {$e^+$};
            \vertex [left=0.4142136em of MuWnuMu] (help1);
            \vertex [above=1em of help1] (eTurn);
            \vertex [above right=3cm of eTurn] (eOut) {\textcolor{blue}{$e^+$}};
            \diagram* {
            (Mu) -- [fermion] (MuWnuMu) -- [fermion] (nuMu);
            (MuWnuMu) -- [boson, edge label'=$W$] (EWnue) -- [anti fermion] (nue);
            (EWnue) -- [fermion] (E);
            (e) -- [anti fermion] (eTurn) -- [anti fermion] (eOut);
            };
            \draw [decoration={brace}, decorate] (Mu.west) -- (e.west)
            node [pos=0.5, left] {$\ammu$};
        \end{feynman}
        \end{tikzpicture}
    }\vspace{1pt}
    \subfloat[$e^+$ energy spectrum.\label{fig:antimuounium-positron-spectrum}]{
        \includegraphics[width=0.3\textwidth]{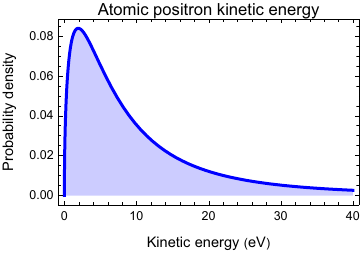}
    }
    \caption{Energy spectrum and the leading-order diagram of antimuonium decay $\overline{\text{M}}\to e^+e^-\bar{\nu}_e\nu_\mu$. The energy spectrum of the fast decay $e^-$ is accurate to next-to-leading-order, atomic shell $e^+$ spectrum assumes $1s$ muonium~\cite{Zhao:2024qjb}.}
    \label{fig:muonium-decay-spectrum}
\end{figure*}

\begin{figure*}[t]
    \nolinenumbers
    \centering
    \subfloat[]{
        \begin{tikzpicture}
            \begin{feynman}
                \vertex (Mu) {$\mu^+$};
                \vertex [right=1.75cm of Mu] (nuWMu);
                \vertex [right= of nuWMu] (EWnu);
                \vertex [right= of EWnu] (E) {$e^{+\,}$};
                \vertex [above= of Mu] (e) {$e^-$};
                \vertex [right=1.75cm of e] (eWnu);
                \vertex [right= of eWnu] (nuWmu);
                \vertex [right= of nuWmu] (mu) {$\mu^-$};
                \diagram* {
                (e) -- [fermion] (eWnu) -- [fermion, edge label=$\nu$] (nuWmu) -- [fermion] (mu),
                (nuWmu) -- [boson, edge label=$W$] (EWnu),
                (Mu) -- [anti fermion] (nuWMu) -- [anti fermion, edge label=$\nu$] (EWnu) -- [anti fermion] (E),
                (nuWMu) -- [boson, edge label=$W$] (eWnu),
                };
                \draw [decoration={brace}, decorate] (Mu.west) -- (e.west) node [pos=0.5, left] {$\mmu$};
                \draw [decoration={brace}, decorate] (mu.east) -- (E.east) node [pos=0.5, right] {$\ammu$};
            \end{feynman}
        \end{tikzpicture}
    }\vspace{1pt}
    \subfloat[]{
        \begin{tikzpicture}
            \begin{feynman}
                \vertex (Mu) {$\mu^+$};
                \vertex [right=1.75cm of Mu] (nuWMu);
                \vertex [right= of nuWMu] (EWnu);
                \vertex [right= of EWnu] (E) {$e^{+\,}$};
                \vertex [above= of Mu] (e) {$e^-$};
                \vertex [right=1.75cm of e] (eWnu);
                \vertex [right= of eWnu] (nuWmu);
                \vertex [right= of nuWmu] (mu) {$\mu^-$};
                \diagram* {
                (e) -- [fermion] (eWnu) -- [boson, edge label=$W$] (nuWmu) -- [fermion] (mu),
                (EWnu) -- [fermion, edge label=$\nu$] (nuWmu),
                (Mu) -- [anti fermion] (nuWMu) -- [boson, edge label=$W$] (EWnu) -- [anti fermion] (E),
                (nuWMu) -- [anti fermion, edge label=$\nu$] (eWnu),
                };
                \draw [decoration={brace}, decorate] (Mu.west) -- (e.west) node [pos=0.5, left] {$\mmu$};
                \draw [decoration={brace}, decorate] (mu.east) -- (E.east) node [pos=0.5, right] {$\ammu$};
            \end{feynman}
        \end{tikzpicture}
    }
    \caption{Muonium conversion to antimuonium by the Standard Model weak interaction. This process is strongly suppressed by the tiny neutrino masses.}
    \label{fig:muonium-conversion-in-SM}
\end{figure*}

The process $\mu^+e^- \to \mu^-e^+$ is generated from pure four-fermion effective couplings, and its Lagrangian is the same as that for muonium-to-antimuonium conversion. Neglecting lepton masses, the differential cross section for the process $\mu^+e^- \to \mu^-e^+$ is given by
\begin{equation}
\begin{aligned}
    \frac{\dd\sigma(\mu^+e^-\to\mu^-e^+)}{\dd\Omega} & = \frac{E_\text{CM}^2}{64\pi^2\Lambda^4}\bigg(c_0^2(1+\cos^2\theta) \\
    & + c_1^2\cos\theta + c_2^2(1-\cos^2\theta)
    \bigg)~,
\end{aligned}
\end{equation}
where $E_\text{CM}=\sqrt{s}$ is the center-of-mass energy and $\theta$ is the angle between the incoming electron and the outgoing muon. The coefficients $c_0^2$, $c_1^2$, and $c_2^2$ are defined as follows,
\begin{equation}
\begin{aligned}
&\begin{aligned}
    c_0^2 =4 \bigg( \norm{C^{LL}_V}^2 & + \norm{C^{RR}_V}^2 + 2\norm{C^{LR}_V}^2 \\
    & + \frac{1}{2}\norm{C^{LR}_S}^2 + \frac{1}{2}\norm{C^{RL}_S}^2 \bigg)~,
\end{aligned} \\
    & c_1^2 = 8 \left( \norm{C^{LL}_V}^2 + \norm{C^{RR}_V}^2 - 2\norm{C^{LR}_V}^2 \right)~, \\
    & c_2^2 = 2 \left( \norm{C^{LR}_S}^2 + \norm{C^{RL}_S}^2 \right)~.
\end{aligned}
\end{equation}
The form of the differential cross section is similar to that for $\ell\bar{\ell} \to \ell'\bar{\ell}''$, but without contributions from $Z$ boson or photon couplings~\cite{Altmannshofer:2023tsa}. For the heavy new physics mediators considered here, a higher center-of-mass energy leads to a larger cross section and increased sensitivity to new physics. The new physics scale that can be probed by $\sqrt{s} = 346$~GeV $\mu^+e^-$ collisions with $16^\circ<\theta_\text{lab}<164^\circ$ in the lab frame is estimated as
\begin{equation}\label{eq:muTRISTAN NP scale}
    \Lambda \gtrsim \frac{(0.25~\text{TeV})^{1/2}}{\sigma_\text{up}(\mu^+e^-\to\mu^-e^+)^{1/4}}~.
\end{equation}
Assuming an integrated luminosity of 1~ab$^{-1}$~\cite{Hamada:2022uyn}, the $\mu^+e^-$ collider could probe new physics at the scale of 10--100~TeV. Operator-dependent new physics scales for experiments are shown in \cref{fig:MACE NP scale}, where the $\mu$TRISTAN result is projected from the background-free upper limit with 90\% confidence level (C.L.).  We observe from this result that MACE, as a low-energy experiment, could probe a new physics scale comparable to or even higher than that of a future muon collider. This finding highlights the significance of low-energy experiments in the search for TeV scale new physics.

\section{$\mmu$-to-$\ammu$ conversion signals and backgrounds}

Investigating and understanding signals and backgrounds is essential in experiments searching for new physics beyond the Standard Model. Backgrounds could affect the MACE sensitivity and a deep understanding of backgrounds allows for the optimization of experimental designs, data analysis and signal selection methods, which can improve the MACE sensitivity to new physics.

\begin{figure*}[t]
    \nolinenumbers
    \centering
    \subfloat[$\mmu$-to-$\ammu$ conversion signal.]{
        \includegraphics[width=0.37\textwidth]{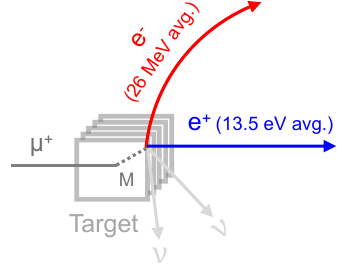}
    }\\
    \subfloat[\label{fig:mace-mu2eeevv-background-topology}Background: SM allow rare decay $\mu^+\to e^+e^-e^+\nu_e\bar\nu_\mu$.]{
        \includegraphics[width=0.27\textwidth]{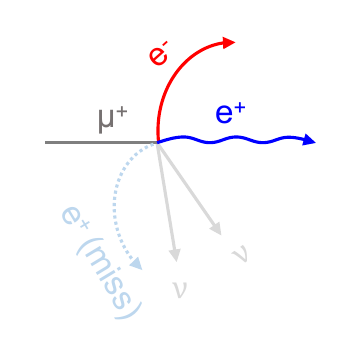}
    }\qquad
    \subfloat[\label{fig:muonium-fsi-background-topology}Background: muonium decay with Bhabha scattering.]{
        \includegraphics[width=0.27\textwidth]{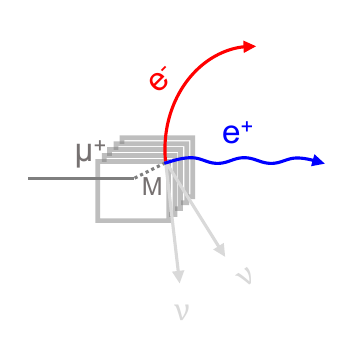}
    }\qquad
    \subfloat[Accidental backgrounds.]{
        \includegraphics[width=0.27\textwidth]{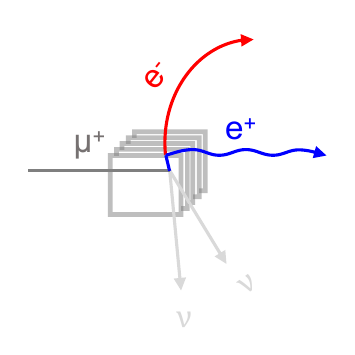}
    }
    \caption{\label{fig:mace-signal-and-background-topology} Event topologies for signals and backgrounds.}
\end{figure*}

\begin{figure*}[t]
    \nolinenumbers
    \centering
    \subfloat[$e^-$ energy spectrum of from $\mu^+$ IC decay.]{
        \includegraphics[width=0.4\textwidth]{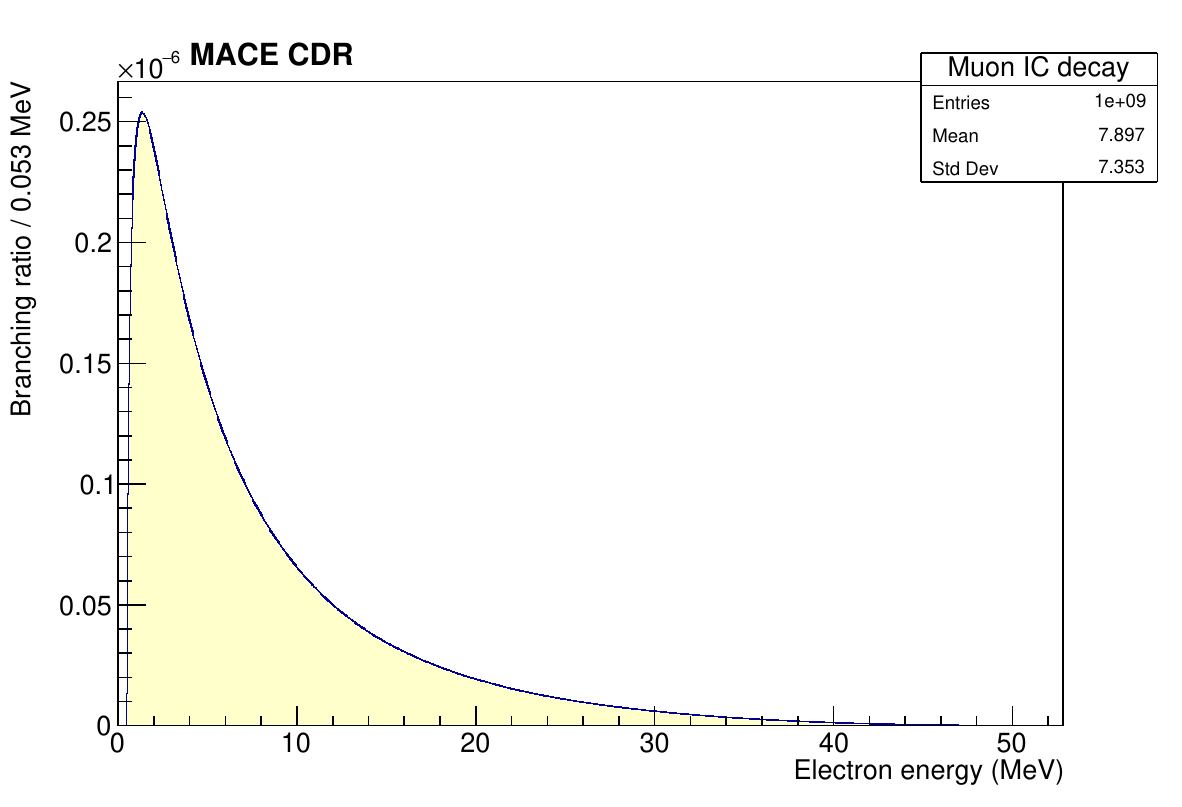}
    }\qquad
    \subfloat[$e^+$ energy spectrum from $\mu^+$ IC decay]{
        \includegraphics[width=0.4\textwidth]{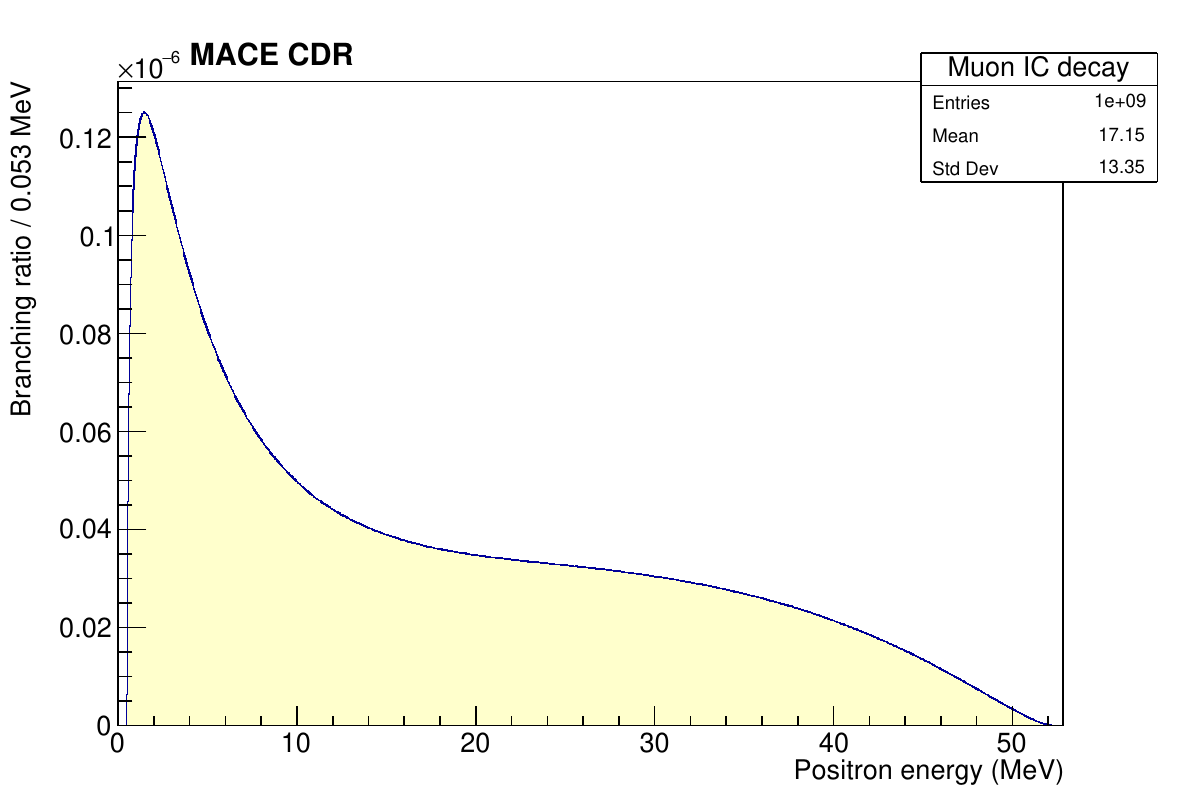}
    }\\
    \subfloat[Joint energy spectrum of $e^+$/$e^-$ from $\mu^+$  IC decay.]{
        \includegraphics[width=0.4\textwidth]{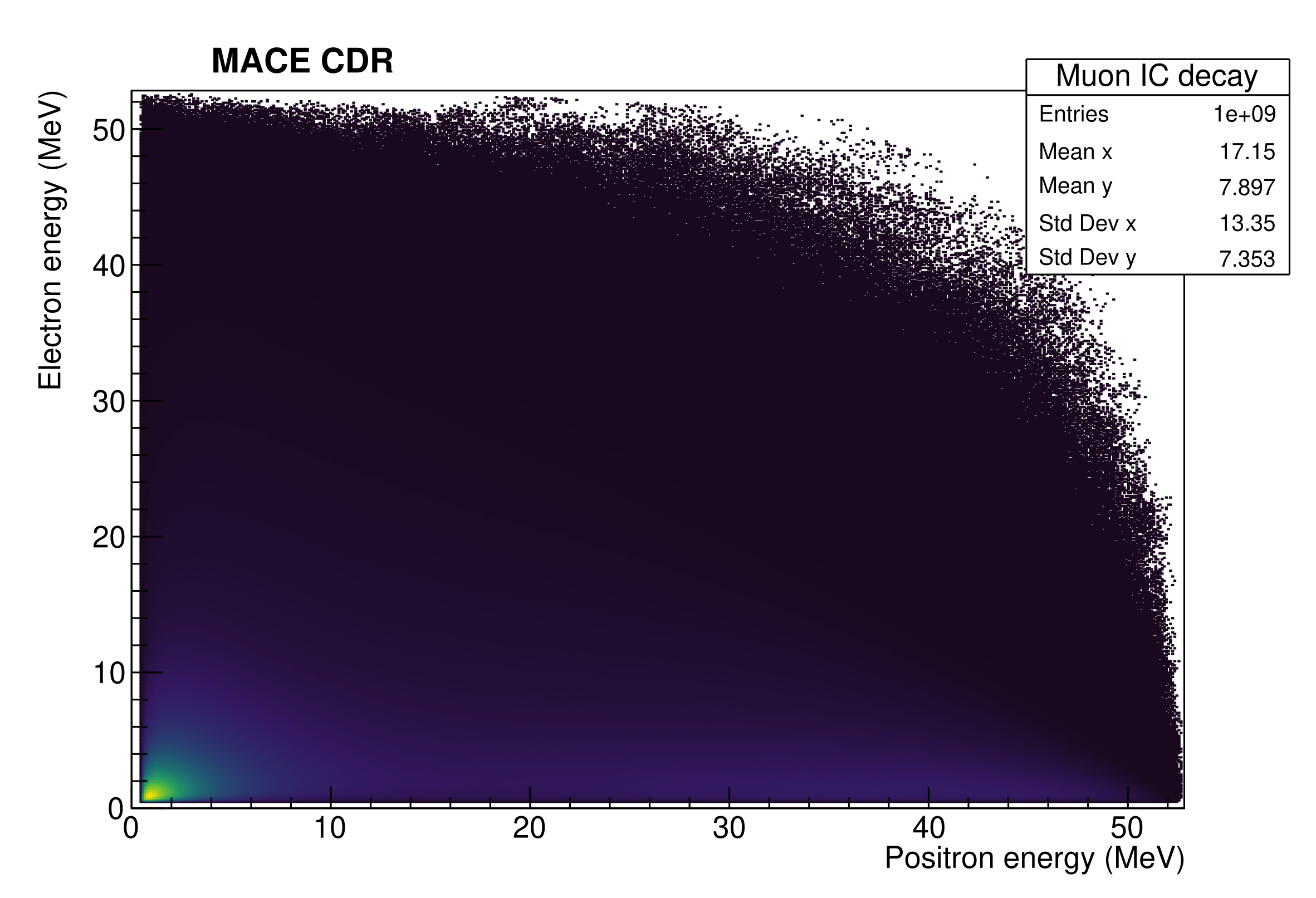}
    }\qquad
    \subfloat[\label{fig:mu2eeevv-spectra-d}Energy spectrum of $e^+$ with the lowest energy from $\mu^+$ IC decay.]{
        \includegraphics[width=0.4\textwidth]{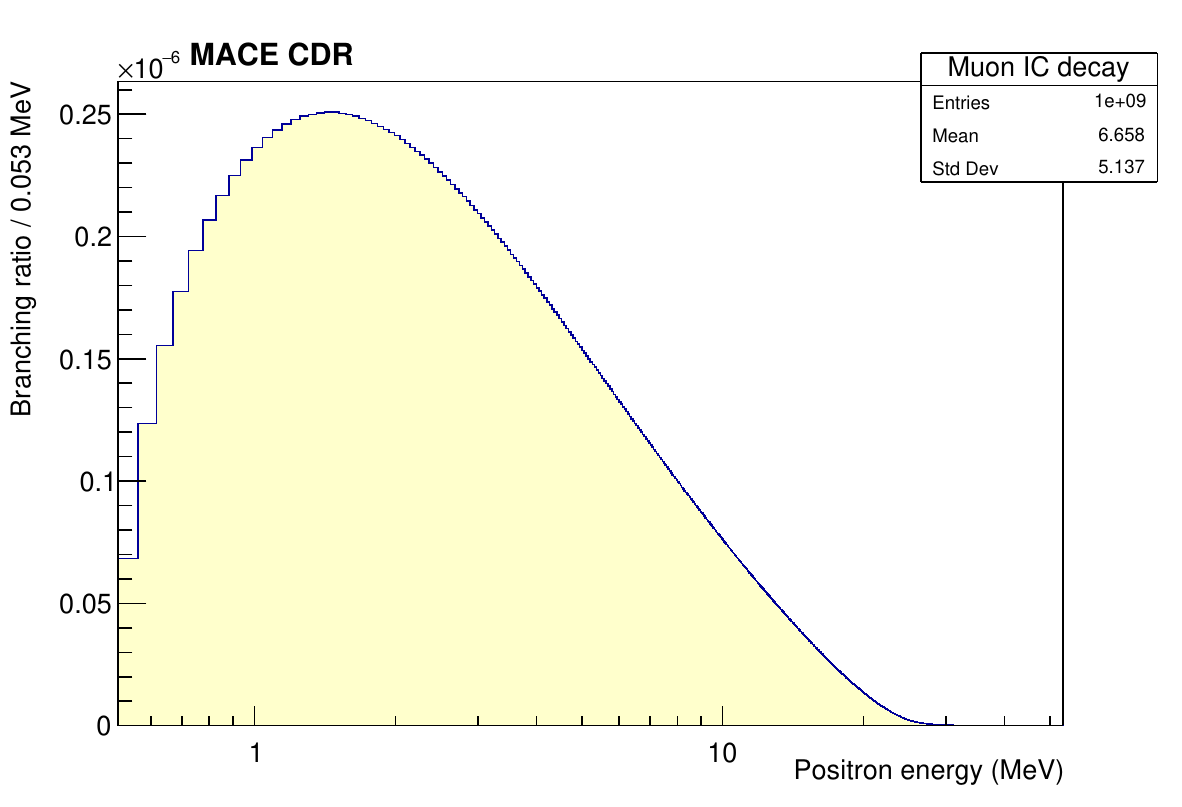}
    }
    \caption{\label{fig:mu2eeevv-spectra}Spectra of muon internal conversion (IC) decay.}
\end{figure*}

\begin{figure}[htbp]
    \nolinenumbers
    \centering
    \includegraphics[width=\columnwidth]{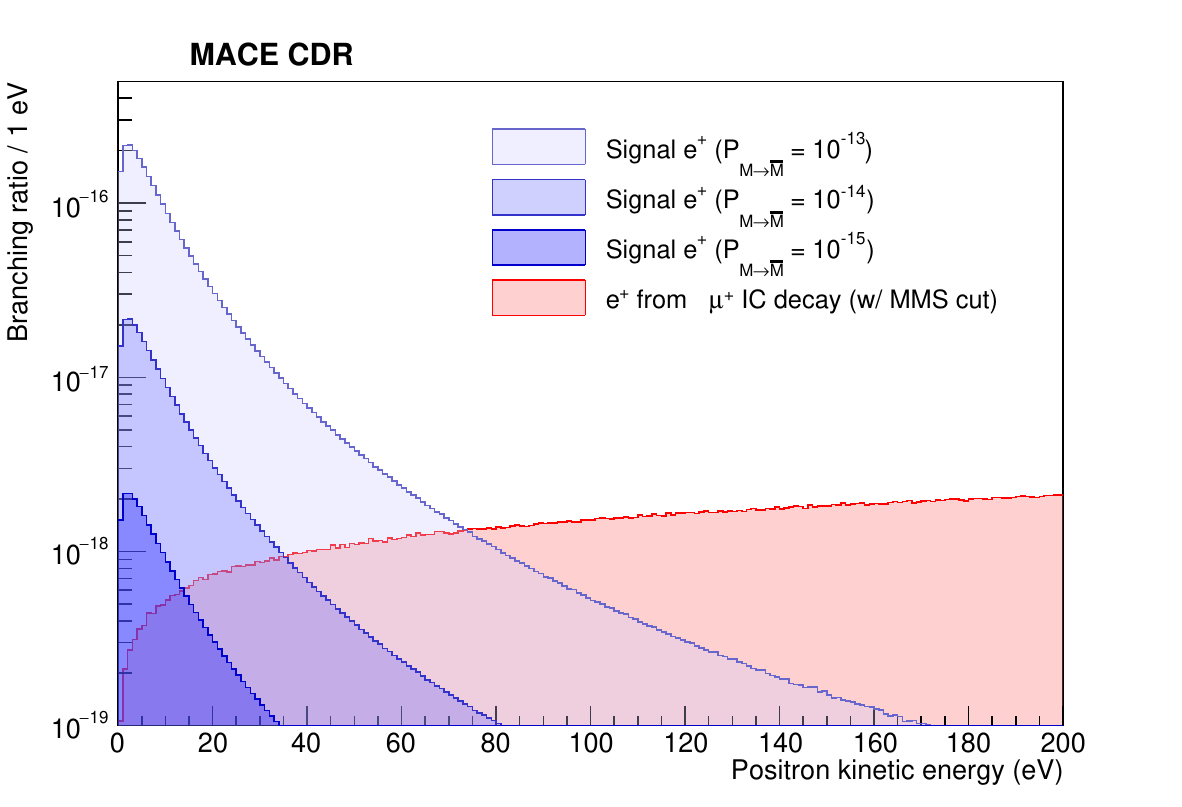}
    \caption{\label{fig:mu2eeevv-bkg-and-signal}$\mmu$-to-$\ammu$ signal $e^+$ kinetic energy spectrum of different conversion probability (blue) and background $e^+$ kinetic energy spectrum from $\mu^+\to e^+e^-e^+\nu_e\bar{\nu}_\mu$ decay process (red).}
\end{figure}

\begin{figure*}[t]
    \nolinenumbers
    \centering
    \subfloat[Transverse momentum spectrum of $e^-$ from $\mu^+$ IC decay (with detector geometric acceptance).]{
        \includegraphics[width=0.4\textwidth]{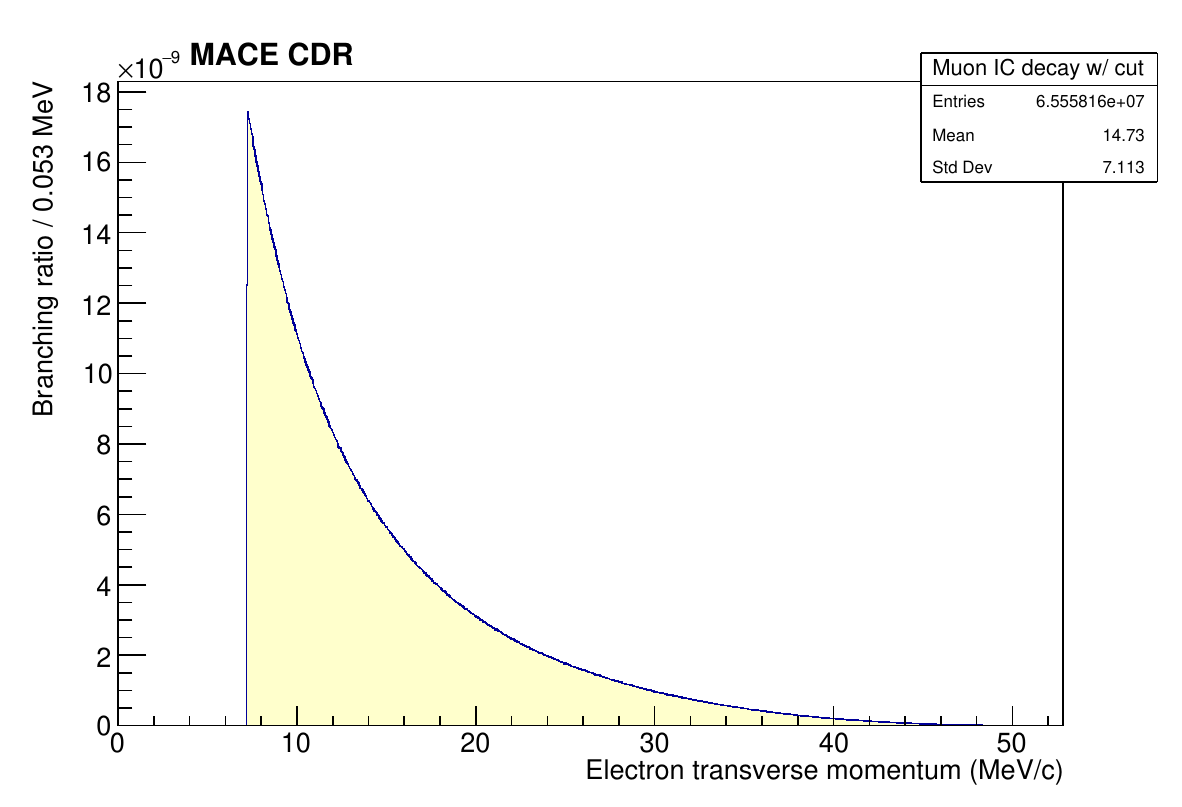}
    }\qquad
    \subfloat[$e^+$ energy spectrum from $\mu^+$ IC decay (with detector geometric acceptance).]{
        \includegraphics[width=0.4\textwidth]{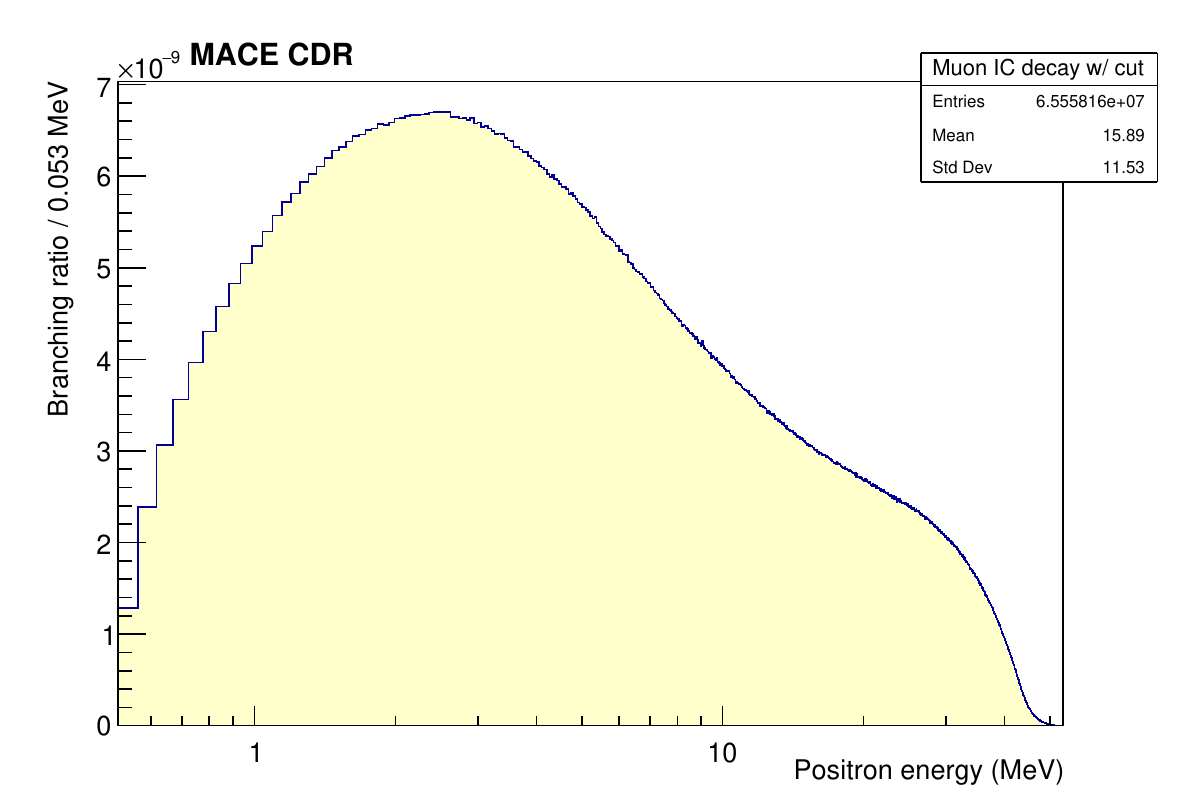}
    }\\
    \subfloat[Joint spectrum of $e^+$ energy and $e^-$ transverse momentum from $\mu^+$ IC decay.]{
        \includegraphics[width=0.4\textwidth]{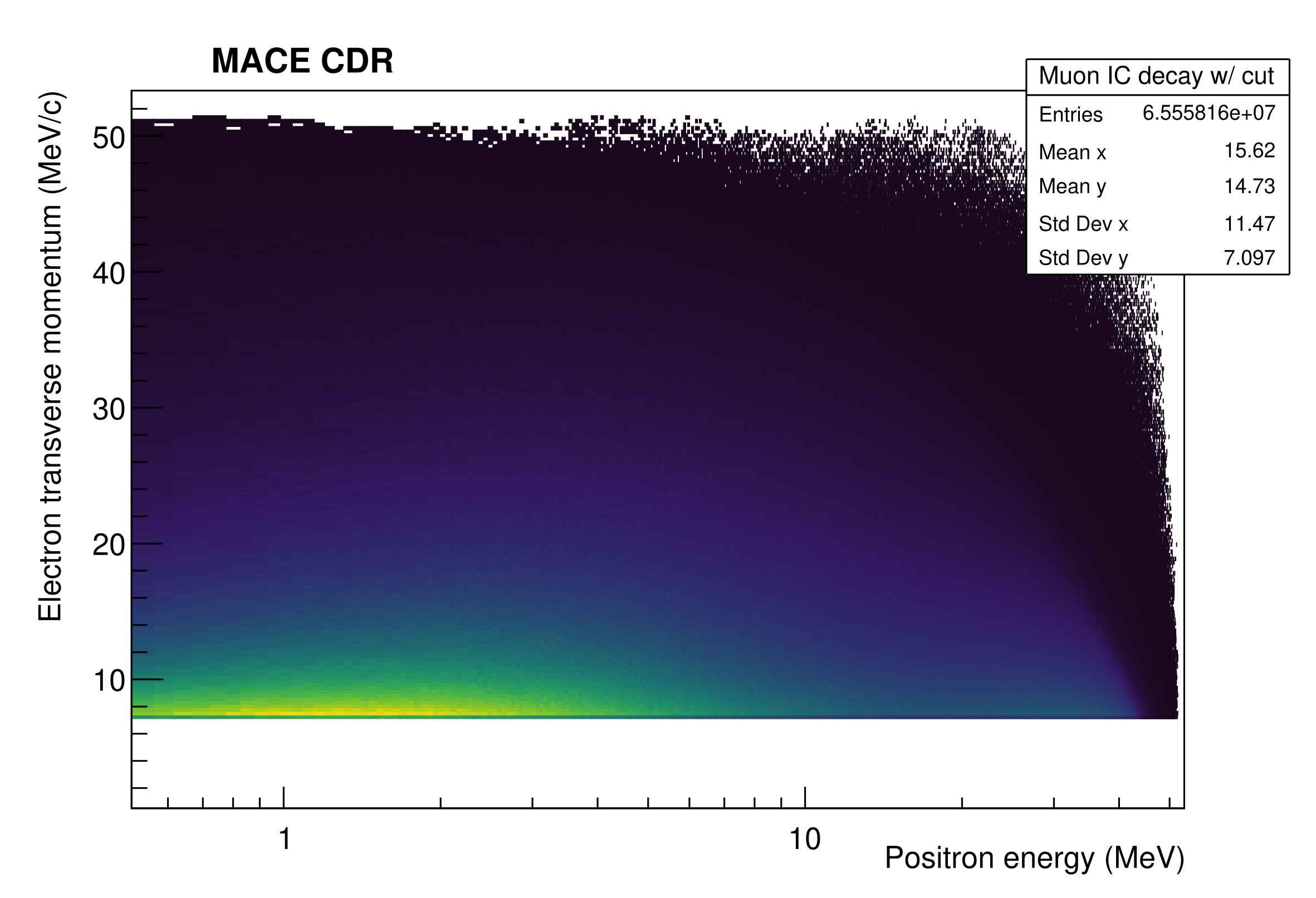}
    }\qquad
    \subfloat[Joint spectrum of $e^+$ energy and $e^-$ energy from $\mu^+$ IC decay.]{
        \includegraphics[width=0.4\textwidth]{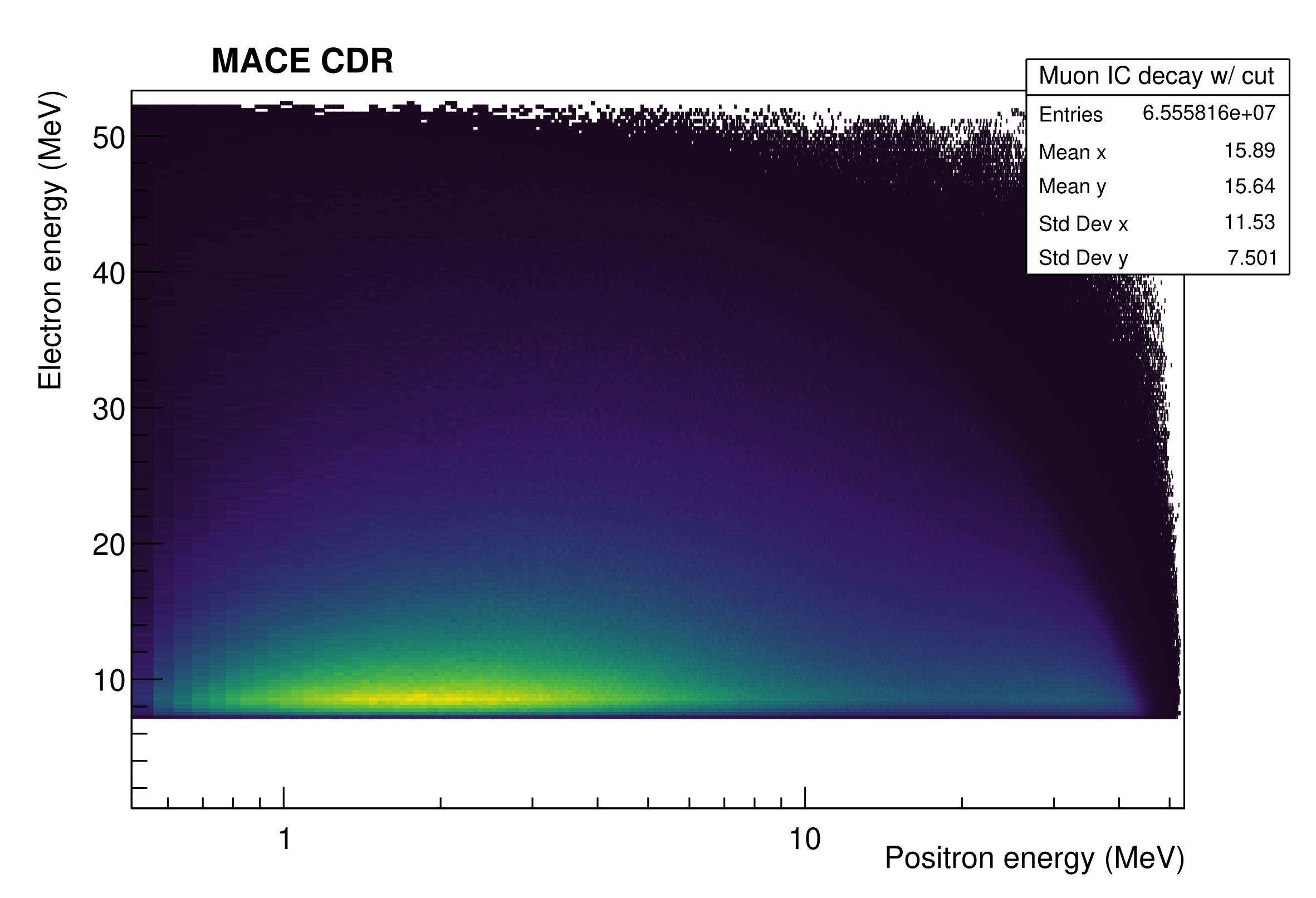}
    }
    \caption{\label{fig:mu2eeevv-spectra-with-cut}Spectra of muon internal conversion (IC) decay with MACE Michel electron magnetic spectrometer (MMS) geometric acceptance.}
\end{figure*}

\subsection{Signal event signature}\label{sec:signal-event-signature}

MACE aims to search for the muonium-to-antimuonium conversion process by detecting antimuonium decay events in a vacuum. The detection scheme is to identify the characteristic decay products of antimuonium. Upon the decay of antimuonium, a fast electron, a slow positron, and two invisible neutrinos are produced. The fast electron originates from the $\mu^-$ decay with a kinetic energy of a few tens of MeV, while the slow positron comes from the atomic shell with a mean kinetic energy of 13.5~eV (considering $1s$ antimuonium, as shown in \cref{fig:muonium-decay-spectrum}). By detecting these two opposite decay products, a few antimuonium decay events can be identified from the vast number of muon and muonium decays.

The signal signature in detection is clear: a signal event involves an energetic electron and a slow positron. The energy spectra of signals are shown in \cref{fig:muonium-decay-spectrum}. The MACE design employs a silica aerogel target that produces muonium by injecting a high-intensity surface muon beam. The target is designed to be porous and perforated to maximize the diffusion of muonium into a vacuum. Following the spontaneous conversion of muonium to antimuonium in a vacuum, the antimuonium decays into an energetic electron and a slow positron. The energetic electron may traverse the charged particle tracker with a sufficient energy threshold, while the positron is guided by the positron transport system towards an position-sensitive detector. Subsequently, the positron may annihilate on the detector, producing a pair of 511~keV gamma rays.

\subsection{Backgrounds}\label{sec:mace-background}

With an signal event involves a energetic electron and a low-energy positron, any events with such signature might be misidentified as signals, making them potential backgrounds. To begin with, we must identify the possible sources of backgrounds. Backgrounds can be grouped into two categories: physical backgrounds and accidental backgrounds. Physical backgrounds may arise from the decay of muons or muonium atoms, while accidental backgrounds could result from two tracks coincident from distinct sources. Our following discussions will focus on these two types of backgrounds.

The primary focus in MACE is on muonium, raising the important question of whether muonium itself could introduce physical backgrounds. In a minimal extension of the Standard Model that simply extending right-handed neutrinos, the $\mmu$-to-$\ammu$ conversion probability will experience GIM-like suppression due to the tiny neutrino masses, similar to cLFV decays such as $\mu\to e\gamma$~\cite{Cheng:1977nv,Lee:1977tib} or $\mu\to eee$~\cite{Hernandez-Tome:2018fbq,Blackstone:2019njl}. As a result, the conversion probability is expected to be considerably low. Consequently, there are no irreducible physical backgrounds, and any background contributions are expected to originate from muon or muonium decay.

MACE will operate at an intensive muon beam, therefore any muon-induced backgrounds should not be overseen. The major decay mode $\mu^+\to e^+\nu_e\bar{\nu_\mu}$ or the radiative decay mode $\mu^+\to e^+\nu_e\bar{\nu_\mu}\gamma$ do not introduce extra electrons, so they are generally safe. However, the SM-allowed internal conversion (IC) decay $\mu^+\to e^+e^-e^+\nu_e\bar\nu_\mu$ introduces an electron in its final state, and its branching ratio is $3.4\times 10^{-5}$ (with transverse momentum cut $p_\text{T}>17~\text{MeV}/c$)~\cite{SINDRUM:1985vbg}. When the positron happens to have a pretty low momentum, and an electron is detected while another is not, the internal conversion decay $\mu^+\to e^+e^-e^+\nu_e\bar\nu_\mu$ can fake signals. The background event topology is shown in \cref{fig:mace-mu2eeevv-background-topology}, and the energy spectra of decay products is shown in \cref{fig:mu2eeevv-spectra}. In fact, the internal conversion decay of muon was identified as one of the major sources of background in the MACS experiment in PSI~\cite{Abela:1996dm,Willmann:1998gd}. To reduce this type of background, we can utilize the difference in kinematics between the antimuonium decay and the muon IC decay.

As shown in the energy spectrum of the lowest positron energy (\cref{fig:mu2eeevv-spectra-d}), one can observe that the branching fraction decreases significantly at low energies. The signal positron typically has a kinetic energy of only a few tens of electronvolts, providing a strong discriminator between the signal and the muon internal conversion decay background, as illustrated in \cref{fig:mu2eeevv-bkg-and-signal}. Selecting low-energy positrons and constraining their kinetic energy to be very low in the MACE detector system can effectively suppress the internal conversion decay background and improve the signal-to-background ratio. This approach calls for the design of a transverse-momentum-selective positron transport solenoid system, as detailed in \cref{sec:mace-magnet-and-solenoid}. On the other hand, the longitudinal momentum can be selected by the time-of-flight of the positron. Furthermore, the transverse momentum of the energetic electron from the same decay event is concentrated in the low momentum region, as shown in \cref{fig:mu2eeevv-spectra-with-cut}), in contrast to the signal electron spectrum shown in \cref{fig:antimuounium-electron-spectrum}. Therefore, selecting the transverse momentum of the energetic electron to be sufficiently also aids in background suppression. With all these signal purify strategies, the phase space of backgrounds can be strongly constrained while keeping the majority of signals, eventually improving the signal-to-background ratio.

There is also a possibility that muonium decay could contribute to physical background in MACE. Muonium shares most of decay properties with the muon, except for the presence of an atomic electron. When a Michel positron from muon decay interacts with the atomic electron and exchanges significant momentum, the muonium decay products behave as a low-energy positron and a high-energy electron, potentially leading to a misidentification as an antimuonium decay. The event topology is shown in \cref{fig:muonium-fsi-background-topology}. This type of background has been discussed in a previous theoretical study~\cite{Feinberg:1961zza}, and the branching ratio with $E_{e^-}>10$~MeV is estimated to be approximately $10^{-10}$. In this scenario, the higher electron energy corresponds to the lower positron energy, indicating a higher momentum exchange and consequently a significantly reduced cross section when the final positron energy is sufficiently low. The background level is expected to be suppressed through the optimized detector design and a similar event selection scheme as discussed above.

Accidental coincidences involving an energetic electron, potentially produced by Bhabha scattering, and a positron from muon or muonium decay, represent another potential source of backgrounds. If a Bhabha scattering event involving a Michel positron and an electron in the porous target material occurs near the material boundary, and the scattered positron is emitted from the material surface with a low energy, it could fake a signal event. The process is similar to the Bhabha scattering of muonium decay final states, and similar strategies can be employed to suppress this type of background.

Furthermore, a selection in an event time window can improve the signal-to-background ratio~\cite{Willmann:2021boq}. All of these three kinds of background source from muon or muonium decay, characterized by an exponentially decreasing time distribution proportional to $\exp(-t/\tau_\mu)$, where $\tau_\mu$ is the muon lifetime. Notably, the muonium-to-antimuonium conversion probability varies with time and is proportional to $t^2\exp(-t/\tau_\mu)$~\cite{Fukuyama:2021iyw,Fukuyama:2023drl}. This time-dependent conversion increases the ratio of antimuonium decay to muonium or muon decay by $t^2$, thereby contributing to the signal-to-background ratio with the accumulated time. The suppression of background can be achieved by selecting a late event time window between the arrival of two beam pulses in a high-repetition-rate muon beam if necessary. A 40--50-kHz muon beam would be potentially available at China initiative Accelerator Driven System (CiADS) in Huizhou (see \cref{sec:beamline}) or Shanghai HIgh repetitioN rate XFEL and Extreme light facility (SHINE)~\cite{Liu:2025ejy}.

In a concise overview, the following potential background contributions should be taken into account in MACE:
\begin{itemize}
    \item \textbf{Internal conversion decay of muon}: The decay $\mu^+\to e^+e^-e^+\nu_e\bar\nu_\mu$ could introduce an energetic electron and a slow positron in its final states simultaneously. If one positron falls within the signal region while the other is undetected, and the electron momentum is sufficiently high to pass the selection, the decay could fake a signal event.
    \item \textbf{Bhabha scattering of muonium decay final states}: Muonium decay final states involve a slow electron from the atomic shell and an energetic positron. Bhabha scattering between these particles can transfer momentum, potentially leading to a background event if the momentum transfer is significant.
    \item \textbf{Accidental coincidences of Michel positron scattered with material electron}: Scattering of positrons generated from muon decay with electrons in the target material could generate an energetic electron and a low-energy positron.
\end{itemize}
The detector design should be optimized to enhance the ability to identify between signal and background events. This includes improving the vertex resolution of the magnetic spectrometer, enhancing the resolution of low-energy positron time-of-flight measurements, and improving the energy resolution in the electromagnetic calorimeter to reduce accidental and physical backgrounds. Utilizing a state-of-the-art positron transport system to select the momentum of transported particles can yield an optimal signal-to-background ratio. Additionally, connecting to a high-repetition-rate muon source can further enhance the signal-to-background ratio. In the subsequent sections, we will discuss the experiment design and necessary specifications.

\section{Beamline}\label{sec:beamline}
The proposal for a high-intensity muon source, driven by the CiADS linac, has been under consideration for several years. From the perspective of beam power, such a muon source has the potential to be one of the state-of-the-art facilities in the world. To meet the challenges posed by the unprecedented beam power, it is essential to explore novel target designs. Solenoid-based capture and transport are also crucial for achieving a higher muon rate. The research and development of the CiADS muon source are currently on the way.

\subsection{Accelerator and proton beam}

\begin{figure}[t]
    \nolinenumbers
    \centering
    \includegraphics[width=\columnwidth]{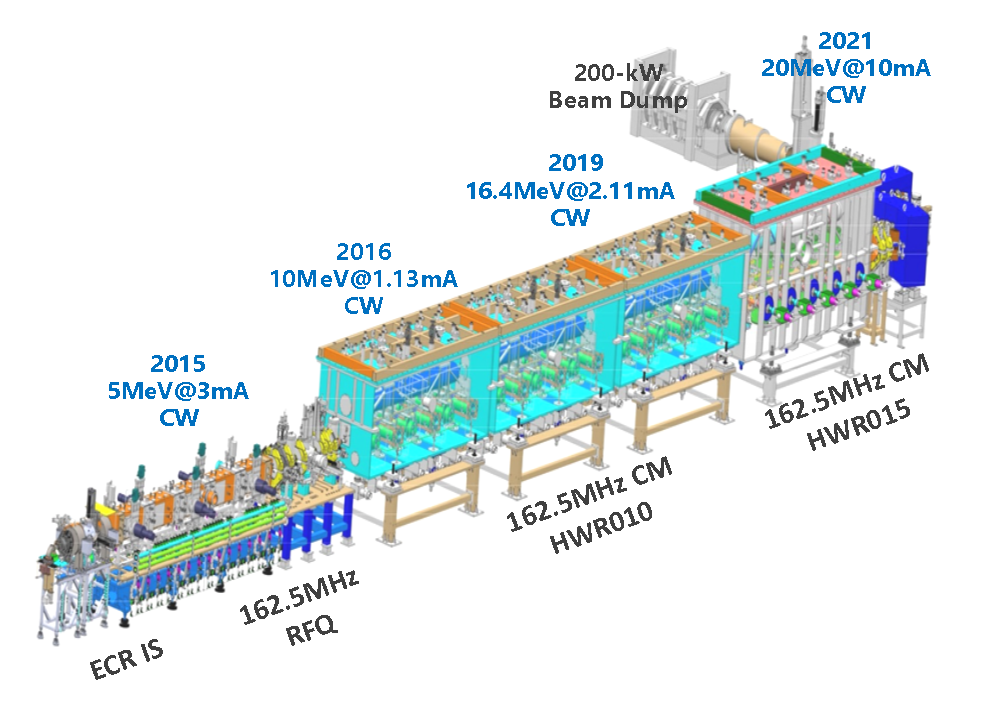}
    \caption{Schematic view of the development and commissioning history of CAFe~\cite{Cai:2023caf}.}
    \label{FIG:CAFe}
\end{figure}

The development of the superconducting linac for ADS in China began in 2011. The prototype front-end linac (CAFe), as shown in \cref{FIG:CAFe}, was developed in stages and includes an Electron Cyclotron Resonance (ECR) ion source, a Radio Frequency Quadrupole (RFQ), a superconducting acceleration section, and a 200-kW beam dump. The commissioning of the hundred kW beam started in 2018 and reached a milestone in early 2021 by producing a 20-MeV proton beam with an average current of 10 mA, demonstrating the feasibility of a superconducting linac in Continuous-Wave (CW) mode.

\begin{figure}[t]
    \nolinenumbers
    \centering
    \includegraphics[width=\columnwidth]{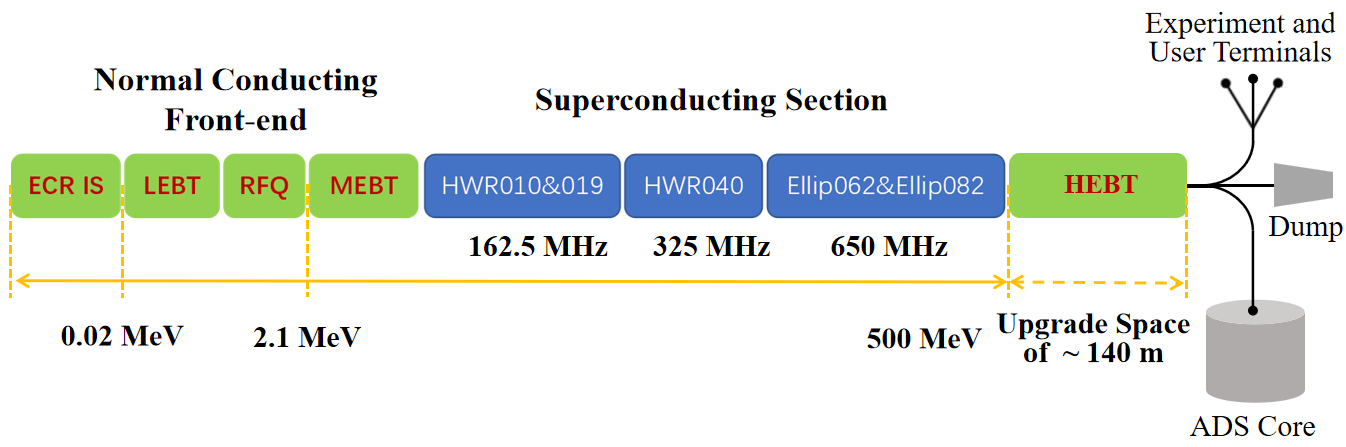}
    \caption{Schematic diagram of the CiADS linac~\cite{Cai:2023caf}.}
    \label{FIG:CiADS Linac}
\end{figure}

The schematic diagram of the CiADS linac, illustrated in \cref{FIG:CiADS Linac}, consists of a normal conducting front-end, a superconducting acceleration section, and several High Energy Beam Transport (HEBT) lines. The civil construction of the linac is complete, while work on the experimental terminals is ongoing. The front-end, integrated in December 2022, includes an ECR ion source, a Low Energy Beam Transport (LEBT) with a fast chopper for beam pulse structuring and machine protection, an RFQ, and a Medium Energy Beam Transport (MEBT). The proton beam from the front-end is a CW beam with a current of 5.2 mA and an energy of 2.18 MeV. The superconducting section accelerates the beam from 2.1 MeV to 500 MeV. With the beam current of 50 $\mu$A on target by 2025, the beam power will be 25 kW at that time, and power ramping to 250 kW and 2.5 MW is expected by 2027 and 2029, respectively. The superconducting section houses three types of Half Wave superconducting Resonators (HWR010, HWR019, and HWR040) and two types of elliptical cavities (Ellip062 and Ellip082) in 32 cryomodules. With the muon source proposal, the superconducting linac can potentially be upgraded to 600 MeV.

\subsection{Muon production and transport}

\subsubsection{Muon production target}

\begin{figure}[htbp]
    \nolinenumbers
    \centering
    \includegraphics[width=\columnwidth]{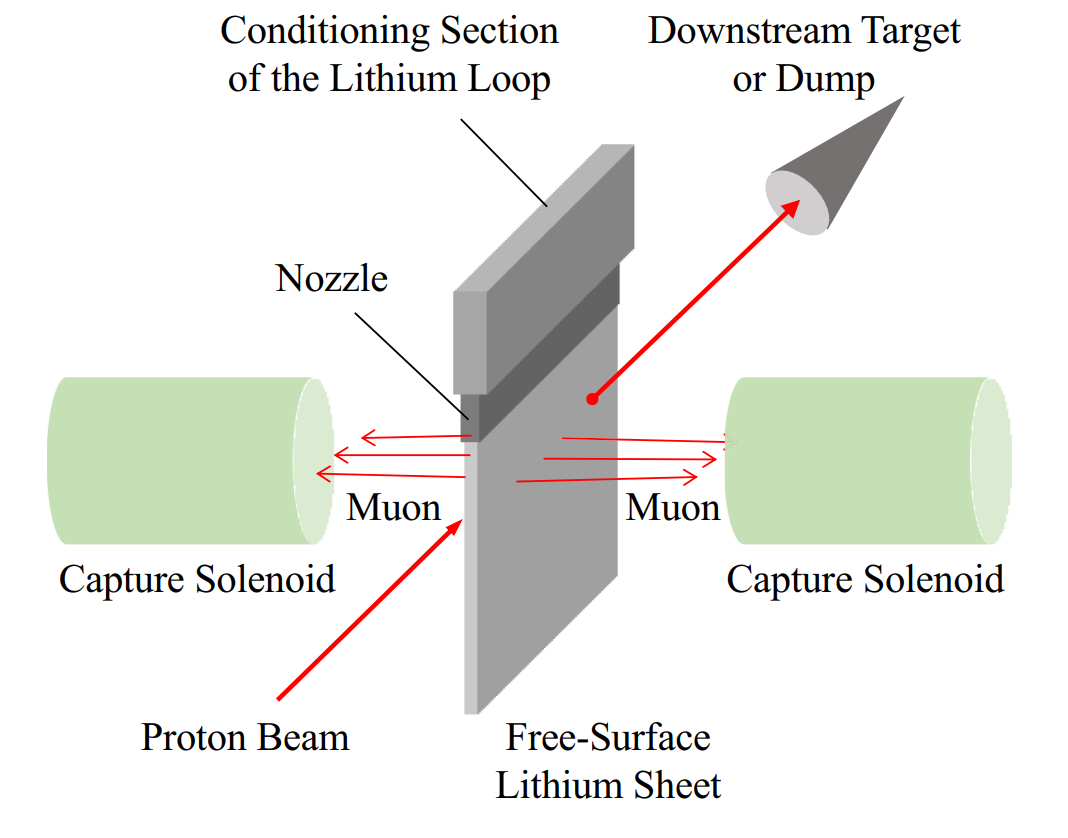}
    \caption{Schematic diagram of the free-surface liquid lithium target~\cite{Cai:2023caf}.}
    \label{FIG:liquid lithium target}
\end{figure}

The production target for a high-intensity muon source is extremely challenging as it requires addressing  high-heat densities and a harsh irradiation environment. Arising from the two-body decay of positive pion stopped close to the surface of the production target, the surface muon escapes from the target with a momentum ranging from 0 to 29.8~MeV/$c$. A new target design, based on a free-surface and sheet-shaped liquid lithium target, is shown in \cref{FIG:liquid lithium target}. Liquid lithium passes through a lithium circuit and forms a sheet jet from the narrow nozzle. The proton beam is collimated to hit the lithium jet at a tiny angle, and the surface muons produced inside lithium then escaping from both sides of the jet are captured by the solenoids.

\begin{figure}[htbp]
    \nolinenumbers
    \centering
    \includegraphics[width=\columnwidth]{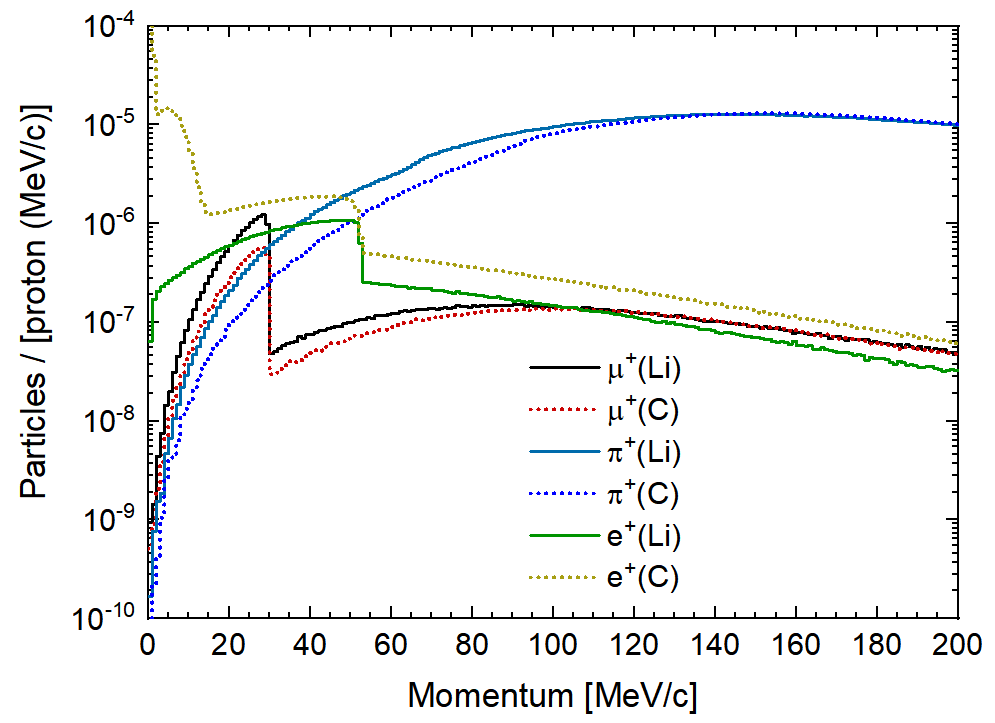}
    \caption{Momentum spectra of \(\mu^+\), \(\pi^+\) and  \(e^+\) recorded by the virtual detector beside the target~\cite{Cai:2023caf}.}
    \label{FIG:Momentum spectra}
\end{figure}

Due to its low melting point, extremely low saturated vapor pressure, high heat capacity, and good compatibility with structural materials, liquid lithium has been used as a neutron production target, a radionuclide production target, and an ion beam charge stripper. Maintaining the stability of free-surface liquid lithium flakes in the jet is a challenging aspect of the design. Extensive research and development works have been carried out to investigate the feasibility of producing the free-surface liquid lithium films or sheets.

A liquid lithium target is particularly well-suited for the high-intensity muon sources driven by the CiADS linac. First of all, lithium has a low atomic number $Z$, and the research performed at PSI has shown that the surface muon production efficiency is roughly proportional to \(Z^{-\frac{2}{3}}\). Second, the lithium target produces more low-energy \(\pi^+\)'s, resulting in more surface muons than graphite target, with a significantly lower rate of positrons. \cref{FIG:Momentum spectra} shows the momentum spectra of the side-leaking \(\mu^+\), \(\pi^+\) and \(e^+\) from lithium and graphite targets recorded by a detector next to the target.

\subsubsection{Muon beamline conceptual design}\label{sec:muon-beamline-design}

\begin{figure}[htbp]
    \nolinenumbers
    \centering
    \includegraphics[width=\columnwidth]{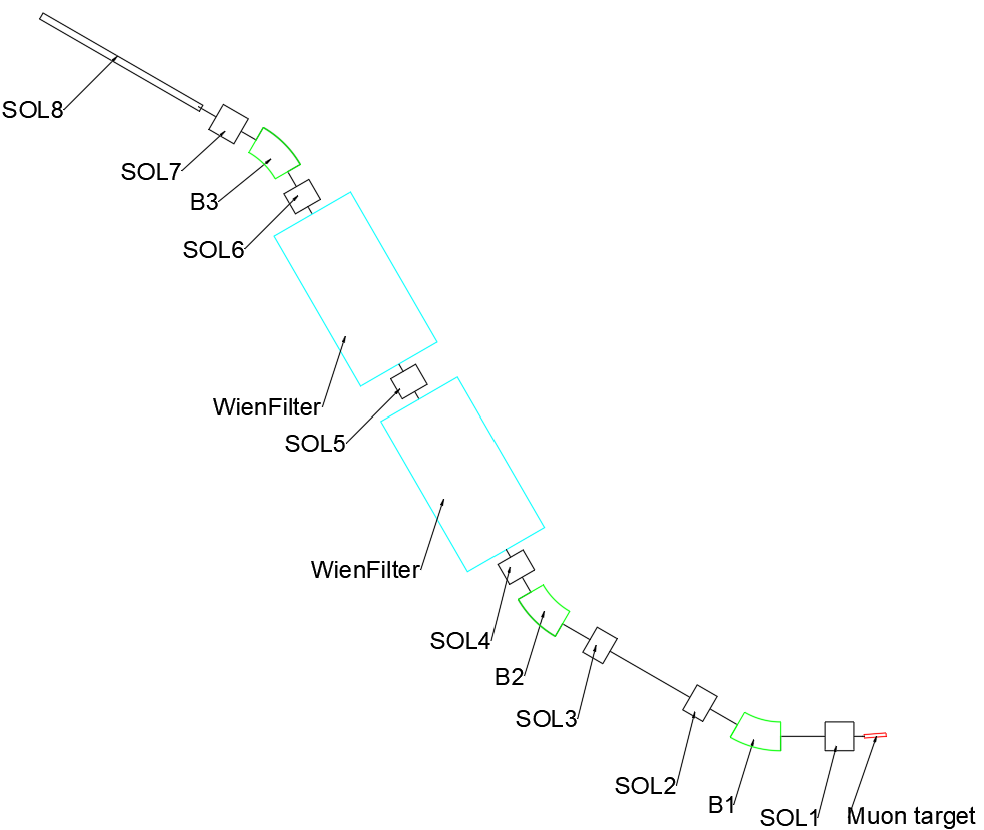}
    \caption{Schematic diagram of the beamline, featuring dipoles, wien filters and solenoids,
    denoted with the labels.}
    \label{FIG:Beamline Schematic}
\end{figure}

Via the interaction of the high-intensity proton beam with the muon target, the generated muons are of large transverse emittance and a significant longitudinal energy spread. To collect the large-emittance muon beam, a solenoid-based beamline concept has been proposed, as depicted in \cref{FIG:Beamline Schematic}.

\begin{figure}[htbp]
    \nolinenumbers
    \centering
    \includegraphics[width=0.5\columnwidth]{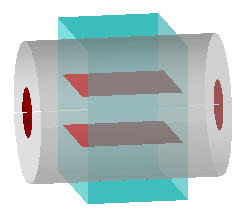}
    \caption{Schematic diagram of a Wien filter.}
    \label{FIG:Wien filter Schematic}
\end{figure}

\begin{figure*}[t]
    \nolinenumbers
    \centering
    \subfloat[Particle distribution at the entrance]{\includegraphics[width=0.33\textwidth]{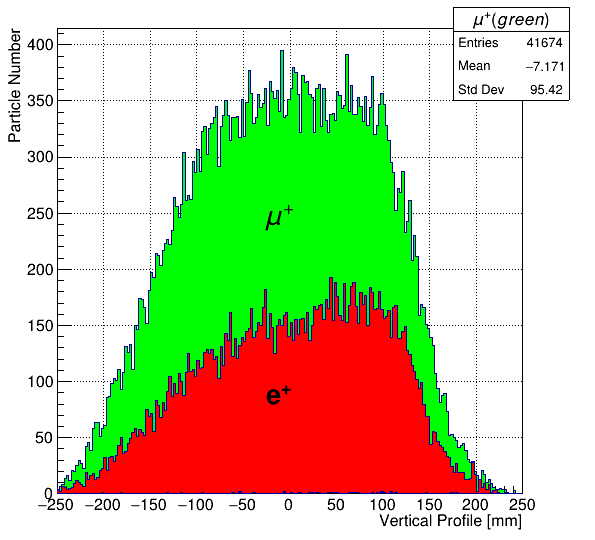}}
    \subfloat[Particle distribution at the exit of a single long Wien filter]{\includegraphics[width=0.33\textwidth]{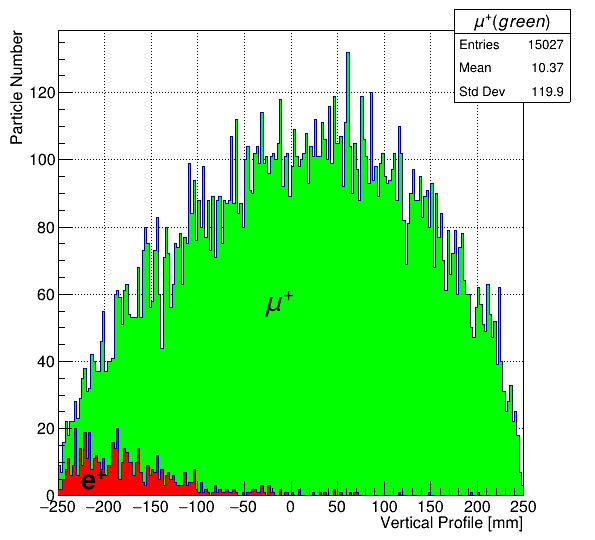}}
    \subfloat[Particle distribution at the exit of the second of two short Wien filter]{\includegraphics[width=0.33\textwidth]{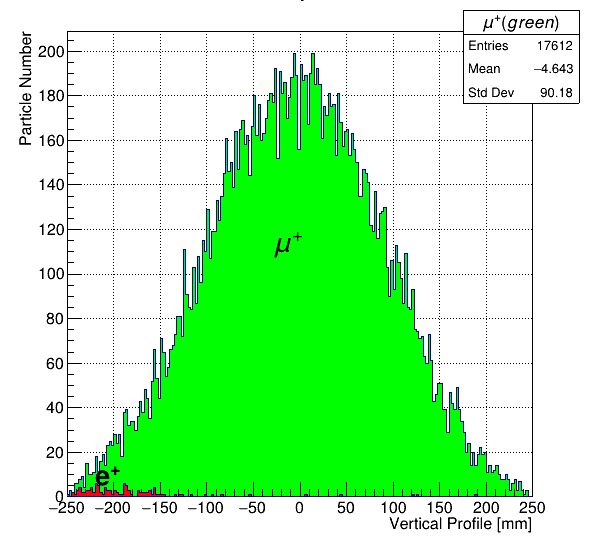}}
    \caption{Comparisons of the positron removal effects using one long Wien filter and two short Wien filters.}
    \label{FIG:Comparison of positron removal effects}
\end{figure*}

The surface muon rate from the target is expected to reach up to $5\times 10^{11}~\mu^+$/s with a 5~mA proton beam. The closer the capture solenoid is positioned to the muon target, the higher the transmission efficiency. However, due to the spatial constraints, the capture solenoid is positioned 200 mm away from the muon target. Drawing on design experience from PSI, we opted for a shorter solenoid with a larger aperture to facilitate particle transport and enhance transmission efficiency. The beamline employs a solenoid with length of 400~mm and diameter of 500~mm. Considering the space utilization and the shielding effectiveness, we set three dipoles to rotate the beam clockwise, clockwise, and counterclockwise.

During the muon collection in the target area and the subsequent beamline transport, a significant number of positrons are generated, which can affect the accuracy of subsequent experimental results. The primary mechanisms for positron production include the decay of $\pi^{0}$ mesons into two photons, leading to the positron-electron pair production, as well as the muon decay. Given that the decay length of muons is 170~m, the number of positrons produced during transport is relatively low, thus with the main source being the $\pi^{0}$ mesons in the target. \cref{FIG:Wien filter Schematic} shows the schematic diagram of Wien filter modeling with G4beamline. A long Wien filter is typically employed to remove positrons from low-momentum muon beams. \cref{FIG:Comparison of positron removal effects} shows the comparisons of positron removal effects using a long Wien filter or two short Wien filters. Simulations using G4beamline indicate that placing two short Wien filters can effectively eliminate positrons and other charged particles while maintaining the transmission efficiency $\mu^+$.

\begin{figure}[htbp]
    \nolinenumbers
    \centering
    \includegraphics[width=\columnwidth]{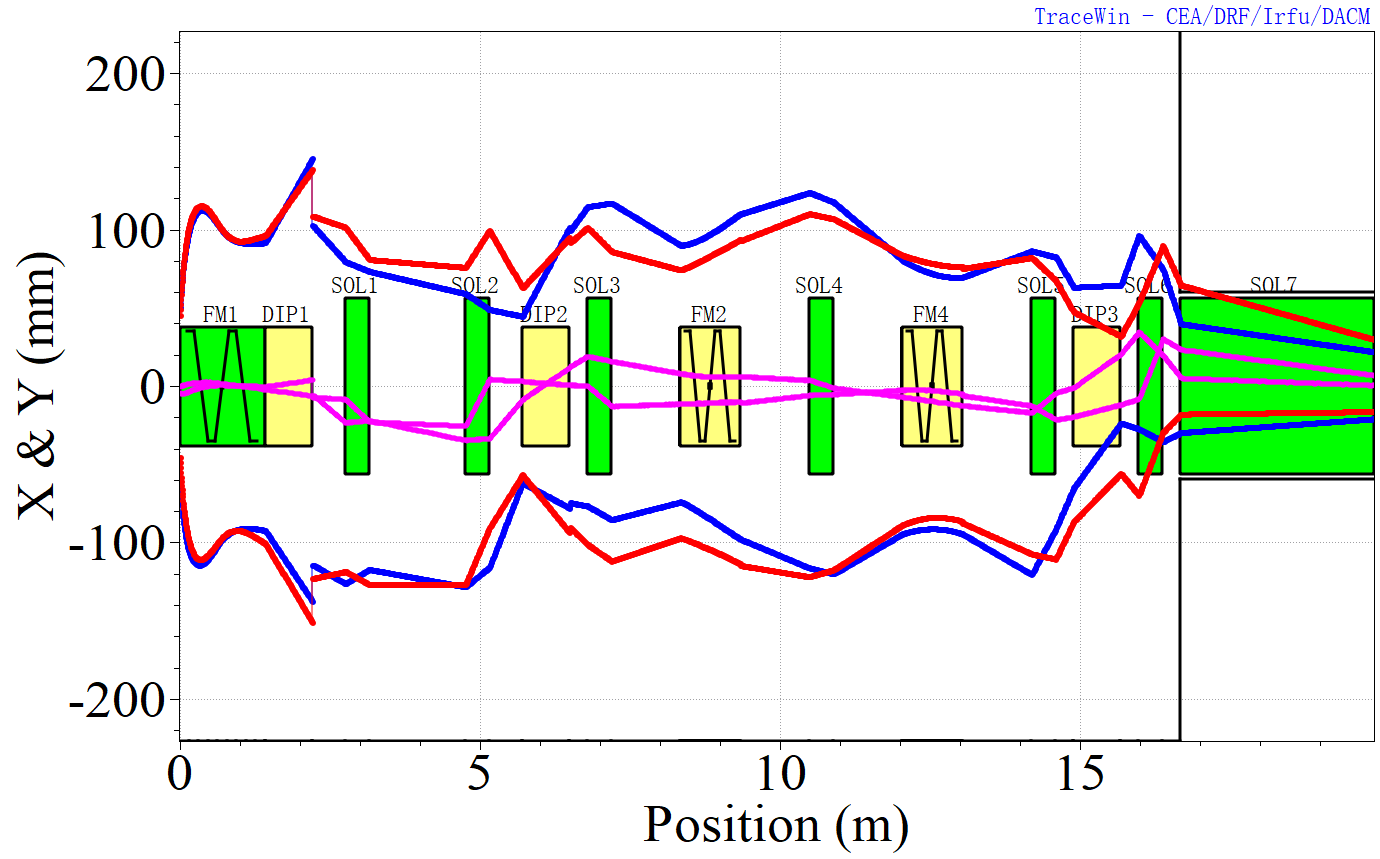}
    \caption{Beam envelope plot from the TraceWin software in the multi-particle mode with hard edge model.}
    \label{FIG:envelope}
\end{figure}

By externally invoking the TraceWin program and utilizing PSO (Particle Swarm Optimization) algorithm, we optimized each component of the beamline (hard edge model) to maximize overall transmission efficiency while minimizing beam spot size at the end. The optimized beam envelope plot is illustrated in \cref{FIG:envelope}. By externally invoking the G4beamline package and utilizing GA (Genetic Algorithm) algorithm, we attempted to optimize each component of the beamline (real magnetic field). The transmission efficiency of the beamline is 3\%, with the transverse RMS size of approximately 20~mm. The surface muon beam rate will be up to $1.5\times 10^{10}~\mu^+$/s with a 5~mA proton beam. Through further optimizations, higher transmission efficiency and surface muon beam rate can be achieved.

\section{Muonium production target}\label{sec:muonium-target}

\subsection{Introduction}

The goal of MACE is to identify spontaneous muonium-to-antimuonium conversion events in a vacuum. MACE cannot detect antimuonium decay events in materials for two reasons. First, atomic positrons will be trapped in the material and not detectable. Second, muonium-to-antimuonium conversion is a coherent process, it is suppressed by decoherence due to interactions of muonium with the material~\cite{PhysRev.123.1439}. Therefore, enhancing the muonium yield is crucial for improving the signal-to-noise ratio. Consequently, increasing sensitivity primarily involves maximizing the muonium yield in a vacuum.

In the production of muonium in a vacuum, a common approach involves directing a surface muon beam into a specific target material. A muon may spontaneously capture an electron to form a muonium atom. Subsequently, these muonium atoms will diffuse within the material, with some escaping into the vacuum. Ideal target materials for this purpose are porous and inert, such as silica powder or silica aerogel. A silica powder target can reach an efficiency of up to 8\% for muonium emission into the vacuum~\cite{SCHWARZ1992244}. It was used in the MACS experiment, the most recent study on muonium-antimuonium conversion, where $5 \times 10^{-3}$ muonium atoms were produced in a vacuum per incident muon~\cite{Willmann:1998gd}. In a more contemporary approach, silica aerogel is employed as the target for muonium production. The self-supporting and porous properties of silica aerogel, together with its relatively high muonium production efficiency, make it suitable for MACE.

However, research has indicated that a significant portion of muonium atoms remains trapped within the target~\cite{Beer:2014ooa,Antognini:2022ysl}, which presents challenges in enhancing muonium emission efficiency. To enhance the diffusion of muonium atoms from the aerogel into the vacuum, prior studies have demonstrated that a perforated surface downstream, created using a pulsed laser, effectively enhances the diffusion of muonium atoms within the target and their emission into the vacuum~\cite{Beer:2014ooa}. The configuration of these perforations is illustrated in~\cref{fig:target-schemantic-diagram}.

The laser ablation technique has been demonstrated to be effective. J. Beare et al. conducted experiments at TRIUMF, investigating various parameters of the ablation structure, and achieved an emission efficiency (the ratio of muonium decay in a downstream vacuum to the total muonium formed) of up to 2\%~\cite{Beare:2020gzr}, representing an order of magnitude enhancement compared to the silica powder target. Additionally, A. Antognini et al. conducted measurements of muonium yields in a vacuum at PSI, employing a slow muon beam with a momentum ranging from 11 to 13 MeV/$c$, achieving a yield (the ratio of muonium decay in a vacuum to the number of incident muons) of 6\%. This yield was enhanced through the utilization of low beam momentum. Furthermore, the J-PARC muon $g-2$ experiment has designed to integrated a perforated silica aerogel target as a crucial component in muon cooling~\cite{Kamioka:2023xob}.

\subsection{Design and optimization of single-layer target}\label{sec:single-layer muonium target}

We have developed a Monte Carlo simulation method for the muonium formation and diffusion in the perforated silica aerogel. The independent simulation results can be well-validated by experimental data. In this section, we summarize the geometrical design and a simulation-guided optimization on the design.

\begin{figure}[htbp]
    \nolinenumbers
    \centering
    \includegraphics[width=\columnwidth]{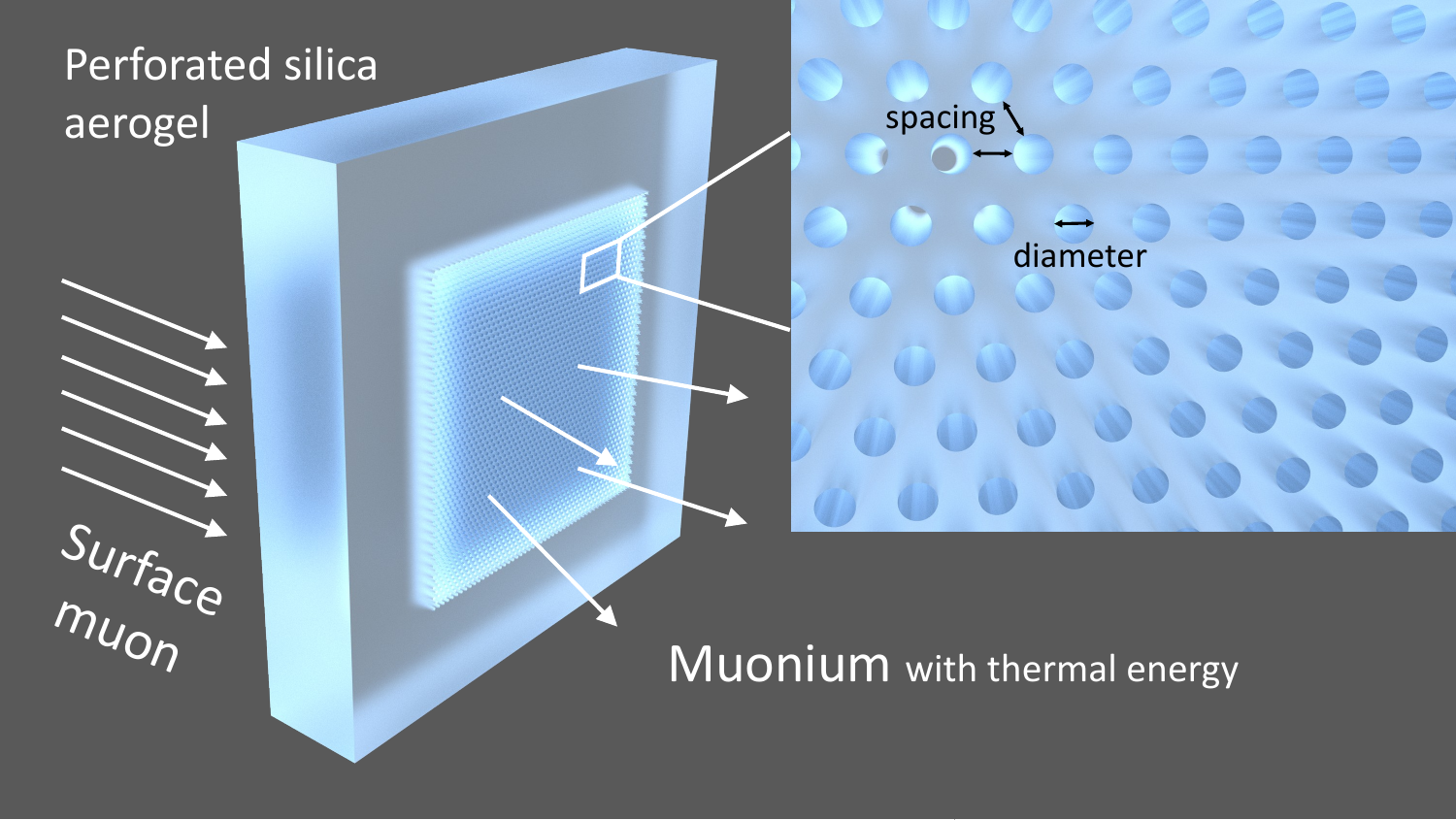}
    \caption{\label{fig:target-schemantic-diagram}The single-layer perforated silica aerogel target concept~\cite{Zhao:2023plv}.}
\end{figure}

\begin{figure}[htbp]
    \nolinenumbers
    \centering
    \subfloat[Total yield $Y_\text{tot}$.]{
        \includegraphics[width=0.6\columnwidth]{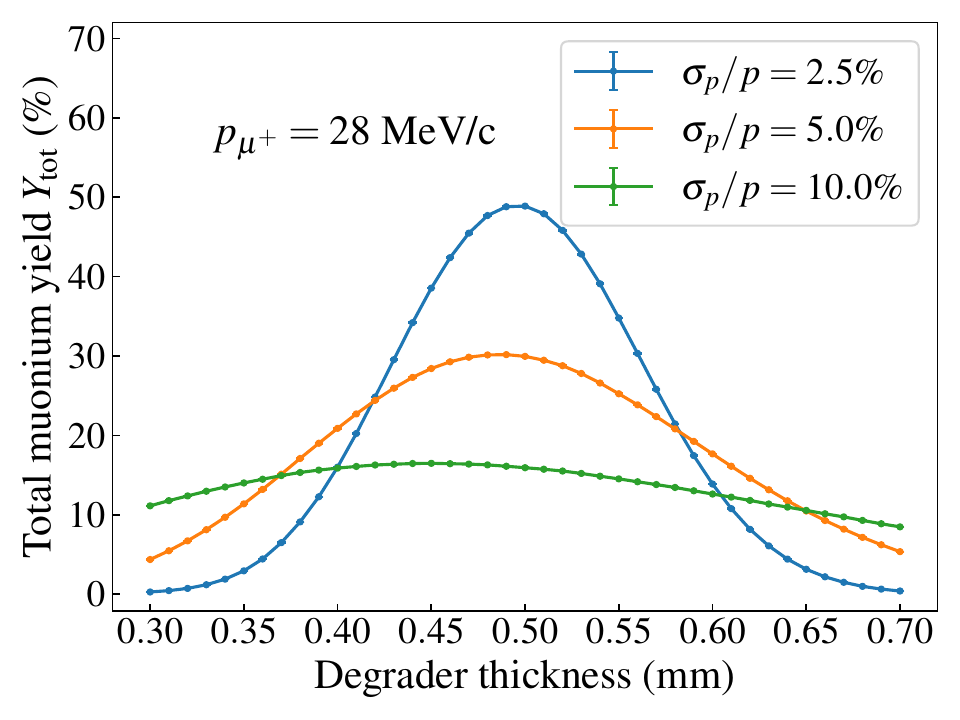}
    }\\
    \subfloat[Emission efficiency $R_\text{vac}$.]{
        \includegraphics[width=0.6\columnwidth]{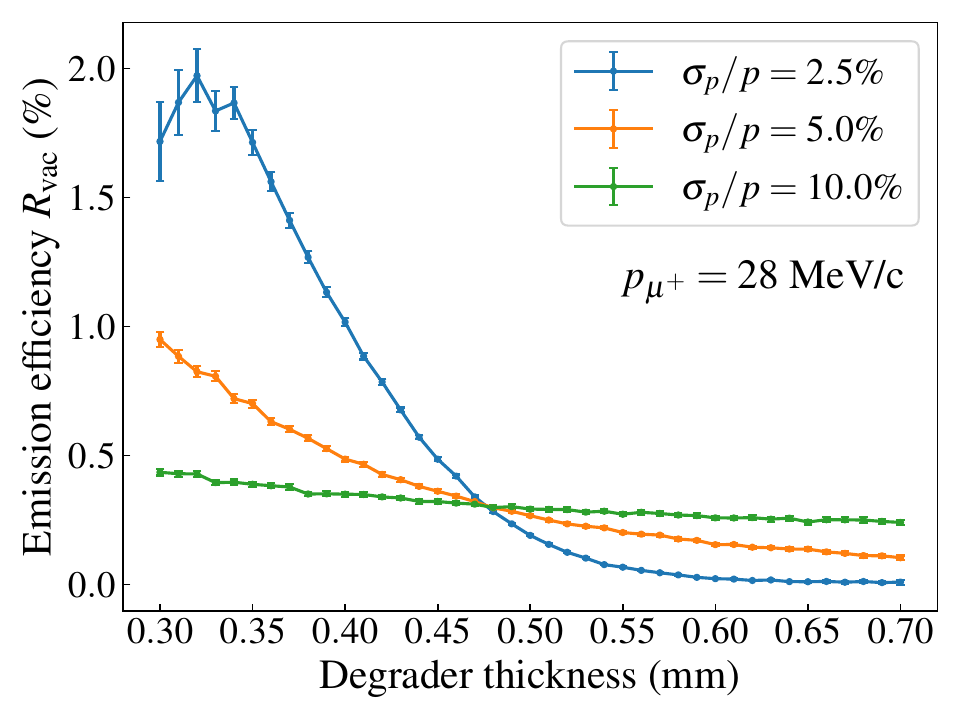}
    }\\
    \subfloat[Vacuum yield $Y_\text{vac}$.]{
        \includegraphics[width=0.6\columnwidth]{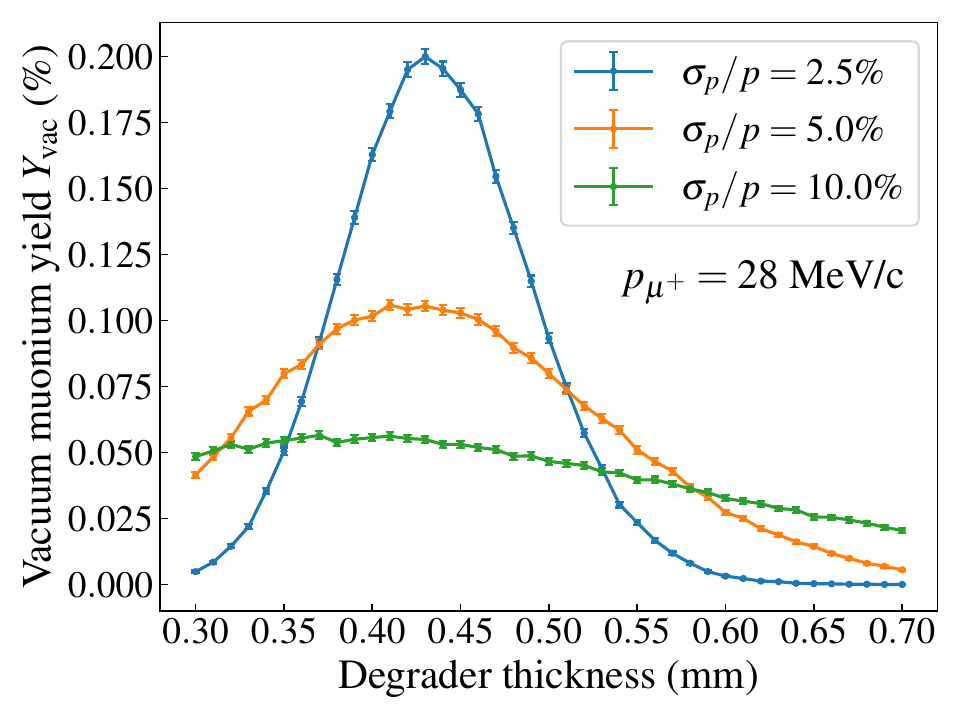}
    }
    \caption{\label{fig:al-degrader-thickness}Muonium production and emission from a flat silica aerogel target with different aluminium degraders ($t_\text{target}=10~\text{mm}, \rho_\text{target}=30~\text{mg}/\text{cm}^3, \lambda=200~\text{nm}$, and $T=322~\text{K}$)~\cite{Zhao:2023plv}.}
\end{figure}

\begin{figure*}[t]
    \nolinenumbers
    \centering
    \subfloat[\label{fig:scan-simulation-a}Total muonium yield $Y_\text{tot}$ ($\sigma_p/p=2.5\%,\ d=1~\mathrm{mm}$)]{
        \includegraphics[width=0.3\textwidth]{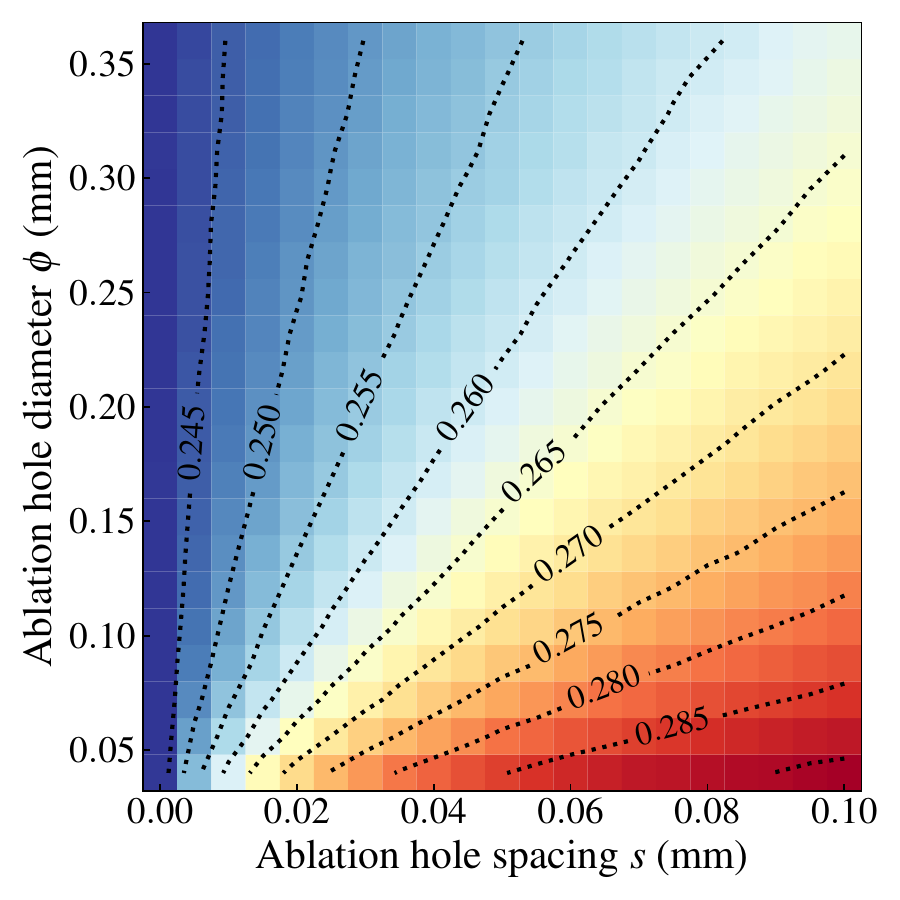}
    }
    \subfloat[Emission efficiency $R_\text{vac}$ ($\sigma_p/p=2.5\%,\ d=1~\mathrm{mm}$)]{
        \includegraphics[width=0.3\textwidth]{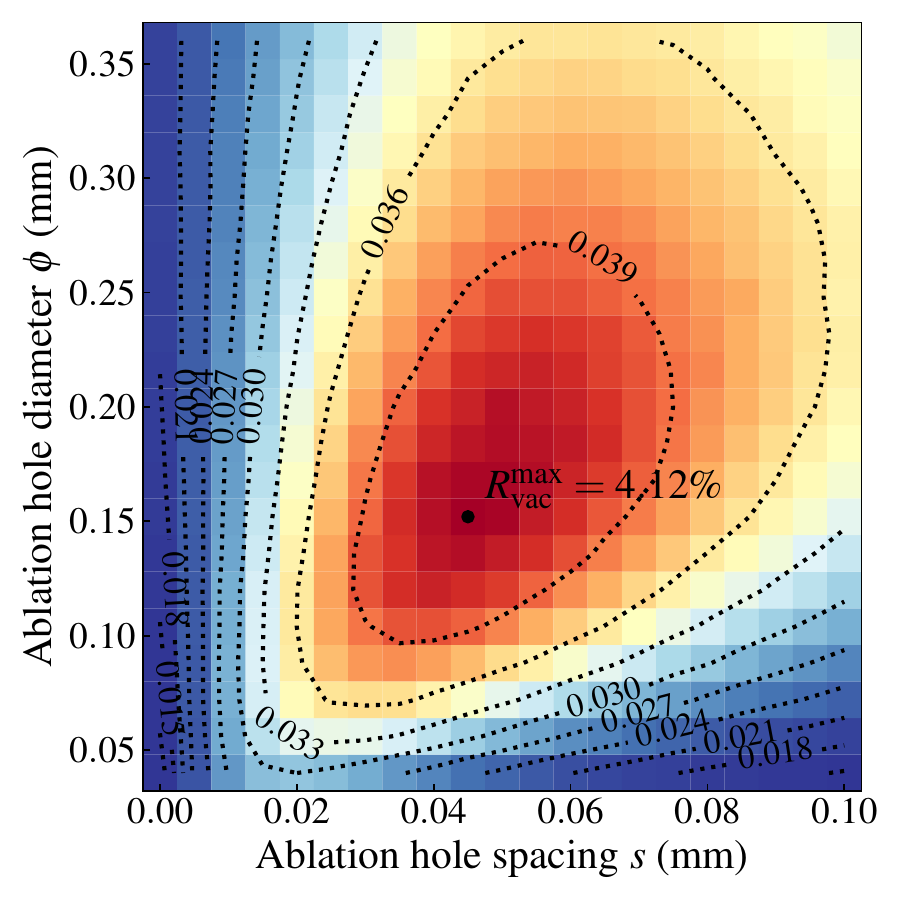}
    }
    \subfloat[Muonium yield in a vacuum $Y_\text{vac}$ ($\sigma_p/p=2.5\%,\ d=1~\mathrm{mm}$)]{
        \includegraphics[width=0.3\textwidth]{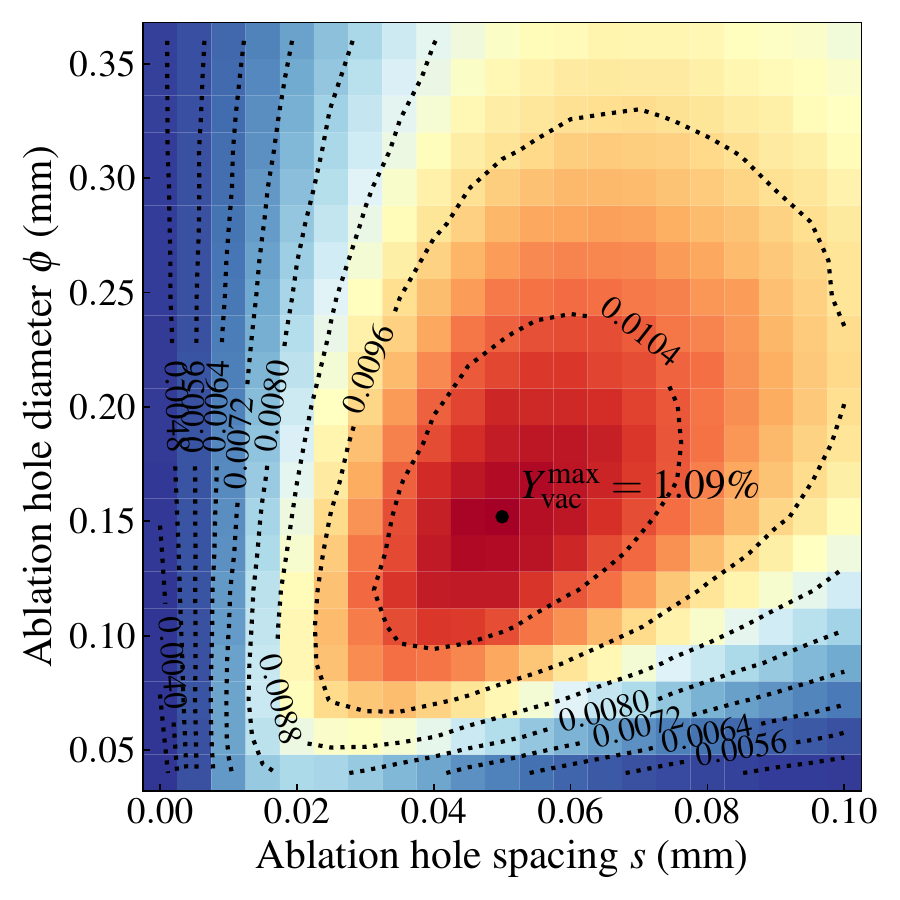}
    }\\
    \subfloat[\label{fig:scan-simulation-d}Total muonium yield $Y_\text{tot}$ ($\sigma_p/p=10\%,\ d=5~\mathrm{mm}$)]{
        \includegraphics[width=0.3\textwidth]{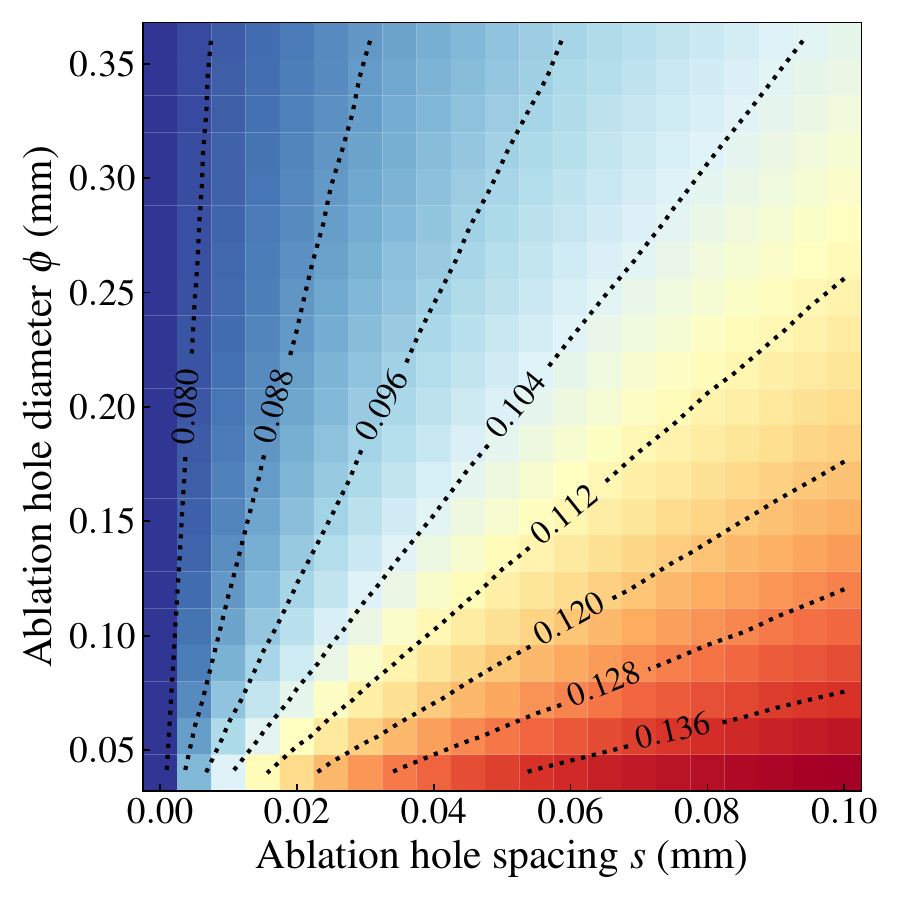}
    }
    \subfloat[Emission efficiency $R_\text{vac}$ ($\sigma_p/p=10\%,\ d=5~\mathrm{mm}$)]{
        \includegraphics[width=0.3\textwidth]{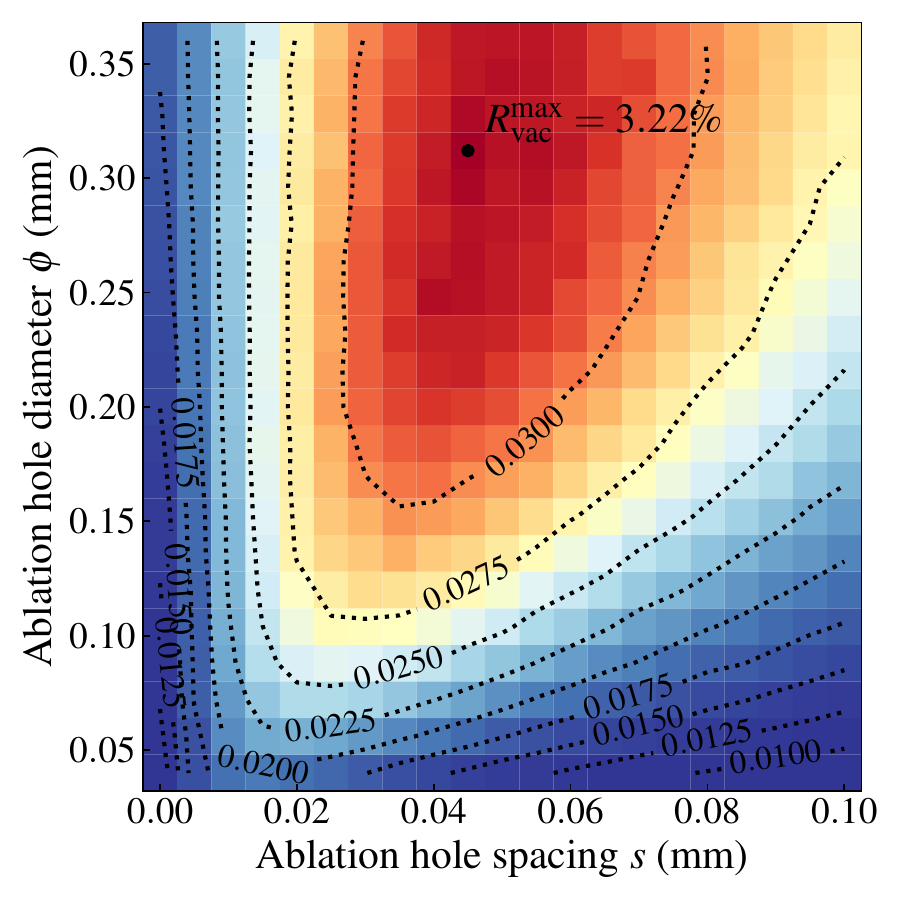}
    }
    \subfloat[Muonium yield in a vacuum $Y_\text{vac}$ ($\sigma_p/p=10\%,\ d=5~\mathrm{mm}$)]{
        \includegraphics[width=0.3\textwidth]{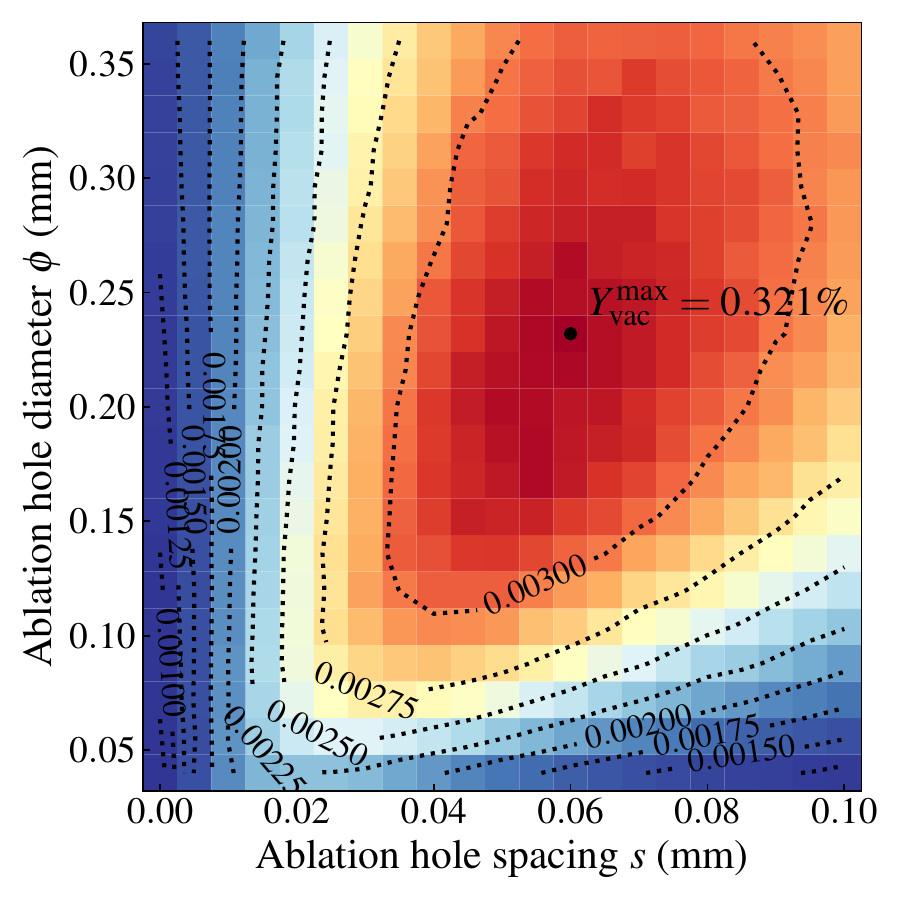}
    }
    \caption{\label{fig:scan-simulation}Projection of total muonium yield, muonium emission efficiency, and vacuum muonium yield under different beam conditions and target geometries~\cite{Zhao:2023plv}.}
\end{figure*}

For the benchmark design, a single-layer target geometry in the shape of a cuboid is selected, with its short edge aligned parallel to the beam direction ($z$-axis), and the remaining two longer edges being of equal length. The downstream face of the target is perforated with blind-ended cylindrical holes perpendicular to the surface, arranged in an equilateral lattice pattern. A schematic diagram is shown in \cref{fig:target-schemantic-diagram}. The geometry is parameterized as follows:
\begin{itemize}
    \item The cuboid geometry: the width and the thickness of the target;
    \item The perforation geometry: the spacing ($s$), the diameter ($\phi$) and the depth ($d$) of the blind-ended holes.
\end{itemize}
The width and thickness of the target are determined by the beam parameters. Specifically, the width is determined by the beam spot size, ensuring it is sufficiently large to cover the entire beam spot for maximum utilization of the beam flux. The target thickness is related to the beam momentum and its momentum spread. It should be adjusted to maximize muonium yield, taking into account any additional materials, such as beam momentum degraders or beam monitors, that are located upstream. In the benchmark design, the target dimensions are 60~mm~$\times$~60~mm (width)~$\times$~10~mm (thickness), with an aluminum beam degrader positioned upstream whose thickness can be tuned.

The perforation geometry entails more considerations. Increasing the spacing between the holes will result in more material, thereby leading to higher rates of muon stopping and increased muonium production. However, these muonium atoms may encounter difficulties in emitting into the vacuum due to obstruction from the material. Conversely, enlarging the hole diameter can enhance emission efficiency but may decrease the overall muonium yield. Therefore, it is likely that a specific combination of geometric parameters will yield the highest vacuum yield when the material and beam conditions are held constant.

It is essential to conduct a combined optimization of the geometric parameters. To perform this process, the first step involves defining the physical quantities relevant to muonium production in a vacuum. Five quantities, $f_{\mu^+}^\text{stop}$, $f_\text{M}$, $Y_\text{tot}$, $R_\text{vac}$, and $Y_\text{vac}$, are utilized to characterize the formation and emission of muonium. These quantities represent the muonium formation fraction, the muon stopping fraction, the total muonium yield, the muonium emission efficiency, and the muonium yield in a vacuum, respectively. They are expressed as
\begin{equation}
\begin{gathered}
     f_{\mu^+}^\text{stop}=\frac{N_{\mu^+}^\text{stop}}{N_{\mu^+}^\text{OT}}~,\quad f_\text{M}=\frac{N_\text{M}^\text{tot}}{N_{\mu^+}^\text{stop}}~,\quad Y_\text{tot}=\frac{N_\text{M}^\text{tot}}{N_{\mu^+}^\text{OT}}~,\quad \\ R_\text{vac}=\frac{N_\text{M}^\text{vac}}{N_\text{M}^\text{tot}}~,\quad Y_\text{vac}=\frac{N_\text{M}^\text{vac}}{N_{\mu^+}^\text{OT}}~.
\end{gathered}
\end{equation}

In these equations, $N_{\mu^+}^\text{stop}$ denotes the number of muons stopped in the target, $N_{\mu^+}^\text{OT}$ denotes the total number of muons on target, $N_\text{M}^\text{tot}$ denotes the total number of produced muonium, and $N_\text{M}^\text{vac}$ denotes the number of muonium in a vacuum. Noteworthy relationships exist among these quantities, including
\begin{equation}
    Y_\text{tot}=f_\text{M}f_{\mu^+}^\text{stop}~,\quad Y_\text{vac}=R_\text{vac}Y_\text{tot}=R_\text{vac}f_\text{M}f_{\mu^+}^\text{stop}~.
\end{equation}
While the value of $f_\text{M}$ is predominantly influenced by material properties and $f_{\mu^+}^\text{stop}$ is associated with both material properties and beam conditions, both $R_\text{vac}$ and $f_{\mu^+}^\text{stop}$ are affected by the aerogel target geometry. A more detailed discussion of the relationships between these quantities can be referred to in reference~\cite{Zhao:2023plv}.

The simulation-guided optimization is conducted as follows. The perforation structure is consistent with that shown in \cref{fig:target-schemantic-diagram}, featuring a target size of $60~\mathrm{mm} \times 60~\mathrm{mm} \times 10~\mathrm{mm}$ and ablation holes with varying geometric parameters within the $40~\mathrm{mm} \times 40~\mathrm{mm}$ region at the center of the downstream surface. The temperature is fixed at 322~K, the aerogel density is set at 27~mg/cm$^3$, and the muonium mean free path is 250~nm, all of which serve as inputs for the Monte Carlo model parameters.

An aluminum degrader is positioned 5~mm in front of the target, where the muon beam traverses and loses energy. A fraction of the muons will stop in the target and form muonium atoms. The degrader, with varying thicknesses, produces different distributions of muon stopping positions, thereby influencing the emission of muonium into the vacuum and resulting in varying vacuum yields. Therefore, optimizing the degrader thickness is crucial for maximizing the yield, as illustrated in the simulation results shown in \cref{fig:al-degrader-thickness}. Based on these results, we find that the optimal degrader thicknesses for this single-layer target with momentum spreads of 2.5\%, 5\%, and 10\% are 430~$\mu$m, 410~$\mu$m, and 370~$\mu$m, respectively.

\begin{table}[htbp]
    \nolinenumbers
    \centering
    \caption{\label{tab:max-Yvac}Simulation of maximum muonium yield in a vacuum and corresponding optimal spacing and diameter with different beam condition~\cite{Zhao:2023plv}. Only statistical errors are shown in this table.}
    \begin{tabular}{c|c|c|c|c|c}
        \hline\hline
        $p_\text{beam}$ & $\frac{\sigma_{p_\text{beam}}}{p_\text{beam}}$ & Depth           & Spacing      & Diameter        & Max M yield in                                         \\
        (MeV/$c$)       & (\%)                                   & $d$ (mm)        & $s$ ($\mu$m) & $\phi$ ($\mu$m) & vacuum $Y_\text{vac}$ (\%)                          \\
        \hline
                        &                                        & 1               & $50 \pm 5$   & $152 \pm 16$    & $1.092 \pm 0.002$                                  \\
                        & 2.5                                    & 2               & $55 \pm 5$   & $184 \pm 16$    & $1.134 \pm 0.002$                                  \\
                        &                                        & 5               & $55 \pm 5$   & $184 \pm 16$    & $1.122 \pm 0.002$                                  \\
        \cline{2-6}
                        &                                        & 1               & $50 \pm 5$   & $152 \pm 16$    & $0.583 \pm 0.001$                                  \\
        28              & 5                                      & 2               & $60 \pm 5$   & $216 \pm 16$    & $0.607 \pm 0.001$                                  \\
                        &                                        & 5               & $50 \pm 5$   & $184 \pm 16$    & $0.604 \pm 0.001$                                  \\
        \cline{2-6}
                        &                                        & 1               & $50 \pm 5$   & $152 \pm 16$    & $0.305 \pm 0.001$                                  \\
                        & 10                                     & 2               & $55 \pm 5$   & $200 \pm 16$    & $0.320 \pm 0.001$                                  \\
                        &                                        & 5               & $60 \pm 5$   & $232 \pm 16$    & $0.321 \pm 0.001$                                  \\
        \hline\hline
    \end{tabular}
\end{table}

Utilizing these optimal degrader configurations, the optimization of perforation geometry can be performed, involving hole spacings ranging from 0 to 100~$\mu$m, diameters ranging from 40 to 360~$\mu$m, and depths of 1~mm, 2~mm, and 5~mm. The simulation results are presented in \cref{fig:scan-simulation}, and the optimal values are shown in \cref{tab:max-Yvac}. As illustrated in \cref{fig:scan-simulation}, maxima in muonium emission efficiency and vacuum yield are observed. When the spacing and diameter deviate from their optimal values, the yield in a vacuum gradually decreases, with significant suppression occurring when the spacing or diameter becomes excessively small.

In summary, for the design of a single-layer target, we recommend an appropriate configuration of perforation parameters that balances the total muonium yield ($Y_\text{tot}$) and muonium emission efficiency ($R_\text{vac}$) to achieve an optimal muonium yield in a vacuum ($Y_\text{vac}$). This optimization can be realized through a combination of the simulation-guided optimization procedure proposed in reference~\cite{Zhao:2023plv} and data-driven experimental optimizations. These approaches are shared between the single-layer and multi-layer target design and will be employed in future technical designs to refine and determine the optimal configurations.

\subsection{Multi-layer target design}
The perforated single-layer target has the potential to achieve a high muonium yield in a vacuum; however, there is still room for improvement. In the single-layer target design, the perforated surface faces downstream relative to the beam, and it is expected that nearly half of the muons will stop within the target to maximize the number density of muons stopped near the perforated surface. This configuration implies that nearly half of the muon beam will penetrate the target without being utilized. Additionally, the single-layer target favors a more concentrated momentum spread to minimize the spread of muon stopping positions; therefore a high muonium yield in a vacuum requires a low beam momentum spread. As observed in \cref{sec:single-layer muonium target}, an increased beam momentum spread significantly suppresses the muonium yield in the vacuum, which is not expected since it amplifies the correlation between muonium yield and beam momentum spread. This reduces design margins and may increase systematic errors associated with the muonium yield. Therefore, we expect a target design that can enhance the utilization of muon beam flux while tolerating a wider beam momentum spread. The design of a multi-layer silica aerogel target has the potential to achieve this goal.

\begin{figure}[htbp]
    \nolinenumbers
    \centering
    \includegraphics[width=\columnwidth]{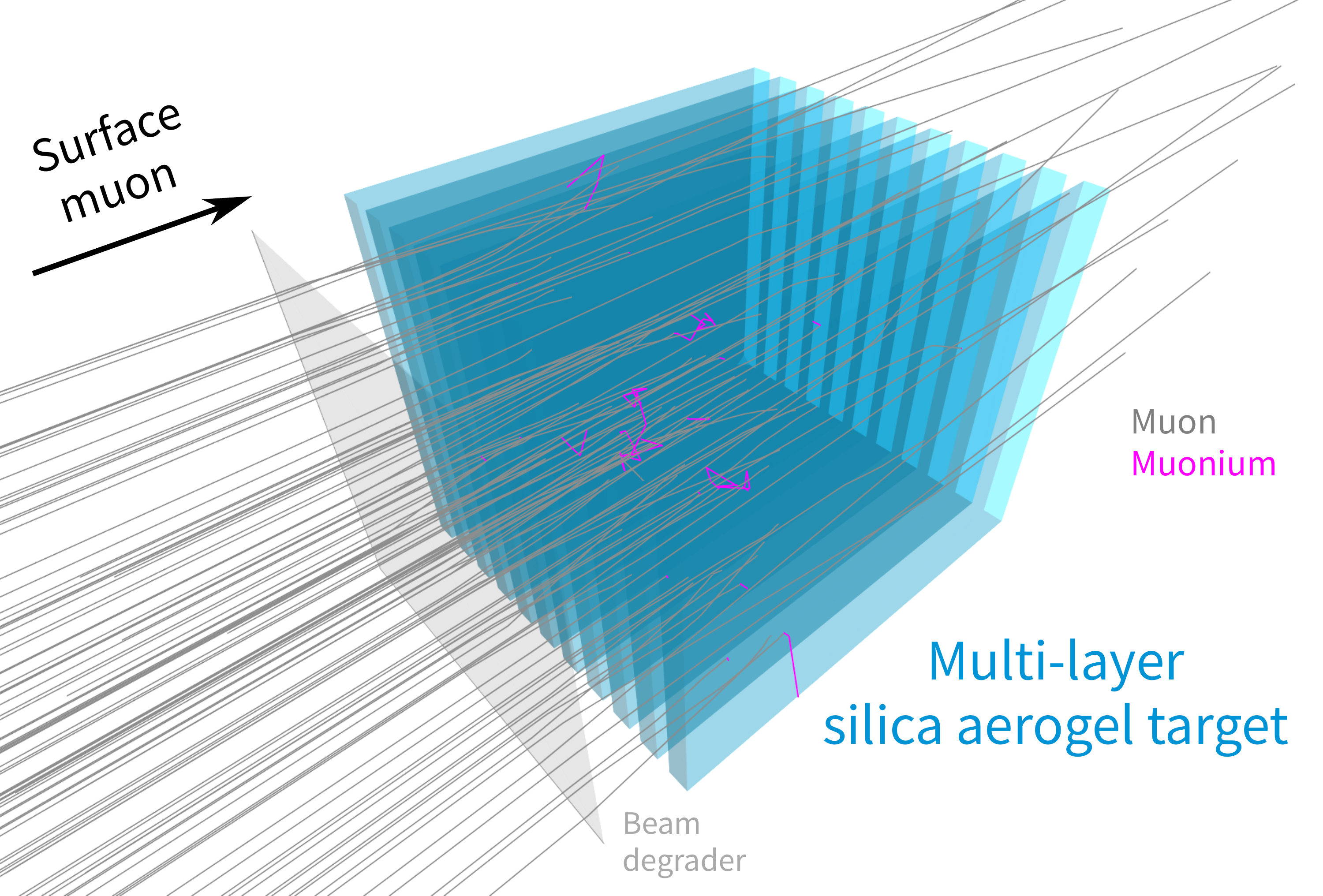}
    \caption{\label{fig:multi layer target concept}The multi-layer silica aerogel target concept.}
\end{figure}

The original concept of a multi-layer silica aerogel target was proposed by Ce Zhang et al.~\cite{Zhang:2022ilj}, with the aim of increasing the total muonium yield in a vacuum to enhance the efficiency of converting a surface muon beam into a thermal muon beam by ionizing the produced muonium atoms in the vacuum. This technology, known as muon cooling, shares the same objective of increasing muonium yield in a vacuum as MACE. In the simulation work by Ce Zhang et al., the multi-layer target design is shown to increase the muonium yield in a vacuum by a factor of 3.45.

\begin{figure}[t]
    \nolinenumbers
    \centering
    \subfloat[Muonium formation vertices.]{\includegraphics[width=0.7\columnwidth]{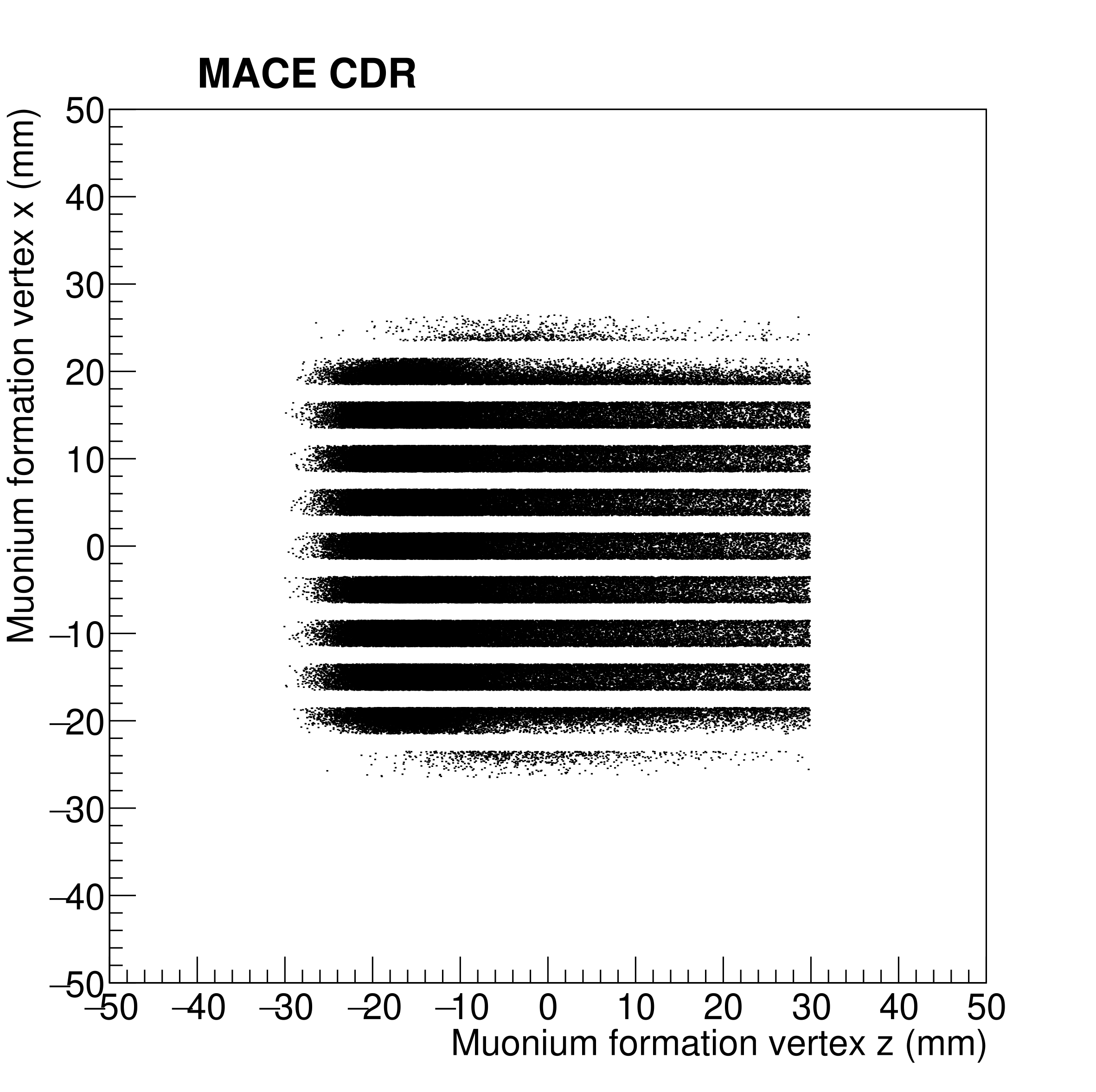}}\\
    \subfloat[Muonium decay vertices.]{\includegraphics[width=0.7\columnwidth]{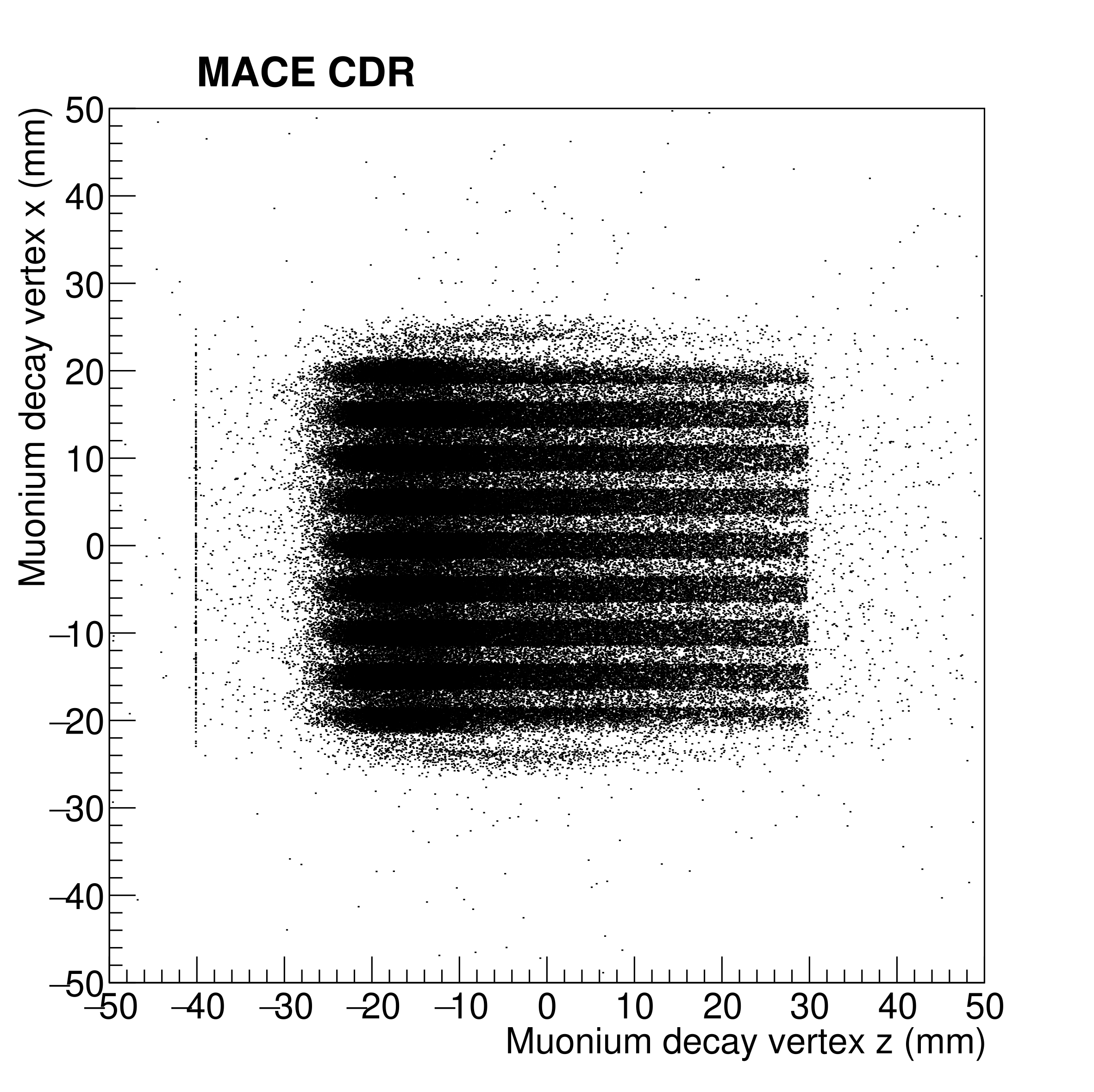}}
    \caption{\label{fig:multi layer target muonium distribution}Simulated distributions of muonium formation and decay vertices. A few percents of muonium atoms will diffuse out of the target before their decay.}
\end{figure}

The design horizontally aligns multiple silica aerogel targets in parallel, with the muon beam directed parallel to the layers. Each silica aerogel layer is perforated, allowing muonium atoms produced within the layer to diffuse and escape into the inter-layer vacuum. If these muonium atoms convert into antimuonium and decay, low-energy positrons will be accelerated parallel to the target layer, exiting the target region and being guided to the positron detection system. This target concept is illustrated in \cref{fig:multi layer target concept}. In scenarios where the beam may be well-collimated, it can pass through the target via the spaces between the silica aerogel layers, potentially leading to reduced utilization of muon beam flux. To address this, a beam degrader may be introduced in front of the multi-layer target to scatter the beam if necessary. By this design, the utilization of muon beam can be effectively enhanced. Furthermore, unlike the single-layer target design, the longitudinal length of the multi-layer target can be extended without significantly compromising muonium emission efficiency, as muonium atoms diffuse and escape primarily in the transverse direction. Consequently, this design achieves the goals of increasing muon beam utilization while tolerating with a wider beam momentum spread.

We conducted simulations of various multi-layer target designs. We utilized a beam momentum of 24~MeV/$c$ and an RMS spread of 1.35~MeV/$c$. The beam is assumed to be collimated, resulting in a beam spot that is a 40~mm$\times$40~mm square, where muons are uniformly distributed. Each layer has a longitudinal length of 60~mm, a height of 50~mm, and varying thicknesses of 2~mm, 3~mm, and 4~mm. The layers are perforated in a 50~mm$\times$40~mm region at their center. The perforation parameters for the aerogel layers are inherited from the optimal parameters of the single-layer target design, i.e. a hole spacing of 55~$\mu$m and a hole diameter of 184~$\mu$m. The spacing between each adjacent layer is kept consistent, varying from 2~mm, 3~mm, to 4~mm. The number of silica aerogel layers is selected to ensure that the overall height and width of the target shape are matched.

In the simulation, $10^6~\mu^+$ are generated for each target design, and the results are summarized in \cref{tab:multi-layer target simulation}. The definitions of the result parameters are consistent with those described in \cref{sec:single-layer muonium target}, with the exception that "in-vacuum muonium" is counted only in the vacuum region of interest, where atomic positrons from antimuonium decay can be accelerated and transported to the positron detection system. From the simulation results, we observed that the muonium yields in the vacuum ($Y_\text{vac}$) are enhanced by factors of 2 to 4 compared to the single-layer target. This enhancement is primarily attributed to the increase in muonium emission efficiency ($R_\text{vac}$).

\begin{table}[htbp]
    \nolinenumbers
    \centering
    \caption{Simulated muonium yield in different multi-layer targets. Only statistical errors are shown in this table.}
    \begin{tabular}{cccccc}
        \hline\hline
        Thickness & Spacing & Aerogel & $Y_\text{tot}$ & $R_\text{vac}$ & $Y_\text{vac}$ \\
         (mm)     & (mm)    & count & (\%)             & (\%)           & (\%)         \\ \hline
        \multirow{3}{*}{2} & 2      & 13 & $20.33(5)$ & $24.3(1)$ & $4.94(2)$ \\ \cline{2-6}
        & 3      & 11 & $16.04(4)$ & $27.6(1)$ & $4.43(2)$ \\ \cline{2-6}
        & 4      & 9  & $13.13(4)$ & $29.3(2)$ & $3.85(2)$ \\ \hline
        \multirow{3}{*}{3} & 2      & 11 & $22.40(5)$ & $17.04(9)$  & $3.82(2)$ \\ \cline{2-6}
        & 3      & 9  & $19.01(5)$ & $19.2(1)$ & $3.66(2)$ \\ \cline{2-6}
        & 4      & 8  & $16.00(4)$ & $20.5(1)$ & $3.28(2)$ \\ \hline
        \multirow{3}{*}{4} & 2      & 9  & $24.14(5)$ & $13.08(8)$  & $3.16(2)$ \\ \cline{2-6}
        & 3      & 8  & $20.80(5)$ & $14.83(9)$  & $3.09(2)$ \\ \cline{2-6}
        & 4      & 7  & $17.44(5)$ & $15.8(1)$ & $2.76(2)$ \\ \hline
        \hline
    \end{tabular}
    \label{tab:multi-layer target simulation}
\end{table}

\begin{figure*}[t]
    \nolinenumbers
    \centering
    \subfloat[Perspective view.]{
        \includegraphics[width=\textwidth]{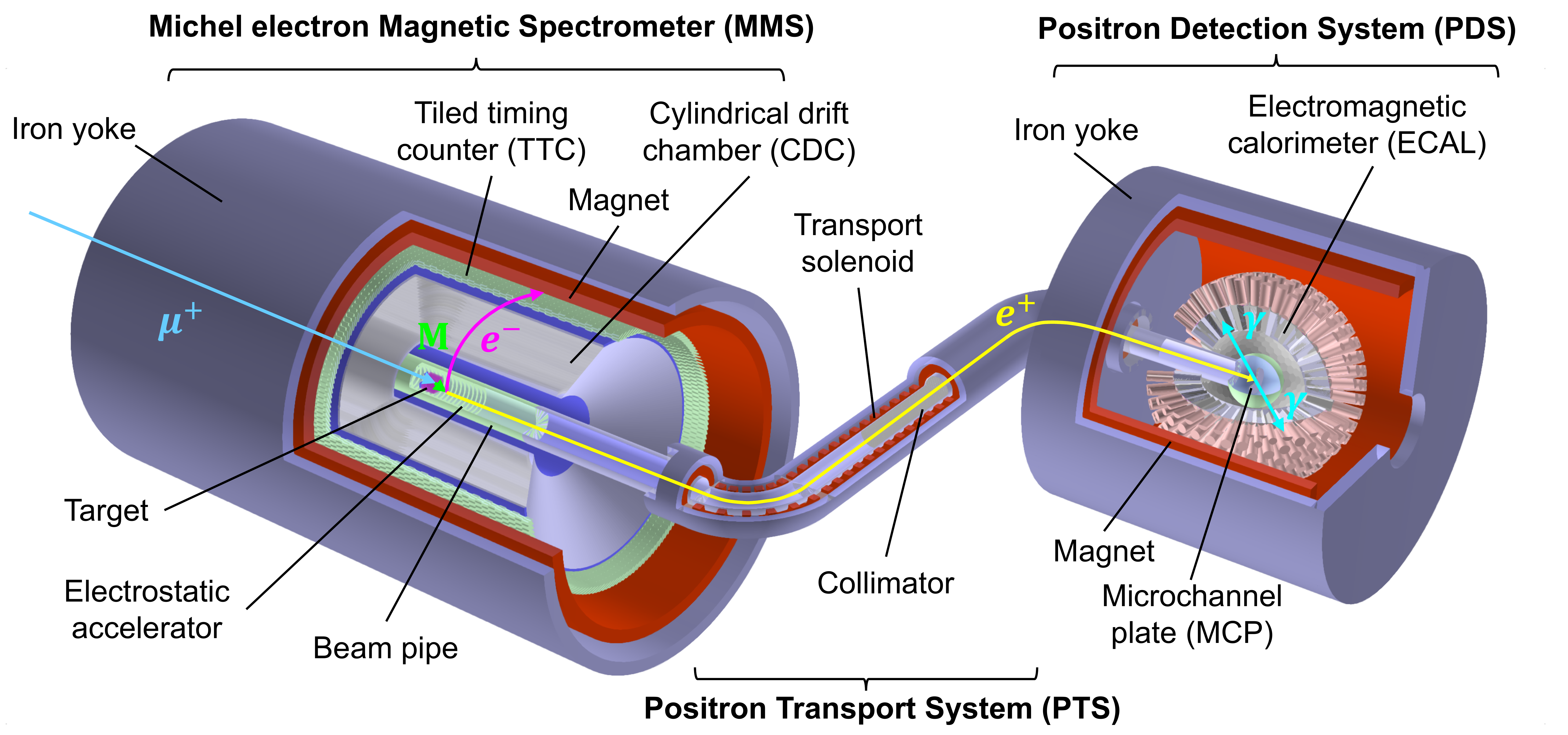}
    }\\
    \subfloat[Top view.]{
        \includegraphics[width=0.8\textwidth]{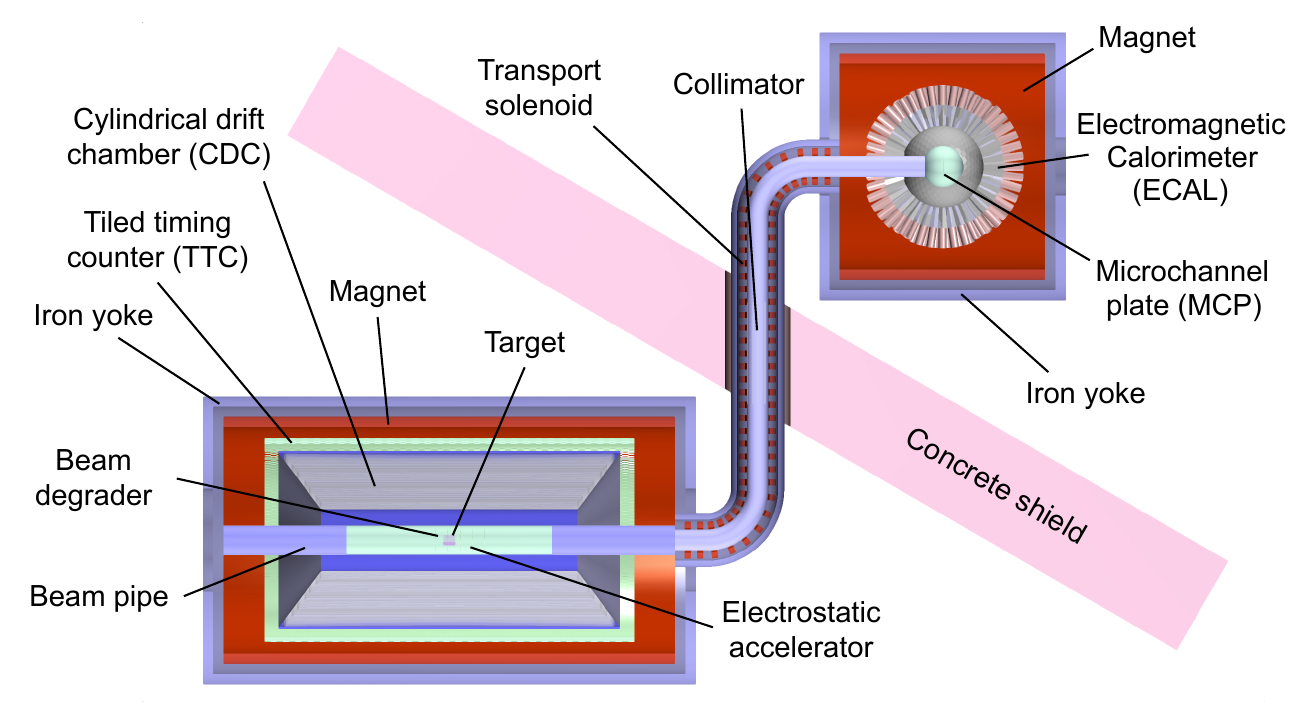}
    }
    \caption{\label{fig:mace-detector-concept}MACE detector concept.}
\end{figure*}

The optimal result is achieved with the smallest single-layer thickness and layer spacing, both at 2~mm, resulting in a muonium yield in the vacuum of 4.94(2)\%. Although the yield number is high, the multi-layer target has its intrinsic limitations. The major concern raised from too much muonium scattering between silica aerogel layers would quench the muonium conversion probability if the layer spacing is too narrow. Therefore, a larger layer spacing should be preferred, and we select a muonium yield in the vacuum of 3.8\% as the benchmark parameter. In conclusion, the results demonstrates the possibility of increasing the muonium yield in vacuum to a few percent if the target geometry is properly designed.

\section{Overview of detector system}\label{sec:mace-detector-overview}
To identify the signal signature demonstrated in \cref{sec:signal-event-signature}, the MACE detector design includes three primary components: a Michel electron magnetic spectrometer (MMS), a momentum-selective positron transport system (PTS), and a positron detection system (PDS). The Michel electron magnetic spectrometer consists of a cylindrical drift chamber (CDC), a set of tiled timing counters (TTC), and a magnet. MMS is used to measure the tracks and momenta of high-energy electrons. The positron transport system includes an electrostatic accelerator and a positron transport solenoid, which transports low-energy positrons to the positron detection system while conserving their transverse position. The PDS includes a microchannel plate (MCP) detector and an electromagnetic calorimeter (ECAL), with the MCP detecting the transverse position of positrons and the ECAL detecting gamma rays produced from annihilation events on the MCP.

MACE utilizes coincident detection from the Michel electron magnetic spectrometer and the positron detection system to discriminate and detect antimuonium signals. Initially, a positron is identified through the coincidence of a hit in the MCP and annihilation signals in the calorimeter. By considering the conservation of the transverse position of positrons in the transport system, the transverse projection of the positron is spatially matched with the track of high-energy electrons measured by the magnetic spectrometer. Most background positrons are filtered out during transportation in the solenoid by a fine collimator. Finally, by comparing the time difference measured by both the PDS and the MMS with the expected signal time of flight, an antimuonium decay signal can be identified.

In summary, the detection scheme involves the primary detector signals as follows:
\begin{itemize}
   \item An energetic electron track in the Michel electron magnetic spectrometer (MMS);
   \item A hit on the microchannel plate (MCP);
   \item Annihilation gamma ray signals in the electromagnetic calorimeter (ECAL).
\end{itemize}
When these signals coincide as follows, they are identified as a muonium-to-antimuonium conversion event:
\begin{itemize}
   \item A positron is identified in the positron detection system (PDS), indicated by the MCP hit coinciding in time with the annihilation signals in the calorimeter;
   \item The positron's transverse position, indicated by the hit on the MCP, matches with the electron track detected in the MMS;
   \item The positron's time-of-flight (TOF) is within the expected signal TOF range.
\end{itemize}

\section{Michel electron magnetic spectrometer}
The Michel electron magnetic spectrometer (MMS) is responsible for detecting and identifying electrons or positrons, measuring their time, track, and momenta for coincidence detection and event selection purposes. The MMS needs to exhibit outstanding spatial resolution to enhance vertex reconstruction precision, thereby allowing MACE to effectively reject accidental backgrounds. An extremely low charge misidentification rate is essential to avoid introducing additional accidental backgrounds. Moreover, it should possess good momentum resolution to facilitate physical analysis and enhance sensitivity to other potential physics subjects.

The primary component of the MMS is a cylindrical drift chamber (CDC) designed to track the decay products of muons, muonium, and antimuonium, specifically positrons and electrons. The CDC should feature a large geometric acceptance to ensure adequate signal efficiency, as well as excellent spatial resolution for optimal momentum and vertex resolution. Surrounding the CDC barrel is a set of timing counters, known as the tiled timing counter (TTC), which functions as the event trigger. Both the CDC and TTC are situated within an axial magnetic field to curve the charged tracks and discriminate between electrons and positrons. The central alignment of the entire MMS system corresponds to the muonium target, the origin point for most tracks.

\subsection{Magnetic field and magnet}
A solenoid magnet surrounds the CDC and TTC to provide an axial magnetic field. Increasing the magnetic flux density can enhance momentum resolution and reduce the charge misidentification rate of the MMS by increasing track curvature. However, the muonium conversion process is influenced by the magnetic field, and since the target is positioned at the center of the MMS, this field affects the conversion process. A strong magnetic field suppresses contributions from certain operators in the conversion process, creating a trade-off between MMS performance and physical sensitivity. As discussed in references~\cite{Hou:1995np,Fukuyama:2023drl}, muonium conversion induced by $(V\pm A) \times (V\pm A)$ and $(S\pm P) \times (S\pm P)$ effective couplings is significantly reduced when the magnetic field reaches 1~T or higher. To maintain MACE's sensitivity to all effective operators and constrain the parameter space, the MMS will operate at $B = 0.1~T$. This magnetic field achieves a balance, ensuring MMS performance while preserving reasonable conversion probabilities associated with $(V\pm A) \times (V\pm A)$ and $(S\pm P) \times (S\pm P)$ couplings. This configuration is also consistent with the choices made in the MACS experiment~\cite{Willmann:1998gd}. Further details on the magnet design are provided in \cref{sec:mace-magnet-and-solenoid}.

\subsection{Cylindrical drift chamber}
\subsubsection{Design objectives}
The cylindrical drift chamber (CDC) is responsible for tracking charged particles. To enhance sensitivity, identify between muonium or muon decay and antimuonium decay, and improve the discrimination between signal and background events, the drift chamber should has high detection efficiency, a large geometric acceptance. High spatial resolution, low charge misidentification rate and high momentum resolution are also essential in enhancing the sensitivity. These criteria guide the design of the MACE cylindrical drift chamber. Therefore, we propose to deploy a light, small-cell cylindrical drift chamber with proper wiring configurations. Design objectives are described as follows.

Reducing the amount of materials is a primary focus on the design of the drift chamber. The tracked particles, electrons and positrons, are disturbed significantly by multiple scattering and Coulomb scattering. These kinds of scattering will affect the resolution of a drift chamber, also known as the material effect. The material budget is mainly built up from three parts: gas, wires, and inner wall. Endplates and outer walls are not taken into account since MACE do not concern the particle momentum after passing the CDC. Among the three parts, gas and wires are the main contributors to the material budget. Therefore, adopting Helium-based gas becomes a optimal solution to improve momentum resolution. This is also the choice of drift chambers constructed in recent years~\cite{BESIII:2009fln,Belle-II:2010dht,Wu:2021ndo,Baldini:2023hxt}. On the other hand, the vertex resolution is a primary focus of MACE to reject accidental backgrounds. A light inner wall will lead to less track distortion, thereby contribute to the decay vertex resolution of a track coincidence with a MCP hit. Thin carbon fiber inner wall is a common solution in collider experiments. However, it is not light enough in muon experiments. In MEG~II, an aluminated Mylar cylindrical foil of 20~$\mu$m thickness is utilized to close the inner surface of chamber~\cite{Baldini:2023hxt}. The aluminated Mylar serves as an option for MACE to further reduces the material budget and improve the vertex resolution. Except for gas and inner wall, wires are another sources of material budget. To reduces the amount of wires, a direct way is to reduce the amount of cells. However, it reduces the rate tolerance of the drift chamber, which is not in favor for MACE due to the high event rate, as to be discussed below. In summary, Helium-based gas and utilizing aluminated Mylar cylindrical foil as the inner wall serves as an optimal solution to minimize the material budget.

\begin{table}[t]
\nolinenumbers
\centering
\caption{Geometric parameters of the cylindrical drift chamber.}
\begin{tabular}{cc}
\hline\hline
Parameter            & Value   \\ \hline
Inner radius         & 150~cm  \\ \hline
Outer radius         & 415~mm  \\ \hline
Inner length         & 1200~mm \\ \hline
Outer length         & 1600~mm \\ \hline
Number of layers     & 21 \\ \hline
Number of cells      & 3536 \\ \hline
Geometric acceptance & 88.8\%       \\ \hline\hline
\end{tabular}
\label{tab:cdc-dimensions}
\end{table}

\begin{figure}[t]
    \nolinenumbers
    \centering
    \includegraphics[width=\columnwidth]{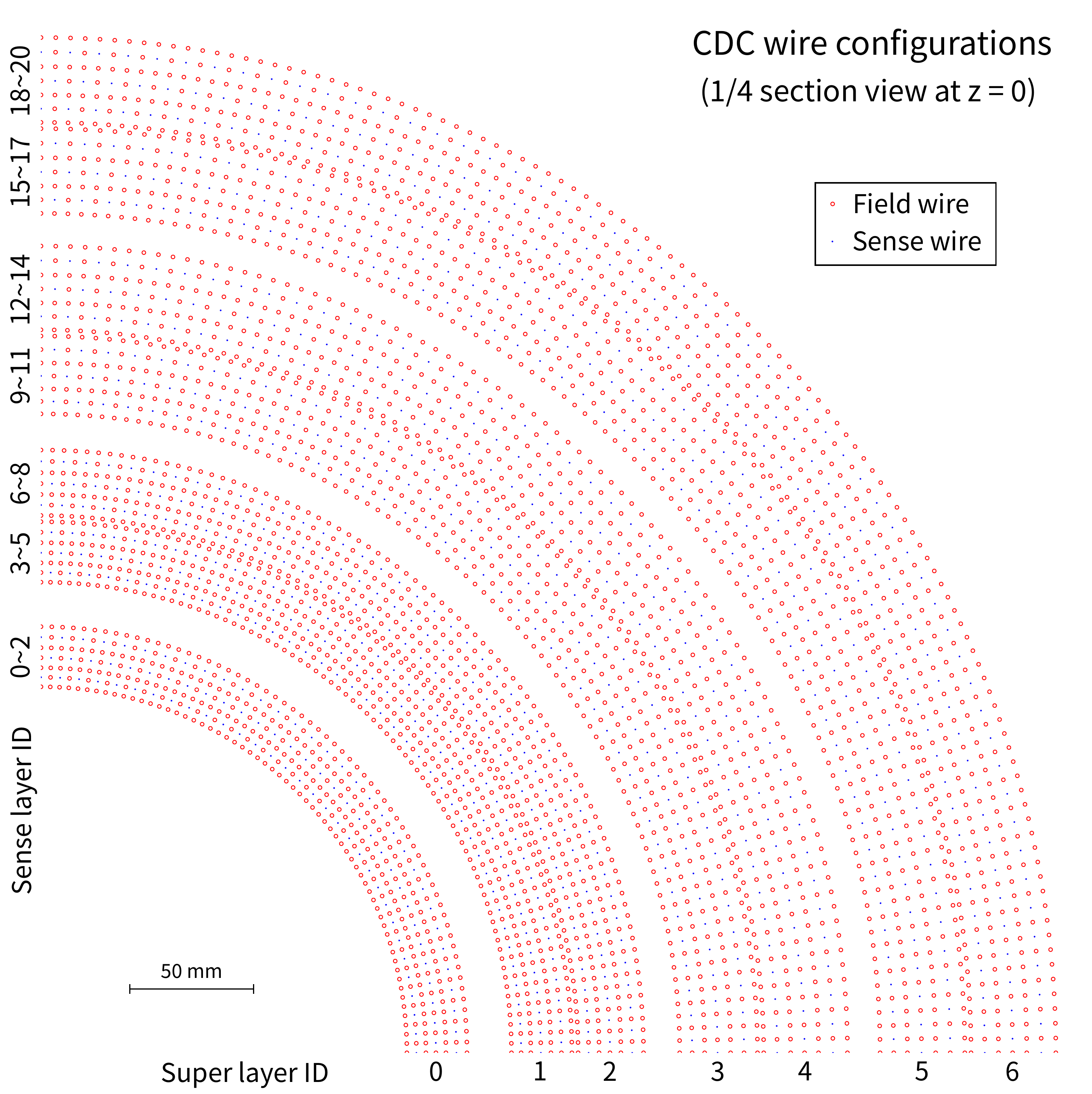}
    \caption{\label{fig:mace-CDC-wiring}A 1/4 section view of CDC wires at $z=0$, where the red and blue circles are the field and sense wires, respectively. The wire radii are scaled by 20 times to be clearly visible. The CDC consists of concentric layers of wires with alternating directions, the odd-numbered super layers being stereo and the even-numbered super layers being axial.}
\end{figure}

\begin{figure}[t]
    \nolinenumbers
    \centering
    \subfloat[1/4 clip-away perspective view]{
        \includegraphics[width=\columnwidth]{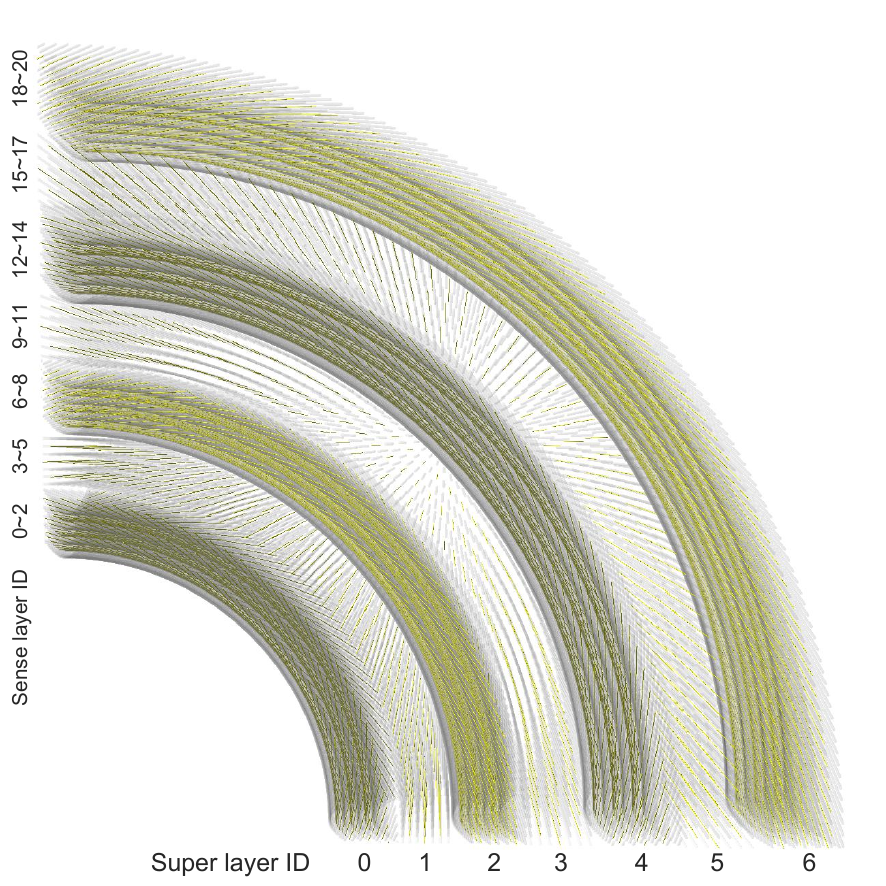}
    }\\
    \subfloat[The $\text{ID}=2$ super layer]{
        \includegraphics[width=0.45\columnwidth]{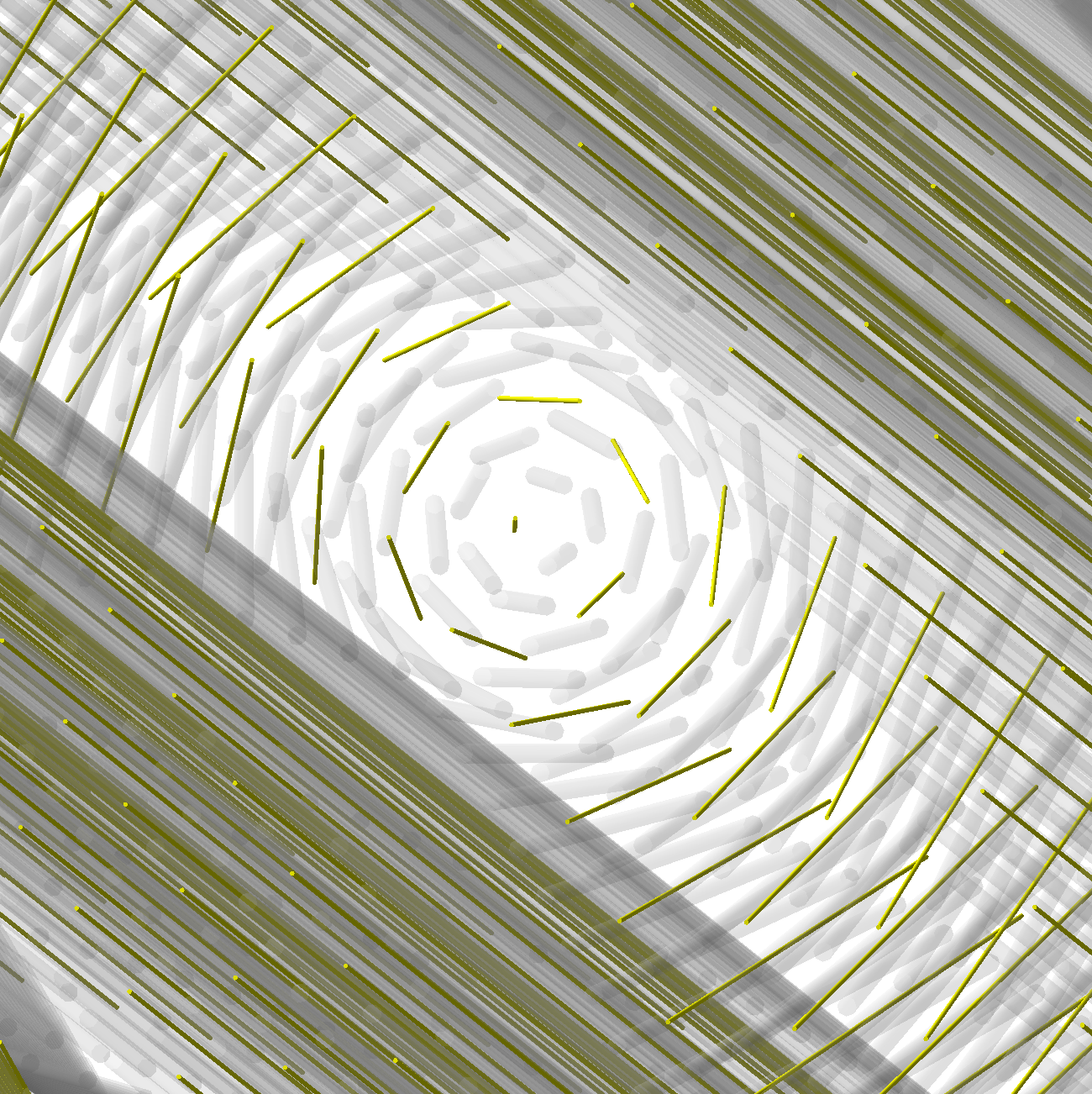}
    }
    \subfloat[The $\text{ID}=3$ super layer]{
        \includegraphics[width=0.45\columnwidth]{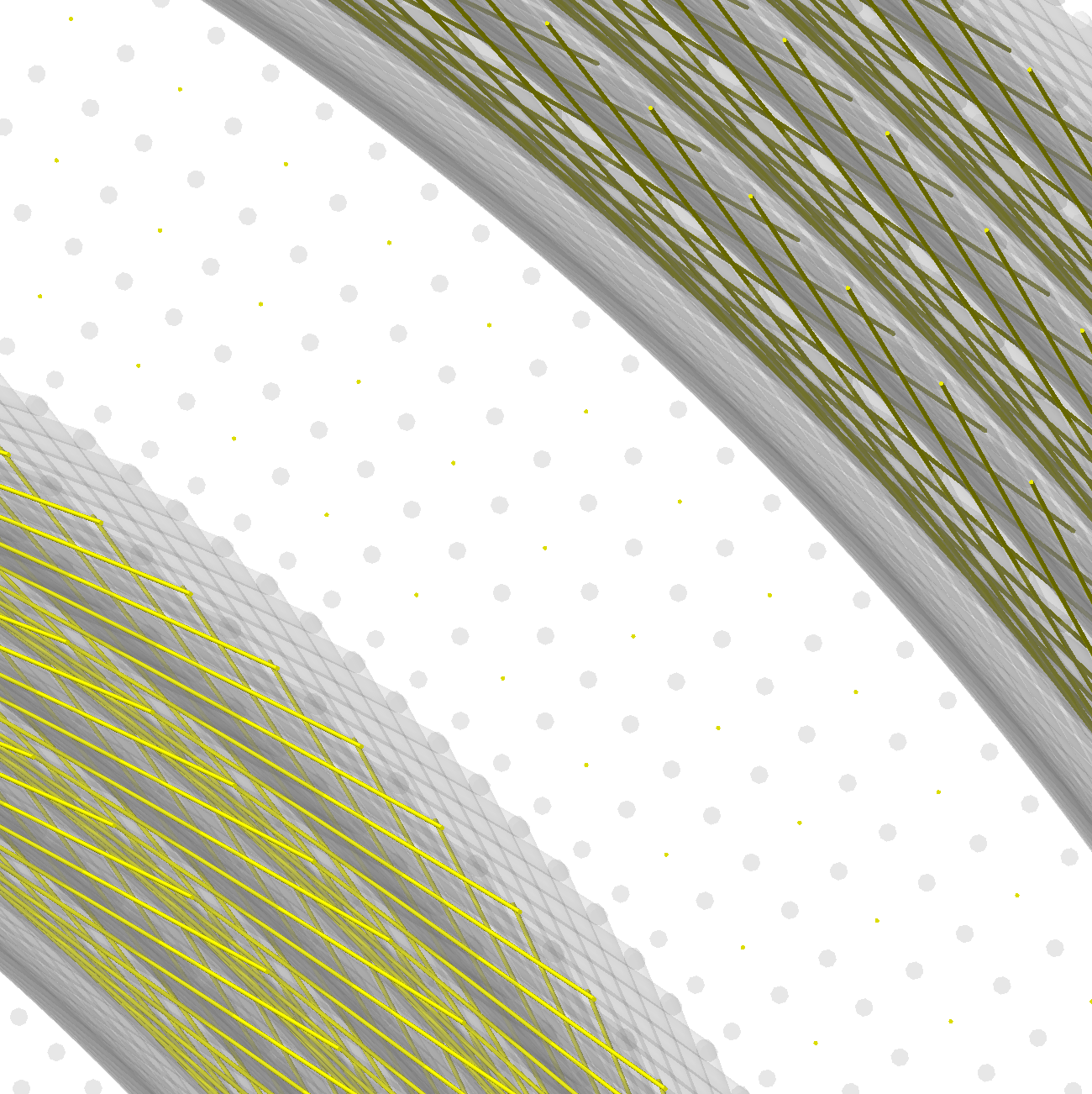}
    }
    \caption{\label{fig:mace-CDC-concept}Perspective view of CDC wires, where golden wires denote sense wires and translucent-gray wires denote field wires. Wire radii are scaled by 20 times to be clearly visible. A hyperbolic profile due to stereo wires in super layer 0, 2, 4, and 6 can be seen in the 1/4 clip-away view.}
\end{figure}

\begin{table*}[t]
\nolinenumbers
\centering
\caption{Wire configurations of MACE cylindrical drift chamber.}
\begin{tabular}{cccccccc}
\hline\hline
\begin{tabular}[c]{@{}c@{}}Super\\ layer ID\end{tabular} & \begin{tabular}[c]{@{}c@{}}Sense\\ layer ID\end{tabular} & Cell ID    & \begin{tabular}[c]{@{}c@{}}Radius at layer\\ center (mm)\end{tabular} & \begin{tabular}[c]{@{}c@{}}$\phi_\text{stereo}$\\ (mrad)\end{tabular} & \begin{tabular}[c]{@{}c@{}}$\phi_\text{1st cell}$\\ (mrad)\end{tabular} & \begin{tabular}[c]{@{}c@{}}Cell width\\ (mm)\end{tabular} & \begin{tabular}[c]{@{}c@{}}Cell length \\ along $z$ (mm)\end{tabular} \\ \hline
\multirow{3}{*}{0}                                       & 0                                                        & 0--123     & 155.992                                                               & 799.799                                                       & 0                                                                       & 7.904                                                     & 1222.530                                                              \\ \cline{2-8}
                                                         & 1                                                        & 124--247   & 164.102                                                               & 807.591                                                       & 25.3354                                                                 & 8.315                                                     & 1235.834                                                              \\ \cline{2-8}
                                                         & 2                                                        & 248--371   & 172.633                                                               & 815.790                                                       & 0                                                                       & 8.747                                                     & 1249.881                                                              \\ \hline
\multirow{3}{*}{1}                                       & 3                                                        & 372--527   & 197.896                                                               & 0                                                             & 20.1384                                                                 & 7.971                                                     & 1265.806                                                              \\ \cline{2-8}
                                                         & 4                                                        & 528--683   & 206.030                                                               & 0                                                             & 0                                                                       & 8.298                                                     & 1277.762                                                              \\ \cline{2-8}
                                                         & 5                                                        & 684--839   & 214.499                                                               & 0                                                             & 20.1384                                                                 & 8.639                                                     & 1290.209                                                              \\ \hline
\multirow{3}{*}{2}                                       & 6                                                        & 840--1003  & 224.193                                                               & -611.622                                                      & 0                                                                       & 8.589                                                     & 1320.833                                                              \\ \cline{2-8}
                                                         & 7                                                        & 1004--1167 & 232.950                                                               & -617.639                                                      & 19.1561                                                                 & 8.925                                                     & 1334.688                                                              \\ \cline{2-8}
                                                         & 8                                                        & 1168--1331 & 242.049                                                               & -623.892                                                      & 0                                                                       & 9.273                                                     & 1349.115                                                              \\ \hline
\multirow{3}{*}{3}                                       & 9                                                        & 1332--1495 & 265.361                                                               & 0                                                             & 19.1561                                                                 & 10.167                                                    & 1365.356                                                              \\ \cline{2-8}
                                                         & 10                                                       & 1496--1659 & 275.726                                                               & 0                                                             & 0                                                                       & 10.564                                                    & 1380.606                                                              \\ \cline{2-8}
                                                         & 11                                                       & 1660--1823 & 286.496                                                               & 0                                                             & 19.1561                                                                 & 10.976                                                    & 1396.451                                                              \\ \hline
\multirow{3}{*}{4}                                       & 12                                                       & 1824--1991 & 298.649                                                               & 501.677                                                       & 0                                                                       & 11.169                                                    & 1428.736                                                              \\ \cline{2-8}
                                                         & 13                                                       & 1992--2159 & 310.031                                                               & 507.616                                                       & 18.7000                                                                 & 11.595                                                    & 1446.392                                                              \\ \cline{2-8}
                                                         & 14                                                       & 2160--2327 & 321.847                                                               & 513.782                                                       & 0                                                                       & 12.037                                                    & 1464.754                                                              \\ \hline
\multirow{3}{*}{5}                                       & 15                                                       & 2328--2519 & 345.728                                                               & 0                                                             & 16.3625                                                                 & 11.314                                                    & 1485.046                                                              \\ \cline{2-8}
                                                         & 16                                                       & 2520--2711 & 357.230                                                               & 0                                                             & 0                                                                       & 11.690                                                    & 1502.017                                                              \\ \cline{2-8}
                                                         & 17                                                       & 2712--2903 & 369.115                                                               & 0                                                             & 16.3625                                                                 & 12.079                                                    & 1519.553                                                              \\ \hline
\multirow{3}{*}{6}                                       & 18                                                       & 2904--3115 & 381.894                                                               & -427.111                                                      & 0                                                                       & 11.318                                                    & 1552.408                                                              \\ \cline{2-8}
                                                         & 19                                                       & 3116--3327 & 393.382                                                               & -431.826                                                      & 14.8188                                                                 & 11.659                                                    & 1570.086                                                              \\ \cline{2-8}
                                                         & 20                                                       & 3328--3539 & 405.217                                                               & -436.683                                                      & 0                                                                       & 12.010                                                    & 1588.319                                                              \\ \hline\hline
\end{tabular}
\label{tab:cdc-wire-config}
\end{table*}

MACE will be connected to an intense muon beam with a flux of $10^8~\mu^+$/s, whose decay products will directly enter CDC, so the event rate and occupancy must be considered. The beam flux results in a physical event rate in the CDC of $7 \times 10^7$/s assuming that 80\% of muons are stopped and decay within the MMS and the MMS acceptance is 90\%. The total event rate within a layer will be less than or equal to the total event rate, with the highest rate expected to be found in the innermost layer due to having the fewest cells. As a gas detector, the CDC has a specific dead time after a trigger in a cell, during which subsequent track passages will not trigger the cell, resulting in missed events. The miss rate can be estimated using the formula $2\tau n^2$, where $\tau$ represents the dead time and $n$ is the event rate. Therefore, the event miss rate in a layer is given by
\begin{equation}
    n_\text{miss} = N_\text{cell} \times 2\tau \left(\frac{n_\text{total}}{N_\text{cell}}\right)^2 = \frac{2\tau n_\text{total}^2}{N_\text{cell}}~,
\end{equation}
where $n_\text{total} = 7 \times 10^7$/s is the physical event rate and $N_\text{cell}$ is the number of cells in a layer. Assuming a dead time of $\tau = 100$~ns, if we want the miss rate in a layer to be less than the threshold of $n_\text{threshold} = 10^7$/s, i.e., $n_\text{miss} < n_\text{threshold}$, we find that
\begin{equation}
    N_\text{cell} > \frac{2\tau n_\text{total}^2}{n_\text{threshold}} \approx 100~.
\end{equation}
In other words, if one layer contains 100 cells, the ratio of the event miss rate to the physical event rate will be slightly more than ten percent. Therefore, a lower bound of at least 100 cells in a layer is established. For a cylindrical drift chamber with a 30 cm inner diameter, this minimum number of cells implies a maximum cell width of 9 mm in the innermost layer, setting an upper limit for the cell width in that layer.

\subsubsection{Wire configurations}
Following the objectives, we considered a CDC geometry consists of near-squared cells. One cell has a 20~$\mu$m gold plated tungsten sense wire at the center, surrounded by eight 80~$\mu$m aluminium field wires, as shown in \cref{fig:mace-CDC-wiring}. Cells are azimuthally arranged in sense layers, in which cells share the same width, length, and stereo angle. Three adjacent sense layers are grouped into a super layer, in which sense layers share the same stereo orientation, number of cells, and cell azimuth angular width. In a super layer, two adjacent layers are staggered by a half cell with to resolve the left-right ambiguity.

The current design contains 7 super layers, with 3 sense layers in each super layer, results in 21 layers in total. Each sense layer contains 124--212 cells, according to the super layer where they are located. This results in 3540 cells in total, made up of 12980 field wires and 3540 sense wires. Among the 7 super layers, 3 super layers are axial layers whose cell orients axially, and remain 4 super layers are stereo layers. The stereo orientations are alternately arranged, with the first super layer being positively twisted, the third super layer being negatively twisted, and so on, forming a UAVAUAV-twisted pattern. Detailed wire configurations are listed in \cref{tab:cdc-wire-config}. Stereo layers have a hyperbolic profile, as shown in \cref{fig:mace-CDC-concept}. This introduces a radially shrink toward the center, leaving a clear space between the next axial super layer in the region close to central $xy$ plane, and also a clear space between the last axial layer in the region close to endplate. Therefore, field wire layer are not shared between two adjacent super layers, results in cell geometries to be nearly identical.

The stereo angle ($\phi_\text{stereo}$) is defined as the the angle at which the projection of the line on the $xy$ plane spans to the origin, ranging from 430--815~mrad. A large stereo angle provides excellent resolutions along the axial direction. However, it might need extra considerations in future technical design. Cell widths range from 8--12~mm, offering good event rate tolerance under the intensive muon decay event rates. The inner radius and length of the CDC are 150~mm and 1200~mm, while the outer radius and length are 415~mm and 1600~mm, respectively. As the result, the geometric acceptance of CDC will be 88.8\%. This also specifies the geometric acceptance of the tiled timing counter (TTC) to be discussed below. The overall acceptance of the MMS system will be the combination of CDC and TTC.

\subsection{Tiled timing counter}
The goal of MACE is to identify spontaneous muonium-to-antimuonium conversion events in a vacuum. The tiled timing counter (TTC) is one of the core components of the MACE detector system, with its primary objective being to provide timing for particle tracks, particularly for the signal electron tracks. The TTC triggers the data acquisition system by detecting positrons or electrons resulting from muon decay or signal events. It is essential for the tiled timing counter to have high time resolution to accurately measure the arrival times of these particles. This capability enhances the performance of the MACE detector system in two key ways: (1) it improves the time-of-flight resolution for signal positrons by providing an accurate decay vertex time, and (2) it increases the resolution of the CDC through enhanced drift time resolution, which assists in precise track reconstruction and minimizes the number of fake tracks.

\subsubsection{Design objectives}
\paragraph{Time resolution.} A time resolution of $\mathcal{O}(100)$~ps should be achieved in TTC. TTC contributes to measuring the slow signal positron time-of-flight (TOF) by providing the anti-muonium decay time. The signal positron TOF is determined by the time difference between the decay time and the hit time of a coincident microchannel plate (MCP) event. Improved time resolution narrows the signal region, enabling better rejection of background events. Based on simulation results presented in \cref{sec:pts electrostatic accelerator}, the intrinsic spread of the signal positron TOF distribution is approximately 7~ns. The MCP can achieve a time resolution of $\mathcal{O}(100)$~ps, which is about an order of magnitude better than the intrinsic TOF spread; thus, if the TTC also meets the $\mathcal{O}(100)$~ps time resolution requirement, it could vanish the detector resolution effect to TOF measurement. To satisfy the $\mathcal{O}(100)$~ps time resolution requirement, fast plastic scintillators such as EJ-228 or EJ-230~\cite{EJ228EJ230}, with reasonably small dimensions of less than 50~mm, are suitable options. These scintillators can achieve a scintillation rise time of 0.5~ns and a fall time of 1.5~ns, as specified in the vendor's product documentation (see \cref{tab:EJ-228-EJ230-properties-table}).

\begin{table}[htbp]
\nolinenumbers
\centering
\caption{Properties of EJ-228 and EJ-230 scintillators~\cite{EJ228EJ230}.}
\begin{tabular}{ccc}
\hline\hline
Properties                                          & EJ-228 & EJ-230 \\ \hline
Light output (\% anthracene)                        & 67     & 64     \\ \hline
Scintillation efficiency (photons/1~MeV $e^-$)      & 10200  & 9700   \\ \hline
Wavelength of maximum emission (nm)                 & 391    & 391    \\ \hline
Light attenuation length (cm)                       & -      & 120    \\ \hline
Rise time (ns)                                      & 0.5    & 0.5    \\ \hline
Decay time (ns)                                     & 1.4    & 1.5    \\ \hline
Pulse width FWHM (ns)                               & 1.2    & 1.3    \\ \hline
H atoms per cm$^3$ ($\times 10^{22}$)               & 5.15   & 5.15   \\ \hline
C atoms per cm$^3$ ($\times 10^{22}$)               & 4.69   & 4.69   \\ \hline
Electrons per cm$^3$ ($\times 10^{23}$)             & 3.33   & 3.33   \\ \hline
Density (g/cm$^3$)                                  & 1.023  & 1.023  \\ \hline\hline
\end{tabular}
\label{tab:EJ-228-EJ230-properties-table}
\end{table}

\paragraph{Geometric configuration.} TTC system is designed as an array of many small scintillator tiles, where the granularity and geometric configuration are essential parameters. In the MEG II experiment, the pixelated timing counter was designed with high granularity, which enhanced overall time resolution through multiple measurements. The entire timing counter successfully completed a test run, incorporating electronic components, cooling systems, bias voltage supply, and calibration systems. During the test run, it achieved a resolution of less than 40~ps with 6--10 hits~\cite{Nishimura:2020hvl}. For MACE, the coincidence of two or three adjacent scintillator tiles by a tilt design can further improve time resolution while also reducing the level of accidental coincidence events. The specific design details will be discussed below.

\paragraph{Spatial resolution.} Improved spatial resolution in TTC can aid track finding in the CDC by providing an approximate yet known hit position. While the CDC can detect hits on wires, the precise spatial locations where tracks pass through are not directly known. The TTC partially compensates for this lack of information at the ends of the tracks. The spatial resolution is influenced by the size of the scintillator tiles: smaller tile sizes enhance spatial resolution, but this increase comes at the expense of a greater number of channels and increased technical complexity, without a significant improvement in physical performance. Furthermore, the requirement for time resolution constrains the scintillator tile size to be less than 50~mm, which is sufficient for providing an initial seed for track finding in the CDC. Consequently, a tile size of 50~mm is deemed appropriate.

\paragraph{Rate capability.} MACE will be exposed to a muon beam with a flux of $10^8~\mu^+$/s, resulting in a total event rate in the Tiled Timing Counter (TTC) of a similar magnitude. A smaller scintillator tile sizes can help reduce the pile-up events. With scintillator tiles sized at $\mathcal{O}(50~\text{mm})$ arranged in a tilting design, there are approximately 10,000 tiles in total, including at least 3,000 scintillator tile sets for multiple-tile coincidences. Consequently, the physical event rate for a single tile sets is estimated to be around $3 \times 10^4$/s, leading to a pile-up event rate of approximately $3\times 10^4$/s in the entire TTC system, assuming a resolving time of 5~ns. This accounts for only 300~ppm of the total physical event rate. Additionally, background particles in the experimental hall will lead to fake events in TTC. A conservative estimate of background hit rate is $10^4$/s per tile, which results in an accidental two-tile coincidence event rate of about $3\times 10^3$/s across the entire TTC system. This is only correspond to 30~ppm of the physical event rate. In conclusion, the rate capability would satisfy the necessary criteria.

\paragraph{Detection efficiency.} The detection efficiency of TTC affects the total efficiency of the MMS. TTC detection efficiency is contributed both intrinsically and from the geometry. The plastic scintillator has good intrinsic detection efficiency to Michel positrons and electrons, and the main efficiency loss is due to the geometry design. The TTC system should covers the whole barrel to ensure the acceptance. Furthermore, the tilt tile design also affect the efficiency to charged tracks of different signs, which will be discussed below.

\subsubsection{Conceptual design}
\begin{figure}[htbp]
  \nolinenumbers
  \centering
  \includegraphics[width=\columnwidth]{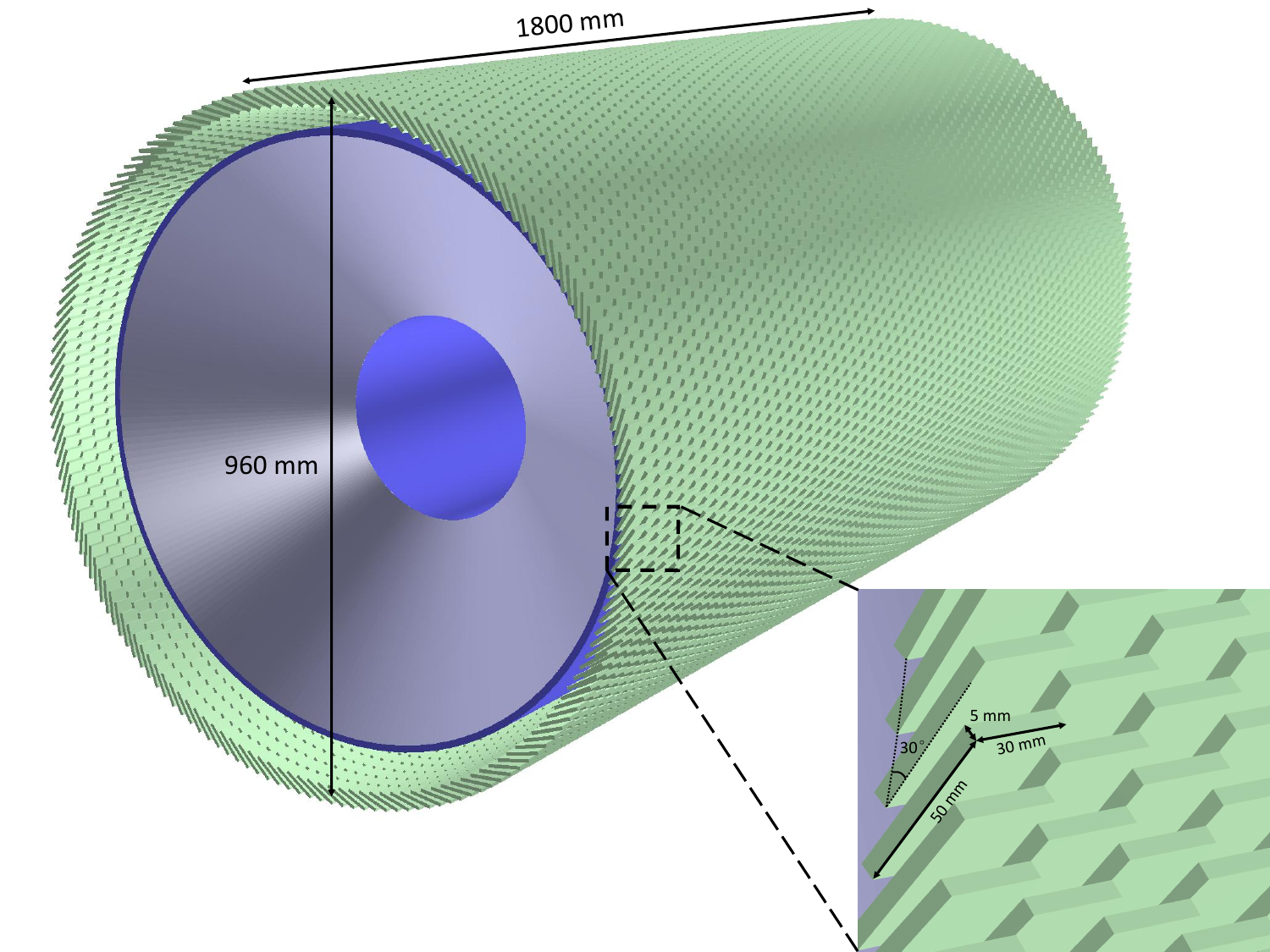}
  \caption{Perspective and zoom-in view of the tiled timing counter (green).}
  \label{fig:TTC-conceptual-design}
\end{figure}

\cref{fig:TTC-conceptual-design} illustrates the geometric design of TTC. Each unit in this design measures 50~mm $\times$ 30~mm $\times$ 5~mm. A total of 9,780 units will be organized in an overlapping arrangement across 60 layers (along the $z$-axis) and 163 columns (along the azimuthal direction) at a tilt of 30 degrees. This overlapping configuration facilitates two-to-three-tile coincidence detection, ensuring that most high-momentum charged tracks traverse at least two scintillators while minimizing the occurrence of accidental coincidences from background particles. Compared to a design with two complete layers of scintillators, this approach reduces the number of channels and reduces cabling complexity. The TTC configuration has a radius of 480~mm and a total length along the $z$-axis of $60 \times 30~\text{mm} = 1.8~\text{m}$. Consequently, the geometric acceptance of the TTC system is calculated to be 88.2\%, which is consistent with that of the CDC.

\begin{table}[htbp]
\nolinenumbers
\centering
\caption{Geometric parameters of the tiled timing counter system.}
\begin{tabular}{cc}
\hline\hline
Parameter             & Value                                 \\ \hline
Layers along the $z$-axis & 60                                \\ \hline
Tiles per layer       & 163                                   \\ \hline
Total number of tiles & 9780                                  \\ \hline
Scintillator size     & 50~mm $\times$ 30~mm $\times$ 5~mm    \\ \hline
Tilt angle            & 30$^\circ$                            \\ \hline
Total length          & 1800~mm                               \\ \hline
Radius                & 480~mm                                \\ \hline
Geometric acceptance  & 88.2\%                                \\ \hline\hline
\end{tabular}
\label{tab:ttc-geometric-parameter}
\end{table}

The tilt of each scintillator tile is intentionally oriented to the direction where the largest surface is more perpendicular to the electron tracks. Since the TTC system will be placed in an axial magnetic field, electrons and positrons will be bent in different directions. This tilt design naturally favors tracks that are more perpendicular to the large surfaces of the tiles, resulting in the positively tilted layers being more sensitive to electrons, which are the decay products of anti-muonium, of particular interest to MACE. The differences in detection efficiency can be observed in the efficiency curves shown in \cref{fig:MMS efficiency}, and its discussion is found in \cref{sec:MMS fast simulation}. The 30-degree tilt angle is calculated to balance the spacing between adjacent tiles within a single layer and the overlap of their projected areas.

Silicon photomultipliers (SiPMs) are known for their excellent time resolution and photon detection efficiency (PDE), making them well-suited for the MACE timing counter. Two SiPMs will be attached to each scintillator tile at its two smallest faces (measuring 10~mm $\times$ 100~mm). This double-ended readout reduces timing bias due to different hit positions and improves the single-tile time resolution. The Hamamatsu S13360 series MPPC is being considered due to its $\sim$\,100~ps time resolution and 40--50\% PDE~\cite{HamamatsuS13360}. The packaging scheme and reflector type for the TTC scintillators are still under investigation. In conclusion, the TTC geometry and detector design are expected to meet the requirements of MACE.

\subsection{Performance}\label{sec:MMS fast simulation}

\begin{figure}[htbp]
    \nolinenumbers
    \centering
    \includegraphics[width=\columnwidth]{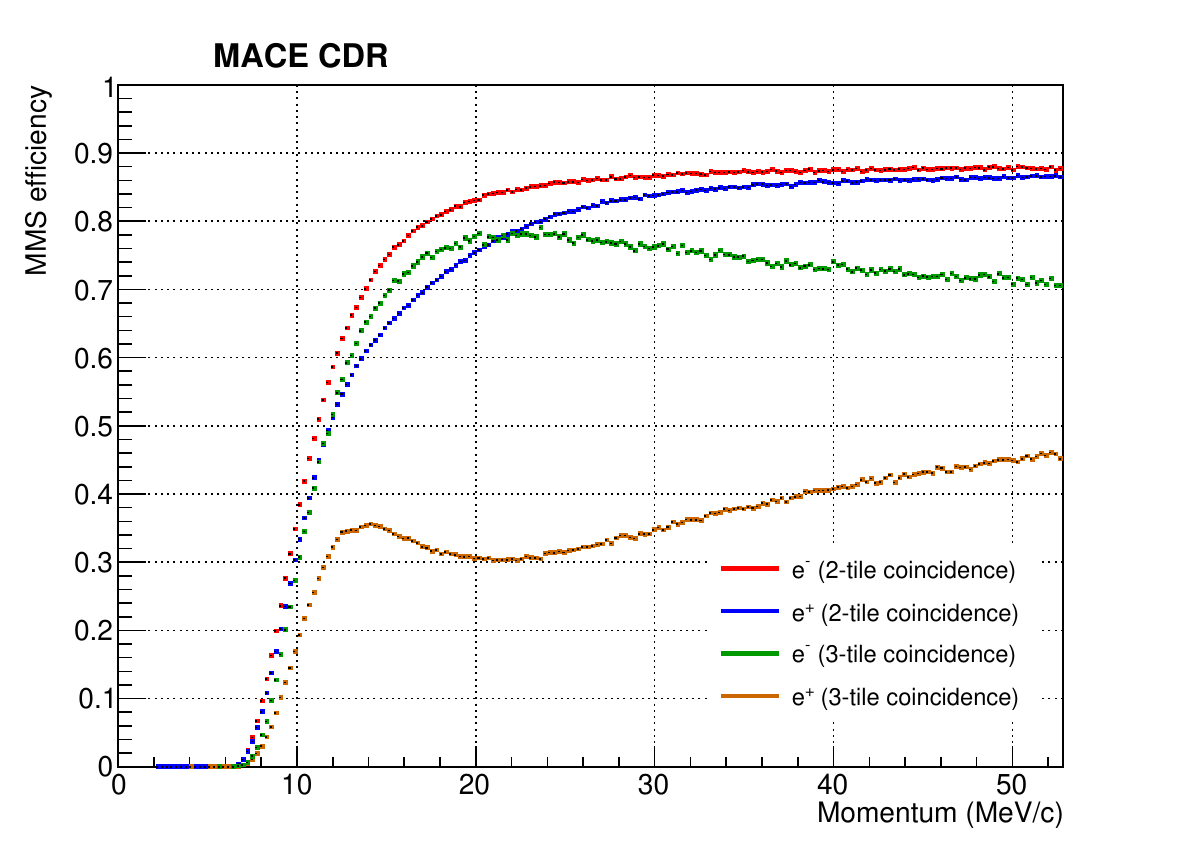}
    \caption{MMS tracking efficiencies for electrons and positrons at two different TTC coincidence thresholds. Reconstruction efficiencies are not included.}
    \label{fig:MMS efficiency}
\end{figure}

\begin{figure}[htbp]
    \nolinenumbers
    \centering
    \subfloat[Electron efficiency.]{\includegraphics[width=\columnwidth]{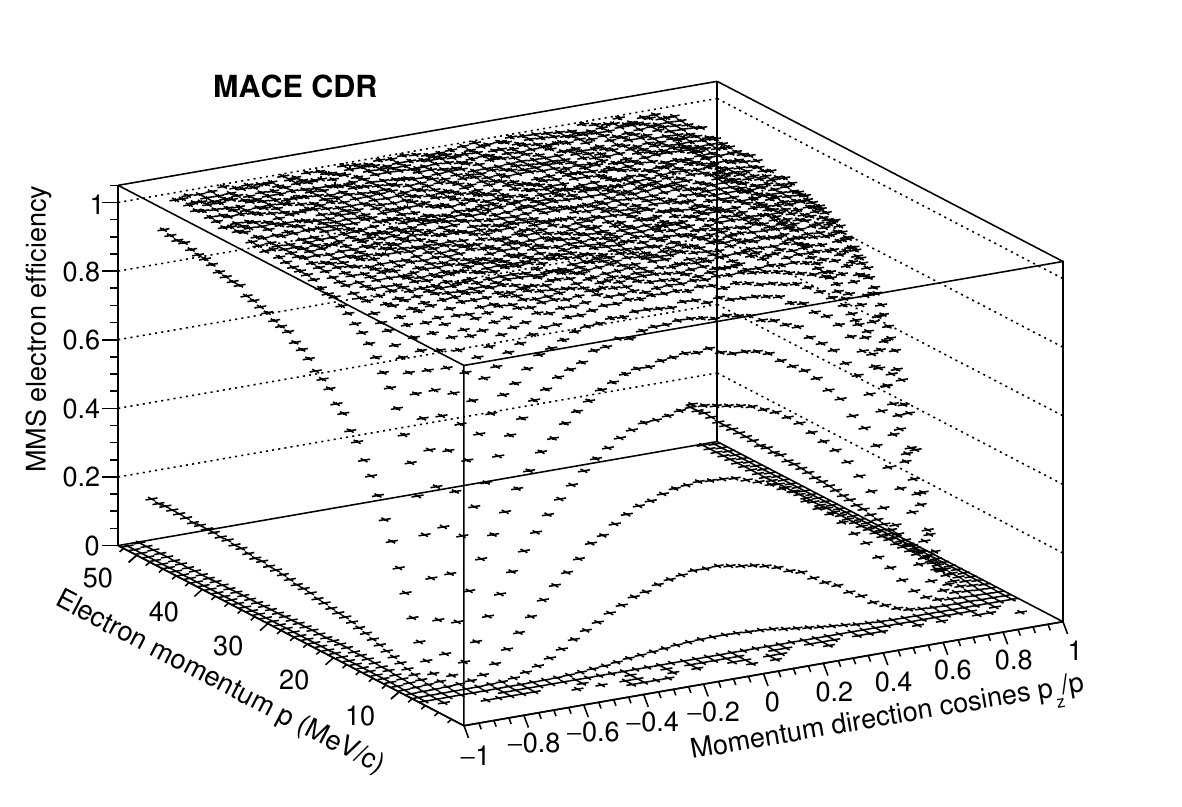}}\\
    \subfloat[Positron efficiency.]{\includegraphics[width=\columnwidth]{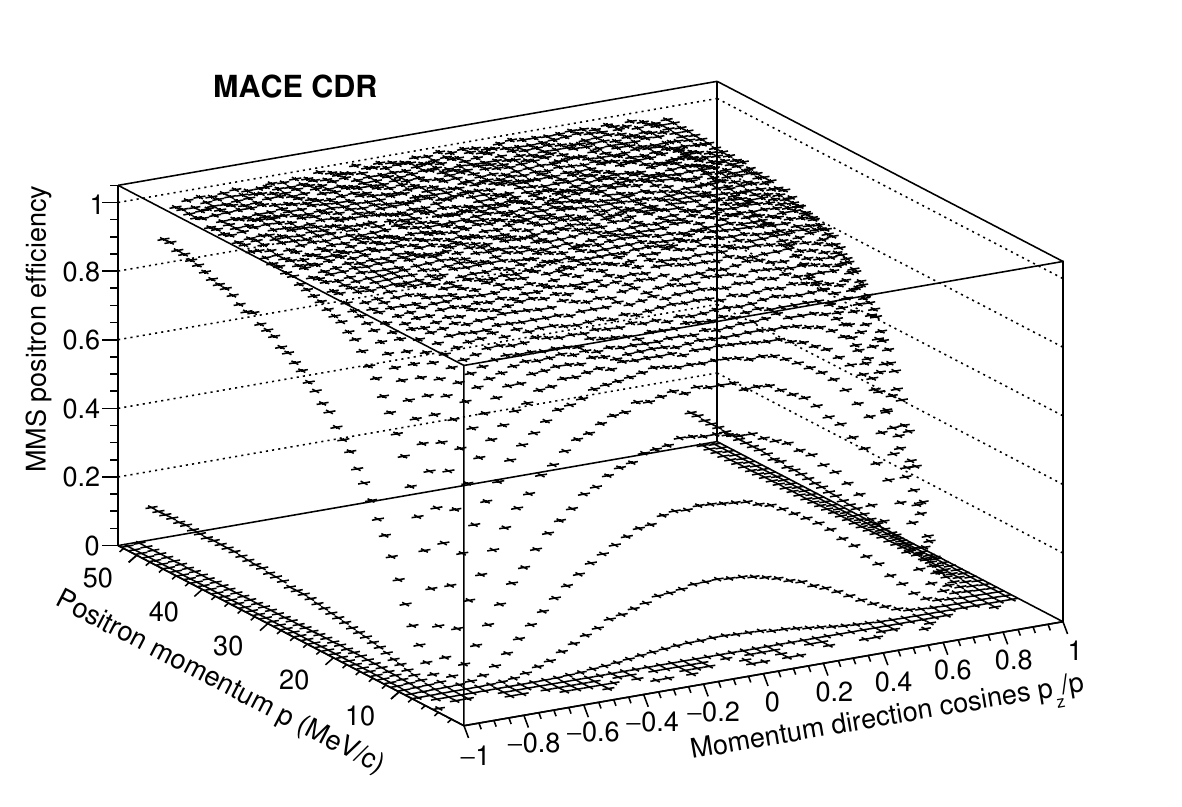}}
    \caption{MMS tracking efficiencies for electrons and positrons with 2-tile TTC coincidence. Reconstruction efficiencies are not included.}
    \label{fig:MMS efficiency 2D}
\end{figure}

In this section, we present the simulation results for the detection efficiency of the MMS system. As discussed earlier, the MMS consists of the CDC and TTC, and their geometric acceptance are consistent by design. A noteworthy characteristic of the TTC design is the overlapping arrangement of scintillator tiles, which results in differing expected coincidence probabilities for electrons and positrons due to the influence of a magnetic field. In the fast simulation, we select tracks that have a number of hit cells equal to or greater than the number of CDC layers, and that hit at least 2 or 3 TTC tiles. The full track reconstruction algorithm is not applied; therefore, the results should be interpreted in terms of the geometric effects on tracking efficiency. The complete tracking efficiency can be considered as this geometric efficiency multiplied by the reconstruction efficiency. Electrons and positrons with isotropic momentum ranging from 1~keV/$c$ to 52.8~MeV/$c$ are generated from the center of the MMS, and the efficiencies for events with 2-tile or 3-tile TTC coincidences are shown in \cref{fig:MMS efficiency}, respectively. Direction-dependent tracking efficiencies for electron and positron are shown in \cref{fig:MMS efficiency 2D}.

\begin{figure}[htbp]
    \nolinenumbers
    \centering
    \subfloat[Momentum spectrum.]{\includegraphics[width=\columnwidth]{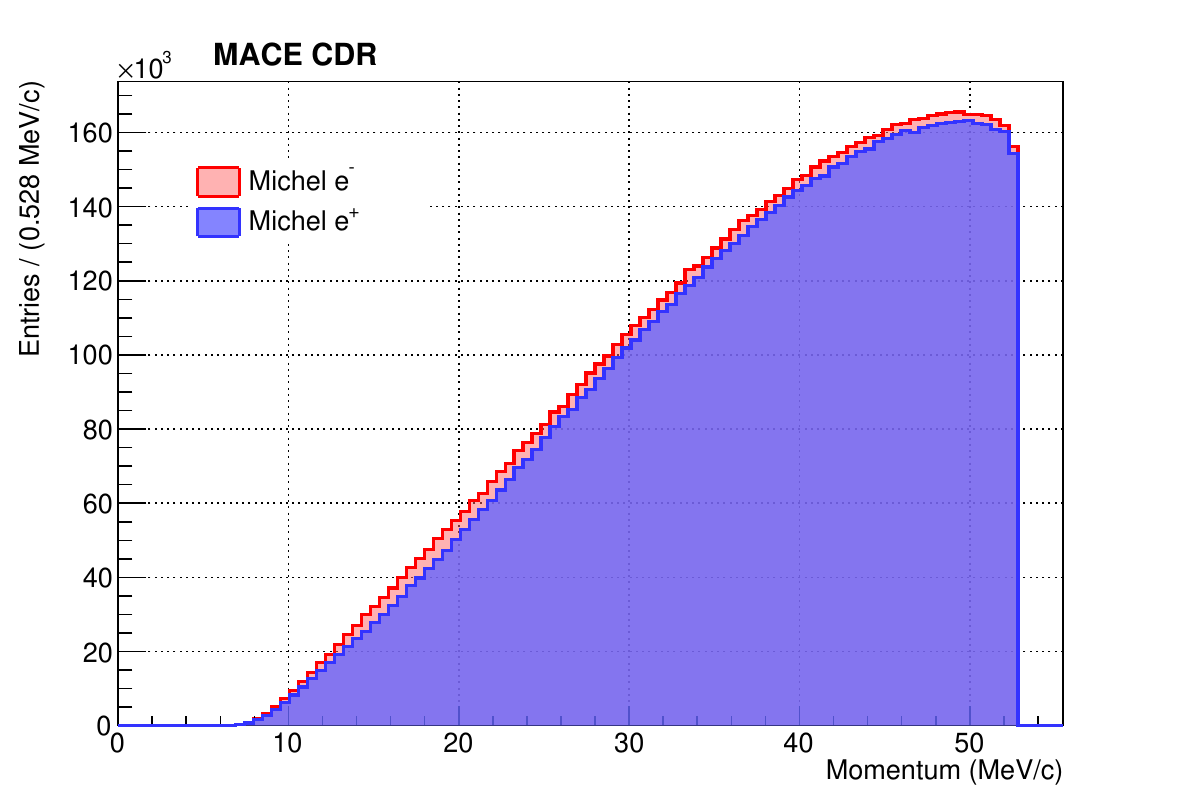}}\\
    \subfloat[Transverse momentum spectrum.]{\includegraphics[width=\columnwidth]{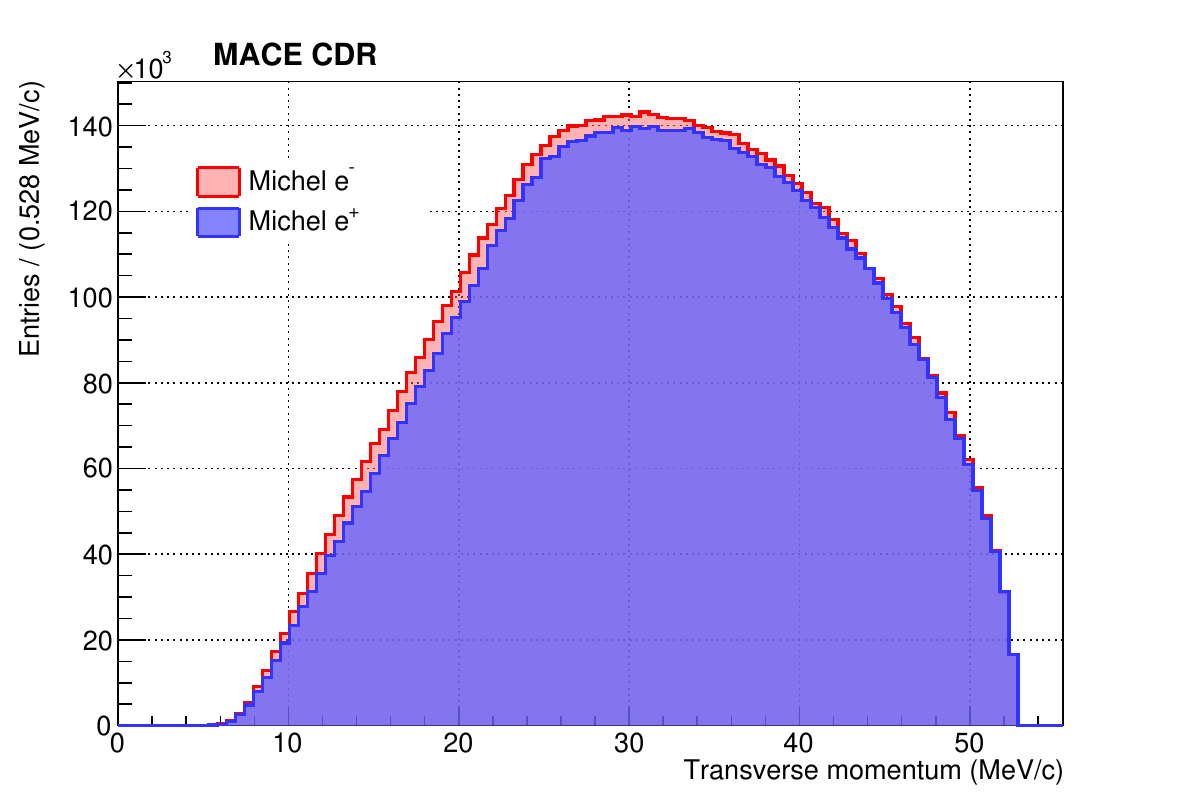}}
    \caption{Michel positron momentum spectrum in MMS without momentum resolution.}
    \label{fig:MMS momentum spectrum}
\end{figure}

\begin{table}[htbp]
\nolinenumbers
\centering
\caption{Tracking efficiency of MMS.}
\begin{tabular}{cccc}
\hline\hline
\multirow{3}{*}{Track}   & \multicolumn{3}{c}{Efficiency} \\ \cline{2-4}
 & Geometry  & Reconstruction  & \multirow{2}{*}{Total} \\
 & $\varepsilon_\text{MMS}^\text{Geom}$ & $\varepsilon_\text{MMS}^\text{Recon}$ &  \\ \hline
$e^-$ from $\ammu$ decay & 84.6\%   & $\sim$\,80\%  & 68\%  \\ \hline
$e^+$ from $\mu^+$ decay & 81.8\%   & $\sim$\,80\%  & 65\%  \\ \hline
\hline
\end{tabular}
\label{tab:MMS efficiency summary}
\end{table}

As expected, we observe that at the 2-tile coincidence threshold, the efficiencies for both electrons and positrons increase with momentum, although the increase in positron efficiency is slower and consistently lower than that of electrons. In the high-momentum region, electron and positron efficiencies converge at maximum values of 88\% and 86\%, respectively, which are limited mainly by the geometric acceptance.

For the 3-tile coincidence threshold, while the positron efficiency remains lower than that of the electron, the efficiencies are no longer monotonic with respect to momentum. For electrons, the 3-tile efficiency initially increases, reaching a maximum around 20~MeV/$c$, before gradually decreasing as momentum increases further. This initial increase can be attributed to the increasing gyration radius, while the subsequent decrease is due to the greater track curvature. If the track curvature becomes too large, an electron track is more likely to traverse 2 TTC tiles instead of 3 due to the overlapping design. In the case of positrons, the trend is similar during the initial increase, but the situation reverses at high momentum. The decrease at high momentum can be attributed to the similar reason as electron. The decrease in efficiency observed between 10 and 20~MeV/$c$ is likely because the track gyration radii become well-matched with the TTC radius and tilt angle, leading to an optimal momentum at which tracks align with the TTC tiles orientations, resulting in a lower coincidence probability.

The MMS is responsible for tracking electrons and positrons from the decays of muon, muonium, or antimuonium. The decay momenta range from 0 to 53~MeV/$c$, with a majority falling within the high-momentum region. The momentum spectra of electrons from antimuonium decay and positrons from muon decay in the MMS are shown in \cref{fig:MMS momentum spectrum}. The momentum spectra for electrons and positrons are slightly different due to differences in tracking efficiency, as discussed earlier. According to simulation results, the tracking efficiency (excluding reconstruction efficiency) for Michel electrons and positrons is 84.6\% and 81.8\%, respectively. These efficiencies are close to the full geometric acceptance, since efficiency losses are mostly contributed from the low-momentum region where the Michel spectrum is low. Assuming a reconstruction efficiency of approximately 80\%, the effective tracking efficiency for Michel electrons and positrons would be about 68\% and 65\%, respectively.

Generally, the 2-tile TTC coincidence is suitable when the accidental coincidence rate is not excessively high, allowing for optimal detection efficiency. Meanwhile, the 3-tile coincidence can be employed if accidental coincidences need to be suppressed, although it results in reduced positron efficiency, while the electron efficiency remains reasonable. In conclusion, the MMS design has been optimized for resolutions and tracking efficiencies, and the system allows for adjustable coincidence configurations for different background rates. The conceptual design of the MMS system is capable of meeting the requirements for the physical goals of MACE.

\section{Positron transport system}
The physical design of the MACE positron transport system (PTS) comprises two main components: an electrostatic accelerator and a solenoid system. This system is designed to transport low-energy atomic positrons produced in antimuonium decays, from the target region to the microchannel plate (MCP) to detect signals. The design aims to minimize background noise from high-energy particles, such as Michel positrons, to the greatest extent possible. Additionally, a position mapping algorithm has been developed to reconstruct the 2D position of the low-energy positron.

\subsection{Magnet and transport solenoid}\label{sec:mace-magnet-and-solenoid}
\begin{figure}[t]
    \nolinenumbers
    \centering
    \includegraphics[width=\columnwidth]{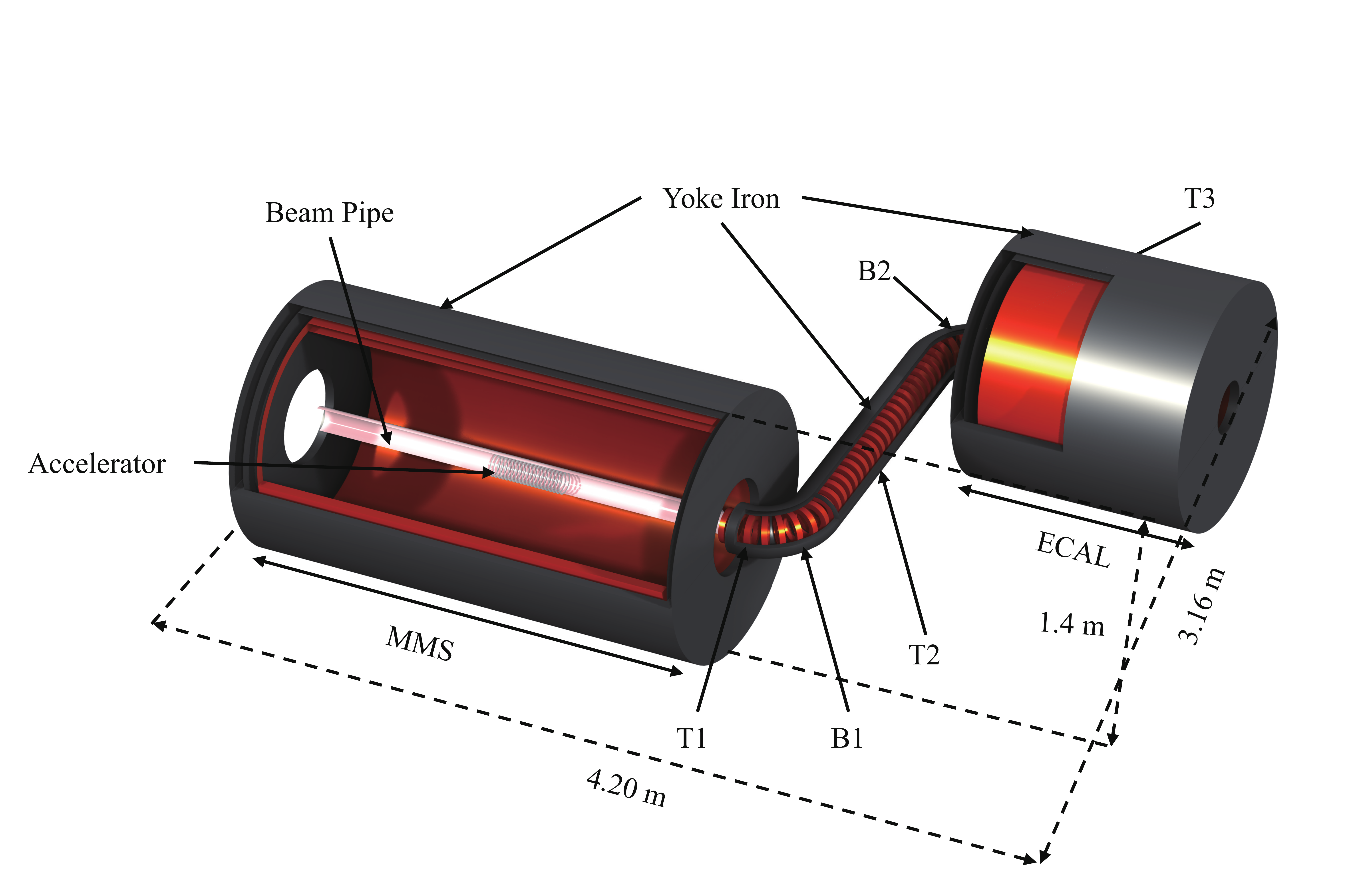}
    \caption{Concept of the MACE positron transport system (PTS). The outer grey shell represents the iron yoke, while the interior copper-colored component denotes the solenoid. Key elements and overall dimensions are indicated in the picture.}
    \label{fig:PTS_solenoid}
\end{figure}

As shown in \cref{fig:PTS_solenoid}, the solenoid system in the PTS has three components: the Michel electron magnetic spectrometer (MMS) solenoid, the transport solenoid, and the electromagnetic calorimeter (ECAL) solenoid. The MMS solenoid is a cylindrical shape, 2400~mm long, and 700~mm in radius, designed to bend charged tracks and confine low-energy positrons, surrounded by a 50~mm thick iron yoke. The S-shaped transport solenoid transports slow positrons to the MCP in the middle of the ECAL. The transport solenoid comprises five sections: three straight solenoids, T1, T2, and T3, and two toroid solenoids, B1 and B2, make up the system. Each solenoid has identical coils with a length of 30~mm, an inner diameter of 120~mm, and an outer diameter of 180~mm. T1 and T3 are 150~mm long, and T2 is 1314.5~mm long. The arcs of B1 and B2 are each 90 degrees, with a radius of 250~mm and opposite rotation directions, forming an S-shape. The transport solenoid is covered by a 30~mm thick iron yoke and connects the centers of the MMS solenoid and the ECAL solenoid. The cylindrical ECAL solenoid is 1200~mm in length and 650~mm in radius, surrounded by a 50~mm thick iron yoke, which can constrain the positrons after transportation.

\begin{figure}[t]
\nolinenumbers
\centering
\subfloat[The distribution of the overall magnetic field across the cross-section of the PTS simulation. Magnetic field lines are also depicted. The corresponding regions in (b) have been marked in the figure.]{\includegraphics[width=\columnwidth]{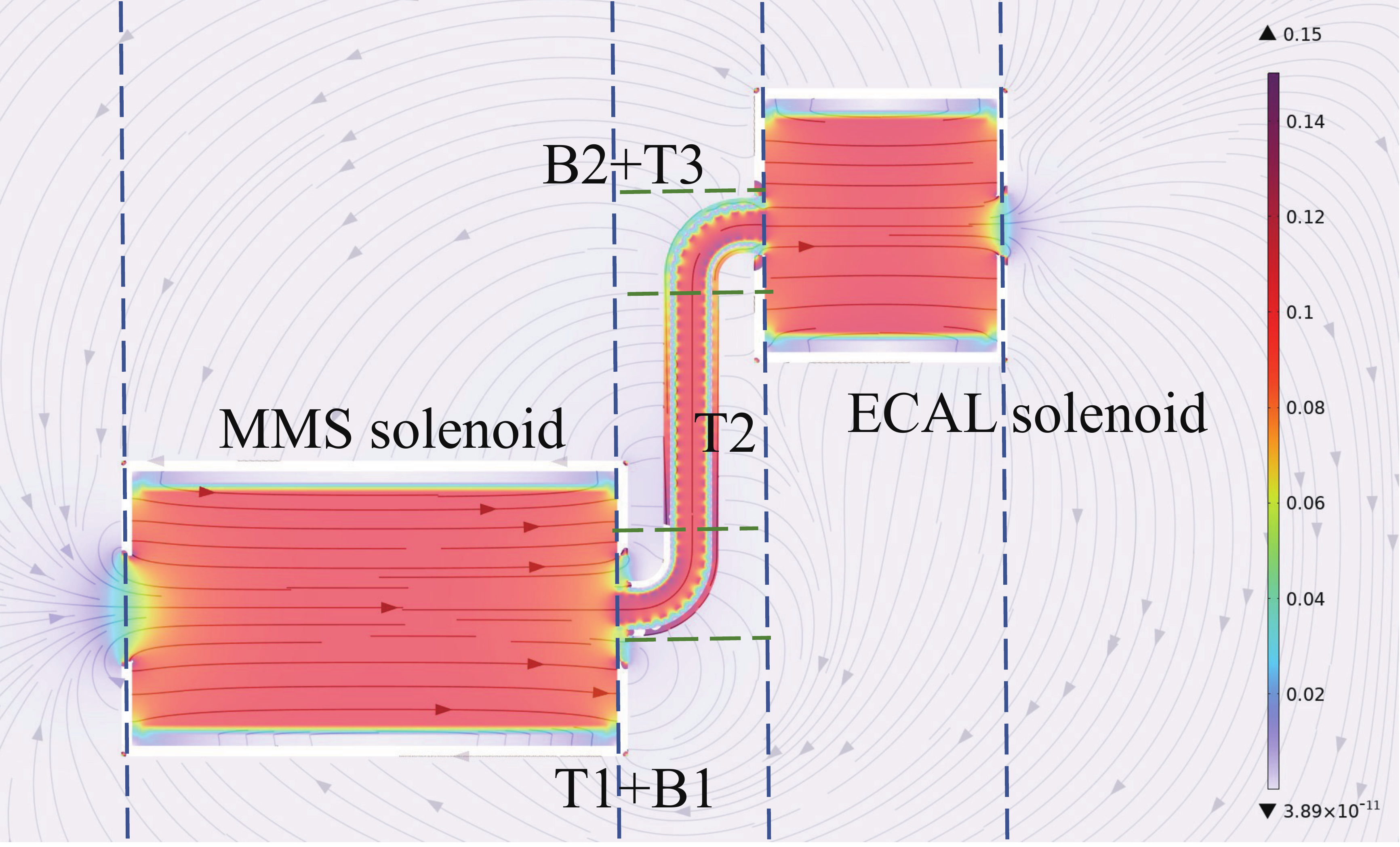}}\\
\subfloat[Distribution of the magnetic field along the central axis of the PTS solenoid. The reference line of 0.1T represents the ideal case.]{\includegraphics[width=\columnwidth]{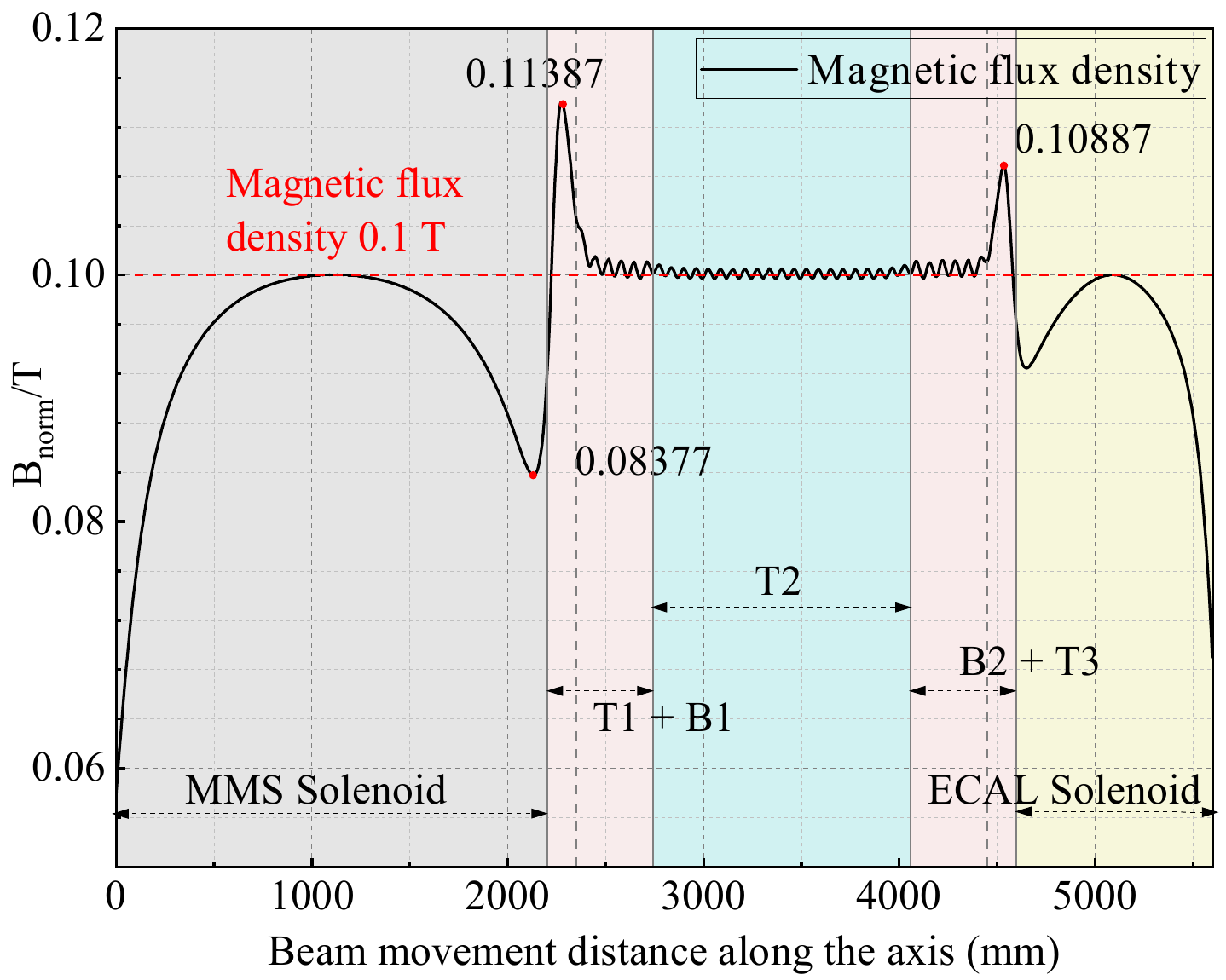}}
\caption{\label{fig:PTS_MEG}The magnetic field distribution.}
\end{figure}

The S-shape of the transport solenoid is designed to reduce background signals. Low-energy positrons with small magnetic stiffness can be considered to follow the magnetic field lines during transport. In contrast, high-energy particles, which have greater magnetic stiffness, tend to collide with the solenoid at the S-shaped turns, effectively acting as a momentum selector. Additionally, as positrons are transported to the MCP, their transverse spatial coordinates and impact times can be measured. A position mapping algorithm can then be employed to reconstruct the decay position of the particles in the target region. This way, the PTS acts as a link between the spatial and timing information of events in the PDS and that in the MMS, demonstrating how the MMS, PTS, and PDS collaborate to coincide with antimuonium events.

To transport positrons with high spatial resolution, it is crucial to maintain a consistent magnetic field in the transportation region of the PTS. This requires that the magnetic fields produced by the MMS solenoid and the ECAL solenoid be of similar magnitude to that of the transport solenoid, as this affects the magnetic field within the transport solenoid. In our experiment, we use a yoke to shield against magnetic field interference between different regions. We then control the magnetic field and its distribution by adjusting the current through the various components.

The magnetic field distribution is shown in \cref{fig:PTS_MEG}. The arrangement of the magnetic flux lines confirms that the magnetic field is uniform. By adjusting the current in the different solenoids, we can control the magnetic field to approximately 0.1~T. The decrease in the magnetic field at the end of the MMS is attributed to the fringe field of the solenoid, whereas the increase is due to the connection between the MMS yoke and T1. The interaction between the MMS and T1, along with the enhancing effect of the yoke, contributes to the magnetic field reaching 0.114 T in this region. The same applies to the T3, but since the ECAL solenoid is smaller than the MMS solenoid, the edge effect is less pronounced. Since the maximum magnetic field intensity caused by the edge effect does not exceed the predetermined 12\%, and the affected region is very short, this edge effect has little impact on the overall transmission performance, according to the simulation results.

\subsection{Electrostatic accelerator}\label{sec:pts electrostatic accelerator}
Signals in MACE are low-energy positrons produced by antimuonium decays. The kinetic energy spectrum is shown in \cref{fig:antimuounium-positron-spectrum}. The momenta of positrons are isotropic, which means they cannot be directed solely by a magnetic field toward the MCP. Additionally, their kinetic energy, typically at a few tens of ev, is insufficient to activate the MCP. However, accelerating the positrons along the axis by a few hundred volts can address both of these challenges.

\begin{figure}[t]
    \nolinenumbers
    \centering
    \includegraphics[width=\columnwidth]{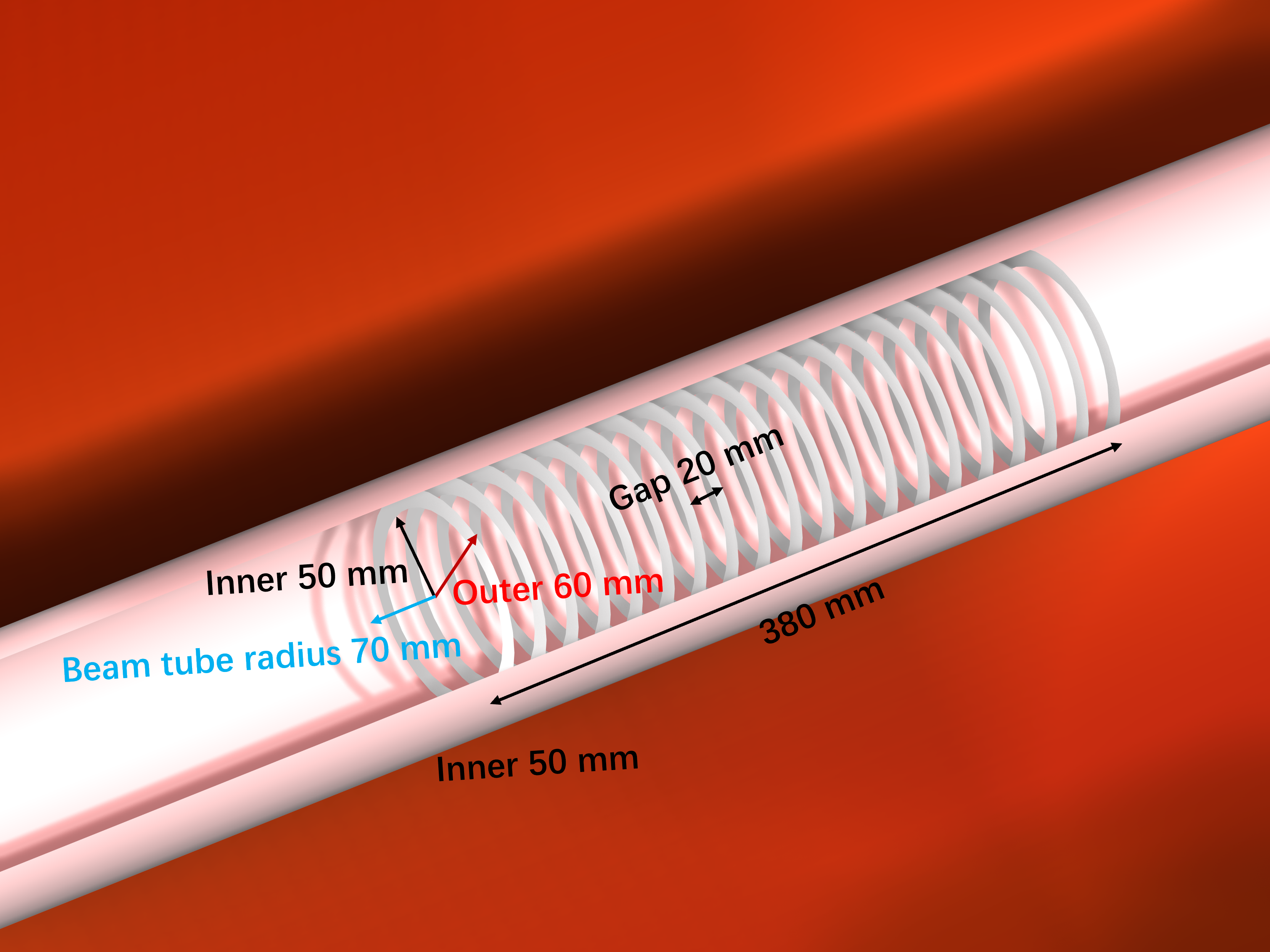}
    \caption{Concept of the electrostatic accelerator.}
    \label{fig:PTS_accelerator}
\end{figure}

In this design, we use an electrostatic accelerator for acceleration. This device generate a uniform electric field in the acceleration section, enhancing the precision of position mapping. Unlike typical designs that place the accelerator outside the beam pipe, MACE's electrostatic accelerator will be positioned inside the beam pipe, which will be grounded to prevent the accumulation of charge on it due to this high-intensity but low accelerator energy situation.

As shown in \cref{fig:PTS_accelerator}, the electrostatic accelerator of PTS consists of multiple circular thin sheets arranged at equal distances. It is situated within the beryllium beam tube along the axis of the MMS solenoid. The electrostatic accelerator has a length of 380~mm an inner radius of 50 mm and an outer radius of 60 mm, enveloping the 60~mm $\times$ 50~mm $\times$ 50~mm target region. The upper end of the accelerator is positioned 110~mm front from the center of the MMS solenoid and the target region. The highest potential is up to 780~V, the distance between adjacent accelerator sheets is 20~mm, and the potential difference is 41.6~V.

\begin{figure}[t]
\nolinenumbers
\centering
\subfloat[Simulated electric field distribution along the axis of the electrostatic accelerator where the blank area represents the region with the reversed electric field.]{\includegraphics[width=\columnwidth]{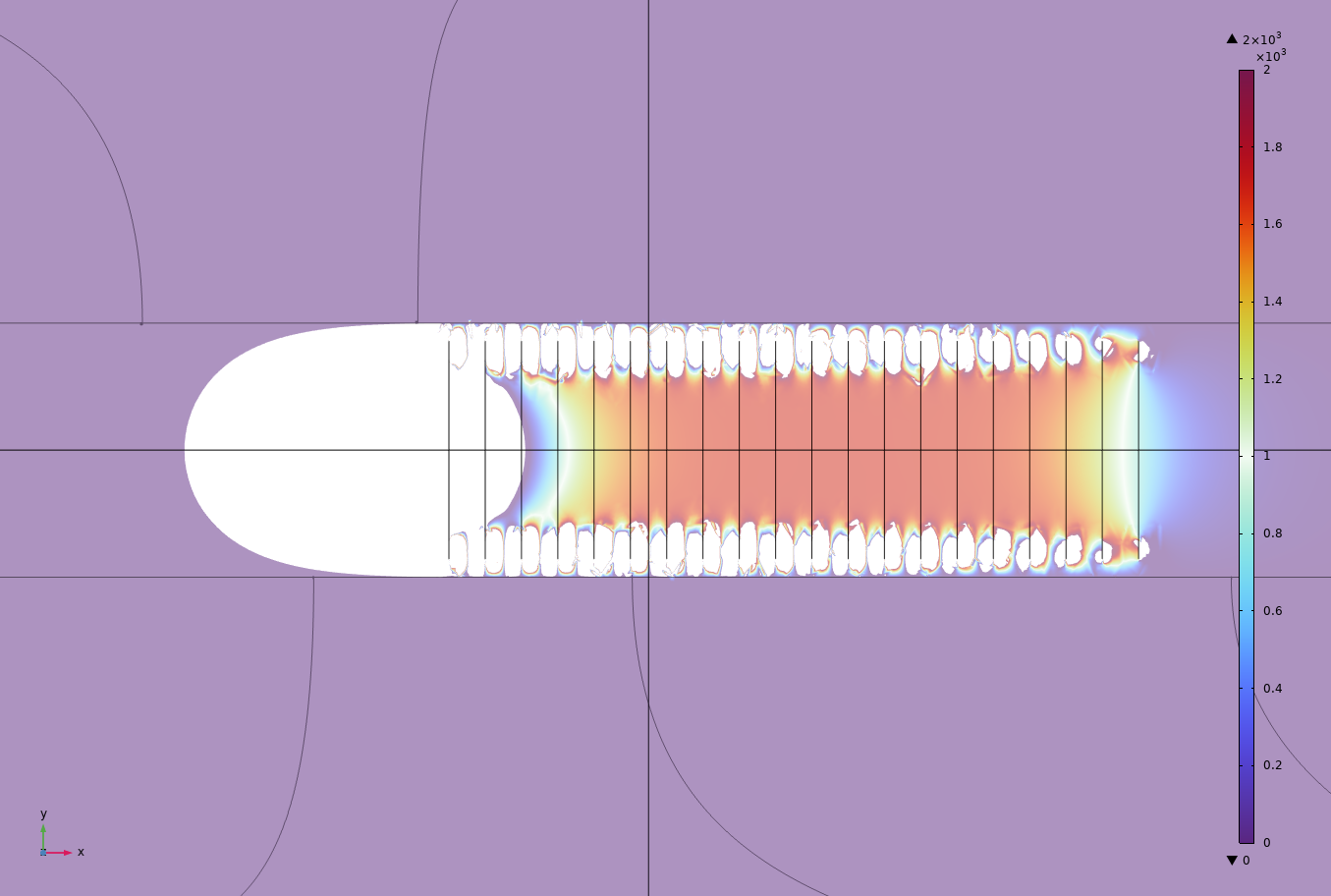}}\\
\subfloat[The electric potential distribution in the axial beam direction. The orange-yellow shaded area is the location of the target area, the black line is the electric field in the beam direction, and the red line is the electric potential.]{\includegraphics[width=\columnwidth]{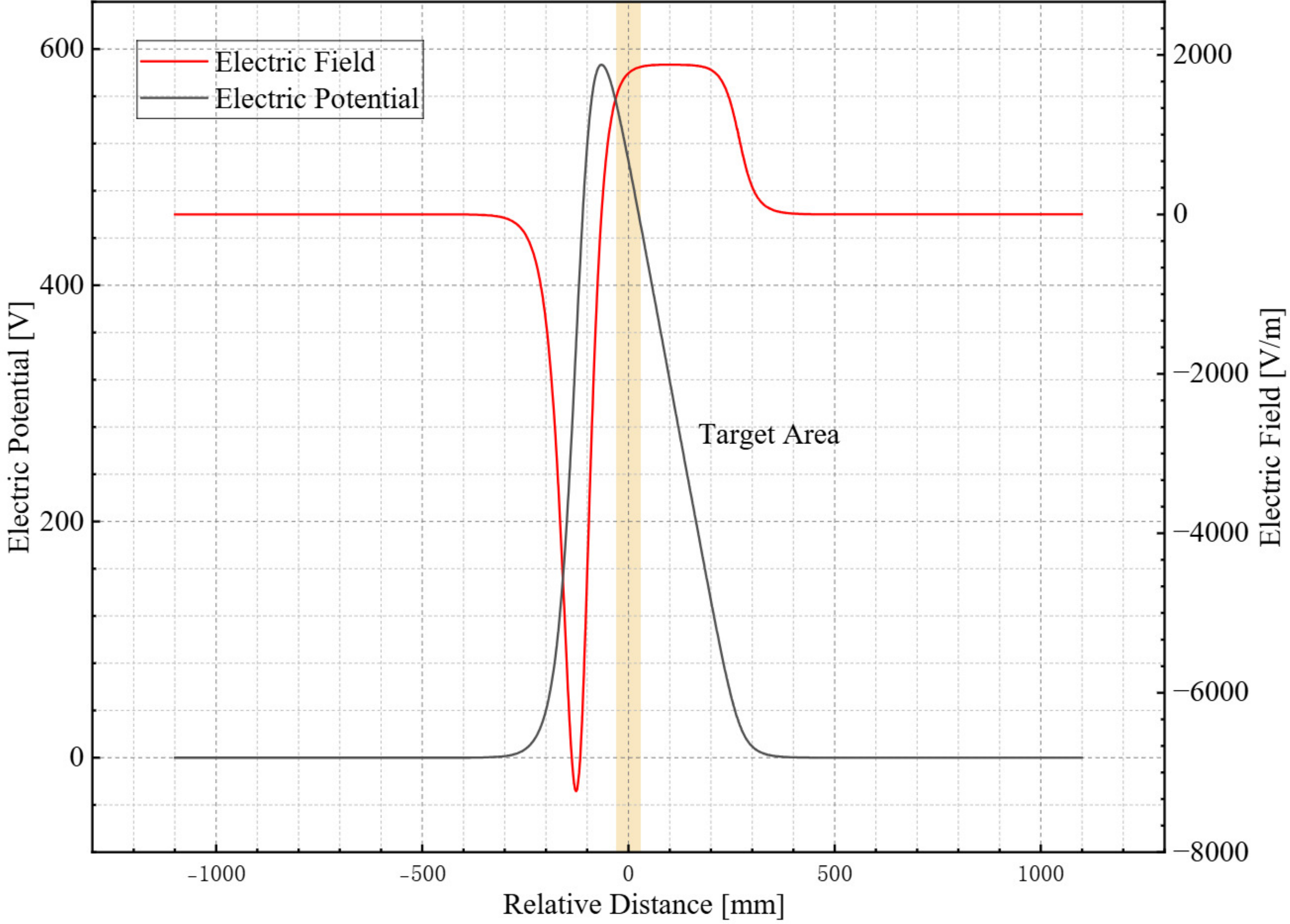}}
\caption{Accelerator electric field distribution}
\label{fig:PTS_ELE}
\end{figure}

\subsection{Performance}
\begin{figure}[t]
    \nolinenumbers
    \centering
    \includegraphics[width=\columnwidth]{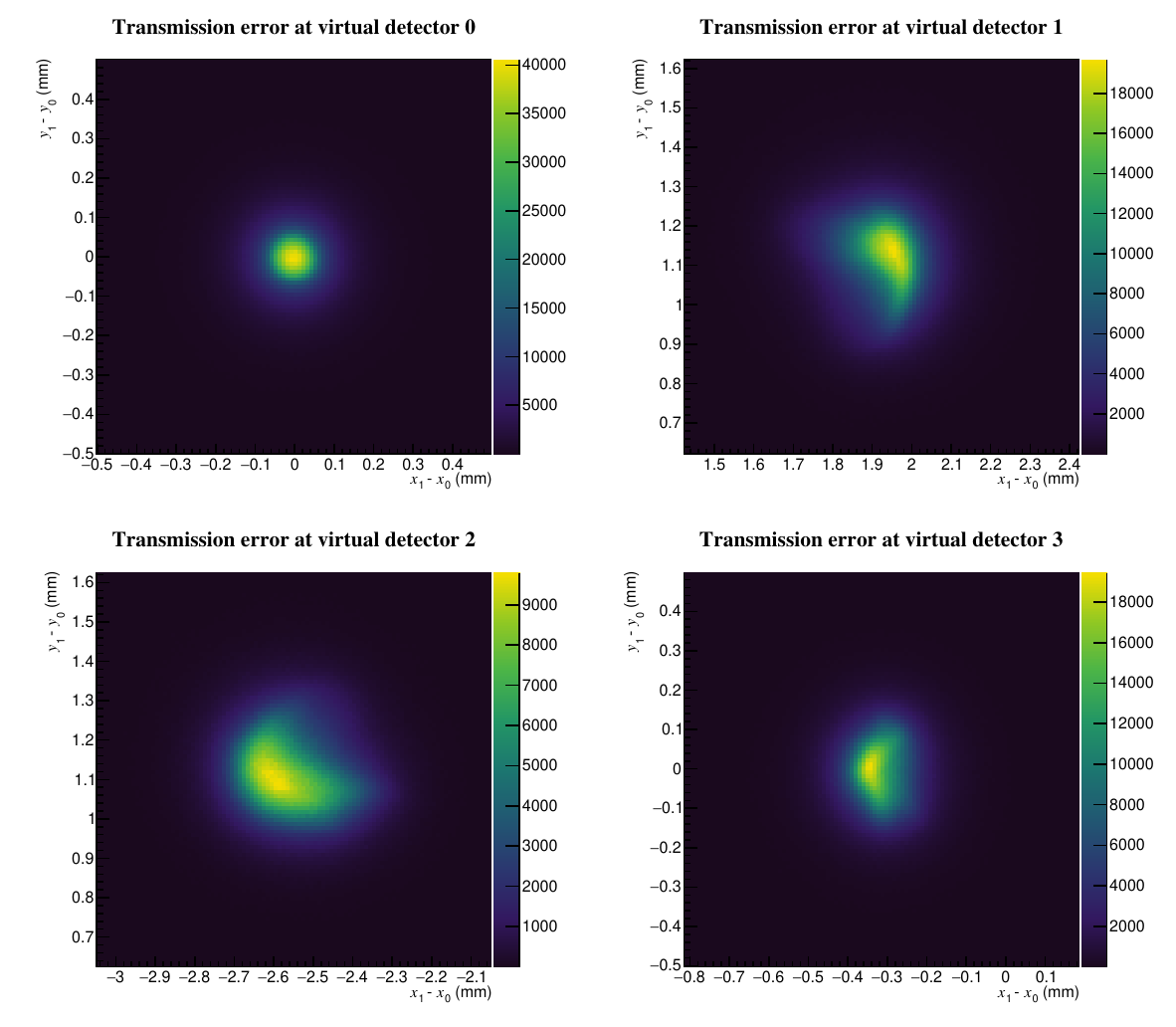}
    \caption{The relative displacement of the positron. }
    \label{fig:PTS_Space_Resolution}
\end{figure}

After the positron is transported to the ECAL solenoid, its transverse position is measured by the MCP. The transverse projection of the positron can be reconstructed using the solenoid position mapping algorithm. Distributions of the difference between the coordinates of the positron transported to the MCP and the coordinates of the antimuonium decay position in the Geant4 simulation results are shown in \cref{fig:PTS_Space_Resolution}. In the plot, $x_1$, $y_1$ represents the coordinates of the particle hitting the virtual detector, $x_0$, $y_0$ represents the initial coordinates of the particle, and the horizontal and vertical coordinates $x$ and $y$ in the picture are the difference between the two, indicating the offset of the particle during transportation, not the beam spot. Virtual detectors 0 to 3 represent the positions of virtual detectors in the beam direction: after acceleration, after B1, before B2, and in the middle of MCP. The average value of the horizontal and vertical coordinates represents the change in relative position during particle transportation. The root mean square represents the spatial resolution of particle transportation, measuring 0.088(1)~mm $\times$ 0.102(1)~mm, and the offset of the transport is 0.315(1)~mm in the $x$ direction and 0.003(1)~mm in the $y$ direction. Our simulation has $10^{7}$ initial incident particles. The signal transmission efficiency is 65.8\%, calculated from the ratio of events detected by each detector to the number of initial incident particles.

\begin{figure}[t]
    \nolinenumbers
    \centering
    \includegraphics[width=\columnwidth]{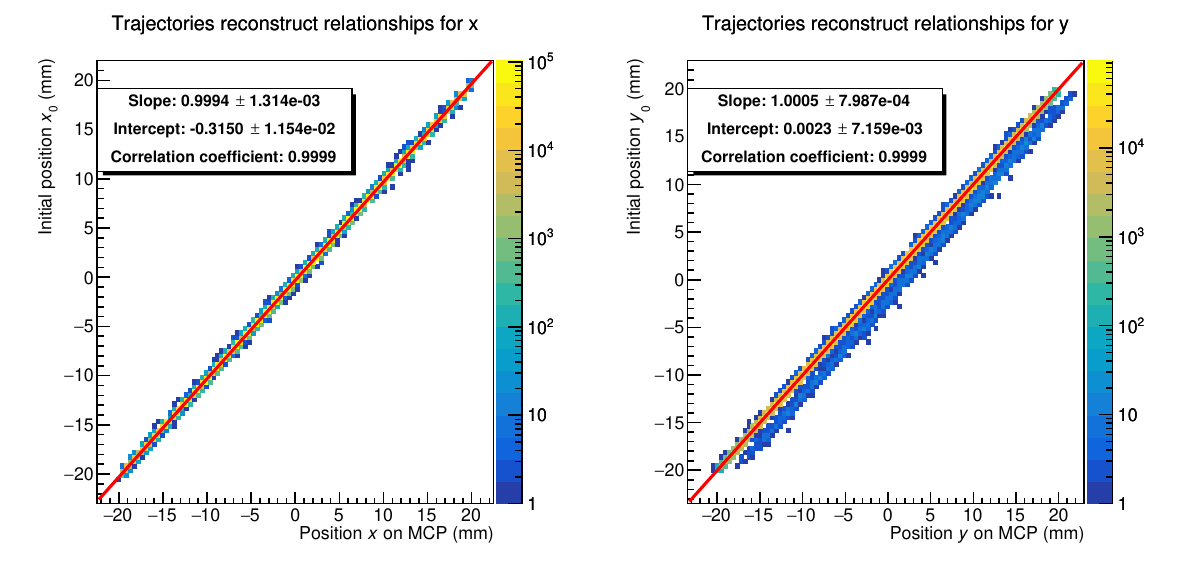}
    \caption{PTS position mapping analysis, the coordinates before transportation are marked as $(x_0, y_0)$, and after transportation are marked as $(x, y)$. The red line is the fitting line, and the parameters are given in the legend.}
    \label{fig:PTS_liner_fit}
\end{figure}

\begin{figure*}[t]
\nolinenumbers
\centering
\subfloat[The simulated particle has an isotropic initial momentum and The upper limit of the simulated energy is 1~keV]{\includegraphics[width=0.5\textwidth]{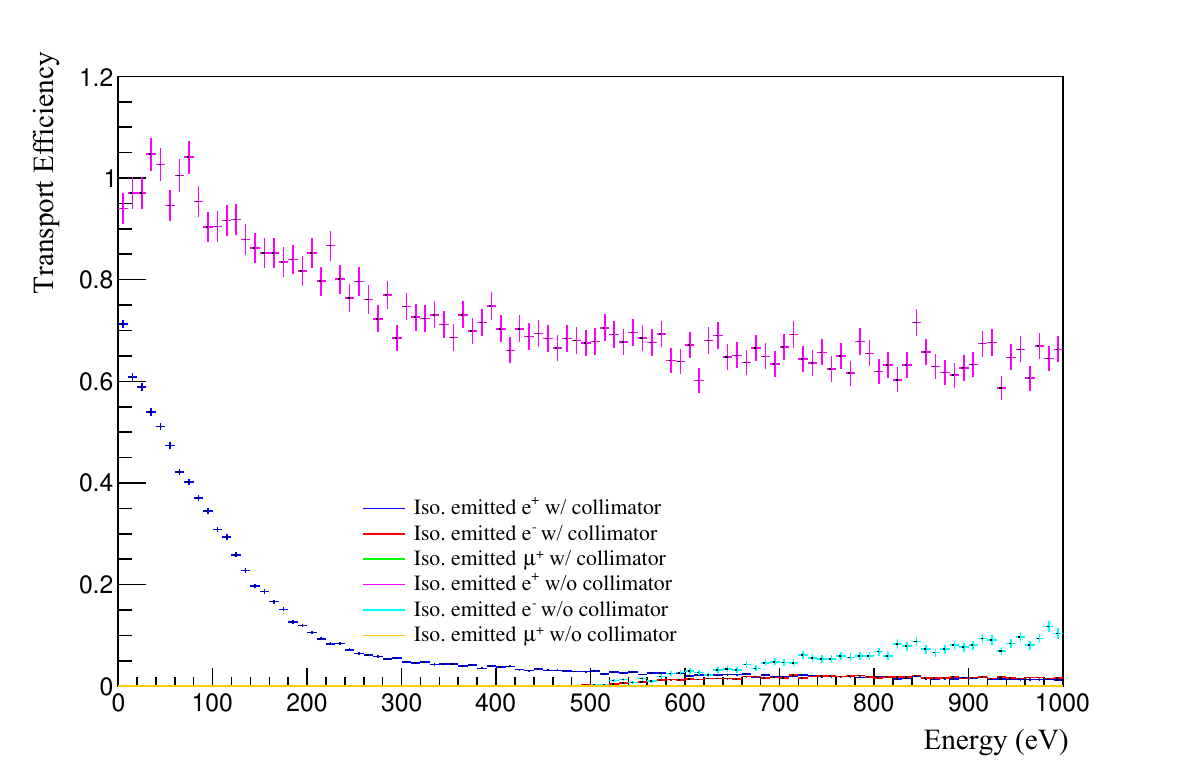}}
\subfloat[The simulated particle has an isotropic initial momentum and The upper limit of the simulated energy is 1~MeV]{\includegraphics[width=0.5\textwidth]{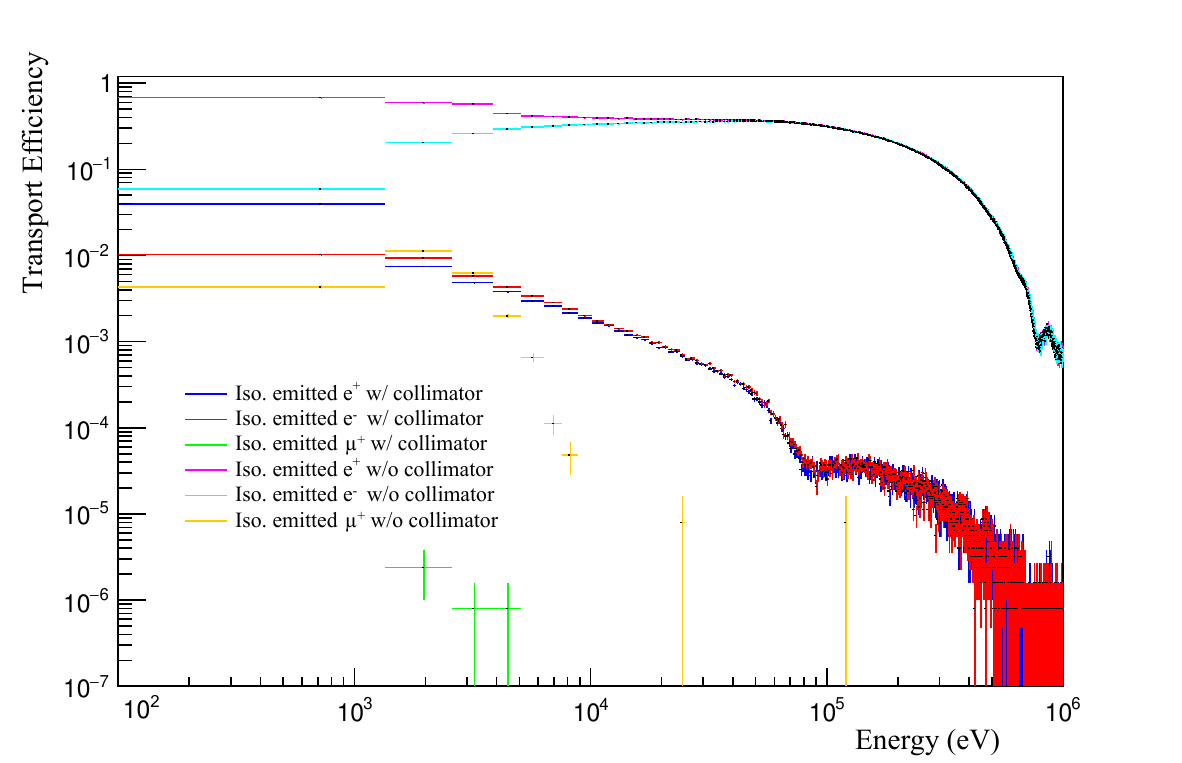}}
\\
\subfloat[The simulated particle has an initial momentum in the axial direction and The upper limit of the simulated energy is 1~keV]{\includegraphics[width=0.5\textwidth]{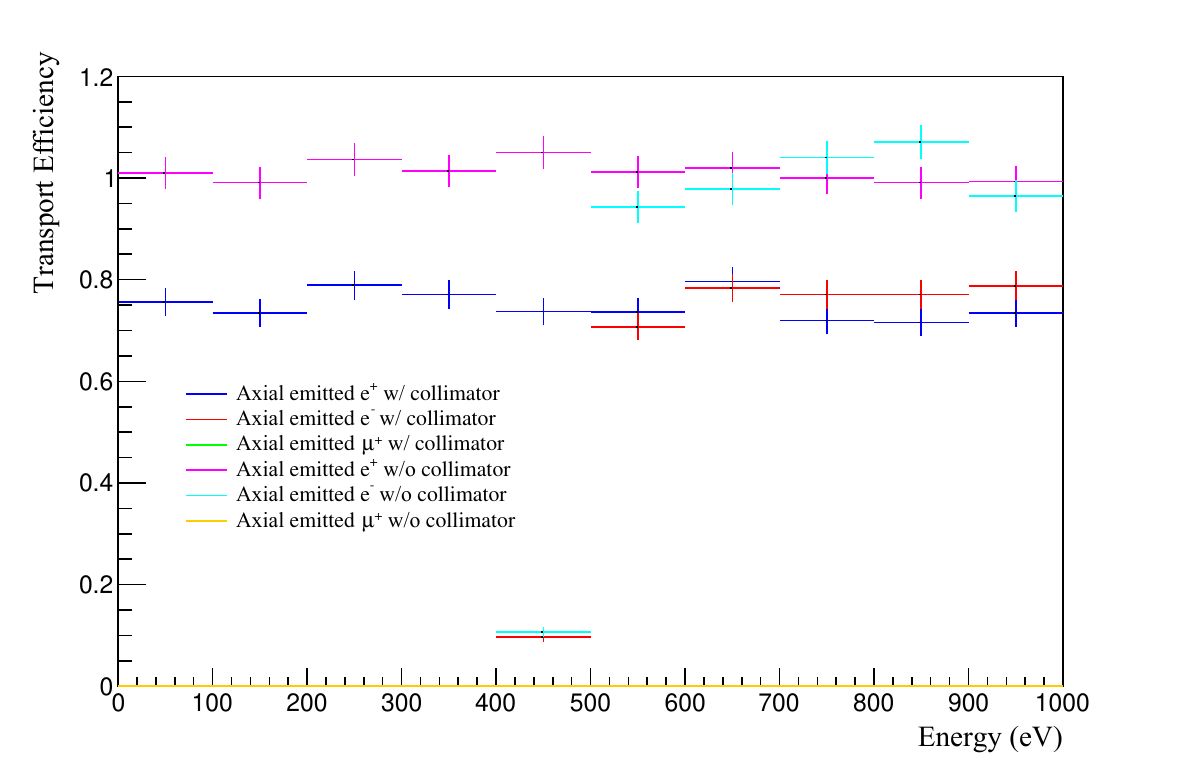}}
\subfloat[The simulated particle has an initial momentum in the axial direction and The upper limit of the simulated energy is 1~MeV]{\includegraphics[width=0.5\textwidth]{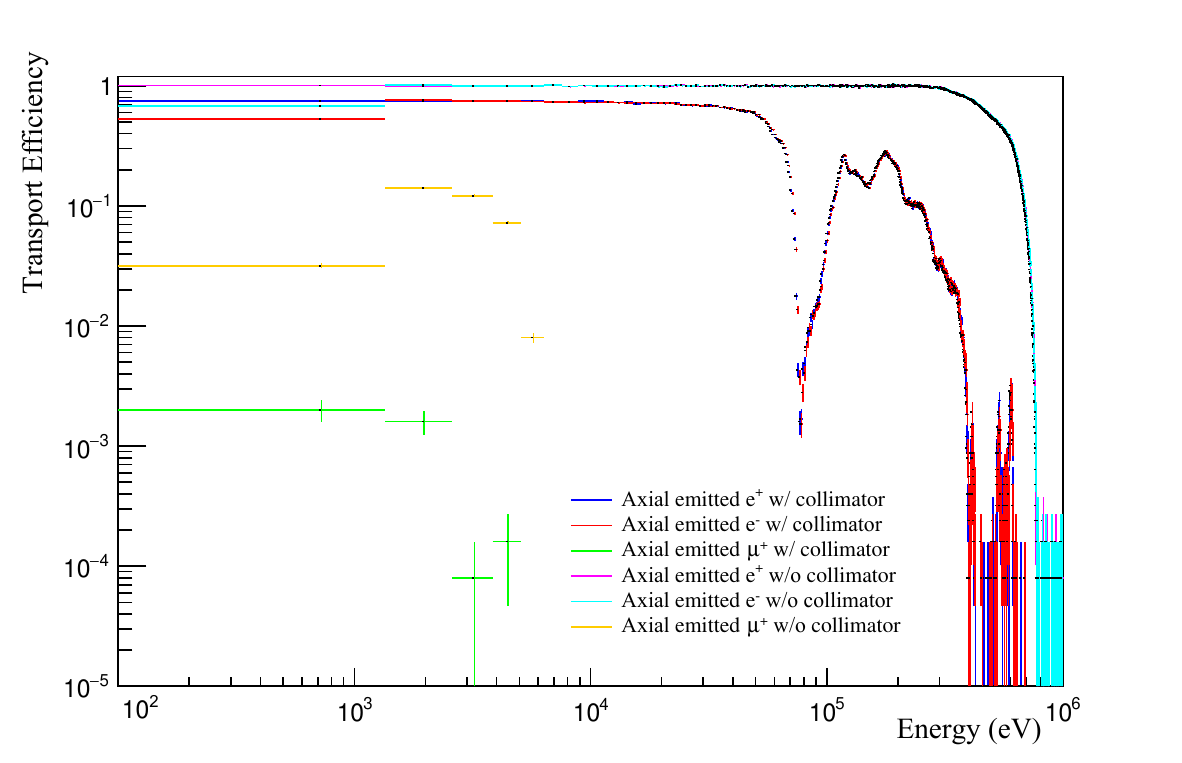}}
\caption{Simulation results of signal transmission efficiency for different charged particles at different energies with and without a collimator.}
\label{fig:PTS_EFF}
\end{figure*}

Beam vertical plane position mapping is performed on the 2D position detected by the virtual detector 3. The position mapping of the positron before and after transportation is shown in \cref{fig:PTS_liner_fit}. In the analysis, the horizontal and vertical coordinates of the particles are treated separately. It can be seen that the coordinates of the transported highly correlate with corresponding initial coordinates, while the calculated coordinates of the transported particles have a low correlation with other initial coordinates, such as $(x, y_0)$, with a correlation coefficient less than 0.01. Therefore, it can be approximately considered that the position mapping function before and after the transportation of the positron is a linear relationship, and the slope is approximately 1, with almost no stretching, and only a partial offset in the vertical beam direction. This offset is consistent with the offset shown in \cref{fig:PTS_Space_Resolution}, so in the current simulation stage, it can be considered that the coordinate transformation before and after the positron transportation is only a position offset, with an offset of 0.315~mm in the $x$ direction and 0.003~mm in the $y$ direction.

Since the low-energy positrons are produced in the whole target region, the acceleration potential and the flight distance vary. Therefore, the overall flight time, the time it takes for the positrons to be transported from the target region to the MCP, is about 322.4~ns, and the flight time width is 6.9~ns.

As shown in \cref{fig:PTS_EFF}, the PTS system simulates the presence or absence of collimator and different charged particles $e^+$, $e^-$, $\mu^-$, $\mu^+$, $\pi^+$, and $\pi^-$, respectively. Since $\pi^+$ and $\pi^-$ promptly decay and cannot be transported to MCP, they are not shown in the figure, and the number of $\mu^-$ and $\mu^+$ events is too small, so only $\mu^+$ is given as a reference. It can be seen that the transmission efficiency is very low for charged particles other than electrons or positrons, and the PTS system is highly selective to energy. Since the high-energy charged particles in MACE are mainly positrons from muon decay and their secondaries, the momentum direction can be regarded as isotropic. The following figure can be used to refer to the PTS system's ability to select momentum.

It can be seen that the transmission efficiency of particles decreases significantly when the collimator exists, especially for high-momentum particles, the filtering capability of the collimator is very outstanding. For positrons with small initial momenta (initial momentum $\leq$~100~eV), the transmission efficiency is $\sim$\,75\%. At this kinetic energy, the transmission efficiency of electrons is extremely low because the momentum cannot overcome the reverse electric field. Low-energy $\mu^+$ cannot be transported either, due to the limited lifetime, so the transmission efficiency is also extremely low. When the initial kinetic energy is greater than 500~eV, the kinetic energy of electrons can overcome the electric field of the electrostatic accelerator, and the transmission efficiency curves of positrons and electrons gradually approach each other. The initial momentum of $\mu$ is sufficient to allow some of them to reach the MCP before decay, but the maximum transmission efficiency does not exceed 1\%. When the initial kinetic energy continues to increase, the transmission efficiency of $\mu$ drops to $10^{-6}$ at 20~keV, and the transmission efficiency of electrons and positrons also drops to 1\%.

In summary, the physical design of the PTS effectively meets the requirements of the MACE system. The PTS achieves a high spatial resolution of 0.088(1)~mm $\times$ 0.102(1)~mm, showcasing excellent transmission capability. In terms of background suppression, the primary sources of background are Michel positrons from muonium decay and muons in the beam. When the kinetic energy exceeds 20 keV, the transmission efficiency for Michel positrons drops to $10^{-2}$, while that for muons is $10^{-6}$. Consequently, the current design demonstrates robust momentum selection capabilities, maintaining a signal transmission efficiency of 65.8\%. The next step will involve exploring the engineering implementation of this design.

\section{Positron detection system}

\subsection{Microchannel plate}

Nowadays, MCP, as a type of electron multipliers based on the secondary electron emission, has been widely used in high energy physics~\cite{PANDACherenkovGroup:2019bmc, Belle-II:2010dht} and other relevant fields~\cite{X-ray-camera:1986, astrophysics:2007, X-ray:1994, LEHMANN2004228, Preston_2011, Williams1998, Choong2010, Heejong2009} due to its excellent spatial and time resolution. MACE, as an experiment searching for the rare process beyond the standard model, also uses MCP as a part of the positron detector, expecting to better suppress the background events using the precise position and time information. After the decay of antimuonium, positrons will be transported along the solenoid to the PDS and annihilate within the material once hitting the MCP with producing a pair of 511 keV gamma rays, which will soon be detected by the electromagnetic calorimeter. Before the annihilation, some initial free electrons will be produced by the ionization of the positron, amplified inside the channels of MCP and finally detected. The energy of the positrons to be detected is several hundred electron volts and the required time resolution is about $100\sim200~{\rm ps}$. The spatial resolution of MCP depends mainly on the distance between the channels, which is commonly $<10~{\rm \mu m}$. The strict requirement of the time resolution excluded most of the mass-produced MCPs, so that further R\&D is needed. To make the design and optimization more convenient, a simulation program for MCP has been developed, which is still independent so far and is possible to be combined into the software system of MACE in the future. However, we are still facing challenges before the combination, such as, the heavy burden to compute the electron avalanche.

The simulation of the electron multipliers like MCP has always been difficult. On one hand, the millions of electrons produced in the electron avalanche will definitely be a challenge for the computing if simulating them just track-by-track. On the other hand, the precision of the simulation cannot be guaranteed if using some phenomenological models due to the necessary approximation and simplification. Furthermore, the mechanism of the secondary electron emission can still not be described so accurately due to the large uncertainty of the dynamics of particles and the fluctuation of statistics. So far, much effort has been put into the simulation of the electron multiplier devices all over the world based on multiple models~\cite{Ozok2017, Wu2008, Cheng2009, Xie2018}. Several commercial softwares are developed for the general simulation of electron multipliers including CST Studio $\rm Suite^{TM}$~\cite{CST}, $\rm SIMION^{TM}$~\cite{SIMION} and so on. However, they are not so easy to be combined into the software system of MACE and difficult to be modified according the real case of the experiment, so that we developed an open-source software based on Geant4~\cite{GEANT4:2002zbu} using Furman model~\cite{Furman:2002du}.

\begin{figure}[htbp]
        \nolinenumbers
        \centering
        \includegraphics[width=0.85\columnwidth]{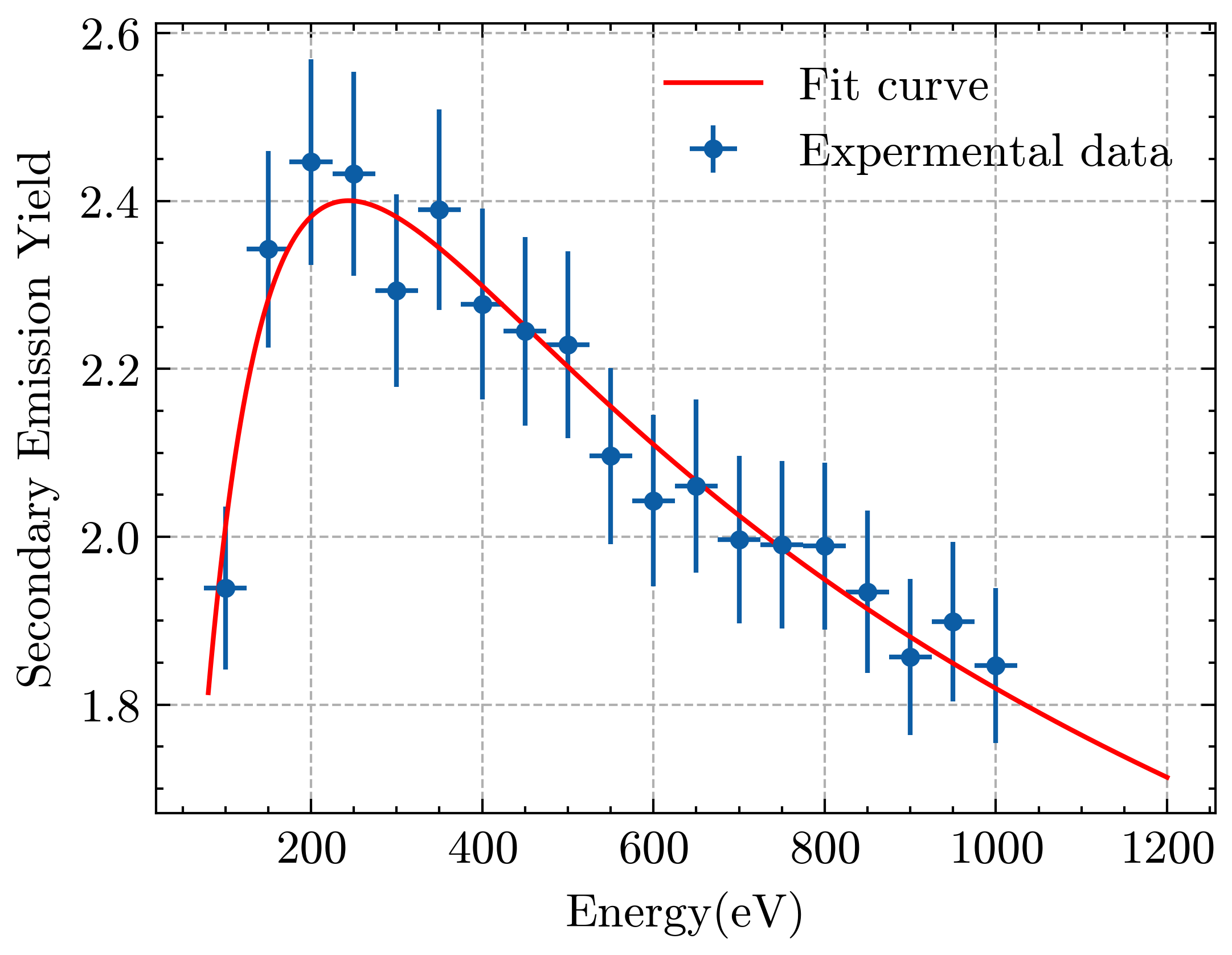}
        \caption{The measured SEY of lead glass overlaid with the best fit.}
        \label{fig:fit_SEY}
\end{figure}

The implementation of the Furman model into Geant4 can be found for detail in reference~\cite{Miao:2023qud, Peng:2023cjp}. The difficulty is how to get the real secondary electron yield (SEY) and the energy spectrum. In the simulation, the SEY distribution with respect to the energy of the incident electrons for lead glass are obtained by fitting to the measured data as shown in \cref{fig:fit_SEY}.

\begin{figure}[htbp]
        \nolinenumbers
        \centering
        \includegraphics[width=\columnwidth]{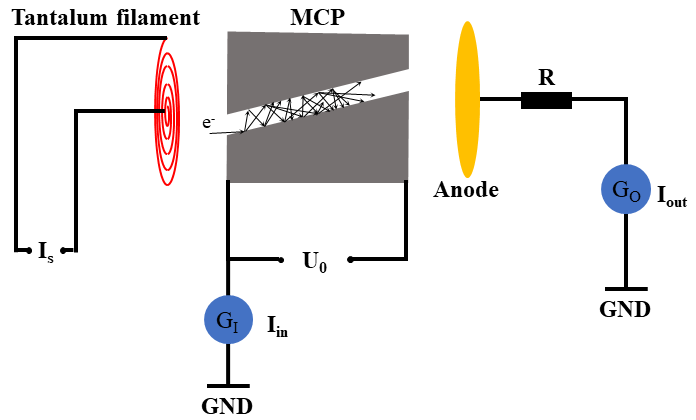}
        \caption{The schematic diagram of the device used to measure the MCP gain.}
        \label{fig:gain_sketch}
\end{figure}

Using the device shown in \cref{fig:gain_sketch}, the MCP gains at various applied voltages are measured. As illustrated in \cref{fig:gain_sketch}, a Tantalum filament, the MCP to be measured and an anode are placed sequentially inside a vacuum device. A voltage of $U_0$ is applied to the MCP. When the system is working, a current $I_{s}$ flows through the tantalum filament and some of the electrons emitted from the filament will enter the input pore of the MCP, of which the input current $I_\text{in}$ will be measured by the galvanometer $G_{I}$. After the avalanche amplification in the MCP channel, the output current $I_\text{out}$ is measured by the anode and the MCP gain at the applied voltage is then calculated by
\begin{equation}
        \label{equ:gain_formula}
        G = \frac{I_\text{out}}{I_\text{in}}
\end{equation}

\begin{figure}[t]
    \nolinenumbers
    \centering
    \subfloat{ \includegraphics[width=\columnwidth]{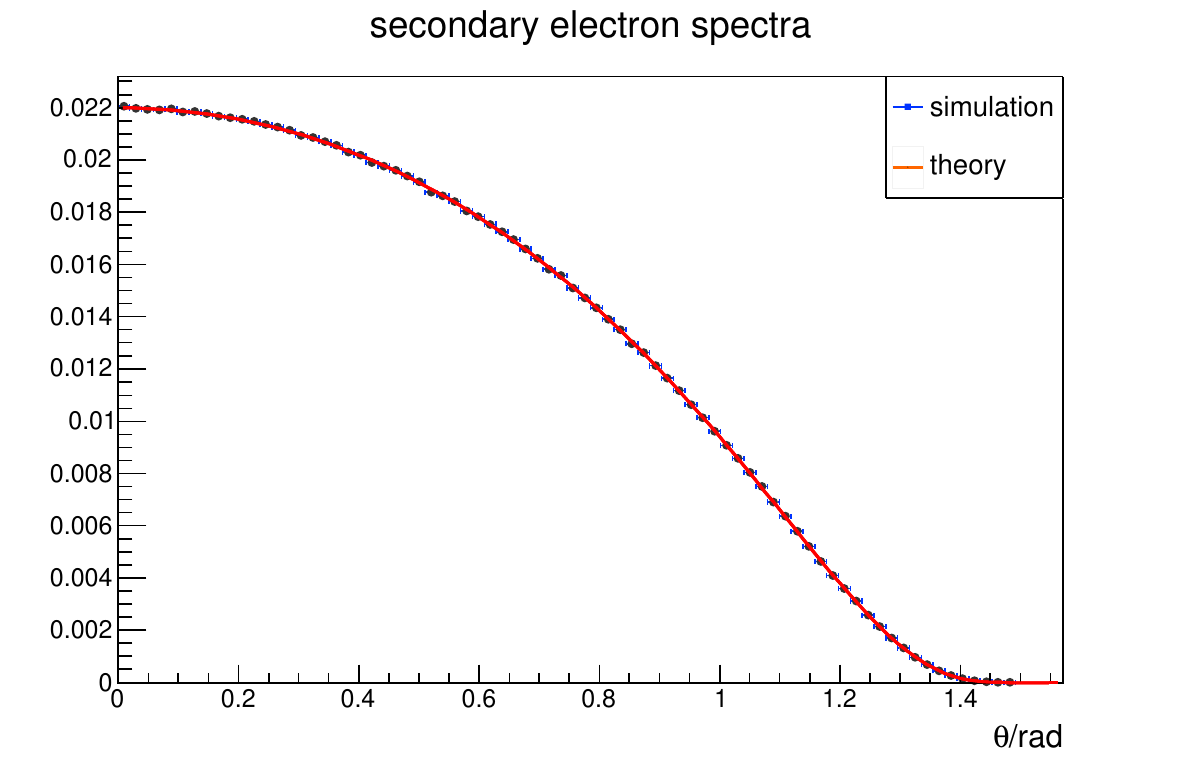} }\\
    \subfloat{ \includegraphics[width=\columnwidth]{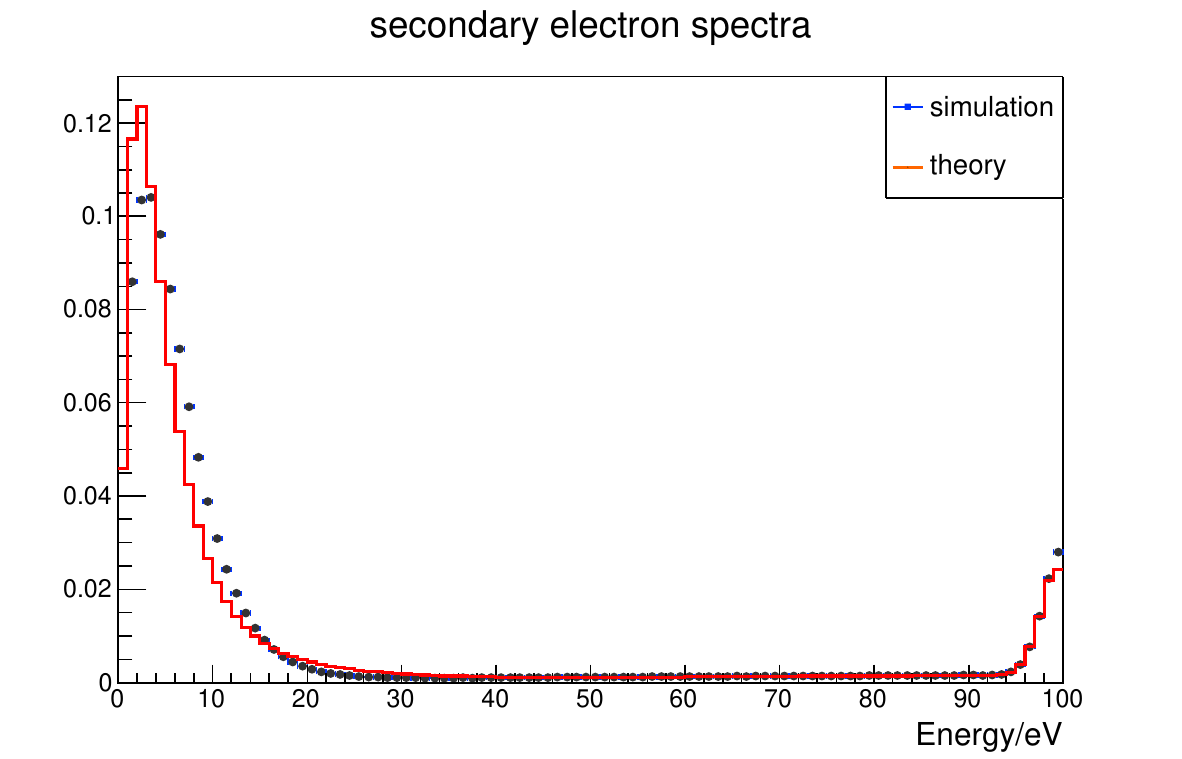} }
    \caption{The angular and kinetic energy distribution of the secondary electrons produced by an incident electron with 100 eV energy. The red solid line is the theoretical prediction and the black points with error bars are the simulation results.}
    \label{fig:comparison_MCP}
\end{figure}

The energy spectrum of the secondary electrons for lead glass is adjusted from the measured spectrum of other materials and iterated in the simulation to get close to the reality. \cref{fig:comparison_MCP} shows the angular distribution and kinetic energy distribution of the secondary electrons when the energy of incident electron is 100 eV.

\begin{figure}[t]
    \nolinenumbers
    \centering
    \includegraphics[width=\columnwidth]{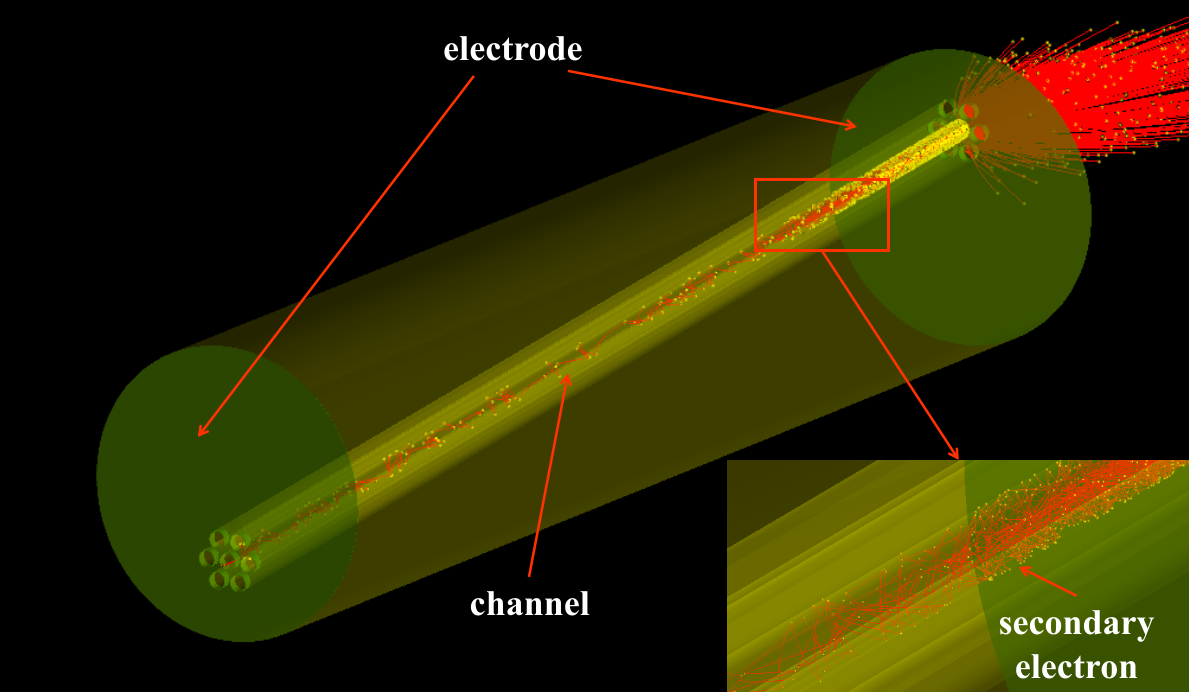}
    \caption{Simulation events using the simple MCP model.}
    \label{fig:sim_MCP}
\end{figure}

After the implementation of the secondary electron emission process, it will be brilliant to study the amplitude and shape of the signal when a certain electron from initial ionization enter a channel, since the response of each channels will be the same or at least similar with each other. Once the signal has been known for detail, the response of a single electron can definitely be accessed by a simple sampling so that the response of the whole MCP will be able to be simulated fast and precisely. We build a simplified geometry of MCP with only 7 channels for the simulation of single-electron response considering the complexity to construct an MCP with millions of channels as shown in \cref{fig:sim_MCP}. Limited by the rendering power of computer, only 5000 tracks are drawn in this figure. However, the amplification of secondary electrons can be clearly seen.

\begin{figure}[t]
    \nolinenumbers
    \centering
    \subfloat[\label{fig:G_V_1}\#1 MCP.]{\includegraphics[height=0.66\columnwidth]{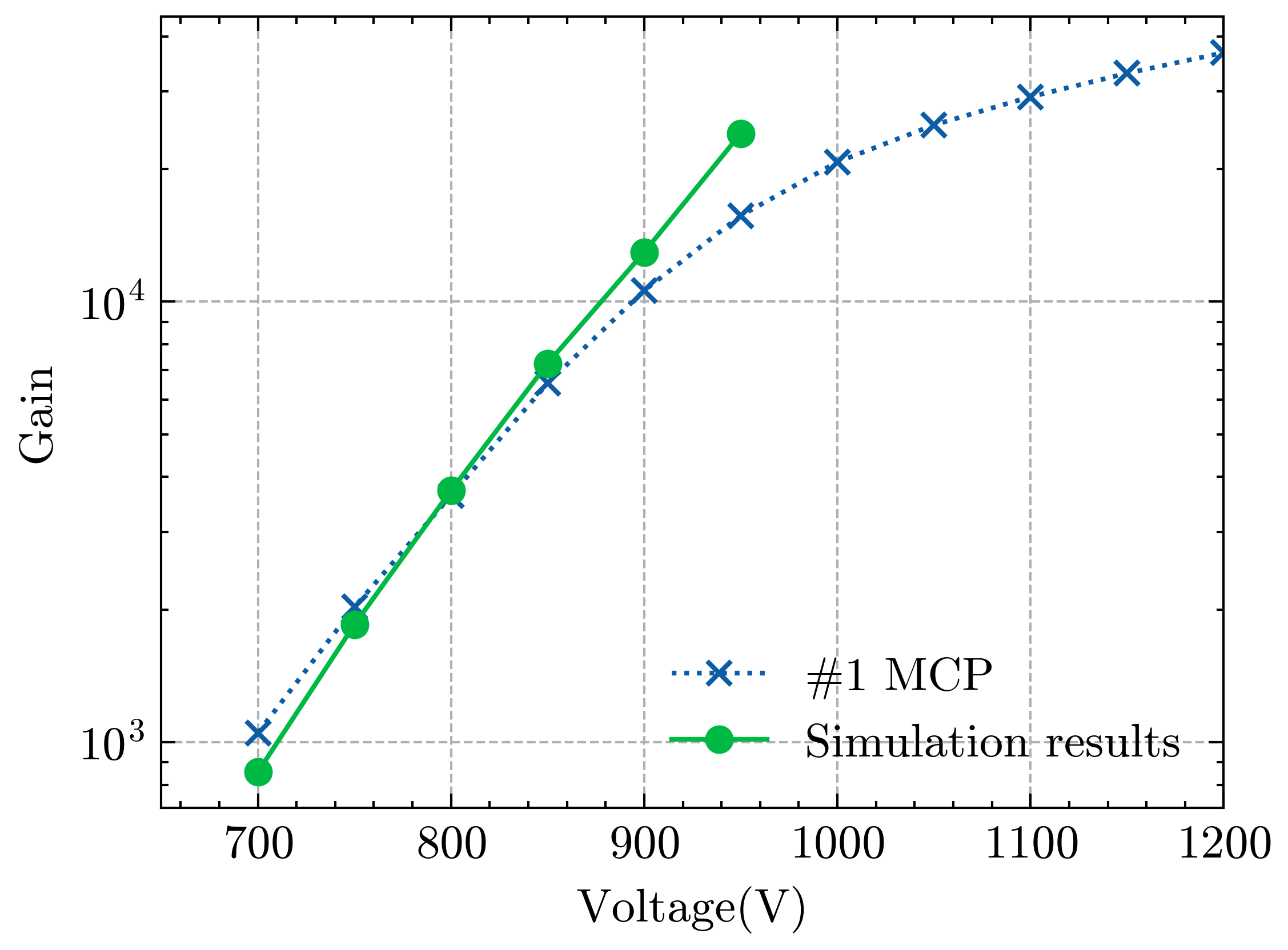}}\\
    \subfloat[\label{fig:G_V_2}\#2 MCP.]{\includegraphics[height=0.66\columnwidth]{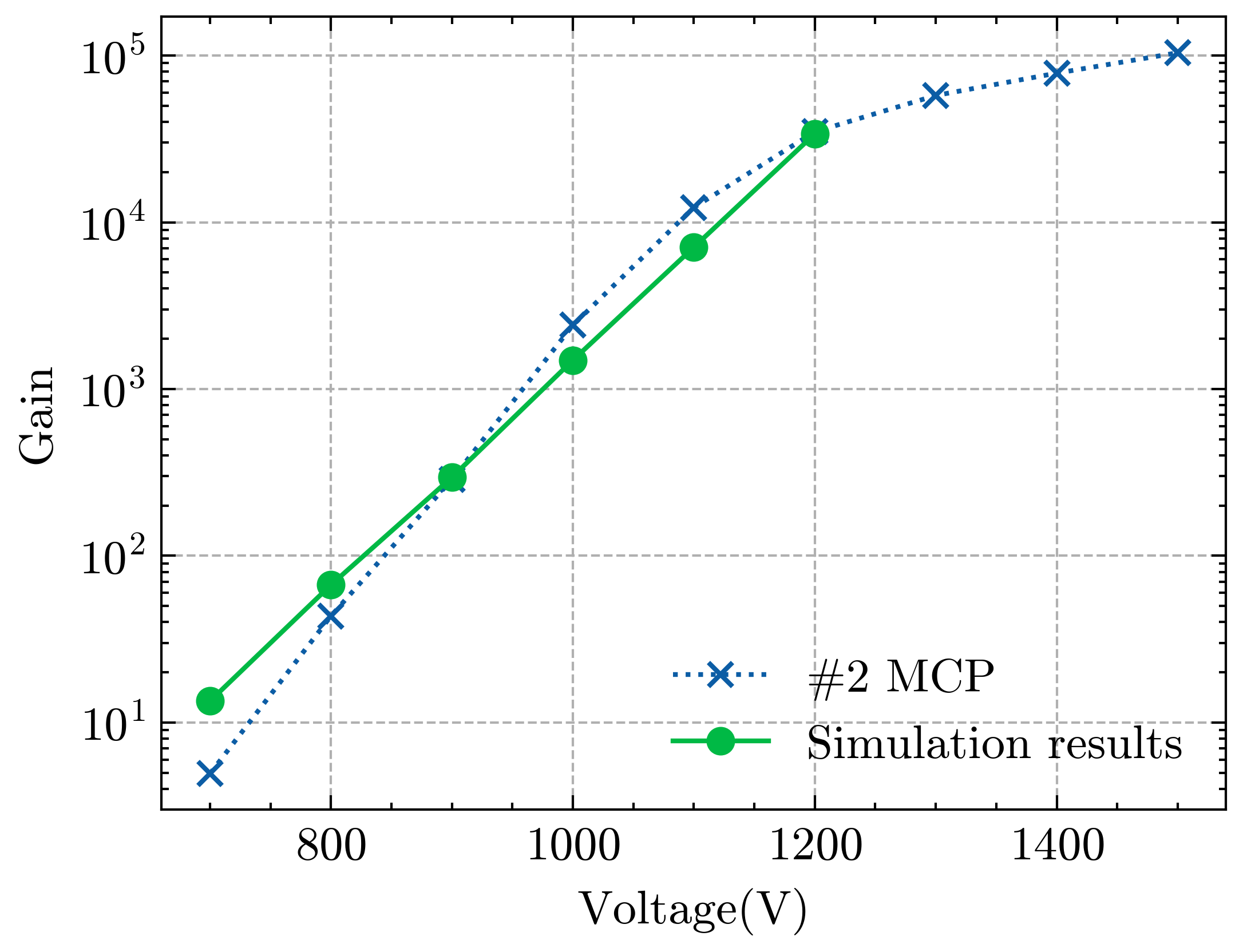}}
    \caption{The gain of two MCPs versus applied voltage. The blue ``$\times$'' markers denote the experimental measurements and the green points denote the simulated results.}
\end{figure}

To further validate our simulation, we simulated the gain of two MCPs versus the applied voltages and compared them with the measured results. The geometries of \#1 and \#2 MCP are established in Geant4 based on the actual measurements and provided in Table~\ref{tab:mcp_1_par}. In the simulation, the kinematic information of all the particles that are able to reach the exit of the channel is recorded without considering the detection efficiency, fluctuation, or other relevant issues of the readout system, so the MCP gain will simply be calculated by dividing the number of the electrons at the exit by the number of the electrons shot towards the MCP. The two pieces of MCPs are simulated at different applied voltages from 700~V to 1200~V with a step of 100~V using the same SEY and energy spectra of the secondary electrons. For each voltage, the gain is simulated for 1000 times with random primary incident electrons and the average is determined to be the gain at this voltage. The gains with respect to the voltages are shown in \cref{fig:G_V_1} and \ref{fig:G_V_2}.

\begin{table}[t]
        \nolinenumbers
        \centering
        \caption{Geometric parameters of the two pieces of MCPs simulated in this work.}
        \begin{tabular}{ccc}
                \hline
                \hline
                Parameter (Unit)   &   \#1 MCP  & \#2 MCP\\
                \hline
                MCP diameter (mm)  &   25 &24.8\\
                MCP thickness (mm) &   0.42 &0.48   \\
                Pore diameter ($\mu$m)    &   10 &6 \\
                Tilt angle (degree)    &   12  &5.5  \\
                Body resistance (M$\Omega$)   &   84  &90\\
                Input current (pA)  &   107 &162  \\
                \hline
                \hline
        \end{tabular}
        \label{tab:mcp_1_par}
\end{table}

\begin{figure}[t]
    \nolinenumbers
    \centering
    \includegraphics[width=\columnwidth]{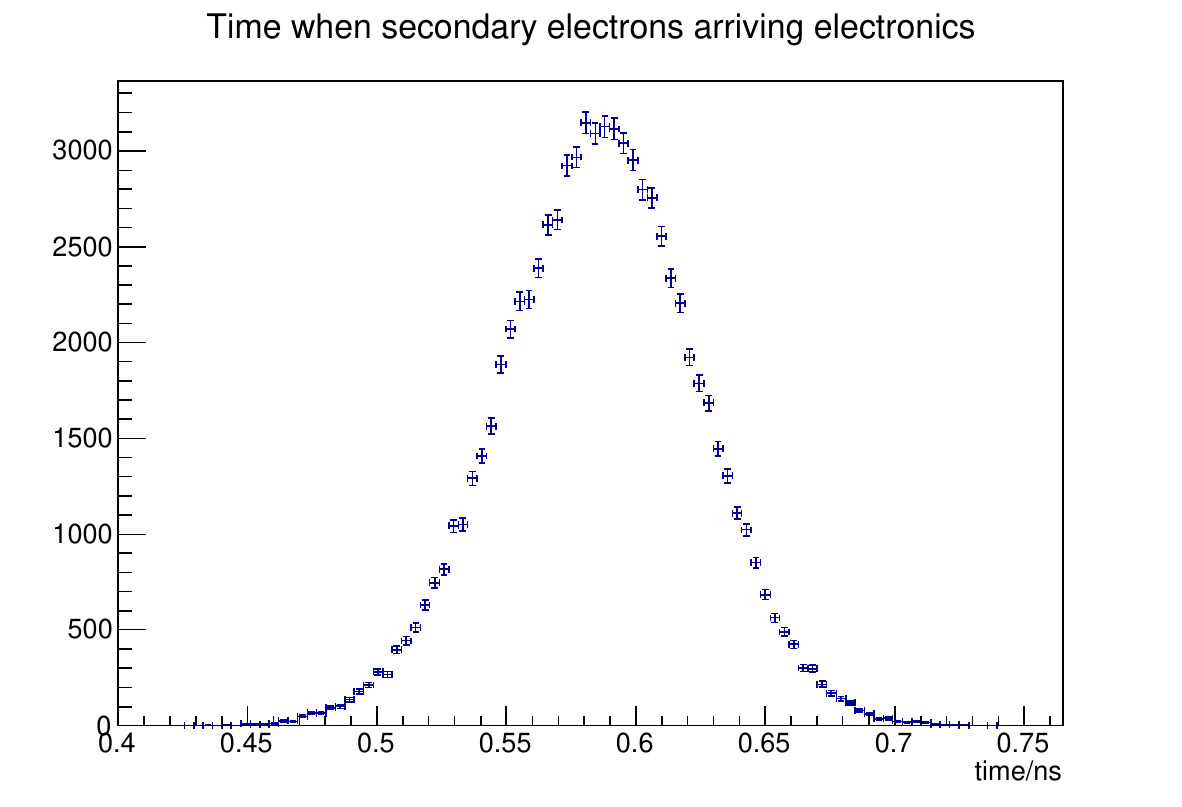}
    \caption{Time distribution of secondary electrons received by electronics. The horizontal axis denotes the time after the initial electron enters the channel.}
    \label{fig:MCP_output}
\end{figure}

According to the simulation results shown in \cref{fig:G_V_1} and \cref{fig:G_V_2}, the gain will increase exponentially with the increasing applied voltage, which is consistent with the experimental measurements and other studies~\cite{evans1965low}, while the measured gain will be obviously lower than the simulated result and tend to stabilize to a constant value when the applied voltages are sufficiently high since the multiplication will be suppressed by the saturation effect~\cite{adams1966mechanism, harris1971saturation} that is not taken into consideration in the simulation at present. The distribution of the time when secondary electrons arriving the electronics for \#2 MCP is shown in \cref{fig:MCP_output}.

The simulation program for the electron multipliers like MCP has been summarized into an independent software named ``MCPSim''~\cite{Miao:2023qud}. In the future, a new MCP will be developed to satisfy the requirement of MACE. MCPSim will be used in the development and optimization. Meanwhile, the simulation of MCP will be implemented into the software system of MACE with more proper consideration of the saturation and space charge effects.

\subsection{Electromagnetic calorimeter} \label{sec:ecal}

\subsubsection{Overview}
The primary purpose of the electromagnetic calorimeter (ECAL) is to enable the annihilation $\gamma$-rays with positron hits on MCP and rejection of various backgrounds.
ECAL is expected to discriminate the annihilation signal from backgrounds and retain all event information as much as possible.
It leads to the baseline performance specifications of ECAL as follows:

\begin{itemize}
    \item An energy resolution of $\sim$\,10\% at 0.511 MeV to perform precise energy reconstruction, rejecting the contribution of natural backgrounds.
    \item A spatial resolution of $\sim$\,1~cm to reconstruct the tracks of two back-to-back $\gamma$-rays, rejecting accidental coincidence.
    \item A timing resolution of $\sim$\,1~ns to ensure that the energy deposits in ECAL modules are in time with MCP response.
    \item A geometrical acceptance of $\sim$\,95\% to prevent leakage of annihilation events.
\end{itemize}

In the next two sections, we first introduce the geometrical layout, module assembly, and event reconstruction of ECAL.
Then, simulation studies are conducted to estimate its performance of $\mmu$-to-$\ammu$ signal.
Moreover, a proposal of MACE \textsc{Phase-I} experiment which consists of the ECAL and an inner tracker system for other muon cLFV processes searching will be further introduced in \cref{sec:phasei}.

\subsubsection{Conceptual design}
\begin{figure}[htbp]
    \nolinenumbers
    \centering
    \includegraphics[width=\columnwidth]{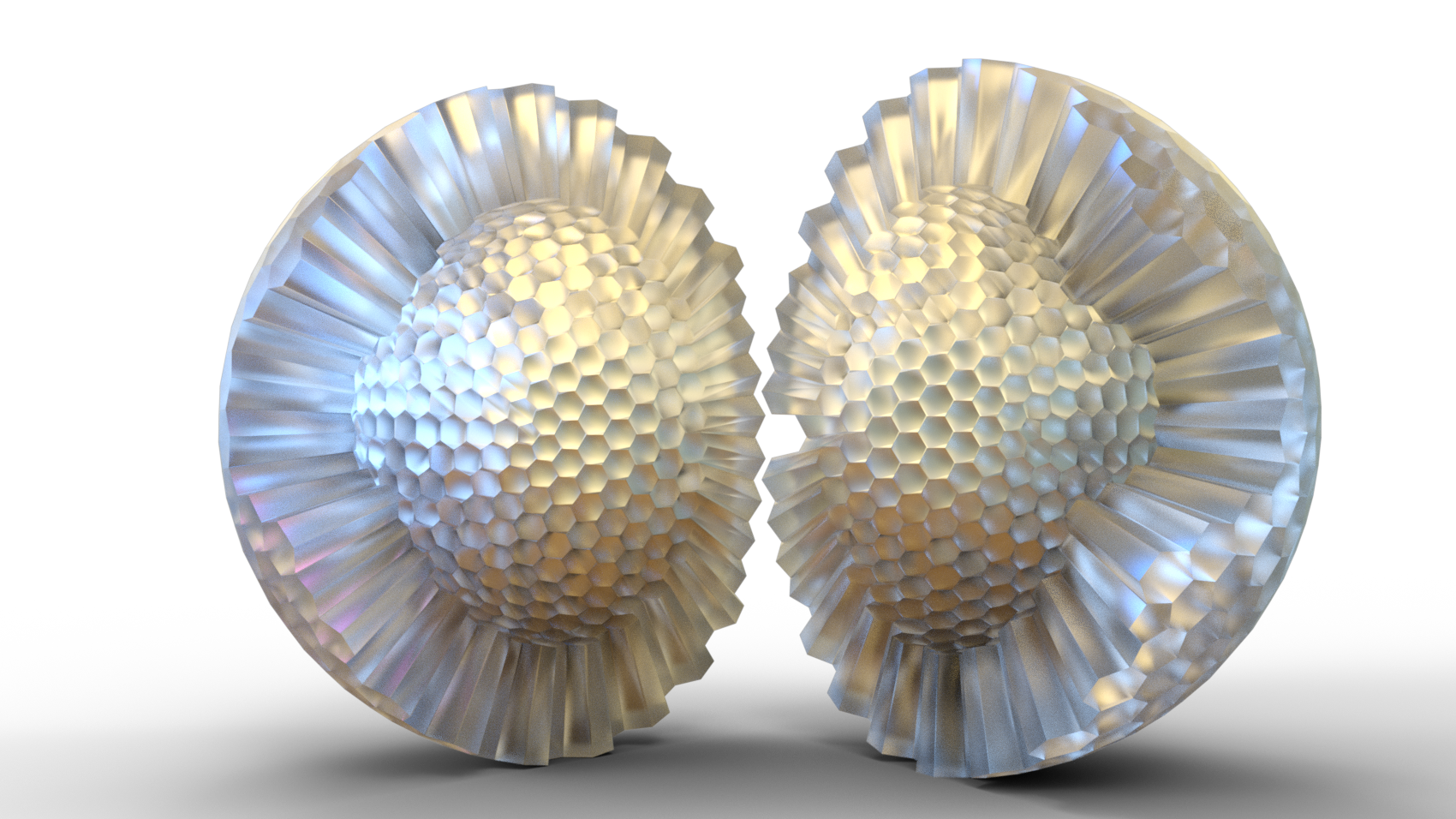}
    \caption{3D rendered image of the ECAL geometry~\cite{Chen:2024jmg}.}
    \label{fig:ecal-3d}
\end{figure}

To meet the mentioned requirements (e.g., excellent energy resolution, high detection efficiency, and good geometrical acceptance), we proposed a Goldberg-polyhedron geometry design of the ECAL (see \cref{fig:ecal-3d}).
The Goldberg polyhedron consists of pentagons and hexagons with every three faces meeting at the same vertex ~\cite{Goldberg1937}.
It is similar in appearance to the sphere as an inpolyhedron and possesses an icosahedral rotational symmetry.
A Class I GP$(8,0)$ Goldberg polyhedron geometry layout is introduced.
It consists of a total of 642 faces, including 12 regular pentagons, and 630 irregular hexagons which can be categorized into 9 types.
Modules created from the different types of faces are identified as Type-PEN and Type-HEX (01--09).
There are 19 modules at the beam entrance and 1 module at the exit removed, resulting in a total of 622 modules and achieving 97\% coverage of the $4\pi$ solid angle.

The ECAL should take account of the particle energy in sub-MeV and several tens of MeV at the same time, due to the different physics goal in MACE \textsc{Phase-I} and MACE. In the 0.511 MeV energy regime of interest, a total absorption homogeneous calorimeter with inorganic scintillators generally provides excellent energy resolution which is expected for MACE.
Three types of crystals have been considered for the MACE calorimeter, including bismuth germanium oxide (BGO), thallium-doped cesium iodide (CsI:Tl) and cerium-doped lutetium-yttrium oxyorthosilicate (LYSO:Ce).
The CsI(Tl) crystal was widely employed by the electromagnetic calorimeter of various experiments due to its high light yield~\cite{Bauer2005,BESIII:2009fln,Belle-II:2010dht}.
However, it is characterized by slow emission time which may cause a loss of efficiency. The short radiation length also increase the cost of materials outside the ECAL.
BGO combines an acceptable light output and a fast decay time, and is widely used in positron emission tomography (PET) which also investigates the signal of annihilation $\gamma$-rays.
Besides, LYSO could be an appealing option as it is brighter and faster than BGO.
The properties of these crystals are summarized in \cref{tab:crystal}.

\begin{table}[htbp]
    \nolinenumbers
    \centering
    \caption{Properties of BGO, CsI(Tl), and LYSO(Ce) scintillation crystal~\cite{ParticleDataGroup:2024cfk,Yeh:2022yog}.}
    \label{tab:crystal}
    \resizebox{\linewidth}{!}{%
        \begin{threeparttable}[b]
            \begin{tabular}{lccccccccc}
                \hline \hline
                Parameter                                                         &
                $\rho$                                                                      &
                $X_0$                                                                       &
                $R_{\text{M}}$                                                                       &
                $\dd E/\dd x$                                                                     &
                $\tau_{\text{decay}}$                                                              &
                $\lambda_{\text{max}}$                                                             &
                $n$                                                                         &
                \multirow{2}{*}{\begin{tabular}[c]{@{}c@{}}Relative \\ output\end{tabular}} &
                \multirow{2}{*}{\begin{tabular}[c]{@{}c@{}}Hygro-\\ scopicity\end{tabular}}                                                                                                       \\
                Unit                                                              & g$/\text{cm}^3$ & cm   & cm   & MeV$/$cm & ns            & nm  &      &                  &        \\ \hline
                BGO                                                                         & 7.13            & 1.12 & 2.23 & 9.0      & 300           & 480 & 2.15 & 21               & no     \\
                CsI(Tl)                                                                     & 4.51            & 1.86 & 3.57 & 5.6      & 1220          & 550 & 1.79 & 165              & slight \\
                LYSO(Ce)                                                                    & 7.40            & 1.14 & 2.07 & 9.6      & 40            & 420 & 1.82 & 85               & no     \\ \hline \hline
            \end{tabular}%
        \end{threeparttable}
    }
\end{table}

The technical specifications of the photonsensor (e.g., active area, photon detection efficiency) determine the efficiency of light collection.
Commonly used photosensors in current particle and nuclear physics experiments include Photomultiplier Tubes (PMTs) and Silicon Photomultipliers (SiPMs).
The active area of PMTs is relatively larger than SiPMs.
However, they are characterized by limited quantum efficiency ($\sim$20\%), high operating voltage, and shielding requirements in magnetic fields.
On the other hand, SiPMs offer excellent photon detection efficiency ($\sim$50\%) and tolerance to magnetic fields.
However, a single SiPM comes with small areas (usually 6~mm at maximum).
Thus SiPM arrays of large active areas are considered to enhance the light collection efficiency.

\subsubsection{Simulation}
\begin{figure}[htbp]
    \nolinenumbers
    \centering
    \includegraphics[width=0.9\columnwidth]{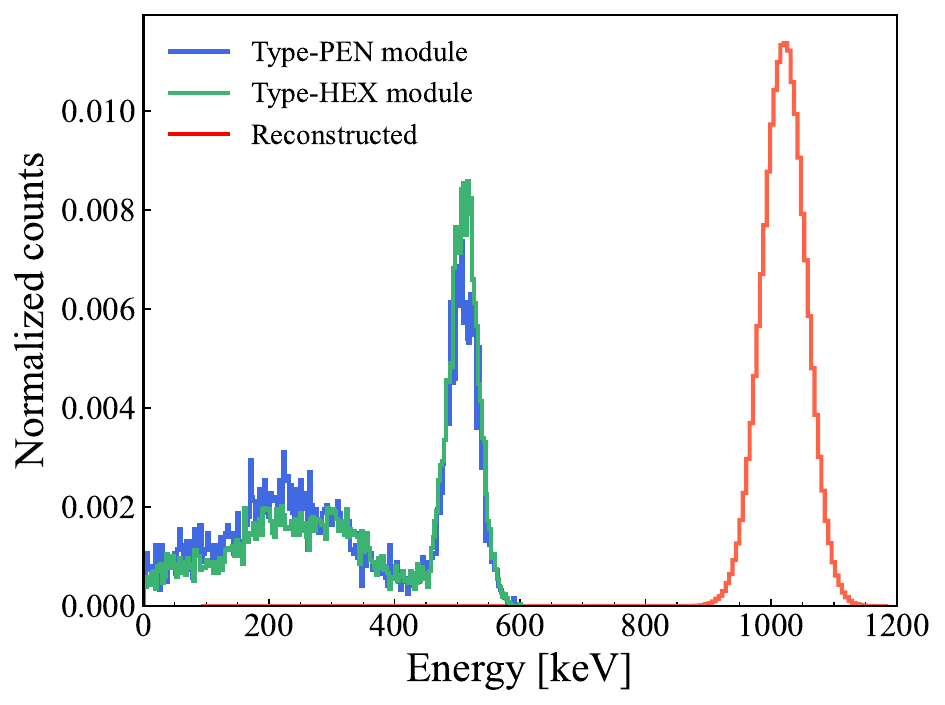}
    \caption{Spectra of a single Type-PEN module (\textit{blue line}), a single Type-HEX module (\textit{green line}) and reconstructed annihilation signals (\textit{red line}) of M-to-$\bar{\text{M}}$ conversion by simulation.}
    \label{fig:ecal-signal}
\end{figure}

A beam of \(10^8\) \(\mu^+\) events with a momentum of 26.3~MeV$/c$ entering the MACE detector has been simulated to evaluate the signal efficiency of ECAL.
Following the coincidence with MCP, a clustering algorithm is applied to search seed modules and their adjacent modules.
The energy of each triggered module is smeared assuming an energy resolution of 10\% at 0.511 MeV.
We then estimate the energy sum and triggered time of the cluster and apply an energy difference cut of $\Delta E_\text{recon}<0.1$ MeV to the reconstructed annihilation events.
The resulting ECAL response spectra are plotted as \cref{fig:ecal-signal}. Events that fall within the \(3\sigma\) interval of the full-energy peak are identified as signal events.

\subsection{Performance}
The overall signal efficiency of the PDS can be derived from the results of the MCP and ECAL. First, the detection efficiency of the MCP depends on the kinetic energy of the incident positron. According to simulation results, the MCP efficiency for positrons accelerated by the PTS is 32.6\%. Second, we found that the primary components of the MCP, lead glass and aluminum, may induce multiple Compton scattering to the 0.511~MeV photons. This effect results in a 36.6\% efficiency loss, leading to a 63.4\% incident efficiency of annihilation events for the ECAL. Additionally, the geometric and reconstruction efficiencies of the ECAL are considered, which are 95.3\% and 94\%, respectively. The total signal efficiency can be estimated as
\begin{equation}
    \varepsilon_{\text{PDS}}=\varepsilon_{\text{MCP}}\varepsilon^{\text{In}}_{\text{ECAL}}\varepsilon^{\text{Geom}}_{\text{ECAL}}\varepsilon^{\text{Recon}}_{\text{ECAL}}~.
\end{equation}
The values of these efficiencies are summarized in \cref{tab:pds-eff}. Consequently, this analysis yields a signal efficiency of 18.5\% for the PDS. A higher efficiency can be achieved using MCP  with a low material budget substrate.

\begin{table}[htbp]
\nolinenumbers
\centering
\caption{Summary of signal eﬀiciencies in the Positron Detection System.}
\label{tab:pds-eff}
\begin{tabular}{ccc}
\hline\hline
Component & Efficiency type                                 & Efficiency value \\ \hline
MCP       & Detection efficiency $\varepsilon_{\text{MCP}}$ & 32.6\%           \\ \hline
\multirow{3}{*}{ECAL} & Incident efficiency $\varepsilon^{\text{In}}_{\text{ECAL}}$          & 63.4\% \\ \cline{2-3}
                      & Geometric efficiency $\varepsilon^{\text{Geom}}_{\text{ECAL}}$       & 95.3\% \\ \cline{2-3}
                      & Reconstruction efficiency $\varepsilon^{\text{Recon}}_{\text{ECAL}}$ & 94.0\%   \\ \hline
\multicolumn{2}{c}{Total signal efficiency}                 & 18.5\%           \\ \hline\hline
\end{tabular}
\end{table}

\section{Offline software}
\subsection{Introduction}
The significance of offline software has grown in high-precision frontier experiments. The MACE offline software is responsible for physical event generation, detector simulation, event reconstruction, offline event display, and analysis. In the conceptual design stage, the offline software supports Monte Carlo simulations of physical and background events to guide detector system design and optimization. During the engineering stage, the offline software plays a crucial role in detector alignment and can enhance the understanding of the real circumstance through detailed simulations. Once the physical run starts, the offline software serves as a bridge that connects the experimental data to physical results. A high-quality offline software system can simplify the workflow for researchers and potentially reducing systematic errors by accommodating more advanced technologies and algorithms, and lead to accelerated research progress and increased physical sensitivities.

There are some objectives to ensure the expected functionality and performance of the offline software.
\begin{itemize}
    \item \textbf{Framework and architecture}: A well-designed architecture helps the development of simulation applications, reconstruction algorithms, and analysis, ensuring code consistency and reusability while enhancing software quality for better maintainability and reduced errors.
    \item \textbf{Performance}: In the case of MACE as a precision frontier experiment, the event rate could reach $10^7$/s, and distributed computing is necessary for comprehensive physical studies, detector optimization, and data reconstruction and analysis in future operations.
    \item \textbf{Validation}: The output of the offline software, especially the physical results, should be continuously checked and compared to experiment data whenever possible. This can minimize any technical or physical errors in the code.
    \item \textbf{Continuous integration (CI)}: CI involves the frequently build, deployment, and testing of the offline software to uphold its availability and quality.
\end{itemize}

\subsection{Framework}
Mustard framework is the most fundamental library of the MACE offline software. Mustard is a generic offline software framework for particle physics experiments~\cite{MustardFramework}, aims to providing a modern, high-performance architecture for distributed computing capabilities, unified geometric interfaces, and high-level abstraction layer of data models out of the box, accelerating the development and performance of applications and libraries. The Mustard framework was formerly developed within the MACE offline software and now it is extracted as an general-proposed framework. It is designed as a set of commonly-used interfaces and functionalities in fixed-target experiments simulation, reconstruction and analysis.

The framework is mainly developed with a concept-based object-oriented paradigm and template metaprogramming paradigm, utilizing the language constructs and standard library features of \texttt{C++20}. This approach enables the development of performance-critical utilities and interfaces without introducing extra runtime overhead. Following the technology roadmap, a low-level component, inspired by the principles of the \texttt{C++} standard library, involves a set of useful concepts, type traits, functions and optimized numeric algorithm. The low-level components are predominantly found in muc~\cite{muc} and in the \texttt{Concept} and \texttt{Utility} components within Mustard.

Built on top of the low-level infrastructure, Mustard provides a set of interfaces and utilities for distributed computing, simulation, customization of data models and distributed data processing, parameterized geometry construction and serialization, command argument parsing, logging, etc. An environment component manages the resources and variables required by these functionalities. Mustard directly depends on MPI~\cite{mpi41}, ROOT~\cite{Brun:1997pa}, Geant4~\cite{GEANT4:2002zbu,Allison:2006ve,Allison:2016lfl}, Eigen~\cite{Eigen}, and other libraries to form a high-level abstraction on these functionalities, while maintaining compatibility with the common workflow of these underlying frameworks at maximum. A detailed description of these functionalities will be provided in the subsequent sections.

\begin{figure}[htbp]
    \nolinenumbers
    \centering
    \includegraphics[width=\columnwidth]{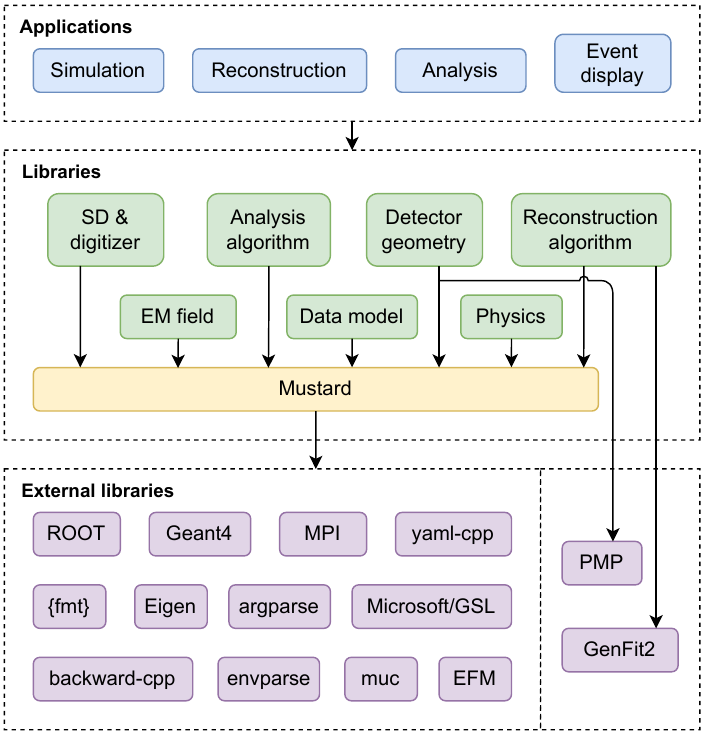}
    \caption{The architecture of MACE offline software.}
    \label{fig:macesw-arch}
\end{figure}

\subsection{Parallel computing}
MACE will be operating with a muon beam with a flux of $10^8$/s, and will be collecting data from $3\times 10^{15}$ muon or muonium decay events. As a result, distributed parallel computing is essential to handle the data processing requirements and to carry out simulations during the design stage. Two parallel computing models have been investigated:
\begin{itemize}
    \item Hybrid model: Multithreading for intra-node parallelism and communication, and message passing for inter-node parallelism and communication.
    \item Pure message passing model: Message passing for both intra- and inter-node parallelism and communication.
\end{itemize}
Both models demonstrate excellent scalability and performance, though they differ in practice. The hybrid model is found to be more memory-efficient and provides greater flexibility, thanks to its capability to finely control the intra-node parallelism. However, it shows slightly lower speedup due to increased thread synchronization and potentially poorer CPU cache locality, as well as increased complexity in coding. On the other hand, the pure message passing model is slightly more memory-consuming due to data (e.g. \textsc{Geant4} data) duplication for each process within a node, but it offers superior speedup and reduced technical complexity. These are results of potentially better CPU cache locality and absence of thread competition. Consequently, the pure message passing model has been selected as the parallel computing model for the MACE offline software.

\begin{figure}[htbp]
    \nolinenumbers
    \centering
    \includegraphics[width=\columnwidth]{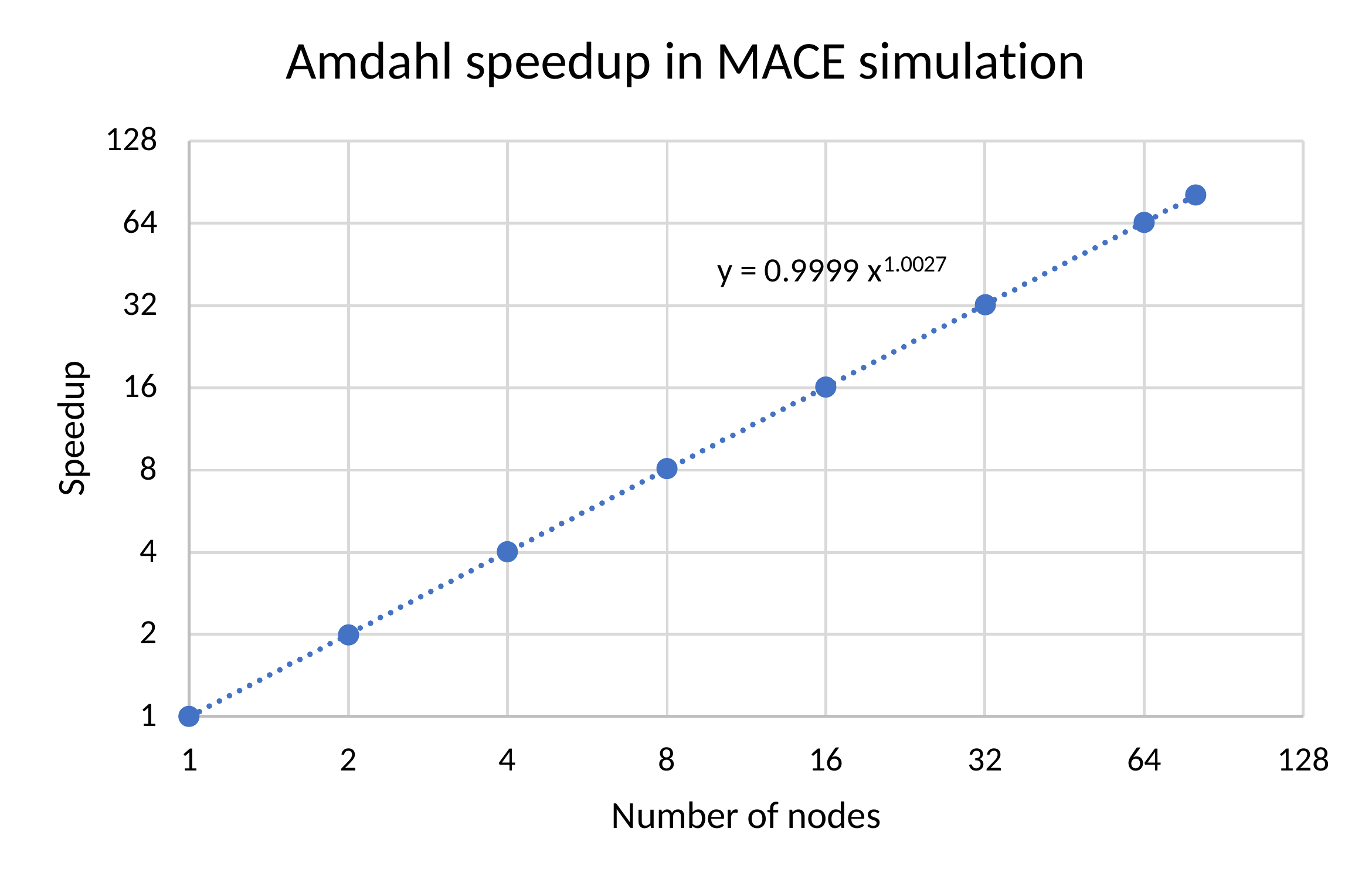}
    \caption{Speedup test at Tianhe-2 supercomputer. Each node has 24 cores. Results shows a linear speedup, and no significant performance degradation is observed.}
    \label{fig:macesw-th2-speedup}
\end{figure}

The message passing model is implemented, based on the Message Passing Interface (MPI)~\cite{mpi41}. MPI offers a scalable interface for distributed computing on both high-performance computing (HPC) cluster and high-throughput computing (HTC) cluster. Within the Mustard framework, a generic executor based on MPI is provided for dynamically scheduling computing tasks among all executing processes. The executor handles scheduling and execution, while users customize the actual tasks. Building upon this extensible functionality are a \textsc{Geant4} simulation run manager and an event-by-event data processor, as solutions to distributed \textsc{Geant4} simulation and distributed event reconstruction or analysis. In the executor, a master/slave pattern communication algorithm is employed for task division and assignment, and the overlap between the computation and communication is maximized to minimize latency and enhance performance. A speedup test is performed and results are shown in \cref{fig:macesw-th2-speedup}. The results demonstrate a linear speedup on 1920 cores (80 computing nodes) in the event loop of a benchmark simulation example, and no significant performance degradation is observed. Thus, it is concluded that the distributed computing infrastructure could meet the requirements for future development of the MACE offline software.

\subsection{Event data model}
Event data model stands in the central position in the offline software system. It conceptualizes how data are produced and stored, the data streamline, and how data should be processed. Therefore, it conceptualizes a large functional parts of the offline software. The Mustard framework provides an interface for building static event data model, encoding the field information in the \texttt{C++} type system with the help of template metaprogramming infrastructure. This enables the check of data model consistency and the usage in user code at the compile time. With all fields' name and type known at the compile time, Mustard also provides a zero-overhead interface mapping the data model to a tuple-like type. This interface serves as a bridge between the data model and the actual data object, which ensures the consistency and is practical in any operation, such as reading/writing data from/to disk or apply algorithm to process the data. Based on the data model interface, the MACE offline software has predefined some commonly used data models for simulation, reconstruction and analysis applications. It is also simple to extend data models for developers or users. However, due to the static nature in the event data model interface and architecture, it remains a challenge in dynamically defining data model via configuration files such as JSON or YAML. This will be addressed in future development.

\subsection{Detector geometry and field}
Geometry system is extensively used in both simulation and reconstruction, to describe and define the geometry, material, and field of the detector system. The task of a geometry framework is to describe the information needed to both human and machine. This involves two aspect. A human-readable geometry description generally is characteristic geometric parameters, such as the inner radius of CDC, the number of collimator foils, etc. While the machine need the full description of the whole geometry, like the position and transformation of every single CDC wires. Therefore, there exists a gap between the real human-readable parameters and the exact geometry definition.

\begin{figure}[htbp]
    \nolinenumbers
    \centering
    \includegraphics[width=\columnwidth]{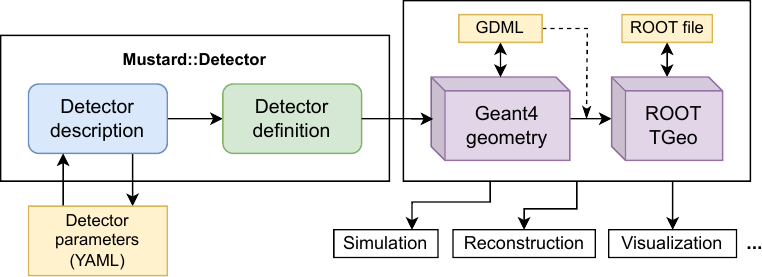}
    \caption{Flow chart of geometry management.}
    \label{fig:macesw-geometry}
\end{figure}

The geometry interface of Mustard framework aims to address this discrepancy. It separates the human-readable detector description from the detailed detector geometry definition. In this architecture, detector definitions is responsible for the exact construction of the detector geometry and material, depending on parameters specified in the corresponding detector descriptions. The detector description does nothing but provides an interface of defining and fetching these parameters. This architecture separates the technical parts from the physical parts during the detector construction. The flow chart of the detector construction is shown in \cref{fig:macesw-geometry}.

The electromagnetic field is another essential subject in the simulation and reconstruction. The Mustard framework provides an interface for users to define a set of reusable field definitions. The defined field can be used in both simulation and reconstruction through the corresponding wrappers. The Mustard framework also supports defining interpolated electromagnetic fields from discrete field data to fulfill the requirement of importing MACE field map.

\subsection{Continuous integration}
Continuous integration (CI) is a practice of frequently integrating the software code, build, deploy and test its functionality. CI can keep the software in an available state and efficiently take newly developed functionalities into practice or production. CI also serves as a practice to observe and fix errors in an early stage. The development branch of the MACE offline software has been continuously building and deploying on the Tianhe-2 supercomputer utilizing the container technique via Apptainer (formerly known as Singularity)~\cite{Kurtzer2017SingularitySC}. An autonomous workflow has been developing to build fully optimized Apptainer images for different MPI implementations. Currently, the image size ranges around 4~GB for image with \textsc{Geant4} data and around 2~GB for images without \textsc{Geant4} data. They are upload to Tianhe-2 supercomputer and updated on demand. The CI workflow is still in investigation and it is planned to deploy a CI workflow on local or cloud servers to automatically perform CI of the MACE offline software in the future.

\subsection{Event display}
For experimental particle and nuclear physics, event display software has become increasingly important. This software is used to visualize the detector response information and the reconstructed information for particles that pass through the detectors. The importance of this software is threefold. First, it is essential for outreach and educational purposes. Students and new researchers can use event displays to gain a more intuitive understanding of how particles interact with detectors and how data is collected and analyzed. This visual approach helps bridge the gap between theoretical knowledge and practical application, making it easier to grasp complex concepts. Second, by providing a clear and detailed visualization of the event data, researchers may spot inconsistencies or errors in the data processing pipeline that might not be obvious through numerical data alone, allowing them to identify potential issues more easily in the data analysis. Third, it is crucial for online monitoring of data taking, ensuring the data quality. Real-time event displays allow researchers to monitor the data as it is being collected, quickly identifying any issues with the detectors or data acquisition system. This immediate feedback is vital for maintaining high data quality and ensuring that any problems are addressed promptly, minimizing the risk of data loss or corruption.

For the MACE, we need to develop event display software for similar purposes. We envision that the software should have the following contents and functionalities:
\begin{itemize}
    \item \textbf{3D Models}: The software should contain 3D models for the main detectors of the MACE experimental apparatus. This will provide a comprehensive visual representation of the experimental setup, aiding in the understanding and analysis of the data.
    \item \textbf{Detector Signals and Reconstructed Hits and Tracks}: For both experimental data and simulated events, the program should display detector signals as well as reconstructed hits and tracks.
    \item \textbf{Online Data Monitoring}: For online data monitoring, the program should automatically take some events from time to time from the data stream, analyze the event online, and display information as described in the previous point. This functionality is crucial for ensuring the continuous monitoring and immediate analysis of data to detect any potential issues promptly.
    \item \textbf{Average Occupancy Display}: The software should display the average occupancy of all detector channels over a period of time. This will help in monitoring the performance of the detectors and ensuring that they are running properly. Any anomalies or irregularities in the occupancy can be quickly identified and addressed.
    \item \textbf{User Interaction with 3D Modules}: The program should allow users to interact with the 3D modules. For instance, when a user clicks on an ECAL module, there should be a pop-up window displaying detailed information such as the channel name, high-voltage setting, and the ADC spectrum over a period of time. This interactive feature will enhance the usability of the software and provide researchers with easy access to detailed detector information.
\end{itemize}

\begin{figure}[t]
    \nolinenumbers
    \centering
    \subfloat{\includegraphics[width=0.48\columnwidth]{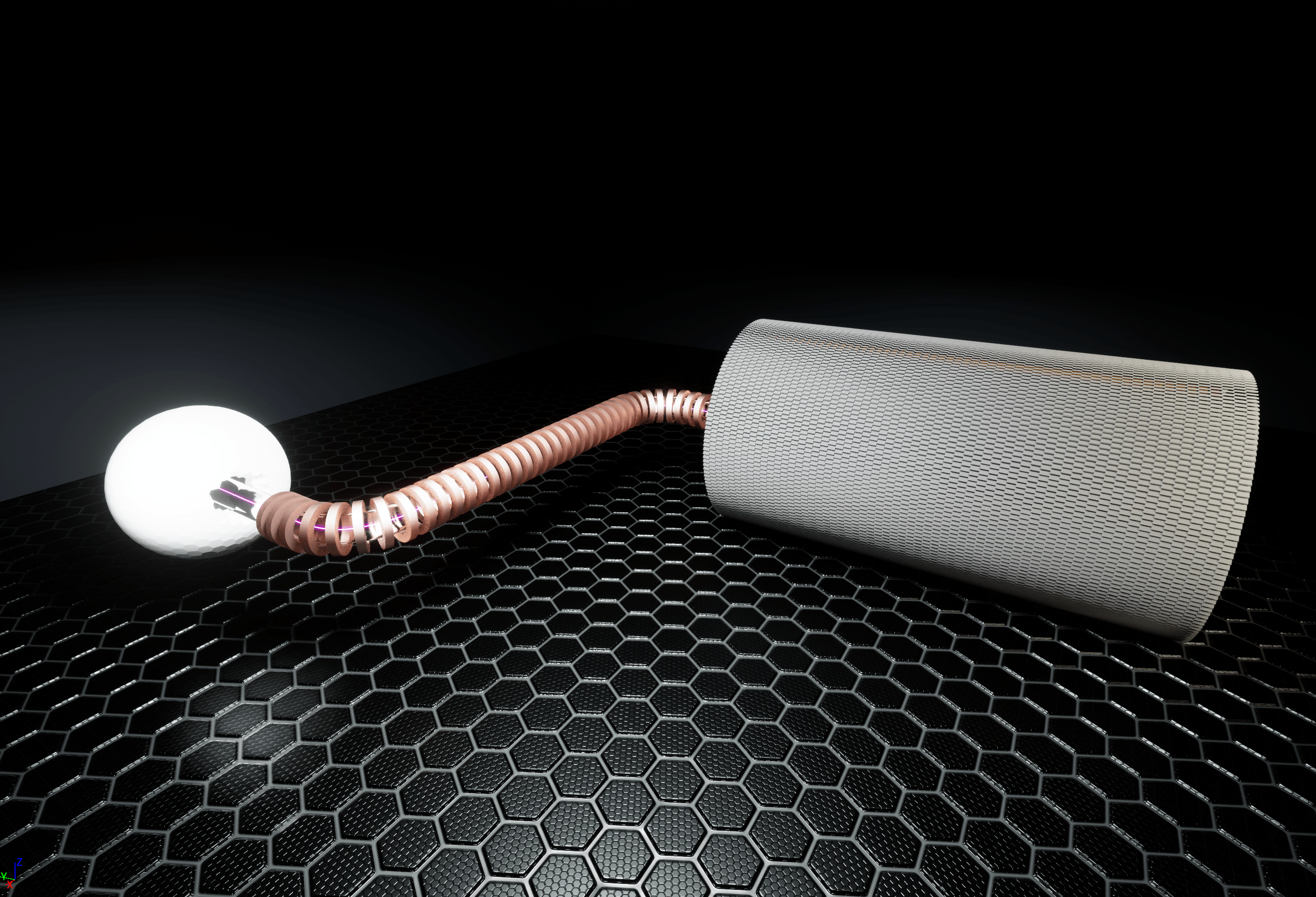}}
    \hfill
    \subfloat{\includegraphics[width=0.48\columnwidth]{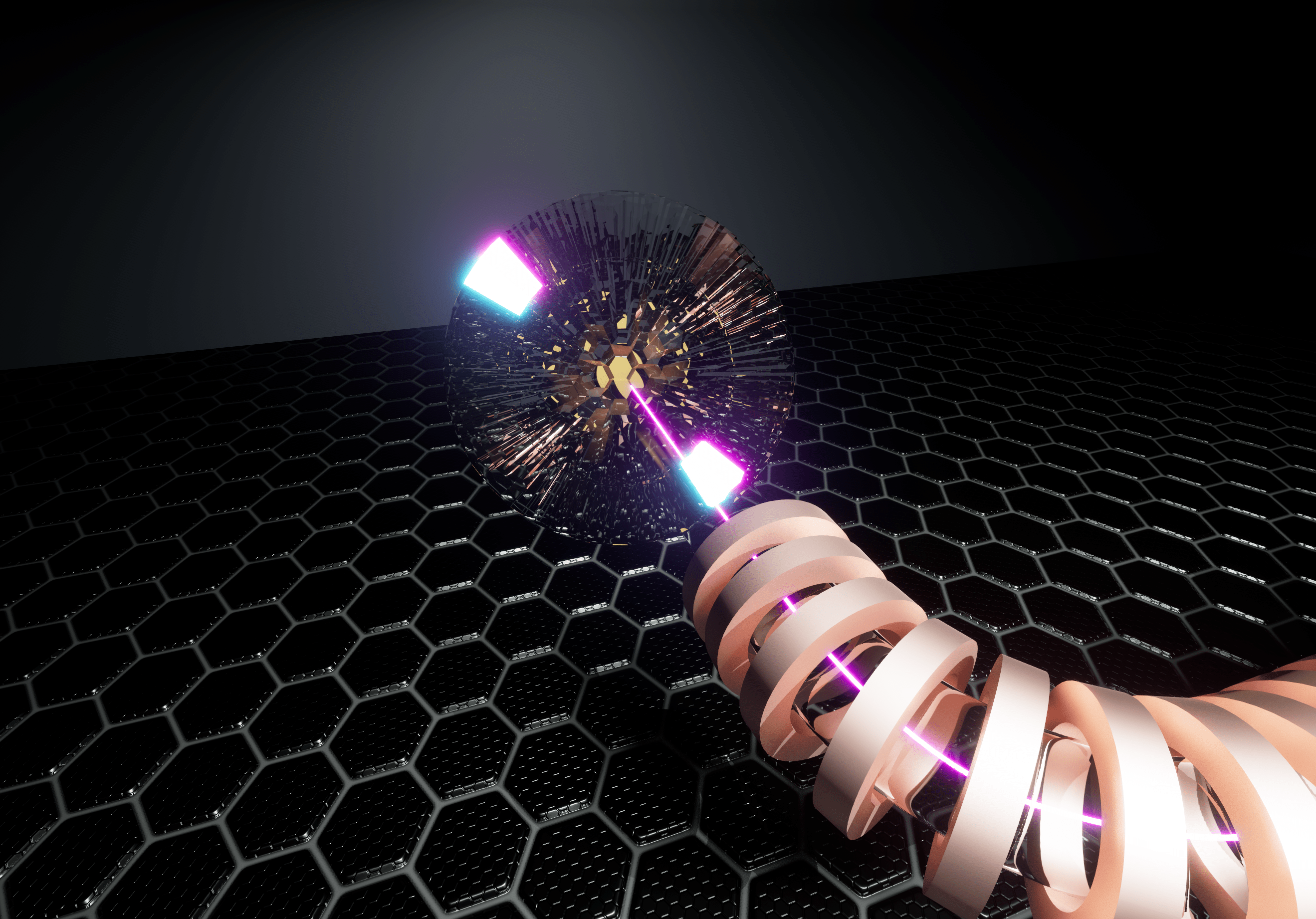}}
    \vskip\baselineskip
    \subfloat{\includegraphics[width=0.48\columnwidth]{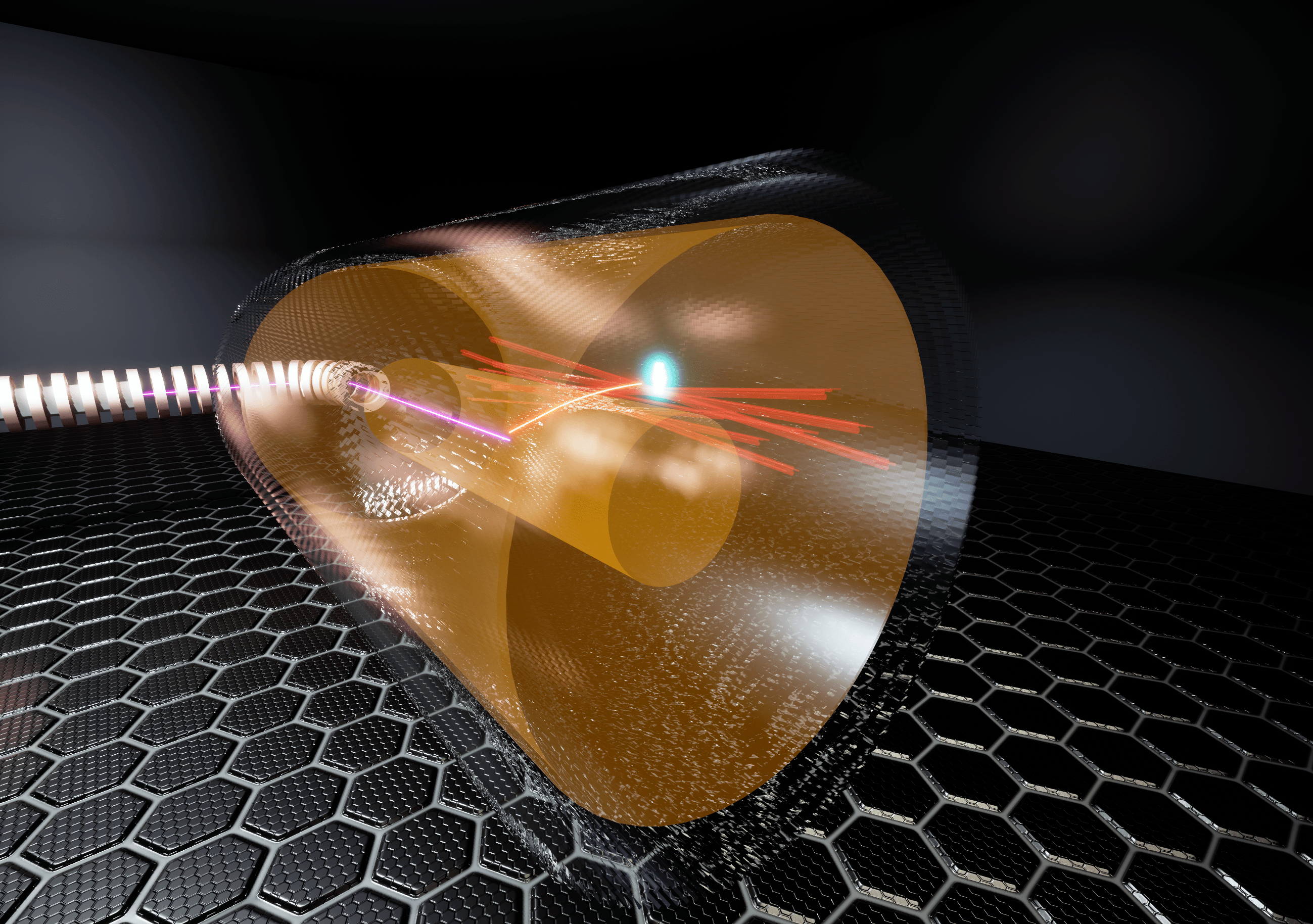}}
    \hfill
    \subfloat{\includegraphics[width=0.48\columnwidth]{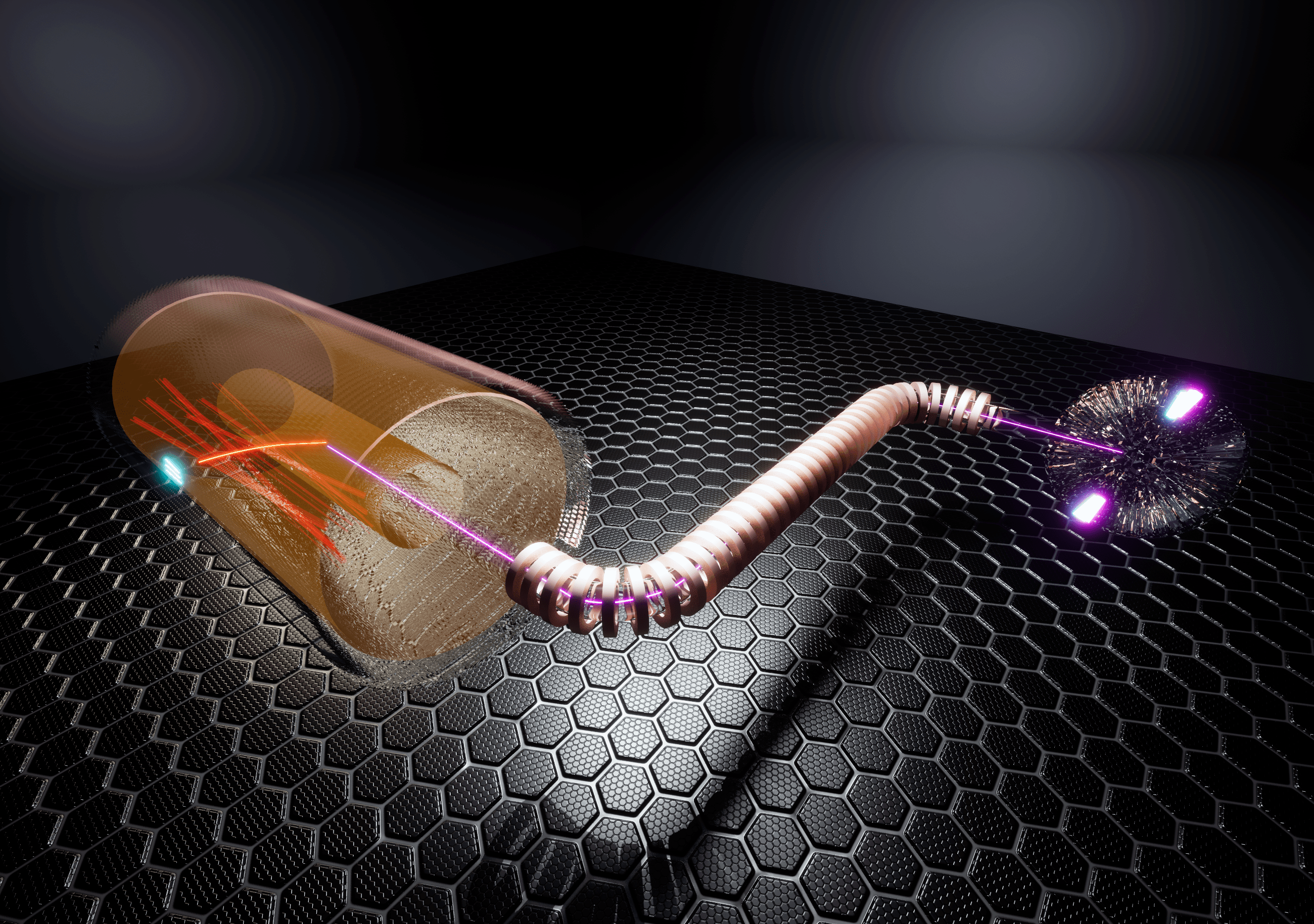}}
    \caption{\textbf{Top-left:} 3D modules imported from Geant4 into the Unreal Engine, including ECAL, MCP, PTS, TTC and CDC. Sensitive detectors will be made semi-transparent during event display for better visualization of signals. \textbf{Top-right:} An event display for the ECAL. Energy deposition in the ECAL modules is displayed using a rainbow color scheme, with purple indicating low energy and red indicating high energy. \textbf{Bottom-left:} A event display for the CDC. The highlighted curved tracks depict the electron trajectory, while the semi-transparent colored cylinders indicate the drift cells that were fired. The radii of these cylinders represent the drift distances. The electron eventually hits the TTC, leaving detectable signals on several adjacent modules. \textbf{Bottom-right}: An overview of the entire event.}
    \label{fig:Event_display}
\end{figure}

To meet these requirements, game engines are suitable candidates, as they are designed to handle complex graphics, real-time rendering, and interactive environments, making them ideal for developing sophisticated event display software. Among the available game engines, Unreal~Engine~5~\cite{unrealengine5} stands out as one of the optimal choices for several reasons.
\begin{enumerate}
  \item It offers unparalleled graphical fidelity. Its advanced rendering capabilities can create highly detailed and realistic 3D models of the MACE experimental apparatus.
  \item It provides powerful tools for real-time data processing and visualization. Its robust framework allows for the integration of live data streams, enabling the software to display detector signals, reconstructed hits, and tracks in real time. This capability is essential for online data monitoring and ensures that researchers can promptly detect and address any issues with the data acquisition system.
\item It supports extensive interactivity. With its user-friendly interface and comprehensive toolset, developers can create interactive 3D modules that allow users to click on specific components. This interactive feature significantly improves the usability of the software, providing researchers with easy and intuitive access to crucial data.
\item It has robust support for Python through various plugins. This allows developers to utilize powerful Python libraries, such as Matplotlib, to create histograms and other visualizations for displaying data. The integration of Python simplifies the development process, making it easier to implement complex data analysis and visualization features directly within the software.
\item Unreal Engine 5 is free software, at least for educational and non-profit purposes, with an extensive user community and strong support network, significantly facilitating the development process.
\end{enumerate}

Currently, the event display software development is in its early stages. We have successfully imported the 3D models for the essential from Geant4 into Unreal Engine 5 (top-left of \cref{fig:Event_display}). For simulated events, the software is already capable of displaying several key elements. With simple keyboard inputs, it can visualize the energy deposition on the ECAL modules for different events, showing how and where particles interact with the calorimeter (top-right of \cref{fig:Event_display}). Additionally, the software can display electron tracks within the spectrometer, allowing researchers to observe the trajectories of electrons as they move through the magnetic field, and eventually produce signals on the TTCs. Furthermore, the software can identify and display the drift cells that are activated by these particles, providing a detailed view of the particle interactions within the detector (bottom-left of \cref{fig:Event_display}).

\section{Background estimation}
In this section, we summarize background simulation results available at present.

\subsection{Physical backgrounds}
\paragraph{Muon internal conversion decay~$\mu^+\to e^+e^-e^+\nu_e\bar{\nu}_\mu$.}
As stated in \cref{sec:mace-background}, the SM-allowed muon rare decay $\mu^+\to e^+e^-e^+\nu_e\bar{\nu}_\mu$ is a major source of physical background. We generated $10^9$ $\mu^+\to e^+e^-e^+\nu_e\bar{\nu}_\mu$ events by Metropolis-Hasting algorithm according to the leading-order squared amplitude of $\mu^+\to e^+e^-e^+\nu_e\bar{\nu}_\mu$~\cite{Djilkibaev:2008jy}. Kinematic cuts are applied correspond to the MMS geometry to reduce the simulation result variance. Assuming a $10^8~\mu\text{OT}/\text{s}$ muon-on-target ($\mu$OT) flux, the generated internal conversion decay events correspond to 114.12 years of data, predicted by the leading-order QED calculations based on McMule~\cite{Banerjee:2020rww}. After the event selection, the background rate is estimated to be $0.29 \pm 0.02/(10^8~\mu\text{OT}/\text{s}\cdot 1~\text{yr})$.

\subsection{Accidental backgrounds}
\paragraph{Beam-related Backgrounds.}
The muon beam produced by stopping \(\pi^+\) on the surface of a target usually contains positrons or other particles of the same momentum~\cite{Xu:2024btf}.
These backgrounds are considered as accidental coincidence.
As shown in \cref{sec:muon-beamline-design}, the ratio of positrons to muons will reach the level of 1\%.
Assuming a $10^8~\mu\text{OT}/\text{s}$ muon-on-target flux, the upper limit of the background rates is estimated to be \(0.07/(10^8~\mu\text{OT}/\text{s}\cdot 1~\text{yr})\) at 90\%~C.L.

\paragraph{Cosmic-Ray Muon Backgrounds.}
The EcoMug generator~\cite{Pagano2021} is employed to generate cosmic-ray muon events in the simulation.
The ECAL is placed at the center of a hypothetical tunnel, surrounded by a copper solenoid and a iron shield.
Cosmic-ray muons are set to emit from a plane source at the sea level.
No event passed the event selection, and the 90\% C.L. upper limit of background rates is estimated to be $2/(10^8~\mu\text{OT}/\text{s}\cdot 1~\text{yr})$.
Generally, a veto system can be installed outside the ECAL to further exclude the background contribution of the cosmic-ray muon~\cite{Adamov:2018vin,Shah2024}.

\begin{table}[htbp]
\nolinenumbers
\centering
\caption{Summary of currently investigated backgrounds.}
\label{tab:mace background}
\begin{tabular}{ccc}
\hline\hline
\multicolumn{2}{c}{\multirow{2}{*}{Background type}}                            & Counts per \\
\multicolumn{2}{c}{}                                                            & $10^8~\mu\text{OT}/\text{s}\cdot 1~\text{yr}$ \\ \hline
Physical                    & $\mu^+\to e^+e^-e^+\nu_e\bar{\nu}_\mu$ & $0.29 \pm 0.02$                                     \\ \hline
\multirow{2}{*}{\quad Accidental\quad} & Beam postiron                          & $<0.07$                                               \\ \cline{2-3}
                                       & Cosmic ray (with veto)                 & $<0.1$                                                \\ \hline
\multicolumn{2}{c}{Total}                                                       & $<1$                                                  \\ \hline\hline
\end{tabular}
\end{table}

\section{Sensitivity}
In the baseline run plan, MACE will be taking data from 1 year data acquisition duration with $10^8$~s$^{-1}$ muon-on-target ($\mu$OT) flux, with beam momentum of 24~MeV/$c$ and an RMS spread of 1.35~MeV/$c$. A ratio of in-vacuum muonium to $\mu$OT ($N_\mu^{\text{vac}}/N_{\mu^+}^{\text{OT}}$) of 3.8\% is possible to achieve in the muonium production target. As a result, during the physical run, MACE is expected to collect data from $N_\mu^{\text{vac}} = 1.20 \times 10^{14}$ muonium decay events in the vacuum of interest.

\begin{table}[htbp]
\nolinenumbers
\centering
\caption{Summary of signal efficiencies in the MACE detector system.}
\resizebox{\linewidth}{!}{
\begin{tabular}{ccc}
\hline\hline
Detector, component or analysis & Efficiency type                                                & Efficiency      \\ \hline
\multirow{2}{*}{Magnetic spectrometer}  & Geometric ($\varepsilon_\text{MMS}^\text{geom}$)       & 84.6\%     \\ \cline{2-3}
                      & Reconstruction ($\varepsilon_\text{MMS}^\text{recon}$)                 & $\sim 80\%$ \\ \hline
Positron transport system                   & Transmission ($\varepsilon_\text{PTS}$)                & 65.8\%     \\ \hline
Microchannel plate                   & Detection ($\varepsilon_\text{MCP}$)                   & 32.6\%     \\ \hline
\multirow{3}{*}{Electromagnetic calorimeter} & Incident ($\varepsilon^{\text{In}}_{\text{ECAL}}$)          & 63.4\% \\ \cline{2-3}
                                                               & Geometric ($\varepsilon^{\text{Geom}}_{\text{ECAL}}$)       & 95.3\% \\ \cline{2-3}
                                                               & Reconstruction ($\varepsilon^{\text{Recon}}_{\text{ECAL}}$) & 94.0\%   \\ \hline
\multicolumn{2}{c}{Total detection efficiency}                                               & 8.25\%     \\ \hline
Analysis                   & Event selection ($\varepsilon_\text{Cut}$)                     & $\sim 80\%$     \\ \hline
\multicolumn{2}{c}{Total signal efficiency}                                               & 6.6\%     \\ \hline\hline
\end{tabular}
}
\label{tab:mace-signal-efficiency-summary}
\end{table}

\begin{figure*}[t]
    \nolinenumbers
    \centering
    \includegraphics[width=0.9\textwidth]{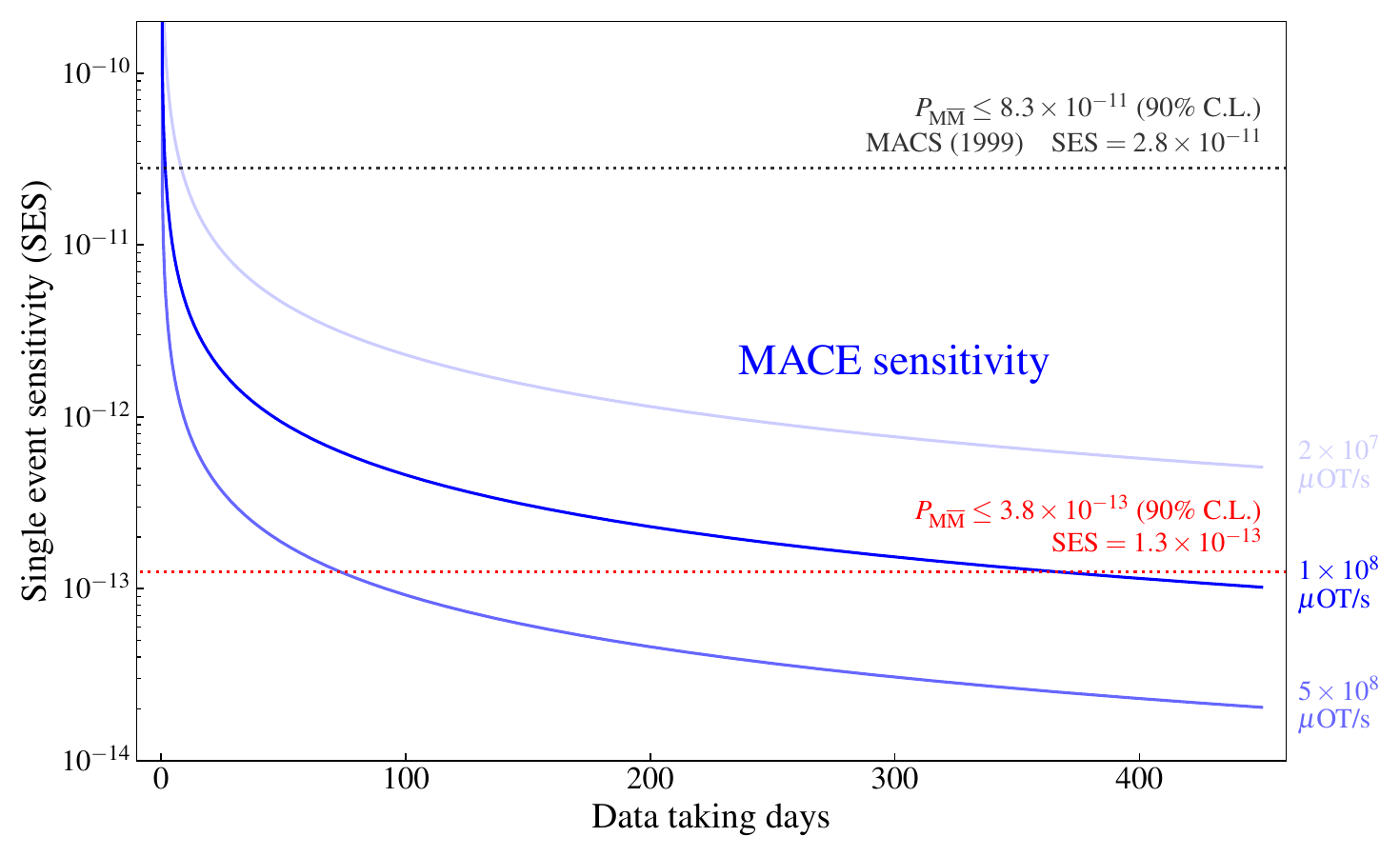}
    \caption{The MACE single event sensitivity (SES) to the muonium-to-antimuonium conversion.}
    \label{fig:MACE sensitivity}
\end{figure*}

Not all antimuonium decay events in the vacuum can be detected due to limitations in detector geometric acceptance and detection efficiency. We summarize the signal efficiency of each detector or component in \cref{tab:mace-signal-efficiency-summary}. Background simulations from the previous section suggest that a near background-free search for muonium-to-antimuonium conversion is possible, and the single event sensitivity (SES) can serve as an indicator of experimental sensitivity. Based on the number of muonium decay events in a vacuum and the signal efficiencies of all detectors or components, the single event sensitivity of the experimental design presented in this article to muonium-to-antimuonium conversion is estimated as
\begin{equation}
\begin{aligned}
    \text{SES} & =\frac{1}{\varepsilon_\text{MMS}^\text{geom}\varepsilon_\text{MMS}^\text{recon}\varepsilon_\text{PTS}\varepsilon_\text{MCP}\varepsilon^{\text{In}}_{\text{ECAL}}\varepsilon^{\text{Geom}}_{\text{ECAL}}\varepsilon^{\text{Recon}}_{\text{ECAL}}\varepsilon_\text{Cut}N_\mmu^\text{vac}} \\
    & =1.3\times 10^{-13}~.
\end{aligned}
\end{equation}
We estimate the upper limit for the muonium-to-antimuonium conversion probability using a Poisson counting statistic~\cite{ParticleDataGroup:2024cfk}. Assuming a conservative background count expectation of $b=1$ according to \cref{tab:mace background}, and an observed event count of $n_\text{obs}=b$, the experimental design presented in this article is expected to attain an upper limit, at a confidence level of $1-\alpha=0.9$,
\begin{equation}
\begin{aligned}
    P(\mmu \to \ammu) & \lesssim\left(\frac{1}{2}F_{\chi^2}^{-1}\left(1-\alpha;2(n_\text{obs}+1)\right)-b\right)\times\text{SES} \\
    & =3.8\times 10^{-13}~.
\end{aligned}
\end{equation}

\section{Proposal of MACE \textsc{Phase-I}} \label{sec:phasei}

\subsection{Introduction}
Taking advantage of the optimized MACE detector system and the construction of high-intensity muon beams in China, a search for other cLFV processes involving muon could be forthcoming.
Therefore, we propose the MACE \textsc{Phase-I} experiment aiming at further physics goals.
And it will help to evaluate the performance of particular sub-detectors before the full construction of MACE. 
Additionally, a measurement of muon polarization, similar to that performed in the MEG experiment~\cite{MEG:2015kvn}, can be conducted to characterize the energy and angular distributions of muon decay products.

\begin{figure}[htbp]
    \nolinenumbers
    \centering
    \includegraphics[width=0.9\linewidth]{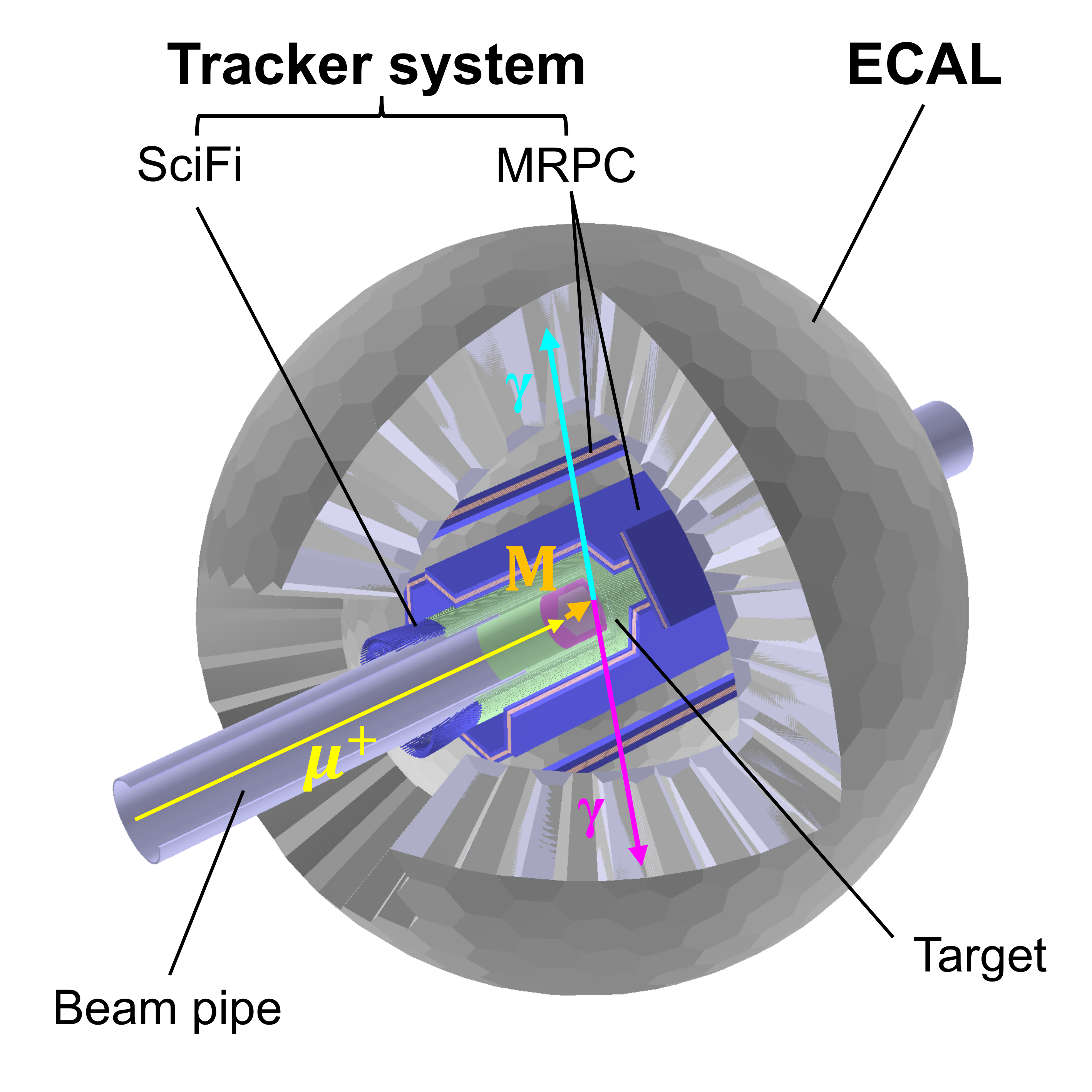}
    \caption{The MACE \textsc{Phase-I} detector concept (an $\mmu\to\gamma\gamma$ event is displayed for example).}
    \label{fig:phasei-3d}
\end{figure}

\cref{fig:phasei-3d} presents the layout of \textsc{Phase-I} detector system.
The detector system will be based on the ECAL and an extra inner tracker system, which consists of a scintillating fiber (SciFi) tracker and a multigap resistive plate chamber (MRPC) hodoscope. The tracker system is design to provide hit position and hit time of $e^+$ which could be signal or background depending on specific processes. The photon barely triggers the SciFi and MRPC, such that it could only be reconstructed by the ECAL.
The cLFV processes in the detector's capacity may include $\mu^+e^-$ scattering, rare decay of muon and muonium.
To better emphasize the necessity of the \textsc{Phase-I} experiment stage, efforts on searching such processes in history is reviewed as follows.

\subsubsection{$\mu^+\to e^+\gamma\gamma$ decay}
The Crystal Box experiment at Los Alamos Meson Physics Facility (LAMPF) firstly constrained the branching ratio of $\mathcal{BR}(\mu^+\to e^+\gamma\gamma)<7.2\times10^{-11}$ at 90\% confidence level in 1988~\cite{Bolton:1988af}. There have been no updated results and dedicated experimental plan on this decay mode. Several New Physics models predict a higher rate or sensitivity of $\mu^+\to e^+\gamma\gamma$ comparing to $\mu^+\to e^+\gamma$~\cite{Fortuna:2023paj,Uesaka:2024tfn}, which offers a solid motivation to search for this process.

\subsubsection{Muonium decays}
The muonium decay, e.g. $\mathrm{M} \to \gamma\gamma$, features a relatively clear final state consisting of two back-to-back 53 MeV photons.
The most recent search for the muonium decay can be traced back to the experiment conducted in 1959 by C. M. York et al.~\cite{PhysRevLett.3.288}. The detector system consists of a muon production target, a stopping target, and two sets of NaI scintillation counters. A 77 MeV $\pi^+$ beam was directed to a carbon block to produce decay muon, which further stopped in a copper target. The experiment merely setup a loose bound of $\mathcal{O}(10^{-5})$ on $\mu^+ e^-\to \gamma\gamma$~\cite{PhysRevLett.3.288,Hughes1975}. With the optimized detector system in \textsc{Phase-I}, a significant improvement of sensitivity is expected for the $\mmu\to\gamma\gamma$ decay.

\subsection{Muon beam}
An estimation of the requirements for the muon beam is listed in \cref{tab:phasei-beam}. The beam intensity is mainly limited by sensitivity and the rate of accidental coincidences. An excessively low intensity is insufficient to meet the sensitivity goals of \textsc{Phase-I}, while an excessively high intensity would lead to an increased rate of accidental coincidences, which would not contribute further to enhancing sensitivity. As a conservative estimation, the optimal beam intensity ranges from $10^6$--$10^7~\mu^+/$s. Within the next few years, the surface muon beam provided by CiADS is expected to achieve the required beam parameters. 

\begin{table}[htbp]
\nolinenumbers
\centering
\caption{Muon beam requirements for MACE \textsc{Phase-I}.}
\begin{tabular}{cc}
\hline\hline
Muon flux                      & $10^6$--$10^7~\mu^+/$s \\\hline
Working mode                    & Continuous wave      \\ \hline
Beam spot diameter                    & $\lesssim 30$~mm \\ \hline
Momentum spreading ($\sigma_p/p$)    & $\lesssim 10\%$ \\\hline
Positron-to-muon ratio              & $\lesssim 1\%$ \\\hline
Data acquisition duration                   & 360 days \\ \hline\hline
\end{tabular}
\label{tab:phasei-beam}
\end{table}

\subsection{Design of the detector system}

\subsubsection{Electromagnetic calorimeter}
The basic design of ECAL in \textsc{Phase-I} is identical with \cref{sec:ecal} but with some modules around the beam window removed. 
Instead of the 0.511~MeV regime in MACE, the signal concerned by MACE \textsc{Phase-I} is up to $\sim50$~MeV which gives a challenge to the current ECAL design.
Both BGO and LYSO crystals are under consideration due to their relatively high density and light yield.
Currently, LYSO crystals ($\sim 12X_0$) are employed in \textsc{Phase-I} simulation studies to achieve better energy resolution and a more compact design.
The fast emission time of LYSO also makes it capable to reject backgrounds and pileup events.

\subsubsection{Tracker system}
The timing and spatial resolution of the ECAL is limited. More importantly, the ECAL can hardly discriminate between $e^+$ and $\gamma$ by itself. An inner tracker system is designed to identify $e^+$s, and measure their arrival time and track. Firstly, Three layers of scintillating fibers is arranged along the surface of the cylinder outside the beam pipe. It reconstruct the track and hit time of an incoming $e^+$. Secondly, the outer layer is an array of multigap resistive plate chamber (MRPC), which offers extreme high timing precision. This combination ensures a great enhancement of sensitivity by reducing any accidental and physical back ground, as a result of increased tracking and timing precision.

\begin{figure}[htbp]
     \nolinenumbers
     \centering
     \subfloat[Overall view of SciFi Tracker]{\includegraphics[width=0.5\columnwidth]{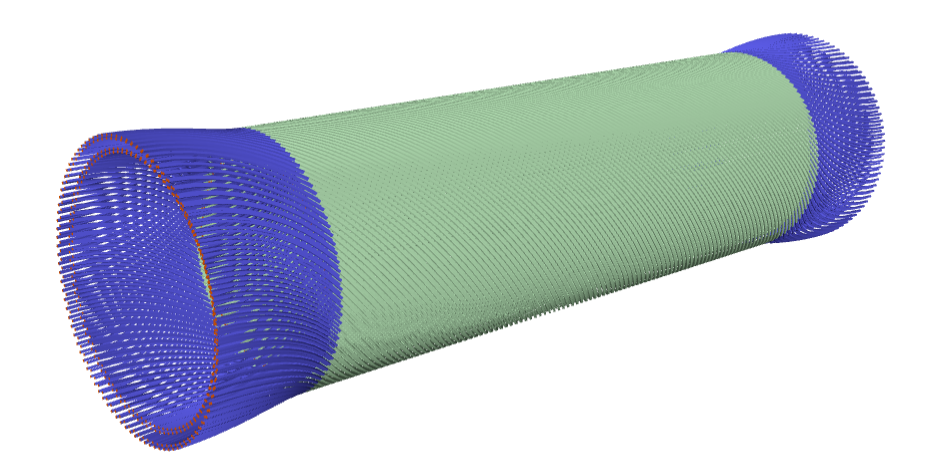}}
     \hfil
     \subfloat[SciFi arrangement]{\includegraphics[width=0.5\columnwidth]{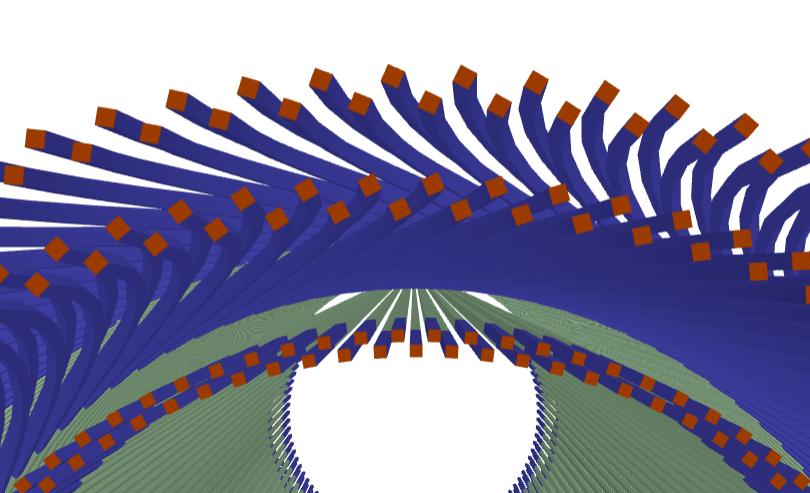}}
     \caption{\textbf{Left:} Overall view of the SciFi Tracker. The radius of the detector is about
 5.5~cm, the length is about 32.5~cm.
 \textbf{Right:} Details of the SciFi tracker. In this figure, One of the three layers is made up of two layers of fibers.Each Scintillating fiber (green) is connected with an optical fiber (blue) and a SiPM (orange) is tied at the end of the optical fiber.}
     \label{fig:sciFi tracker}
 \end{figure}

\paragraph{Scintillating fiber tracker}
A scintillating fiber (SciFi) tracker for high-precision timing and tracking is designed. It is expected to have a time resolution of $\mathcal{O}(1)$~ns and a high detection efficiency to distinguish the signal from the background.
In the current design, the SciFi tracker consists of three layers: two helical layers and a transverse layer. The helical fibers are arranged at a tilt angle to ensure that they rotate 360$^\circ$ in the azimuthal angle $\phi$ from one side of the veto to the other side, while the transverse fibers extends parallel to the axial direction. Each layer contains 200 fibers and extends 325.4 mm along the axial direction with an average thickness of 1.3 mm, as shown in \cref{fig:sciFi tracker}. The radii of each layer are 45~mm, 50~mm and 55~mm, respectively. This design leads to a geometry acceptance of 95.6\%. The 1~mm width single cladding square scintillating fibers produced by Kuraray are considered in the benchmark design~\cite{kuraray_psf}. Each fiber is mechanically held in place independently. 
The scintillation light emitted in the fibers will be transported by light guide fibers and read out by SiPMs, where signals of each fiber output individually. The number of channels is 600 in total under the current design.

\paragraph{Multigap resistive plate chamber hodoscope}
Although the SciFi tracker can provide good spatial resolution, better timing capability is required to further suppress the potential physical and accidental backgrounds. The MRPC is typically used in high energy physics experiments as a time of flight detector due to its high performance, simplicity, and low cost. The detector is mainly made by glass and plastic, finally housed in an aluminum alloy gas chamber filled with working gas. The MRPC can achieve a detection efficiency exceeding 95\% for charged particles, along with an excellent timing resolution of approximately 50~ps and $\mathcal{O}$(1) mm spatial resolution~\cite{Wang:2020iwn}. The specification perfectly match the requirement of MACE \textsc{Phase-I}. Under the current design, eight MRPC modules are situated between the SciFi tracker and the ECAL, fully covering the radial direction. The MRPC will tag the charged particle along with the SciFi, and provide high precision time information for event reconstruction.

\section{Summary}
In this study, we have reported the conceptual design of MACE in theoretical and experimental aspects.
With the current MACE detector concept, including muonium production target, Michel electron magnetic spectrometer (MMS), positron transport system (PTS), and positron detection system (PDS), two orders of magnitude or better improvement for muonium-to-antimuonium process comparing to the current limit is forthcoming.
Moreover, we intend to utilize a portion of the MACE detector to search for the muonium rare decay process before the full construction of MACE.
Future works will be continuously conducted in areas such as full simulation and reconstruction, software development, technical design and prototyping of each sub-detector, etc.

The observation of charged lepton flavor violation processes will clearly indicate the presence of new physics beyond the Standard Model. Charged lepton flavor violation in processes like $\mu^-N\to e^-N$ conversion, muon decays such as $\mu^+\to e^+\gamma$ and $\mu^+\to e^+ e^- e^+$, and the conversion of muonium to antimuonium ($\mu^+ e^-\to \mu^- e^+$) may exhibit significant effects that could be within the reach of future experiments. The proposed MACE concept holds promise for exploring new physics through the muonium-to-antimuonium conversion process.
In conjunction with other flavor and collider searches, MACE will illuminate the mysteries of new physics beyond the Standard Model.

\begin{acknowledgments}
We express our gratitude to Klaus~Jungmann, Yoshitaka~Kuno, Kim-Siang~Khaw, Lorenzo~Calibbi, Ce~Zhang, Tsutomu~Mibe, Yuichi~Uesaka, Yu~Bao, Ningqiang~Song and many other colleagues for their insights and expertise during discussions.
We also appreciate the professional and constructive comments provided by Xin~Chen, Kim-Siang~Khaw, Teng~Li, Yi~Liao, Yong~Liu and Xiang~Xiao in reviewing this article.
This project is supported by
National Natural Science Foundation of China under Grant Nos. 12075326, 11535014, 11975017, 12475191, 11905092, 12105132 and 12175039;
Guangdong Basic and Applied Basic Research Foundation under Grant No. 2025A1515010669;
Natural Science Foundation of Guangzhou under Grant No. 2024A04J6243;
Fundamental Research Funds for the Central Universities (23xkjc017) in Sun Yat-sen University;
Basic Research Conditions and Major Scientific Instrument and Equipment Research and Development Projects of the Ministry of Science and Technology under Grant No. 2022YFF0705602;
the State Key Laboratory of Particle Detection and Electronics, SKLPDE-ZZ-202412;
Natural Science Foundation of Shandong Province under Grant No. 2023HWYQ-010;
the ``Fundamental Research Funds for the Central Universities" at Southeast University;
the National Development and Reform Commission of China (Large Research Infrastructures of 12th Five-Year Plan: China initiative Accelerator Driven System, Grant No. 2017-000052-75-01-000590);
and Innovation Training Program for bachelor students in Sun Yat-sen University.
The simulation benefited greatly from the provision of computing resources by the National Supercomputer Center in Guangzhou.
The Feynman diagrams shown in this article are created using TikZ-Feynman~\cite{Ellis_2017}.
\end{acknowledgments}

\end{document}